\documentclass[twocolumn]{aa} 
\usepackage{natbib}
\usepackage{url}
\PassOptionsToPackage{hyphens}{url}
\usepackage{graphicx}
\usepackage{subfigure}
\usepackage{amsmath}
\usepackage{txfonts}
\usepackage{xcolor}
\usepackage{listings} 
\usepackage{times}
\usepackage{multirow}
\usepackage{float}
\usepackage{epstopdf}
\usepackage{longtable}
\usepackage{soul}
\usepackage[colorlinks=true,linkcolor=red,citecolor=blue, breaklinks=true]{hyperref}
\restylefloat{figure}

\newcommand{\hi}{\ion{H}{i}}

\newcommand{\planck}{\textit{Planck}}  
\newcommand{\disperse}{{\tt  DisPerSE}}  
\newcommand{\herschel}{{\it Herschel}}
\def\NHUNIT{\ifmmode {\rm \,cm^{-2}} \else $\rm \,cm^{-2}$ \fi} 
\def\nhh{\ifmmode N_{\rm H_{2}}\else $N_{\rm H_{2}}$\fi} 
\def\nhhc{\ifmmode N_{\rm H_{2}}^0\else $N_{\rm H_{2}}^0$\fi} 
\def\nhhbg{\ifmmode N_{\rm H_{2}}^{\rm bg}\else $N_{\rm H_{2}}^{\rm bg}$\fi} 
\def\nh{\ifmmode N_{\rm H}\else $N_{\rm H}$\fi}
\def\ml{\ifmmode M_{\rm line}\else $M_{\rm line}$\fi}  
\def\sunpc{\ifmmode \rm M_\odot/\rm pc\else $\rm M_\odot/\rm pc$\fi}  
\def\rout{\ifmmode R_{\rm out}\else $R_{\rm out}$\fi}  
\def\av{\ifmmode A_{\rm V}\else $A_{\rm V}$\fi}   
\def\fwhmdec{\ifmmode FHWM_{\rm dec}\else $FWHM_{\rm dec}$\fi}   
\def\rflat{\ifmmode R_{\rm flat}\else $R_{\rm flat}$\fi}   

\def\arcm{\ifmmode {^{\scriptstyle\prime}}
          \else $^{\scriptstyle\prime}$\fi}
\newdimen\sa  \newdimen\sb
\def\parcs{\sa=.07em \sb=.03em
     \ifmmode \hbox{\rlap{.}}^{\scriptstyle\prime\kern -\sb\prime}\hbox{\kern -\sa}
     \else \rlap{.}$^{\scriptstyle\prime\kern -\sb\prime}$\kern -\sa\fi}
\def\parcm{\sa=.08em \sb=.03em
     \ifmmode \hbox{\rlap{.}\kern\sa}^{\scriptstyle\prime}\hbox{\kern-\sb}
     \else \rlap{.}\kern\sa$^{\scriptstyle\prime}$\kern-\sb\fi}

\def\rev{}
\def\Newrev{}
\hypersetup{draft}
\begin{document} 

\title{Characterizing the properties of nearby molecular filaments observed with {\it Herschel} }
  \titlerunning{Properties of nearby filaments observed with  {\it Herschel}}

   \author{
     D. Arzoumanian\inst{1,2}
      \and
            Ph. Andr\'e\inst{2}
                 \and
           V. K\"onyves\inst{2,3}          
         \and
               P. Palmeirim\inst{4}          
         \and
	A. Roy\inst{2,5}          
        \and
         N. Schneider\inst{5,6}          
\and
M. Benedettini\inst{7} 
\and 
P. Didelon\inst{2} 
\and
J. Di Francesco\inst{8} 
\and
J. Kirk\inst{3} 
\and
B. Ladjelate\inst{2,9} 
            }

   \institute{$^1$Department of Physics, Graduate School of Science, Nagoya University, Furo-cho, Chikusa-ku, Nagoya 464-8602, Japan\\
                \email{doris.arzoumanian@nagoya-u.jp} \\ 
                $^2$Laboratoire AIM, CEA/DRF--CNRS--Universit\'e Paris Diderot, IRFU/D\'epartement d'Astrophysique, C.E. Saclay,
              Orme des Merisiers, 91191 Gif-sur-Yvette, France \\
                  $^3$Jeremiah Horrocks Institute, University of Central Lancashire, Preston PR1 2HE, UK\\
                  $^4$Instituto de Astrof\'isica e Ci{\^e}ncias do Espa\c{c}o, Universidade do Porto,
CAUP, Rua das Estrelas, PT4150-762 Porto, Portugal  \\
                  $^5$Universit\'e de Bordeaux, LAB, UMR 5804, 33270 Floirac, France \\
                  $^6$I. Physik. Institut, University of Cologne, Z\"ulpicher Str. 77, 50937 Koeln, Germany \\
              $^7$INAF-Istituto di Astrofisica e Planetologia Spaziali, via Fosso del Cavaliere 100, 00133 Roma, Italy\\
                  $^8$National Research Council Canada, 5071 West Saanich Road, Victoria, BC V9E 2E7, Canada\\
                 $^9$Institut de RadioAstronomie Millim\'etrique (IRAM), Granada, Spain \\
                  }
     \date{}

\abstract{
{\it Context.}  
Molecular filaments have received special attention recently, thanks to new observational results on their properties. 
In particular, our early analysis of filament properties from 
{\it Herschel} imaging data 
in three nearby molecular clouds 
revealed a narrow distribution of  
{\rev median} inner widths centered at a characteristic value of about 0.1\,pc.\\ 
{\it Aims.} Here, we extend and complement our initial study 
with a  detailed analysis of the filamentary structures identified with  \herschel\ in eight nearby molecular clouds  (at distances\,$<$500\,pc). 
Our {\rev main} goal is to establish 
statistical distributions of 
{\rev median} properties {\rev averaged along the filament crests } 
and to compare the results with our earlier work based on a smaller number of filaments. \\
{\it Methods.} We use the column density ($\nhh$) maps derived from  \herschel\  data  
and the DisPerSE algorithm 
to trace a network of individual filaments in each cloud. 
We analyze the 
density structure along and across the main filament axes in detail. 
We build synthetic maps of filamentary clouds to assess the completeness limit of our extracted filament sample 
and validate our measurements of the filament properties. 
These tests  also help us to select the best choice of parameters to be used for 
tracing filaments  
with DisPerSE and fitting their radial column density profiles.\\ 
{\it Results.} Our analysis yields an extended sample of 1310 filamentary structures and 
a selected sample of {\rev 599 filaments} with aspect ratios larger than 3 and column density contrasts larger than 0.3. 
We show that our selected sample of filaments is more than $95\,\%$ complete for column density contrasts larger than 1,  
{\rev with only $\sim 5\,\%$ of spurious detections}.   
{\rev On average}, more than  $15\,\%$  of the total gas mass in the clouds, and more than
$80\,\%$ of the dense gas mass {\rev(at $\nhh>7\times10^{21}$\,cm$^{-2}$)}, is found to be in the form of  filaments, respectively.  
Analysis of the radial column density profiles of the 599 filaments in the selected sample 
indicates a narrow distribution of 
{\rev crest-averaged} inner widths,  
with a median value of  $0.10\, $pc and an interquartile range of $0.07\, $pc. 
In contrast, the extracted filaments span wide ranges in length, central column density, column density contrast, and mass per unit length. 
The characteristic filament width is well resolved by \herschel\ observations, and a 
median value of $\sim$0.1\,pc is consistently found using three distinct estimates based on (1) a direct measurement of the width at half power 
after background subtraction, as well as (2) Gaussian and  (3) Plummer
fits.
The existence of a characteristic filament width is further supported by the presence of a tight correlation between mass per unit length 
and central column density for the observed filaments.\\
{\it Conclusions.} Our detailed analysis of a large filament sample confirms our earlier result that nearby molecular filaments 
share a common {\rev mean} inner width of  $\sim$0.1\,pc, {\rev with typical variations along and on either side of the filament crests of about $\pm 0.06$\,pc around the mean value}. 
This observational result sets strong constraints on possible models for 
the formation and evolution of filaments in molecular clouds. It also provides important hints on the initial conditions of star formation.
}


\maketitle

\section{Introduction}\label{intro}

Both the atomic and the molecular phase of 
the Galactic interstellar medium (ISM) 
have been known to be filamentary for a long time. 
Interstellar filaments 
have initially been detected in dust extinction  \citep[e.g.,][]{Schneider1979,Myers2009}, 
dust emission \citep[e.g.,][]{Abergel1994}, \hi\   \citep[e.g.,][]{Joncas1992,McClure-Griffiths2006}, 
and CO emission from both diffuse molecular gas \citep{Falgarone2001,HilyBlant2009} and dense star-forming gas \citep[e.g.,][]{Bally1987,Cambresy1999}.  
It is only recently, however, that the ubiquity of filamentary structures in the cold ISM 
and their importance for the star formation process have been revealed thanks to 
the unprecedented quality and sky coverage of {\it Herschel} dust continuum  
images at far-infrared and submillimeter wavelengths.
Prominent filamentary structures have been observed with {\it Herschel} in both star-forming and 
non-star-forming low-mass clouds in the solar neighborhood \citep[e.g.,][]{Andre2010,Men'shchikov2010}, 
as well as in massive star-forming complexes  throughout the Galactic Plane 
at distances from a few kpc up to the central molecular zone  
\citep[][]{Molinari2010,Hill2011, Hennemann2012, Schneider2012,Schisano2014,Wang2015}. 
Filaments are also striking features in numerical simulations of molecular cloud formation and evolution
\citep[e.g.,][]{MacLow2004,Vazquez-Semadeni2007,Hennebelle2008,Nakamura2008}, even if
the resulting filament properties do not always match the observed ones \citep[e.g.,][]{Hennebelle2013,Federrath2016,Ntormousi2016,Smith2016}. 

Quite unexpectedly, our early analysis of the radial column density profiles  
observed with \herschel\ 
for 90 filaments in three nearby clouds (IC5146, Aquila, and Polaris) 
suggested that molecular filaments share a common inner width of about 0.1\,pc despite a wide range of central column densities 
\citep[][]{Arzoumanian2011}. 
Moreover, the results of the {\it Herschel} Gould Belt survey  \citep[e.g.,][]{Andre2010,Konyves2015,Marsh2016} indicate that most prestellar cores form in dense, ``supercritical'' filaments for which the mass per unit length exceeds the critical line mass 
of nearly isothermal, long cylinders  \citep[cf.][]{Inutsuka1997},  $M_{\rm line,crit}=2 c_{\rm s}^{2}/G \sim16\,\sunpc$, where $c_{\rm s} \sim0.2$\,km\,s$^{-1}$ 
is the isothermal sound speed for cold molecular gas at $T_{\rm gas} \sim 10$\,K.
Based on  {\it Herschel} results in nearby clouds, it has also been argued that filaments may help to regulate the star formation efficiency in the dense molecular 
gas of galaxies, and may be responsible for a quasi-universal star formation law in the dense ISM of galaxies \citep[cf.][see also \citet{Lada2012,Toala2012}]{Shimajiri2017}. 
These findings support a paradigm for star formation in which the formation and fragmentation of molecular filaments play a central role \citep[cf.][]{Andre2014,Inutsuka2015}. 
To  improve further our understanding of the initial conditions and ``microphysics'' of star formation in the cold ISM of galaxies, 
characterizing the detailed properties of nearby molecular filaments is thus of paramount importance.

One of the cornerstones of the proposed filamentary paradigm for solar-type star formation \citep[][]{Andre2014} is the existence of a characteristic filament width $\sim 0.1\,$pc suggested 
by the early analysis of {\it Herschel} filament properties by \citet[][]{Arzoumanian2011}. 
Recently, \citet{Panopoulou2017}  challenged the conclusion of \citet[][]{Arzoumanian2011}.  
Noting the tension between the presence of a characteristic filament width and the absence of any characteristic scale in the power spectrum of interstellar cloud images \citep{Miville2010,Miville-Deschenes2016},
they discussed potential biases in measurements of filament widths 
based on simple Gaussian fitting of the radial column density profiles.

Here, we extend our previous study of the radial density structure and basic properties of molecular filaments  
to a much broader sample of 
filaments observed with {\it Herschel} in eight nearby clouds, using an improved and more automated method for filament identification, extraction, and characterization.  
We complement our analysis of the {\it Herschel} images with multiple tests performed on 
synthetic maps to estimate the completeness level of the extracted filament sample 
and the reliability of the derived filament properties. 
In particular, we address the concerns raised by \citet{Panopoulou2017} on possible measurement biases. 
Our results essentially confirm and strengthen our earlier findings on filament properties \citep[e.g.,][]{Arzoumanian2011,Peretto2012,Palmeirim2013,AlvesdeOliveira2014,Benedettini2015,Cox2016}.
In a parallel paper \citep{Roy2018}, we also show that the essentially scale-free power spectra of {\it Herschel} images are consistent 
with the presence of a characteristic filament width $\sim 0.1\, $pc and do not invalidate the conclusions drawn from the analysis of filament profiles.

The present paper is organized as follows. 
Section\,\ref{FilIdentification} describes the different steps of the method employed to identify/extract filamentary structures in molecular clouds. 
Section\,\ref{FilMeasure} details the  steps followed to measure the properties of the filament sample. 
Three Appendices (\ref{App1}, \ref{App2},  and \ref{App4}) complement these two sections by summarizing multiple analyses performed on synthetic maps to estimate  the completeness limit of the filament sample and the reliability of measuring their properties. 
Section\,\ref{FilStat} presents the statistics of the measured filament properties.
In Sect.\,\ref{Discussion}, we discuss the implications of our results for our theoretical understanding of filament formation and evolution in the cold ISM, and the link with the star formation process.    
Finally, Sect.\,\ref{Summary} summarizes our main findings. 

\begin{table*}[!ht]  
\label{Table1}    
\centering
 \caption{Summary and properties of the nearby molecular clouds analyzed in this paper.} 
\begin{tabular}{|c|c|ccccccc|}   
\hline\hline   
Field & Color &Distance & $\tilde{A}^{\rm cloud}_{\rm tot}$  & $A^{\rm cloud}_{\rm tot}$ & $M^{\rm cloud}_{\rm tot}$& $M^{\rm cloud}_{\rm dense}/M^{\rm cloud}_{\rm tot}$ & $A^{\rm cloud}_{\rm dense}/A^{\rm cloud}_{\rm tot}$ & Reference \\
& &  [pc] & [deg$\,\times\,$deg] & [pc$^{2}$]& [$10^3$M$_\odot$] &[$\%$] &[$\%$]  &  \\
  (1)&(2)& (3) & (4) &(5) &(6)&(7)&(8)&(9) \\
\hline
IC5146&{\color{black} $\bullet$} &460&3.2$\times$2.5&145&3.7&9&1&1 \\
Orion B&{\color{cyan} $\bullet$} &400&6.8$\times$8.6&803&26.7&12&1 &2,3\\
Aquila&{\color{orange} $\bullet$} &260&4.4$\times$4.7&230&25.5&19&9 &4,5,6,7\\
Musca&{\color{yellow} $\bullet$} &200&3.1$\times$4.2&33&1.0&1&<1&8 \\
Polaris&{\color{blue} $\bullet$} &150&5.4$\times$5.2&35&0.5&0&0 & 4, 9,10,11\\
Pipe&{\color{green} $\bullet$} &145&7.5$\times$3.3&53&2.0&1&<1&12 \\
Taurus L1495&{\color{violet} $\bullet$} &140&5.4$\times$3.7&48&2.4&11&2&13,14,15,16 \\
Ophiuchus&{\color{red} $\bullet$} &140&4.8$\times$5.0&59&3.4&17&3 & 17\\
                           \hline  \hline
                  \end{tabular}
\begin{list}{}{}
 \item[]{{\bf Notes:} 
The fields analyzed in this paper correspond only to part of the regions covered by the {\it Herschel} Gould Belt survey (HGBS -- see http://gouldbelt-herschel.cea.fr). 
 {\bf Col. 1:} Cloud name. 
 {\bf Col. 2:} Color code used to represent each cloud in the plots shown throughout the paper.  
 {\bf Col. 3:} Default distance adopted for each region. See references for details. 
  {\bf Col. 4:} Apparent surface area in square degrees of (the portion of) each cloud analyzed in the paper.
 {\bf Col. 5:} Physical surface area in pc$^2$ of (the portion of) each cloud analyzed in the paper.
 {\bf Col. 6:}  Total gas mass derived from integrating the \herschel\ column density map of each cloud over the surface area given in Col.\,4 and Col.\,5, {\rev adopting a mean molecular weight per hydrogen molecule $\mu_{\rm H_{2}}=2.8$.} 
  {\bf Col. 7:} Fraction of dense gas mass in each cloud, where dense gas is defined based on column density, $\nhh>7\times10^{21}$\,cm$^{-2}$ \citep[see][]{Konyves2015}. 
 {\bf Col. 8:} Fraction of cloud area occupied by dense gas with $\nhh>7\times10^{21}$\,cm$^{-2}$ in the plane of the sky. 
 {\bf Col. 9:} References: 1 = \citet{Arzoumanian2011}, 2 = \citet[][]{Schneider2013}, 3 = K\"onyves et al. 2018, in prep., 4 = \citet{Andre2010}, 5 = \citet{Konyves2010}, 
 6 = \citet{Bontemps2010}, 7 = \citet{Konyves2015}, 8~=~\citet{Cox2016}, 9 = \citet{Men'shchikov2010}, 10 = \citet{Miville2010}, 11 = \citet{Ward-Thompson2010}, 12 = \citet{Peretto2012}, 
 13 = \citet{Palmeirim2013}, 14 = \citet{Kirk2013}, 15 = \citet{Marsh2014}, 16 = \citet{Marsh2016}, 17 = Ladjelate et al., in prep.
   }
 \end{list}      
  \end{table*}


\section{Identifying filaments in nearby molecular clouds}\label{FilIdentification}

In this paper, we make use of the optically thin sub-millimeter dust continuum emission imaged with  the \herschel\  Space Observatory \citep{Pilbratt2010} 
to study the column density  structure of eight nearby molecular clouds (distance\,<\,500\,pc) observed as part of the \herschel\ Gould Belt survey \citep[HGBS,][]{Andre2010}.
The clouds analyzed here span a wide range of physical conditions, from active cluster-forming intermediate- to high-mass star-forming regions such as Ophiuchus, Aquila, Orion, to relatively low-mass star-forming regions like the Pipe nebula, Taurus L1405, IC5146, Musca, to a quiescent and non-star-forming region like the Polaris Flare. 
Table\,\ref{Table1} summarizes the properties of the target clouds and 
gives the distance, projected size (on the plane of the sky), and mass of each region, along with appropriate references.

To take a census of individual filaments in each of these clouds, 
the first step 
is to follow the crests of filamentary structures 
in the corresponding \herschel\  column density maps. 
For this purpose, we used the \disperse\ algorithm which is a powerful tool to trace filament networks \citep{Sousbie2011}. 
While other methods have  been recently developed  
\citep[e.g.,][and others]{Men'shchikov2013,Clark2014,Schisano2014,Koch2015}, we have chosen \disperse, which 
has been  successfully used since 2011 \citep{Arzoumanian2011}  to trace the filamentary web of the  ISM revealed by \herschel\ observations of 
star-forming clouds \citep[e.g.,][]{Hill2011,Peretto2012,Schneider2012,Palmeirim2013}, and non  \herschel\  observations \citep[e.g.,][]{Panopoulou2014,LiGX2016}, 
as well as in numerical simulations  \citep[e.g.,][]{Smith2016,Federrath2016}.

In the following subsections, we describe our methodology and the successive steps taken 
to build a filament sample from the  \herschel\ observations of the target clouds.

\subsection{Working definition of  filament}

We define a molecular filament 
as any elongated structure detected in a 2D column density map of a molecular cloud, 
which has a minimum aspect ratio and a minimum column density excess 
over the local background. 

The aspect ratio of a filamentary structure is defined by 
\begin{equation}
AR=l_{\rm fil}/W_{\rm fil}\,, 
\end{equation}
\noindent
where $l_{\rm fil}$ and $W_{\rm fil}$ are the length and the width of the structure, respectively. 
The length $l_{\rm fil}$ scales with the number of pixels tracing the crest of the structure in the input column density map.
An elongated structure with wriggles is longer than a straight one,
for the same origin and end points. 

The intrinsic column density contrast of a filament is defined as 
\begin{equation}
C^0=(\nhh^{\rm fil} - \nhh^{\rm bg})/\nhh^{\rm bg}=\nhh^0/\nhh^{\rm bg}\,, \label{contEq}
\end{equation}
\noindent
where $\nhh^{\rm fil}$ and $\nhh^{\rm bg}$ are the column densities observed along the crest of the filament 
and toward the local background, respectively. 
The column density amplitude of the filament is 
$\nhh^0=\nhh^{\rm fil}-\nhh^{\rm bg}$. 
The column densities appearing in Eq.\,\ref{contEq} are estimated at each map pixel along the filament crest 
and then averaged along the crest.

\subsection{Tracing networks of filaments with the \disperse\ algorithm}\label{SkelDisperse}

\begin{figure*}[!ht]
   \centering
  \hspace{0.1cm}
         \resizebox{8.cm}{!}{
     \hspace{0.1cm}
        \vspace{-0.9cm}
\includegraphics[angle=0]{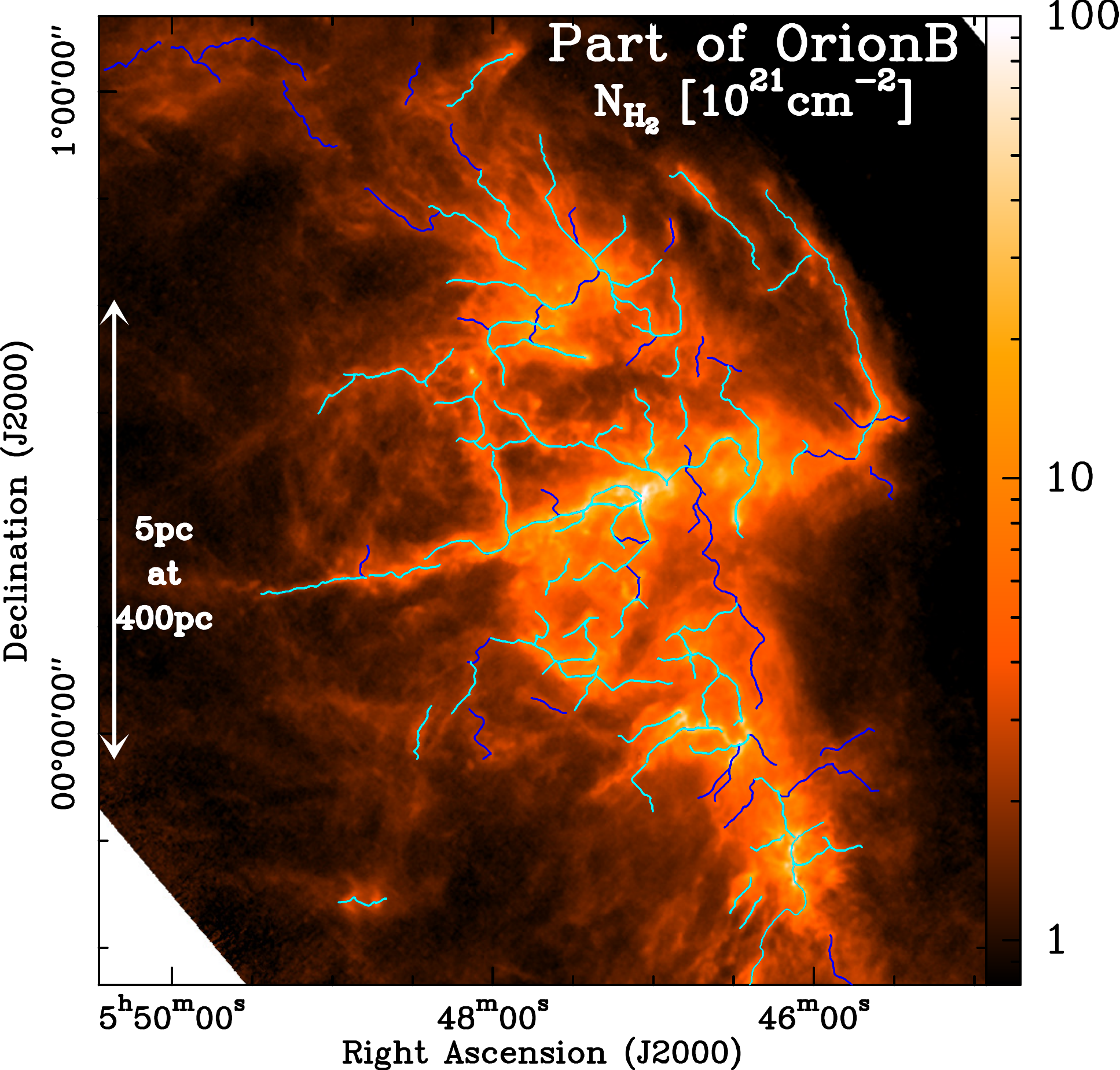}}
    \resizebox{9.9cm}{!}{
\includegraphics[angle=0]{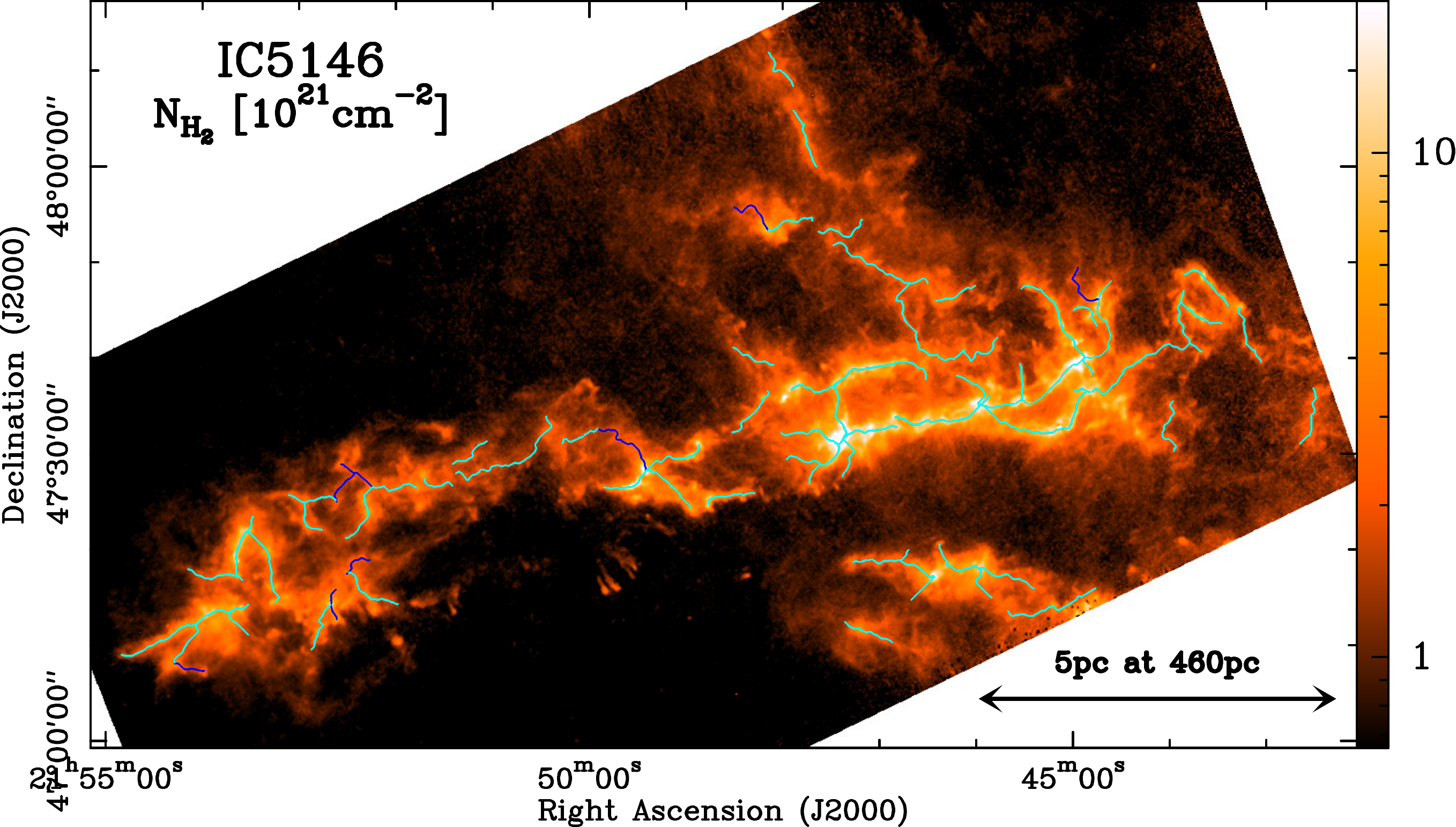}}
 \caption{{\it Herschel} column density maps of a portion of the analyzed region in Orion~B (left) 
and of the entire field analyzed here in IC5146 (right), 
as derived from HGBS data 
  \citep[http://gouldbelt-herschel.cea.fr/archives -- cf.][K\"onyves et al. 2018]{Andre2010,Arzoumanian2011}. 
The effective HPBW resolution of these maps is $18\parcs2$.    
The crests of the filamentary structures traced in the two clouds using \disperse\ as explained in Sect.\,\ref{SkelDisperse} are overlaid as solid curves.
{\rev The cyan blue curves trace the filament crests of the selected sample, and the dark blue curves trace the crests of additional filaments  
from the extended sample (cf. Sect.\,\ref{SelecSample} and Table\,\ref{tab:SumParamDisp}).}
The maps of the other fields analyzed in this paper are shown  in Appendix\,\ref{App4}.
}          
  \label{ColdensMapsSkel}
    \end{figure*}

To trace filament networks, we adopted \disperse\ \citep[][\url{http://www2.iap.fr/users/sousbie/web/html/indexd41d.html}]{Sousbie2011},  
 an algorithm designed to identify persistent topological features such as peaks, voids, and filamentary structures in astrophysical data sets. 
\disperse\ stands for "Discrete Persistent Structures Extractor" and was initially developed 
to analyze large-scale filamentary structures in the galaxy distribution, i.e., the cosmic web 
\citep{Sousbie2011b}.

The \disperse\ algorithm traces filaments by connecting critical points (e.g., saddle points and maxima) with integral lines, following the gradient in a map. 
Critical points are the positions where the gradient of the map is zero. We refer to the unique integral line that joins two connected critical points as an ``arc''. 
The absolute difference between the two map values at a pair of critical  points is called the {\it persistence} of the pair. 
In \disperse, the concept of persistence is used to select the pairs of critical points  that have a {\it persistence}, i.e., difference, 
larger than a minimum threshold value: the {\it persistence threshold}. 
This approach selects  topological features that are robust with respect to  data noise and ``background fluctuations''. 

In a first  step, \disperse\  builds a ``skeleton'' of crests tracing  all of the arcs with a {\it persistence} larger than 
the specified {\it persistence threshold}. 
In a second step,  \disperse\ 
1) removes  the arcs of the skeleton based on a {\it robustness} criterion and 
2) assembles  aligned arcs into longer filaments whenever the relative orientation between two neighboring arcs is smaller than a pre-defined assembly angle. 
The {\it robustness} parameter can be understood as a 
measure of how much contrast an arc has with respect to the local background.  
In practice, this parameter is measured by \disperse\ by comparing the mean map value along an arc with the local background values 
at the scale of the arc. 
The derived skeleton is then  
reprojected onto the same grid as the input image, 
after smoothing the arcs over a given number of pixels. 
The result of these two steps is a skeleton tracing the crests of the network of filaments 
in the input map
for the specified  {\it persistence} and {\it robustness} thresholds.  

The free parameters of the \disperse\ run are thus the following: the {\it persistence threshold} ($PT$), the  {\it robustness threshold} ($RT$), 
the maximum assembly angle ($AA$) between neighboring connected arcs  at the assembling step, 
and the number of pixels ($N_{\rm pix}$) used to smooth the skeleton 
before reprojection onto the same grid as the input image. 

\subsection{Using \disperse\ on \herschel\ column density maps}\label{SkelDisperseHerschel}  

In the analysis presented in this paper, we used column density maps   
derived from \herschel\  imaging data taken as part of the HGBS (cf., http://gouldbelt-herschel.cea.fr/archives.html).
The \nhh  column density maps of the eight regions discussed here were produced following the same procedure 
as described in Sect.\,4.1 of \citet{Konyves2015} for the Aquila cloud. 
These \nhh\ maps {\rev were calculated  adopting a mean molecular weight per hydrogen molecule $\mu_{\rm H_{2}}=2.8$ \citep[e.g.,][]{Kauffmann2008} and } 
 have an estimated accuracy of better than $\sim 50\%$ \citep[see][]{Konyves2015,Roy2013,Roy2014}.
We used both standard \nhh\ maps at the $36\parcs3$ (half-power beam width -- HPBW) resolution of \herschel/SPIRE  500\,$\mu$m data 
and ``high-resolution''  \nhh\ maps at  the $18\parcs2$ resolution of \herschel/SPIRE  250\,$\mu$m data.
The multi-scale decomposition method used to derive \herschel\ 
column density maps at $18\parcs2$ resolution is described in detail in Appendix~A of \citet{Palmeirim2013}. 
As examples, Fig.~\ref{ColdensMapsSkel} shows  the ``high-resolution''  column density maps of the fields analyzed here in 
the Orion~B and IC5146 clouds (only a portion of the Orion~B field is displayed in Fig.~\ref{ColdensMapsSkel} -- see Fig.~\ref{Maps1} in Appendix~\ref{App4} for the whole field).
The column density maps of the other regions discussed in this paper are shown in Appendix~\ref{App4}.

As part of the present study, 
we ran \disperse\ on column density maps at the standard resolution of $36\parcs3$ 
and subsequently measured the properties of the extracted filamentary structures in the corresponding high-resolution column density maps (equivalent $HPBW = 18\parcs2$).
Running  \disperse\ on the $36\parcs3$ resolution column density maps  produces skeletons that are smoother (less wriggly) than when \disperse\ is run on the $18\parcs2$ high resolution maps. 
In this context, 
an appropriate 
choice for the {\it persistence threshold} ($PT$) to be used when running \disperse\  
is on the order of  the minimum root mean square (${\rm rms}_{\rm min}$) level of 
the ``background cloud fluctuations'' in the input column density map. 
An appropriate value for the  {\it robustness threshold} ($RT$) is on the order of the minimum filament \nhh\ amplitude to be 
detected 
in the maps. %
The robustness threshold is thus
linked to the minimum column density contrast of the filaments to be extracted, $RT\,\sim\,C^0 \nhh^{\rm bg}$ (see  Appendix\,\ref{App1}  for details).
The filament background column density $\nhh^{\rm bg}$ is a local quantity for each filament and varies through the \herschel\ column density maps. 
Ideally, therefore, the $RT$ parameter should be given a local value, varying as a function of position in the input column density map. 
However, this consideration cannot easily be taken into account with the current version of the \disperse\ code. 
The $RT$ parameter was thus chosen to be a scaled version of
the minimum background column density value, $\nhh^{\rm bg,min}$, observed in each region. 
In practice, 
we derived  $\nhh^{\rm bg,min}$ and ${\rm rms}_{\rm min}$ using the median and the standard deviation of all values 
in the first bin of the column density histogram of each field, adopting a bin size of  $10^{21}\, {\rm cm}^{-2}$ 
(see Appendix\,\ref{App1}  and Fig.\,\ref{SyntMap}b for details). 

As a final post-treatment step (outside \disperse), the filament skeletons generated by \disperse\ 
were ``cleaned'' by removing features shorter than $10\times$ the HPBW resolution of the input images.

\subsection{Completeness and reliability of the extracted filament sample}\label{comp}

To select optimum parameters for our method of tracing filamentary structures, 
we performed several tests on synthetic maps including well-defined populations of mock filaments. 
These tests are described in detail in Appendix\,\ref{App1}. 

The optimum choice of \disperse\ parameters results from a compromise between two conflicting requirements:
maximizing the completeness of the extracted filament sample and maximizing the reliability of the sample, i.e.,
minimizing the level of contamination 
by spurious structures. 
For a given choice of  \disperse\ parameters, the completeness and reliability of the output sample   
depends on the column density contrast ($C^0$) and aspect ratio ($AR$) of the filaments to be extracted.
To select optimum parameters and 
estimate the completeness and reliability of our census of filaments in the observed clouds, 
we therefore applied the method outlined above to synthetic maps and 
investigated the variations of the 
fractions of extracted synthetic filaments and spurious structures
as a function of 
the persistence ($PT$) and robustness ($RT$) thresholds of the \disperse\ runs on one hand, 
and the contrast $C^0$ and aspect ratio $AR$ of the injected synthetic filaments on the other hand.
{\rev Another method of choosing appropriate \disperse\ parameters and tracing filament crests 
has been presented by \citet{Green2017} }
 
The results of our tests, {\rev described  in Appendix\,\ref{App1}}, indicate that the extracted filament sample 
is more than $95\%$ complete to filaments with intrinsic contrast $C^0=2$
for wide ranges of the  \disperse\  parameters $PT$ and $RT$ : $0.5\, {\rm rms}_{\rm min}\le PT \le 4\, {\rm rms}_{\rm min}$ and $0.75\, \nhh^{\rm bg,min}\le RT \le 2.25\, \nhh^{\rm bg,min}$, respectively. 
For synthetic filaments with intrinsic contrast  $C^0=1$, 
the completeness of the extracted sample is about  $95\%$ for $RT \le1.5\nhh^{\rm bg,min}$ and larger than $80\%$ for $RT=2\nhh^{\rm bg,min}$.
{\rev The fraction of spurious detections  decreases when $RT$ increases. 
For $C^0\ge1$ and $RT\ge1.5\,\nhh^{\rm bg,min}$, the fraction of spurious extracted structures is  only $\sim5\%$, 
but it increases rapidly for $RT<1.5\,\nhh^{\rm bg,min}$.}

For a given robustness threshold $RT$, 
the fraction of extracted synthetic filaments depends mainly on filament contrast $C^0$, 
and is less affected by the aspect ratio $AR$ of the filaments (see Fig.\,\ref{Comp_randW_AR_C}). 
For reference, the column density contrast of isothermal model filaments in pressure equilibrium with 
the ambient cloud is 
$<\Sigma_{\rm fil}>/\Sigma_{\rm cloud}\approx 1.18\times \sqrt{f_{\rm cyl}/(1-f_{\rm cyl})}$, 
where $\Sigma_{\rm fil}$ and $\Sigma_{\rm cloud}$ are  the  gas surface densities of the filament and the cloud, respectively, and 
$f_{\rm cyl} \equiv M_{\rm line}/M_{\rm line,crit} < 1 $ \citep[cf.][]{Fischera2012}.\footnote{Equilibrium model filaments exist only for subcritical masses per unit length, i.e.,  $f_{\rm cyl} \leq 1$.}
Thus, thermally transcritical filaments with $M_{\rm line,crit}/2  \la M_{\rm line} < M_{\rm line,crit} $ (i.e., $f_{\rm cyl} \ga 0.5 $) 
are expected to have column density contrasts $C^0 \ga 1$, while thermally supercritical filaments with well-developed power-law density profiles may have $C^0 >> 1$.

While the completeness of the extracted filament sample does depend on the adopted robustness threshold, 
it is not very sensitive to the exact choice of the persistence threshold 
within a factor of two around the minimum ${\rm rms}$ value in the map (${\rm rms}_{\rm min}$).
The fraction of extracted synthetic filaments is larger for $RT\le 1.5$--2$\, \nhh^{\rm bg,min}$, but the fraction of spurious detections also increases. 
To maximize the fraction of ``true'' detections while maintaining the  fraction of ``spurious'' detections to a reasonably low level, 
the best compromise for the choice of the robustness parameter appears to be $RT \sim 1.5\nhh^{\rm bg,min}$.  

In the following analysis, based on the multiple tests of Appendix\,\ref{App1}, we set  $PT={\rm rms}_{\rm min}$, $RT=1.5\nhh^{\rm bg,min}$, the assembling angle $AA$ to 50$^\circ$,  and 
the number of pixels of the smoothing kernel $N_{\rm pix}$ to $2\times HPBW/pix$, where $pix$ is the pixel size.  
Table\,\ref{tab:SumParamDisp} gives the absolute values of 
the persistence and robustness thresholds used to trace filamentary structures in each of the 8 fields, 
as well as  the number of extracted filaments.  
The previous discussion and the tests of Appendix\,\ref{App1} suggest that our census 
of filamentary structures is more than $95\% $ complete to transcritical and supercritical filaments, 
{\rev with a possible contamination from spurious structures of about $\sim5\%$. 
Our sample of subcritical filaments is admittedly less complete and 
may also contain a larger number of spurious structures.}  
 
 The filament skeletons derived as explained above are overlaid as  solid curves in Fig.\,\ref{ColdensMapsSkel} 
 for part of the Orion~B field (left panel) and the IC5146 cloud (right panel), respectively 
 (see Fig.~\ref{Maps1} to Fig.~\ref{Maps4}  in Appendix\,\ref{App4} for the whole Orion~B skeleton 
 and the skeletons derived in the other regions).


 \begin{figure*}[]
   \centering
     \resizebox{19cm}{!}{
\includegraphics[angle=0]{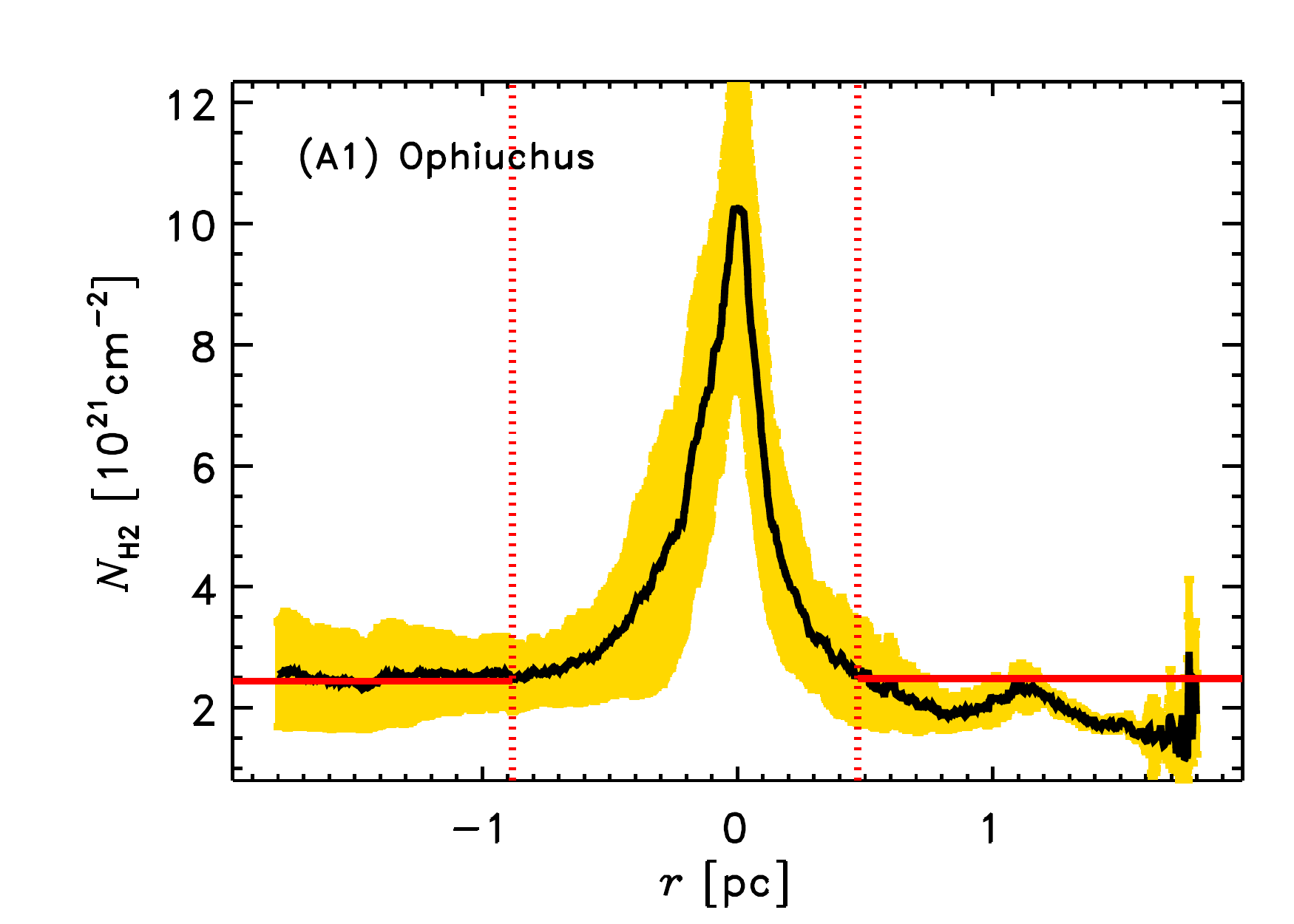}
\includegraphics[angle=0]{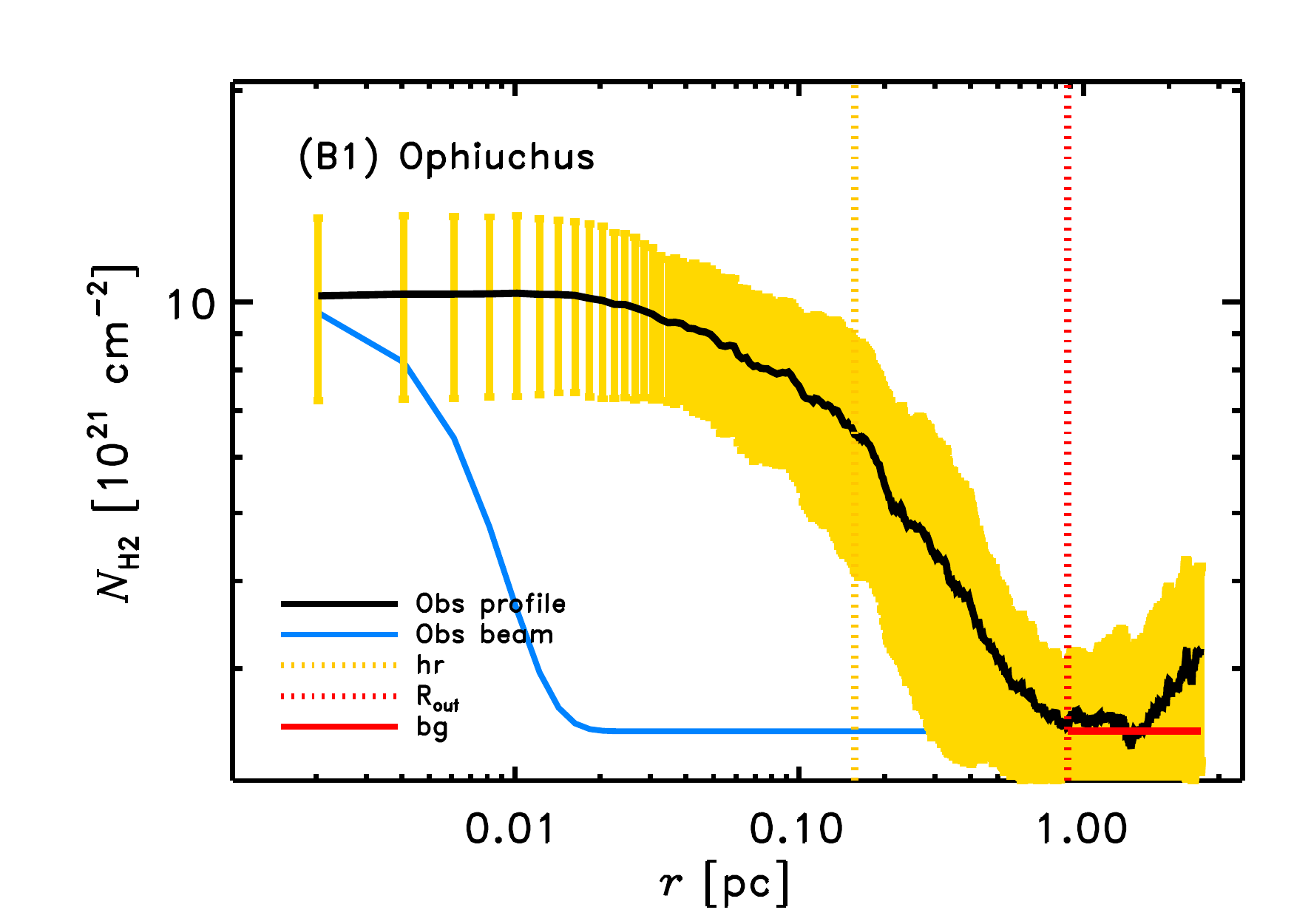}
\includegraphics[angle=0]{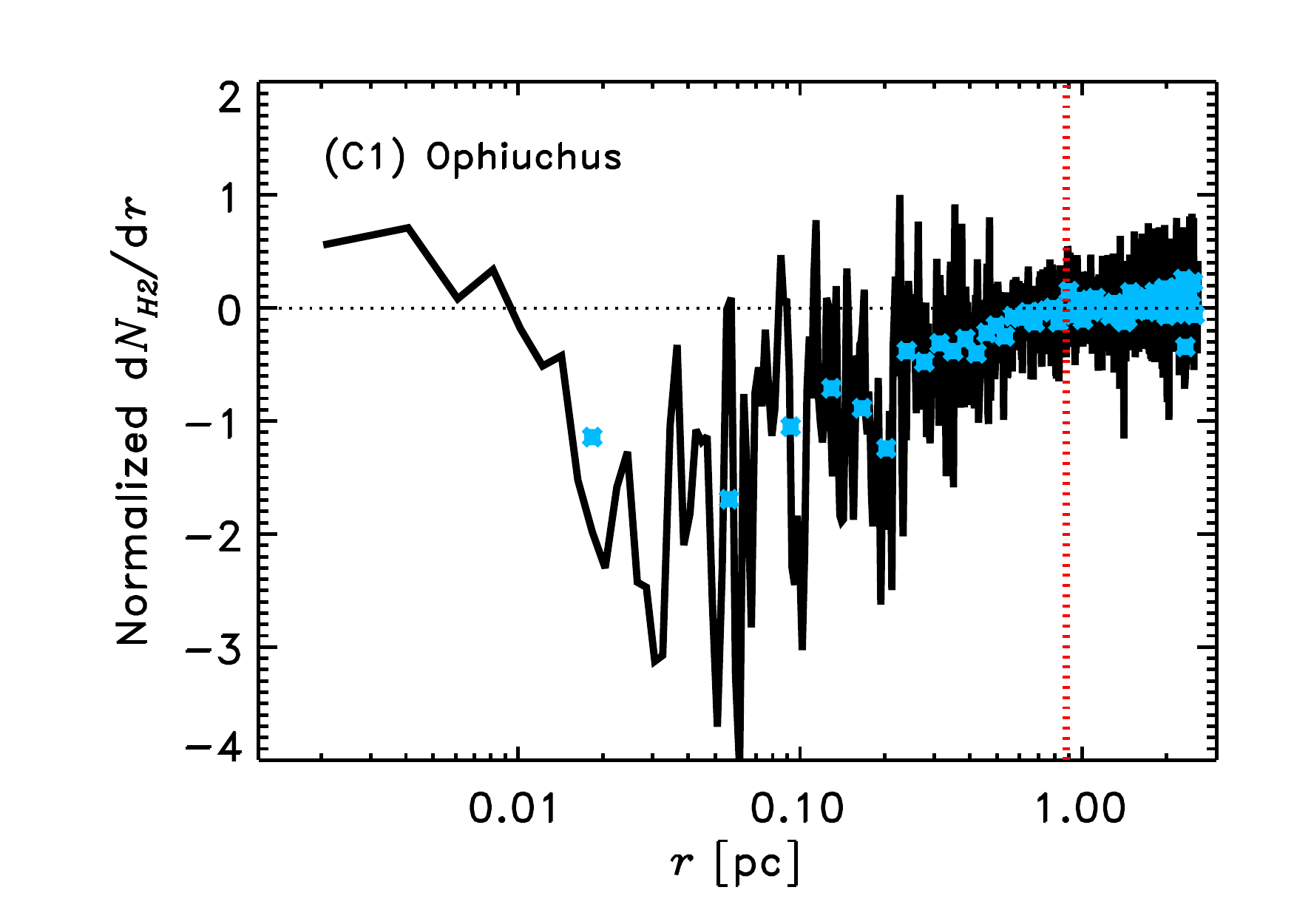}
}
     \resizebox{19cm}{!}{
     \hspace{-0.3cm}
\includegraphics[angle=0]{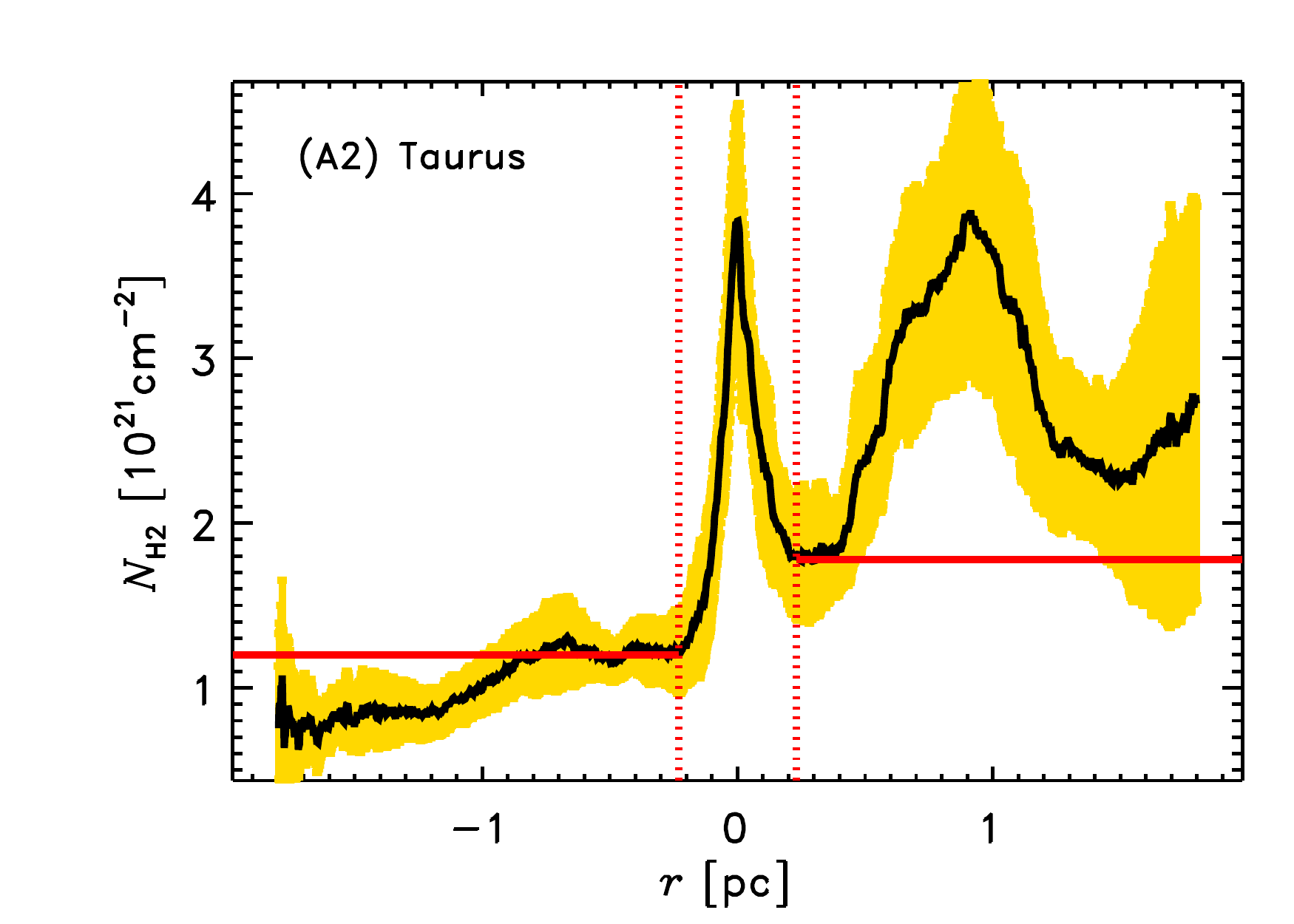}
\includegraphics[angle=0]{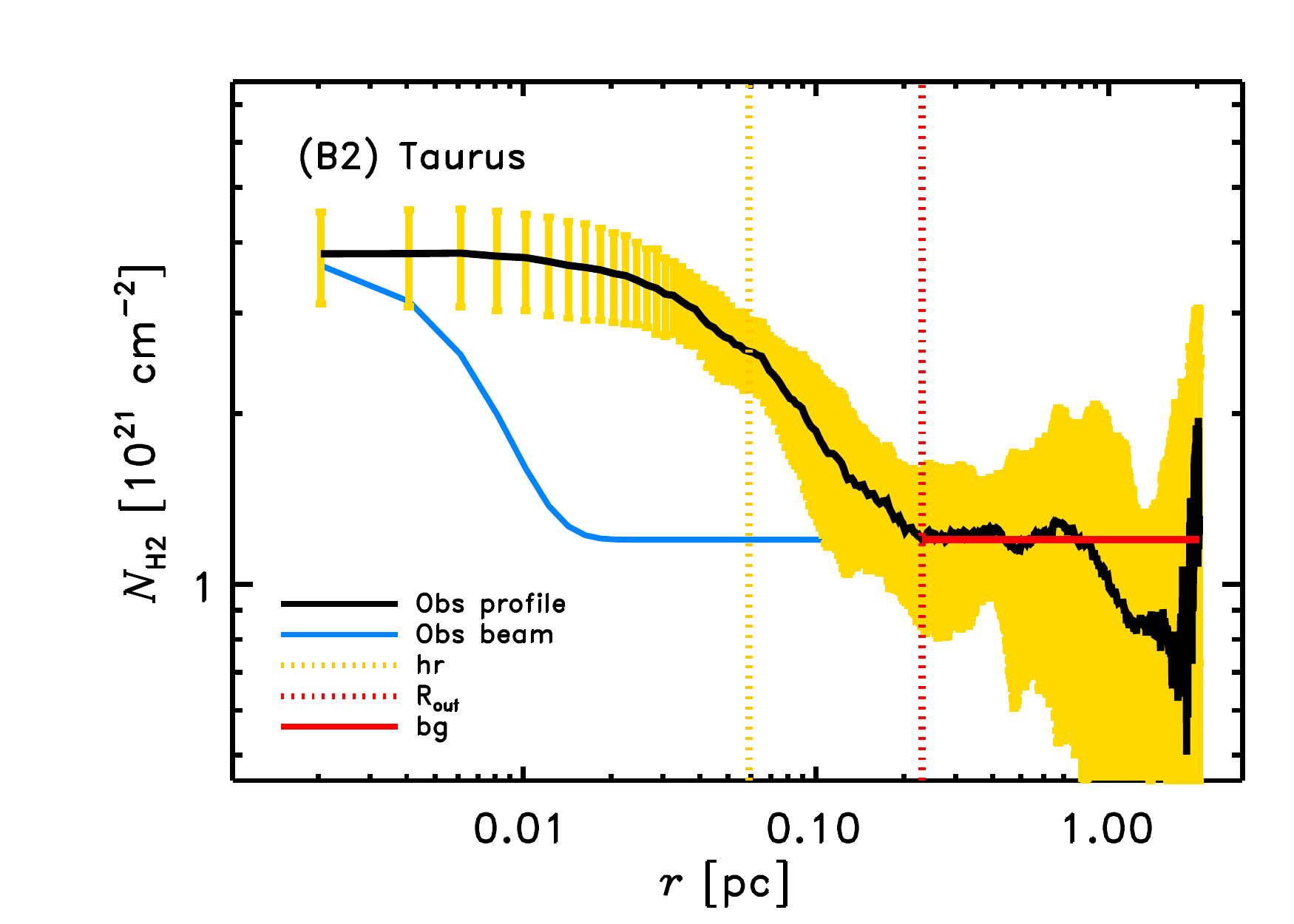}
\includegraphics[angle=0]{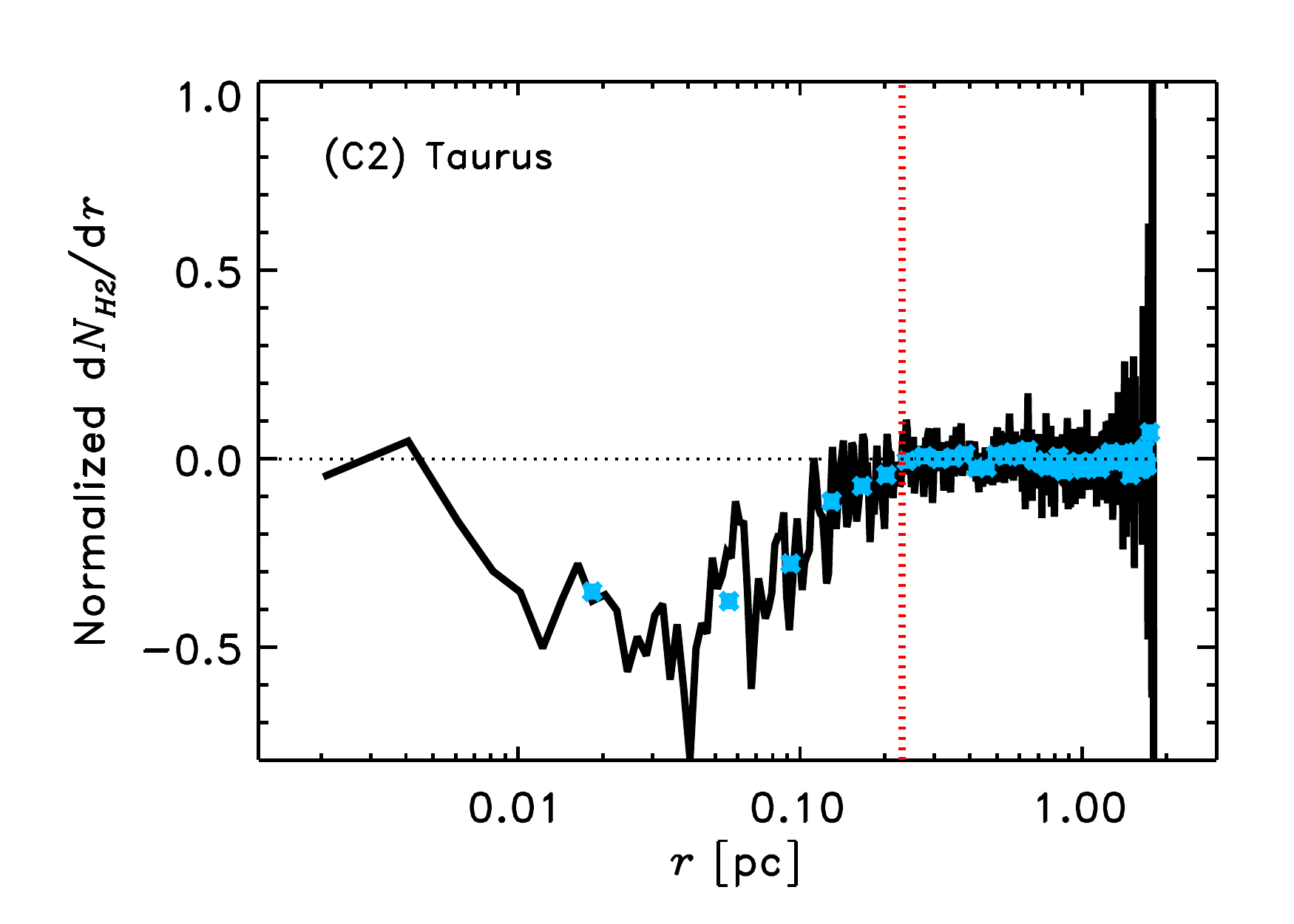}
}
  \caption{Columns from left to right for both rows: 
 {\bf (A)} Median radial column density profile observed on either side of the filament axis in lin-lin format, 
  {\bf (B)}  median radial column density profile on the left-hand side of the crest [i.e., $r<0$ in  {\bf (A1)} and  {\bf(A2)}]  in log-log format, and 
 {\bf (C)}  logarithmic slope of the column density profile (${\rm d\, ln} \nhh/{\rm d\, ln}r$) as a function of radius from the central crest,    
  for a filament in the Ophiuchus cloud (top three panels, {\bf A1, B1, C1}) and a filament in the Taurus cloud (bottom three panels, {\bf A2, B2, C2}).
  In {\bf (A)} and {\bf (B)}, the yellow area/error bars correspond to the median absolute deviation ($\rm {mad}(r)$) of the distributions of independent cuts  taken perpendicular to the filament crest (see Sect.\,\ref{RadProf}).  
  {\rev In {\bf (A)},  the outer radii $R_{\rm out}^\pm$ (vertical dotted red line) and the background column densities $  \nhh^{\rm bg \pm}$ (horizontal red line) are marked.} 
In {\bf (B)}, the observational beam of $18.2\arcsec $ (blue curve), the outer radius $R_{\rm out}^{-}$ (vertical dotted red line),  the background column density $\nhh^{\rm bg-}$ (horizontal red line), and the half power radius $hr^-$ (vertical dotted yellow line) are  shown.
As explained in Sect.\,\ref{FilBack}, the values of $R_{\rm out}^\pm$, $\nhh^{\rm bg\pm}$, and $hr^{\pm}$ were derived from the data {\rev independently on either side of the filament crest}, 
using the logarithmic slope plot {\bf (C)} smoothed over 3$\times HPBW$. 
In {\bf (C)}, the black solid curve represents the logarithmic slope profile at the original resolution,  
while the blue symbols show the logarithmic slope profile smoothed over 3$\times HPBW$; the red dotted vertical line marks the derived outer radius $R_{\rm out}^-$, 
and the black dotted horizontal line marks a zero slope (flat column density profile).
}          
  \label{2FilProf}
    \end{figure*}

\section{Measurements of filament properties }\label{FilMeasure}

In this section, we provide details on how we derive radial column density profiles for the extracted  filaments and 
how we estimate properties such as
filament width, outer radius, profile shape at large radii, mass per unit length, etc. 
The filament properties presented in this paper are derived from measurements performed on 
{\it Herschel}  column density maps at $18\parcs2$ resolution.

\subsection{Construction of radial profiles for the extracted filaments}\label{RadProf}

To construct radial profiles perpendicular to the long axis of a given filament, we first determine the 
normal direction to the filament at each point along its crest. 
To do so, we compute the Hessian matrix  $H$  of  second-order partial derivatives  for all pixels in the corresponding ``high-resolution'' column density map.  
The angle $\alpha$ between the $x$-axis of the Cartesian coordinate system  and the local tangential direction to the filament crest is  given by:
 \begin{equation}
 \alpha = \frac{1}{2}\, \arctan\left( \frac{2H_{xy}}{H_{x^{2}} - H_{y^{2}}}\right),$$ \label{Eq:Angle}
\end{equation}
where the $x$ and $y$ axes are the horizontal and vertical axes  of the column density map 
in  Cartesian coordinates, 
and  $H_{x^{2}}=\partial ^{2}\nhh/\partial x^{2}$,  $H_{y^{2}}=\partial ^{2}\nhh/\partial y^{2}$ and  $H_{xy}=H_{yx}=\partial ^{2}\nhh/\partial x\partial y$. 
The angle $ \alpha+90^\circ$ then gives the normal direction to the filament crest at each position  \citep[see also][]{Arzoumanian2012}. 
From the $2\times2$ Hessian matrix, 
the minimum curvature of the column density field can also be estimated at each point (as the smaller of the two eigenvalues of the 
matrix), which may then be used to enhance the contrast of filamentary structures in the map.
Indeed, a filament is an elongated structure which corresponds to a relatively small curvature of  the column density field along its main axis, 
and a significantly stronger curvature along its short axis, i.e., the normal direction to the filament crest.  
Based on this idea, the Hessian matrix has been used to identify filaments in \herschel\ and \planck\ maps  \citep[][]{Schisano2014,planck2016-XXXII}. 
Here, we only use the Hessian matrix to calculate the normal direction to the filament crest, using Eq.\,\ref{Eq:Angle}  on the original \nhh\ map.\\

Using the above approach, we thus construct radial profiles 
at each pixel position along the filament crest 
in the high-resolution column density map.
We then build the median radial column density profile of the filament by computing the median of all radial cuts along the  crest.    
We also derive a set of spatially independent radial profiles for the filament by dividing the crest into  consecutive segments of $2\times HPBW$ length 
and averaging the radial cuts obtained in each segment of adjacent pixels along the filament crest. 
{\rev As mentioned in Sect.\,\ref{intro}, prestellar and protostellar cores are observed along supercritical filaments. 
The presence of these dense cores may affect estimates of the filament properties locally and should ideally be removed. 
However, since cores contribute only a small fraction of the mass of dense filaments, 
typically $15\%$ on average \citep[e.g.][]{Konyves2015},  
they are not expected to alter significantly the median profile and median properties of a filament. 
For simplicity, 
in this paper, we thus present results derived from an analysis of original \nhh\ maps without subtracting dense cores.
}

Figure\,\ref{2FilProf} shows  two examples of median radial column density profiles, obtained 
for a supercritical filament in Ophiuchus and a subcritical filament in Taurus.
The local dispersion of the radial \nhh\ profiles 
(shown in yellow) is estimated as the median absolute deviation   
${\rm mad}(r)={\rm median}[|\nhh^{pix}(r)-{\rm median}(\nhh(r))|]$, where $\nhh^{pix}(r)$ is the value of the $\nhh(r)$ radial profile 
at each pixel position along  the filament crest.
Radial dust temperature profiles can be constructed in a similar way 
from the line-of-sight dust temperature maps 
derived from \herschel\ data at $36\parcs3$ resolution  \citep[cf.][]{Konyves2015}.

In the following subsections,  we describe how filament properties can be derived from the 
radial profiles.
All of the properties  discussed below were first measured independently on either side of the filament crest. 
The values corresponding to the two sides were then averaged.

\subsection{Estimating the local background and outer boundary  of each filament}
\label{FilBack}

Molecular filaments are not  isolated structures but are embedded in parent 
interstellar clouds which often exhibit complex, multi-scale internal structure.
To {\it measure} the properties of an individual filament, 
one should first estimate its outer boundaries and characterize the local background, 
i.e., the emission properties of the parent cloud in the immediate vicinity of the filament.
Especially in the case of a low-contrast filament, an accurate determination of the local background 
is key for  deriving intrinsic filament properties 
but 
this is not straightforward.
Even the B211/B213  filament in Taurus, which has a very clean and very symmetric radial \nhh\ profile, 
exhibits some differences in background properties on either side of its long axis \citep{Palmeirim2013}. 
Here, we thus perform separate measurements on either side of the filament axes. 

At each point $s$ along the crest, the outer radius \rout\ $(s)$ of a filament may be defined 
as the radial distance from the filament axis at which the amplitude  $\nhh^0=\nhh^{\rm fil}-\nhh^{\rm bg}$ 
of the radial column density profile becomes negligible compared to the column density $\nhh^{\rm bg}$ of the background cloud 
(without beam deconvolution). 
In practice, $R_{\rm out}(s)$ can be estimated from the observed radial column density profile 
as the closest point to the filament crest for which the 
logarithmic slope of the profile ${\rm d\, ln} \nhh/{\rm d\, ln}r$, smoothed over $3\times$HPBW,  is consistent with zero 
(see Fig.\,\ref{2FilProf}).
Strictly speaking, two outer radii $R_{\rm out}^\pm (s)$ are defined at each point $s$ on either side of the filament axis. 
The column density  $\nhh^{\rm bg \pm} (s)$ of the background cloud  on either side of the filament crest is assumed to be constant: 
$\nhh^{\rm bg\, \pm}  (s) = \nhh(r=R_{\rm out}^\pm) $ (see Fig.\,\ref{2FilProf}). 

Once $R_{\rm out}^\pm$, $\nhh^{\rm bg \, \pm}$, and $\nhh^0$ have been estimated, 
the half-power radius $hr^\pm$ of the filament (on either side of the main axis) may be derived, without any fitting of the radial profile, 
as the radial distance from the filament crest where the background-subtracted amplitude of the observed profile is half of the value on the crest, i.e.,  $hr^\pm=r^\pm(\nhh^0/2)$.
The quantity $hd \equiv 2hr$ or $hr^+ + hr^-$ provides a first estimate of the filament width (see Fig.\,\ref{2FilProf}). 
We  deconvolve the measured $2hr$ width from the equivalent $HPBW\,=\,18\parcs2$ beam of the high-resolution column density map 
using the following approximation: $hd_{\rm dec} = \sqrt{(2hr)^{2}- HPBW^{2}}$, where $hd_{\rm dec}$ is the estimated half-power diameter after deconvolution. 

\begin{figure*}
   \centering
     \resizebox{16cm}{!}{
   \includegraphics[angle=0]{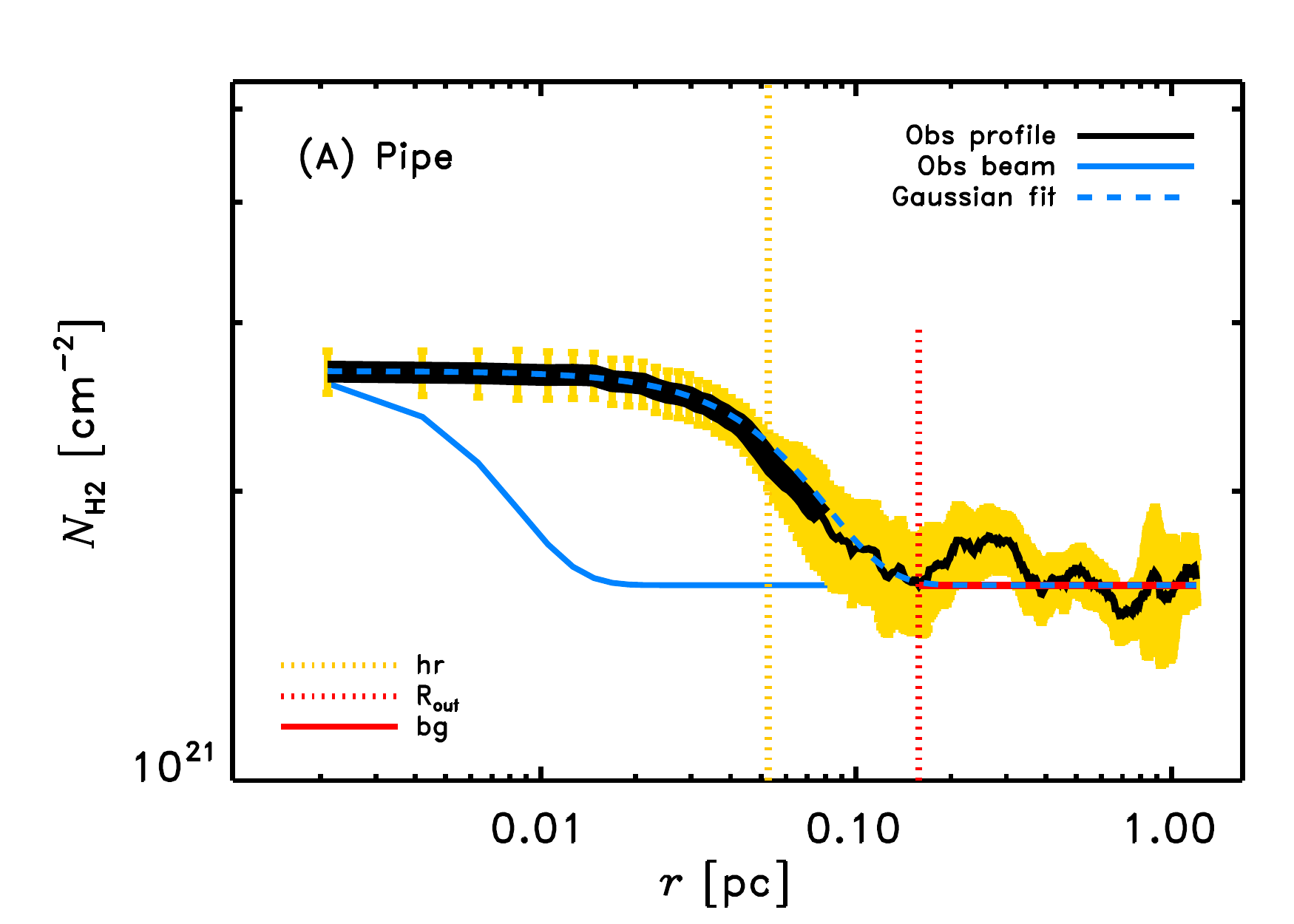}
      \includegraphics[angle=0]{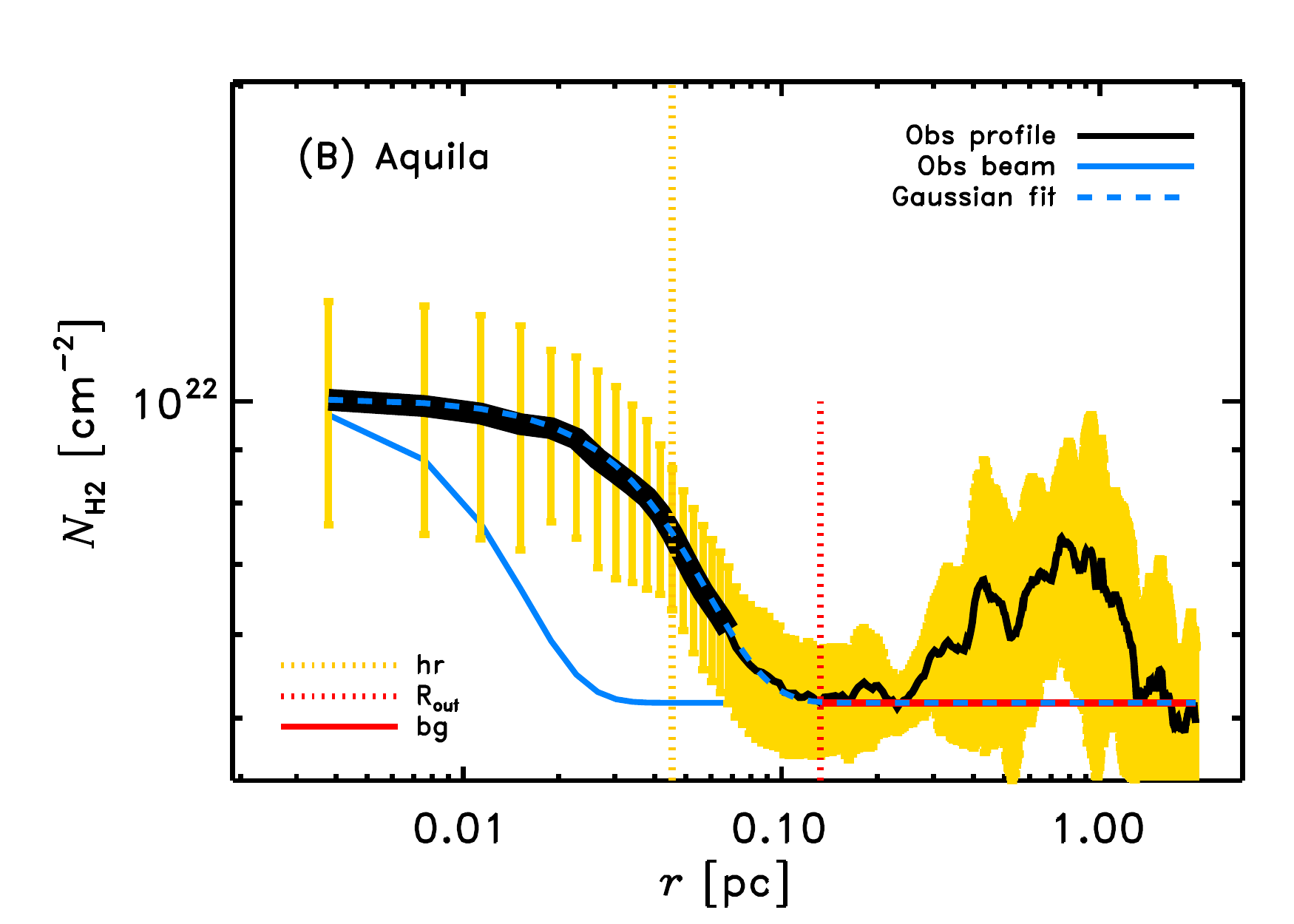}
   }   
 \resizebox{16cm}{!}{
 \includegraphics[angle=0]{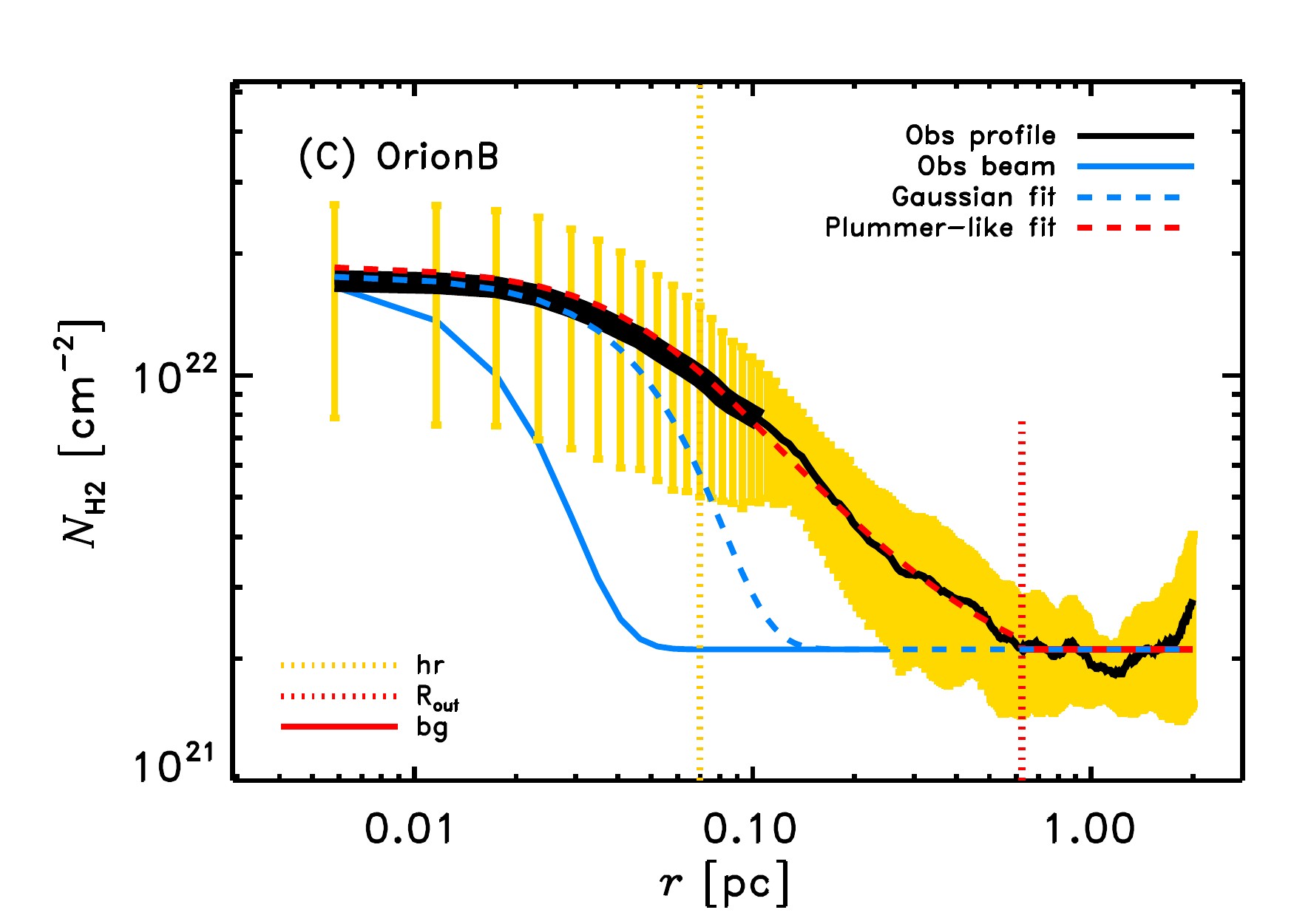}
\includegraphics[angle=0]{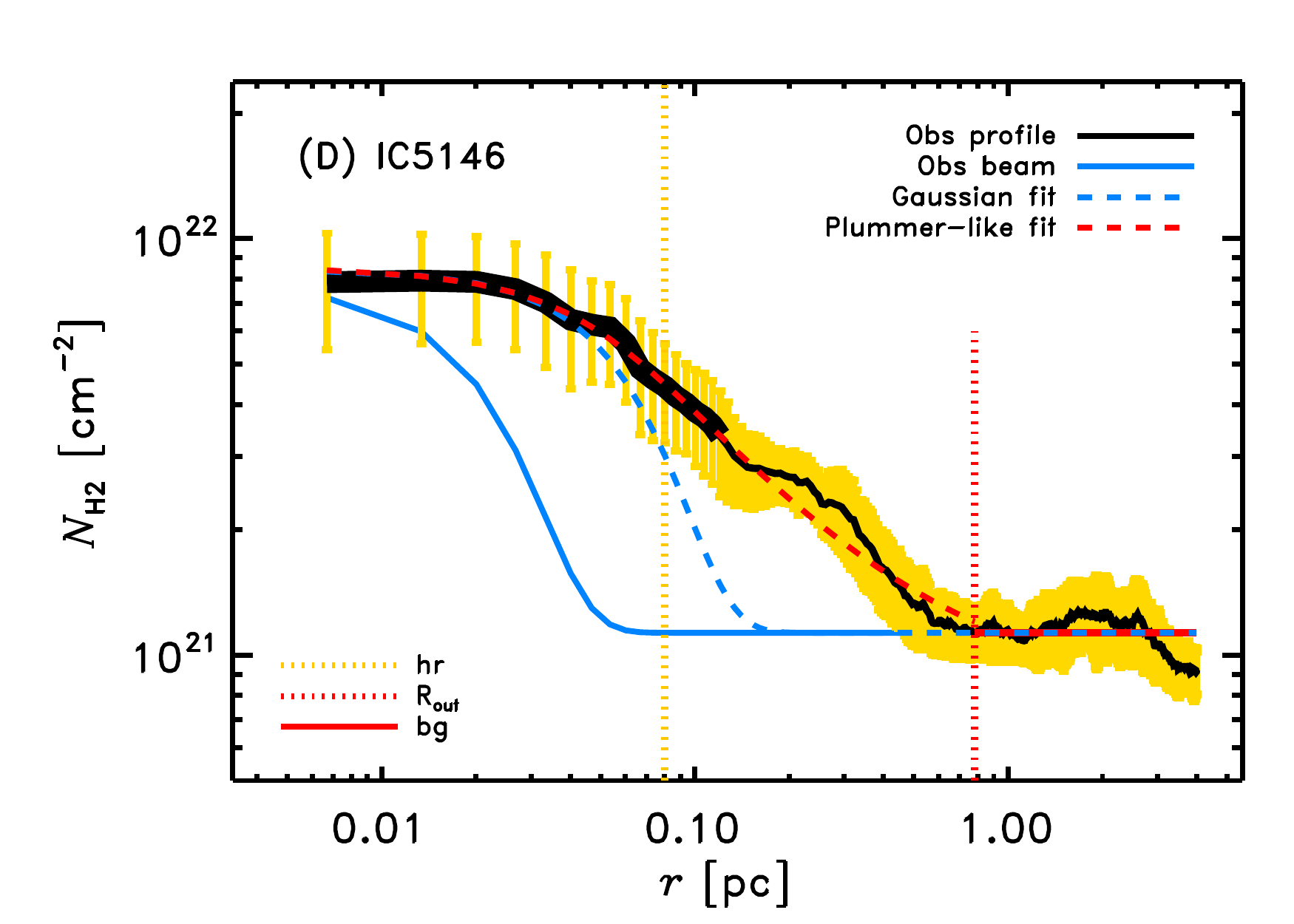}}
  \caption{  {\rev  Radial column density profiles (in log-log format) observed perpendicular to, and averaged along, the crests of four filaments  in four different clouds. 
The cloud name is indicated at the top left of each panel.  
The blue and red dashed curves show the best Gaussian and Plummer fits, respectively (see Sect.\,\ref{filfit}). 
In panels (A) and (B), the observed profiles are Gaussian-like and the derived ($hd_{\rm dec}$, $FWHM_{\rm dec}$) widths are (0.11\,pc, 0.12\,pc) and (0.09\,pc, 0.09\,pc), respectively. 
In panels (C) and (D), the inner parts of the profiles are well reproduced by Gaussian functions, but the outer pars show power law wings, which are  much better fitted 
with Plummer functions. The derived ($hd_{\rm dec}$, $FWHM_{\rm dec}$,  $D_{\rm flat}$, and $p$) parametes are (0.14\,pc, 0.10\,pc, 0.09\,pc, 2.2)  and  (0.16\,pc, 0.11\,pc, 0.08\,pc, 2.2), 
for  panels  (C) and (D),  respectively.
On each plot, the blue solid curve represents the effective beam resolution of the column density map ($18.2\arcsec $). 
The half-power radius $hr$, outer radius $\rout$, and background column density (bg), 
are also indicated (see legend on the bottom left of the panels and Sect.\,\ref{FilBack}). 
The black thick section of each profile  indicates the fitting range of the Gaussian fits (i.e., $r\le1.5hr$). 
The Plummer fits were performed for $r\le\rout$.
The yellow area/error bars correspond to the median absolute deviations [${mad}(r)$] of the distribution  
of independent cuts taken perpendicular to the filament crest (see Sect.\,\ref{RadProf}). 
  } }
  \label{fig:Multiprof}
\end{figure*}

To summarize, estimates of $R_{\rm out}(s)$, $\nhh^{\rm bg}(s)$, and $hd(s)$ are obtained at each point along, and on either side of, the crest of each filament. 
In practice, the quantities used in Sect~\ref{FilStat} and Sect~\ref{Discussion} below
correspond to the medians of 
the $R_{\rm out}(s)$, $\nhh^{\rm bg}(s)$, and $hd(s)$ values 
derived from the set of spatially independent profiles for each filament. 
The profiles shown in Fig.\,\ref{2FilProf} and Fig.\,\ref{fig:Multiprof} correspond to
the median radial column density profiles of four filaments taken as examples.

\subsection{Fitting the filament radial profiles}\label{filfit}

The radial column density profiles of each filament were fitted 
with both Gaussian and Plummer-like functions.

For each filament, separate fits were derived on either side of the main axis using 
1) the median radial column density profiles  
and 2) the set of spatially-independent radial profiles constructed along the filament crest. 
In the latter case, median parameter values were calculated from the distribution of fitting parameters obtained along the crest.
In both cases, the parameters of the fits found on either side were also averaged. 
This approach resulted in two values for each fitting parameter, one derived from the median radial profiles and 
the other corresponding to the median parameter value along the filament crest.

In the next two subsections, we describe the Gaussian and Plummer fitting procedures, respectively.
Section\,\ref{Linkfit} gives details on the  link between the filament 
widths derived from the two fitting methods. 
The reliability of the derived parameter values is discussed 
in Sect.\,\ref{SelecSample}.

\subsubsection{Gaussian function fitting}\label{Gaussfit}

Each $\nhh (r)$ profile was fitted for $r \le 1.5\,hr$ with a one-dimensional Gaussian function of the form,
\begin{equation}\label{Eq:EqGauss}
\nhh^{\rm G}(r) = \nhh^{0,{\rm G}}\exp\left[-4\ln2\,(r/FWHM)^2\right]+ \nhh^{\rm bg,G}\,,
\end{equation}
where the column density amplitude, $\nhh^{0,{\rm G}}$, 
the $FWHM$ width, and {\Newrev the (uniform) background column density 
$\nhh^{\rm bg,G}$ estimated from the observed column density at $r=1.5hr$, $\nhh(r=1.5\,hr)$, are the three free parameters of the fit. 
 The free parameters of the fit were bounded as follows: $0.7\nhh^{0} \le\nhh^{0,{\rm G}}\le1.3\nhh^{0}$,  $0.7\nhh(r=1.5\,hr)\le\nhh^{\rm bg,G}\le1.3\nhh(r=1.5\,hr)$, and  $0<FWHM\le\rout$.}
Each fitted data point was weighted by its median absolute deviation $mad$ (see yellow error bars in  Fig.\,\ref{2FilProf} -- see also Sect.\,\ref{RadProf}).  
The fitted $FWHM$ width was then deconvolved from the observational $HPBW$ beam  as  $FWHM_{\rm dec} = \sqrt{FWHM^{2}- HPBW^{2}}$, where 
$HPBW\,=\,18\parcs2$ is the effective resolution of the corresponding high-resolution column density map. 
 
\subsubsection{Plummer-like function fitting}\label{Plumfit}

The radial column density profiles of observed filaments often exhibit power-law wings 
which cannot be well represented by a Gaussian function and are better reproduced by a Plummer function 
\citep[see Fig.\,\ref{fig:Multiprof} and e.g.,][]{Arzoumanian2011,Palmeirim2013,Andre2016,Cox2016}. 
Using a Plummer-like function (Eq.\,\ref{Eq:EqPlum}),  one can in principle reproduce the behavior of the radial  column density distribution for both $r\le R_{\rm flat}$ and $ r>>R_{\rm flat}$, which is not possible with a Gaussian fit. 
Each  median  $\nhh(r)$ profile was thus fitted with the following function:
\begin{equation}\label{Eq:EqPlum}
\nhh^{\rm Pl}(r) = \nhh^{0,{\rm Pl}}/\left[1+\left(r/R_{\rm flat}\right)^2\right]^\frac{p-1}{2}+ \nhh^{\rm bg,Pl}\,,
\end{equation}
where $R_{\rm flat}$ is the radius of a flat inner region with approximately constant (column) density, 
$p$ is the power-law exponent of  the corresponding density profile for $r>>R_{\rm flat}$, 
$\nhh^{0,{\rm Pl}}$ is the central column density of the Plummer-like model filament, 
and $\nhh^{\rm bg,Pl}$ is the background column density (here assumed to be independent of $r$).
{\rev The fitting was performed up to $r=\rout$. The free parameters of the fit were $\nhh^{0,{\rm Pl}}$, $\nhh^{\rm bg,Pl}$,  $R_{\rm flat}$, and $p$. 
{\Newrev  The free parameters were bounded as follows: 
$0.7\nhh^{0}\le\nhh^{0,{\rm Pl}}\le1.3\nhh^{0}$,  $0.7\nhh(r=\rout)\le\nhh^{\rm bg,Pl}\le1.3\nhh(r=\rout)$, 
$0<R_{\rm flat}\le\rout$, and $1.2\le p\le3.5$.}
 The Plummer-like model function described by Eq.\,\ref{Eq:EqPlum} was first convolved with the $18\parcs2$  
Gaussian beam of the corresponding column density map prior to comparison with the observed profile.

{\rev
Recovering the intrinsic $R_{\rm flat}$ value when fitting $R_{\rm flat}$ and $p$ 
simultaneously 
is difficult in the case of low-contrast filaments 
due to 
a partial degeneracy between the two parameters  
whose errors are 
anti-correlated \citep[as discussed by, e.g.,][]{Malinen2012,Juvela2012,Smith2014}. 
The results are nevertheless satisfactory for high-contrast filaments $C^0 \ge 1$, 
such as the Taurus B211/B213 filament \citep[][]{Palmeirim2013}. 
We thus derived  two estimates of $R_{\rm flat}$: $R_{\rm flat}$(1) from fit 1, obtained by fixing the power-law exponent to $p=2$ and leaving %
$R_{\rm flat}$ as a free parameter, 
and  
$R_{\rm flat}$(2) from fit 2, leaving both $R_{\rm flat}$ and $p$ as free parameters. }\\

As an illustration, 
Fig.\,\ref{fig:Multiprof} shows the median radial \nhh\ profiles of 
{\rev four filaments observed in different regions}
along with the corresponding best Gaussian and Plummer
fits. 

 \begin{figure}[!ht]
   \centering
     \resizebox{8.5cm}{!}{
\includegraphics[angle=0]{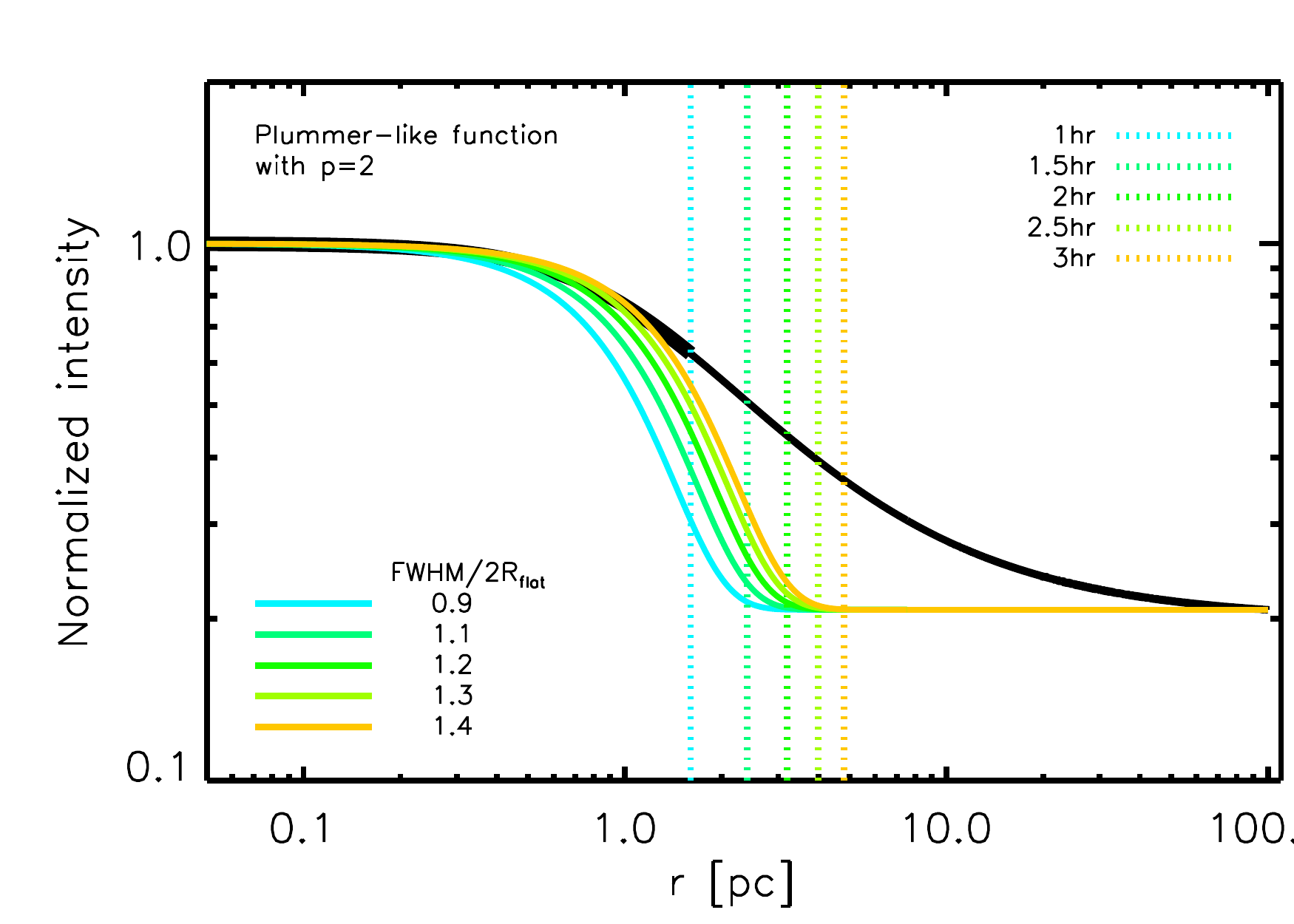}}
  \caption{Gaussian fits (colored curves) to the radial column density profile $\nhh (r)$ of a model filament with a Plummer profile  (black curve)
  described by Eq.\,\ref{Eq:EqPlum} with $p=2$,  $\nhh^{\rm bg}=0.2\nhh^0$, and  $R_{\rm flat}=1$\,pc,
  for five choices of the fitting range: $r \in [0,hr]$, $[0,1.5hr]$, $[0, 2hr]$, $[0,2.5hr]$, and $[0, 3hr]$ (see vertical dotted lines  and color coding at the top right of the plot), {\rev where $hr=1.6\,R_{\rm flat}$}.  
  The $FWHM$ width values obtained from the Gaussian fits are given in units of $D_{\rm flat} = 2\, R_{\rm flat}$ at the bottom left of the plot. 
}          
  \label{ProfFit}
    \end{figure}

\subsubsection{Filament width}\label{Linkfit}

For each radial column density profile constructed from {\it Herschel} data, 
we derived three estimates of the corresponding filament width: a deconvolved half-power diameter 
$hd_{\rm dec}$ (from the deconvolved $2hr$ value) without any fitting of the profile, a deconvolved  $FWHM_{\rm dec}$ width resulting from a Gaussian fit for $r \le 1.5\,hr$, 
and a (deconvolved) flat inner diameter $D_{\rm flat}=2R_{\rm flat}$ resulting from a Plummer-like function fit for $r \le R_{\rm out}$. 
These three different estimates of the 
inner width of each filament are discussed further in Sect.\,\ref{FilStat}. 

We stress that, in the presence of power-law wings, both the {\rev measured half-power diameters $hd$} and the results of Gaussian fits must be interpreted with caution. 
In particular, as pointed out by, e.g., \citet[][]{Smith2014,Panopoulou2017}, 
the $FWHM$ widths derived from Gaussian fitting are  affected by the range of radii used in the fits.
Figure\,\ref{ProfFit} 
illustrates  the link between the  {\rev  half-power diameter $hd$}, the  $FWHM$ width derived from a Gaussian fit,  
and the intrinsic flat inner diameter $2\, R_{\rm flat}$ of a Plummer-like model profile with $p=2$, for several fitting ranges.
In this case, {\rev  $hd=1.6\,(2R_{\rm flat}^{p=2})$} and $FWHM=1.1\,(2R_{\rm flat}^{p=2})$ when the profile is fitted for radii $0 \leq r \leq 1.5\,hr$. The relation between $FWHM$ and $2R_{\rm flat}^{p=2}$ 
changes by $\pm 20\%$ when 
the fitting range varies between $hr$ and $3hr$. 
For Plummer-like model profiles with $p=1.5$, $p=2.5$, and $p=3$,
$FWHM=1.6\,(2R_{\rm flat}^{p=1.5})$, $FWHM=0.9\,(2R_{\rm flat}^{p=2.5})$, and $FWHM=0.8\,(2R_{\rm flat}^{p=3})$, when the profile is fitted for $r \leq 1.5\,hr$, 
with uncertainties of $\pm40\%$, $\pm11\%$, and $\pm8\%$ respectively, when the fitting range varies between $hr$ and $3hr$.

\subsection{Filament mass per unit length}\label{MlineFilAnalysis}

As introduced in Sect.\,\ref{intro}, the mass per unit length (\ml) of a filament is a very important parameter which,  
by comparison with the critical line mass  $M_{\rm line,crit}$, may be used to diagnose whether the filament 
is unstable to gravitational collapse as a cylindrical structure.  

When the filament profile is well approximated by a Gaussian function, \ml\ can be estimated by multiplying the central  surface density 
of the filament by the $FWHM$ width %
\citep[cf. Appendix~A of][]{Andre2010}. 
When the \nhh\ profile is not Gaussian-like and/or includes a significant power-law wing, however, 
the latter approximation of \ml\ may underestimate the true \ml. %
To investigate the relevance of the shape of the radial column density profile for estimating the filament \ml,
we derived and compared three (partly independent) estimates of \ml, making use of our detailed analysis of the radial profiles:\\
\phantom{x}- $M^{\rm w}_{\rm line}=\Sigma_{\rm fil}^0 \times W_{\rm fil} $, where $\Sigma_{\rm fil}^0 = \mu_{\rm H_2}m_{\rm H}  \nhh^0 $ is the central gas surface density of the filament, 
$W_{\rm fil}=2hr$ is the filament  width estimated without any fitting, 
$\mu_{\rm H_2}$ the mean molecular weight per hydrogen molecule, and $m_{\rm H}$ the mass of a hydrogen atom.\\
\phantom{x}- $\ml^{\rm int} = 2\,\int_{0}^{\rout}\  \mu_{\rm H_2}m_{\rm H}(\nhh(r)-\nhh^{\rm bg})\,{\rm d}r$, integrating the column density profile over radius up to \rout,  
after background subtraction.\\%
\phantom{x}- $M_{\rm fil }/l_{\rm fil }$, where $M_{\rm fil }$ is the total mass of the filament calculated by summing the background-subtracted 
column density over the pixels contained in the area around the filament crest bounded by the outer radius values \rout\  measured along the filament crest (see Sect.\,\ref{FilBack}), 
and $l_{\rm fil }$ is the total length of the filament.

\subsection{Reliability of derived filament properties}\label{SelecSample}

Before discussing the statistical properties of the extracted filament sample in Sect.~\ref{FilStat} below, 
the reliability of our method of measuring filament properties must be assessed. 
To this end, we tested the measurement procedures 
described in the previous subsections 
using  synthetic maps. 
{\rev These tests are described in detail in Appendix\,\ref{App2a}  
and Appendix\,\ref{App2b}.}

Briefly, we distributed several sets of synthetic filaments with both Gaussian and Plummer density profiles 
within a realistic ``background'' column density map. 
{\Newrev The synthetic filaments were given  uniform, Gaussian, or power-law distributions of input properties ($FWHM$, $D_{\rm flat}$, $p$) along their crests.}
{\Newrev  The various measurements/fitting steps described in Sects.\,\ref{RadProf} to\,\ref{MlineFilAnalysis} above were then applied to the synthetic maps.}
For synthetic filaments with Gaussian profiles, {\rev the measured  $hd=2hr$ values, derived without any fitting, reproduce the input 
$FWHM$ widths. }
Moreover, the resulting distributions of  $FWHM$ widths estimated from Gaussian fitting 
are in excellent agreement with 
the input distributions of $FWHM$ values for any fitting range between [0,$1.5hr$] and [0,$3hr$] (Appendix\,\ref{App2a}). 
In particular, a peaked distribution of measured 
$FWHM$ widths is recovered solely when 
the input synthetic filaments have a constant  width {\Newrev along their crests or a narrow Gaussian distribution around the mean width value} 
(see Figs.\,\ref{histo_width_mock}A and\,\ref{histo_GaussW_mock} in Appendix\,\ref{App2a}). 
In addition the distribution of measured $FWHM$ widths is flat or {\rev has a power-law distribution down 
to the resolution limit of the column density map} when the input filaments have a flat or {\rev power-law distribution of widths {\Newrev (constant along their crest)}, respectively,}  
(see Figs.\,\ref{histo_width_mock}B,C 
in Appendix\,\ref{App2a}). 
The values of the input filament lengths, column density contrasts $C^0$, and masses per unit length $\ml$, 
are also well recovered.

In the case of synthetic filaments with Plummer-like profiles, 
the median value of the $FWHM$ widths derived from Gaussian fitting 
is significantly affected by the fitting range \citep[see Fig.\,\ref{ProfFit}, and also][]{Smith2014,Panopoulou2017}.  
{\rev However, for a given fitting range (e.g., for $0\le r\le1.5hr$), a peaked distribution of measured $FWHM$ values 
is obtained only when the input $R_{\rm flat}$ values are the same for all mock filaments, 
while a power-law distribution down to the resolution limit of the synthetic map is recovered 
when the input mock filaments have a power-law distribution of $R_{\rm flat}$ values (see Figs.\,\ref{histo_widthPlum_mock} in Appendix\,\ref{App2b}). 
Moreover, both the $R_{\rm flat}$ and the $p$ parameter of the synthetic Plummer filaments 
are reasonably well recovered by our Plummer-fitting method, 
independently of the input distributions of $R_{\rm flat}$ values (uniform, flat, or power law)
and $p$  values (uniform, flat, or Gaussian), {\Newrev 
and whether the input $R_{\rm flat}$ values were constant along the filament crests (see Figs.\,\ref{histo_widthPlum_mock} to\,\ref{fig:histo_mockParamPlumFlatp}) or not (see Fig.\,\ref{fig:histo_mockParamPlum_plw} in Appendix\,\ref{App2b})}.
{\rev The measurements however become increasingly more uncertain for lower-contrast ($C < 1$) filaments, 
as can be seen for instance by comparison of Fig.\,\ref{fig:histo_mockParamPlum_Cont05} with Fig.\,\ref{fig:histo_mockParamPlum_Cont1}.
}

{\rev For Plummer-like input filament profiles, both 
the $FWHM$ widths  derived from Gaussian fitting and the $hd$ values (derived without any fitting) 
are affected by changes in the input power-law slope $p$. 
{\rev The derived $FWHM$ widths are closest to the input $D_{\rm flat} = 2R_{\rm flat}$ values when the fitting range is $0\le r\le 1.5hr$. 
If this fitting range is adopted, the derived $FWHM$ widths provide estimates of the $D_{\rm flat} $ diameters 
that are accurate to better than $\sim 50\% $ for high-contrast ($C^0 > 1$) filaments when $1.5 < p < 3$. 
The uncertainties in the derived $FWHM$ widths are also smaller 
than those in the derived $hd$ values. 
}

Based on the reliability tests and uncertainty assessments of Appendix\,\ref{App2}, 
we present and discuss 
the statistical distributions of our three estimates of the filament inner width ($hd_{\rm dec}$, $FWHM_{\rm dec}$,
and $D_{\rm flat}$) in Sect.~\ref{FilStat} below.
We adopt the $FWHM_{\rm dec}$ estimates as our reference measurements of the  inner width,  
giving satisfactory results for both Gaussian-like filaments and Plummer-like filaments. 
We stress, however, that the $D_{\rm flat}$ estimates are more appropriate and more accurate 
in the case of high-contrast, dense filaments with power-law wings.
}

     \begin{table}[!ht]  
\centering
\caption{Absolute extraction thresholds and numbers of extracted filaments.}
 \label{tab:SumParamDisp}
\begin{tabular}{|c|ccc|cc|}   
\hline\hline   
Field & $\nhh^{\rm bg,min}$& PT&RT&$N^{\rm fil}_{\rm tot}$&$N^{\rm fil}_{\rm select}$\\
&   \multicolumn{3}{|c|}{[$10^{21}\,\rm cm^{-2}$]}  &$\#$&$\#$\\ %
  (1)&(2)& (3) & (4) &(5)&(6) \\
\hline
IC5146&0.66&0.17&1.00&67&59 \\
Orion B&0.72&0.17&1.09&410&234 \\
Aquila&2.71&0.20&4.06&137&71 \\
Musca&0.86&0.09&1.28&47&10 \\
Polaris&0.56&0.17&0.84&32&20 \\
Pipe&0.83&0.12&1.24&148&38 \\
Taurus L1495&0.92&0.07&1.38&266&110 \\
Ophiuchus&0.99&0.01&1.48&204&57 \\
\hline
All&&&&1310 &599 \\
                           \hline  \hline
                  \end{tabular}
\begin{list}{}{}
 \item[]{{\bf Notes:} {}
  {\bf Col. 1:} Field name. 
  {\bf Col. 2:} Minimum background column density in each field (see Sect.\,\ref{SkelDisperseHerschel}).  
  {\bf Col. 3:} Persistence threshold used in \disperse ,  corresponding to the minimum rms level of the background column density fluctuations in each field.
  {\bf Col. 4:} Robustness threshold used in \disperse , corresponding to 1.5$\nhh^{\rm bg,min}$, where $\nhh^{\rm bg,min}$ is the minimum background column density in each field.
  {\bf Col. 5:} Total number of filaments extracted in each field following the procedure explained in Sect.\,\ref{SkelDisperse}. 
  {\bf Col. 6:} Number of selected filaments with aspect ratio $ AR>3  $ and column density threshold $ C^0>0.3$ in each field (see Sect.\,\ref{SelecSample}).
  }
 \end{list}      
  \end{table}

\subsection{Filament samples}\label{FilSample}

Table\,\ref{tab:SumParamDisp} gives the absolute values of the persistence and robustness thresholds used to trace filamentary structures in  
each of the eight target fields, for $PT=rms_{\rm min}$ and $RT=1.5\nhh^{\rm bg,min}$. 
The entire set of filaments resulting from the extraction method described in Sect.\,\ref{FilIdentification} 
is referred to as the  ``total filament sample'' and comprises a total number of $N^{\rm fil}_{\rm tot} = 1310$ filaments.
After measuring the properties of each of these 1310 filamentary structures, 
we selected a subset of more robust filaments satisfying the following two additional conditions: 
$$ AR>3 \,\,{\rm and} \,\, C^0>0.3,$$
where  $AR=l_{\rm fil}/W_{\rm fil}$ with $W_{\rm fil}=FWHM$, and  $C^0=\nhh^0/\nhh^{\rm bg}$. 
This selection discards a fraction of filamentary structures which may be strongly contaminated by spurious features (see Appendix\,\ref{App1} and Fig.\,\ref{Comp_randW_AR_C}) 
or may be associated with elongated cores rather than filaments (for $AR\lesssim3$).  
The number of filaments in this selected sample is $N^{\rm fil}_{\rm select}=599$.

\section{Statistical properties of the extracted filament sample}
\label{FilStat}
  
This section presents the statistical properties of the filament sample extracted as explained in Sect.\,\ref{FilIdentification} in the eight fields of Table\,\ref{Table1}, 
all imaged with {\it Herschel} as part of the HGBS. 
In all of the following plots, the filaments detected in each region are represented by specific colored dots (cf. Table\,\ref{Table1} for the color coding). 
In Sect.\,\ref{AveProp}, we first discuss the distributions of  {\rev median properties 
resulting from ``averaging''  about 10 to 30 independent  measurements along and on either side of each filament crest.}
Properties such as filament column density, dust temperature, width, outer radius, length, mass per unit length, and column density contrast, 
estimated using the measurement steps described in Sect.\,\ref{FilMeasure} and {\rev averaged along the filament crests}, 
are discussed. 
We also investigate possible correlations between these various {\rev filament-averaged} properties. 
Tables\,\ref{tab:Tab_MassFrac}--\ref{tab:PlummerFit}  summarize the global properties of the subsets of filaments identified in each field. 
In particular, the mean value and the standard deviation of each fitted parameter are provided in Table\,\ref{tab:table_stat} (for Gaussian fits)
and Table\,\ref{tab:PlummerFit}  (for Plummer fits) for the eight fields. 
The median value of each parameter and the equivalent standard deviation estimated from the interquartile range (IQR) assuming Gaussian statistics 
are also provided. For a Gaussian distribution, the standard deviation $\sigma $ corresponds to the IQR  divided by a factor 1.349. 
{\rev Finally, in Sect.\,\ref{AlongFilProp}, we discuss the filament properties derived prior to any averaging along the filament crests.}

   \begin{figure*}[h!]
   \centering
     \resizebox{14cm}{!}{
\includegraphics[angle=0]{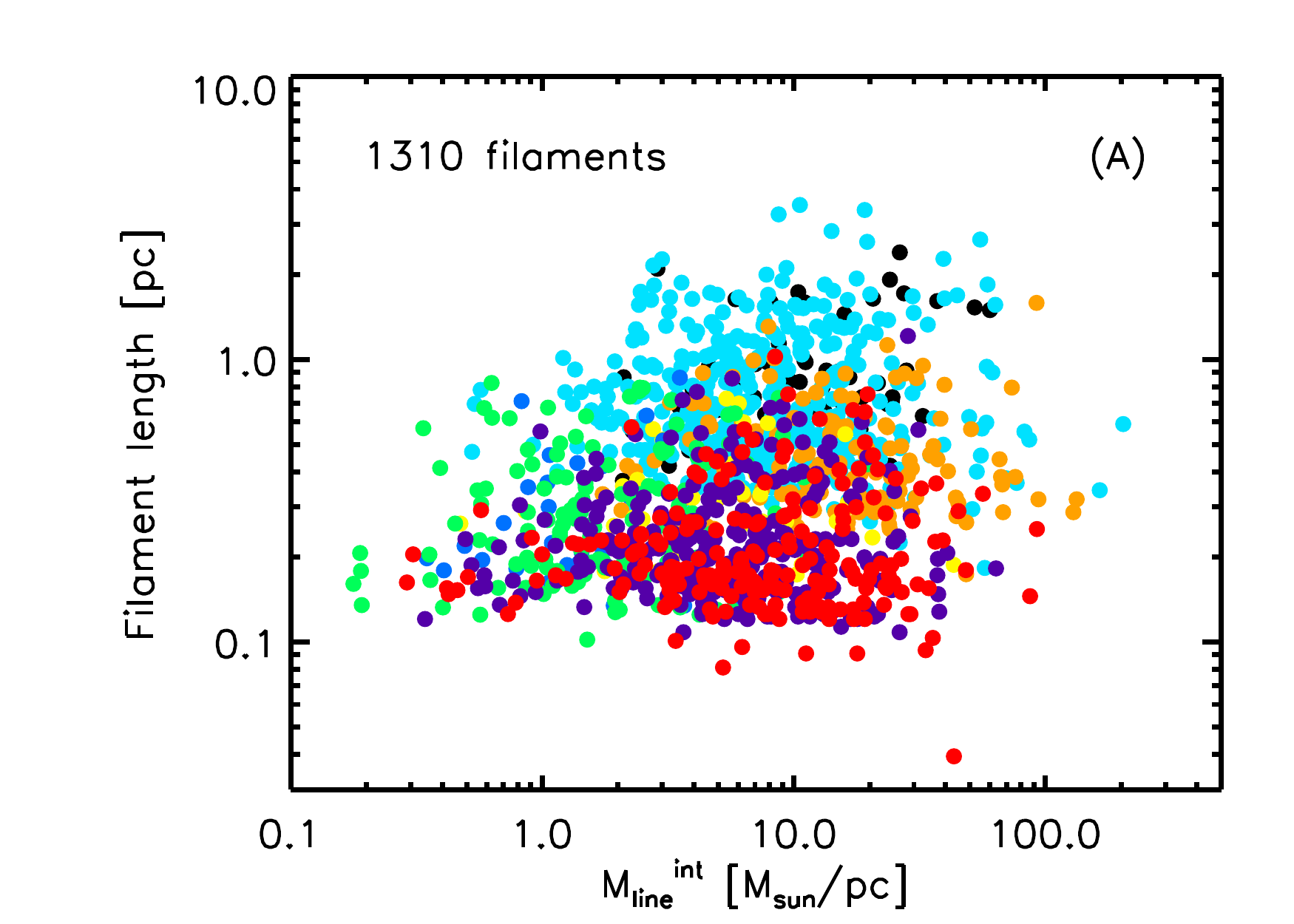}
  \includegraphics[angle=0]{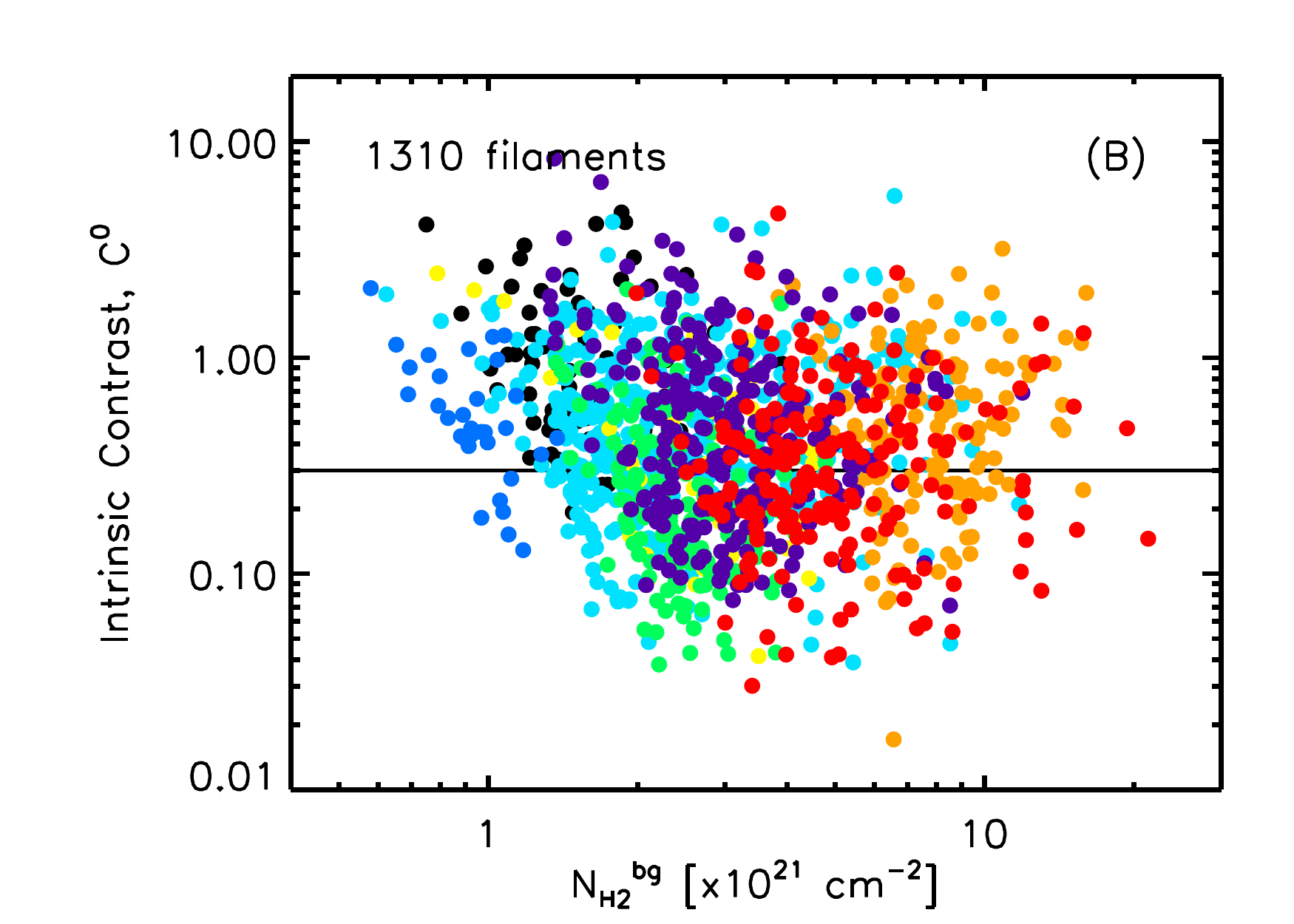}}
      \resizebox{14cm}{!}{
 \includegraphics[angle=0]{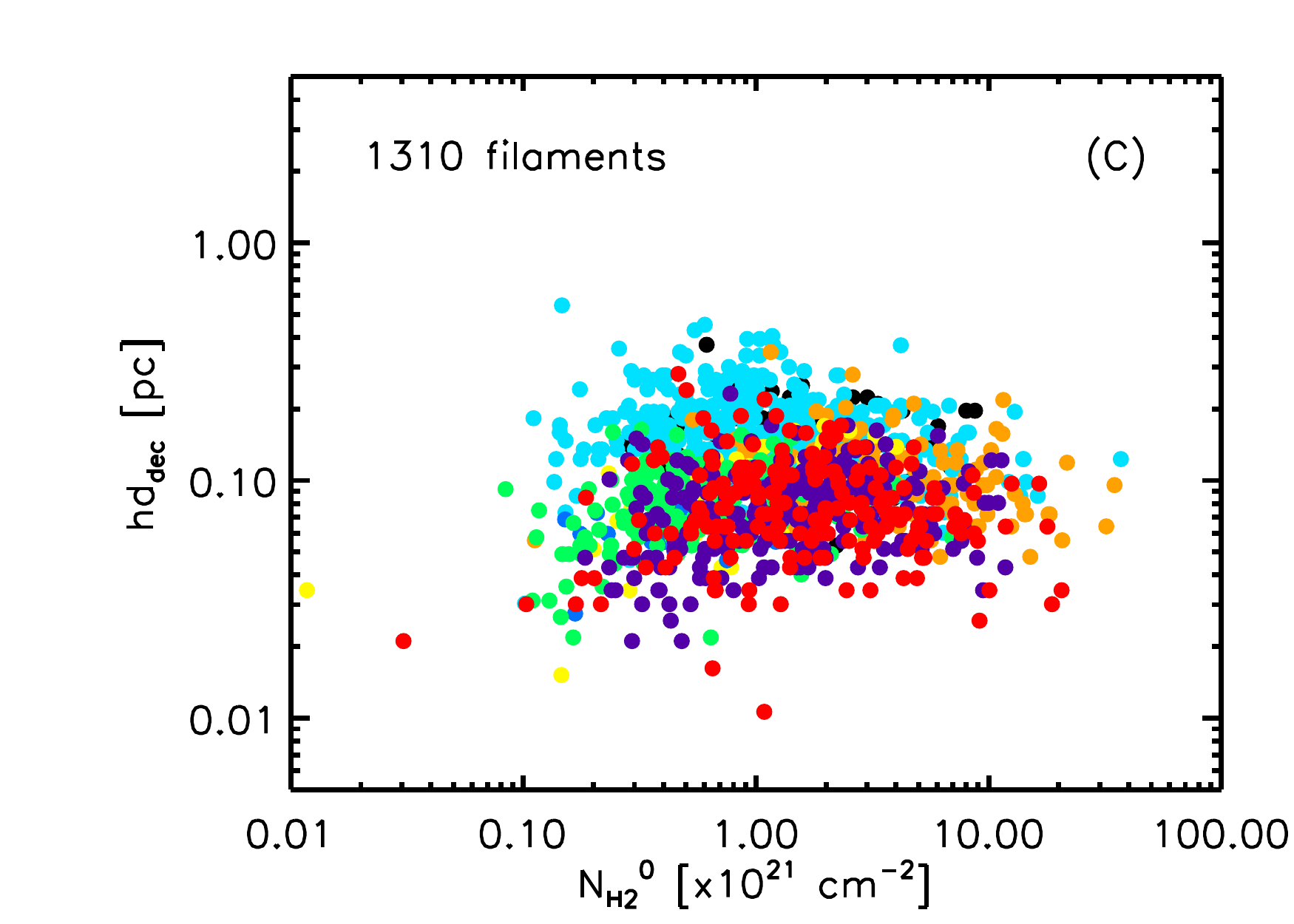}
 \includegraphics[angle=0]{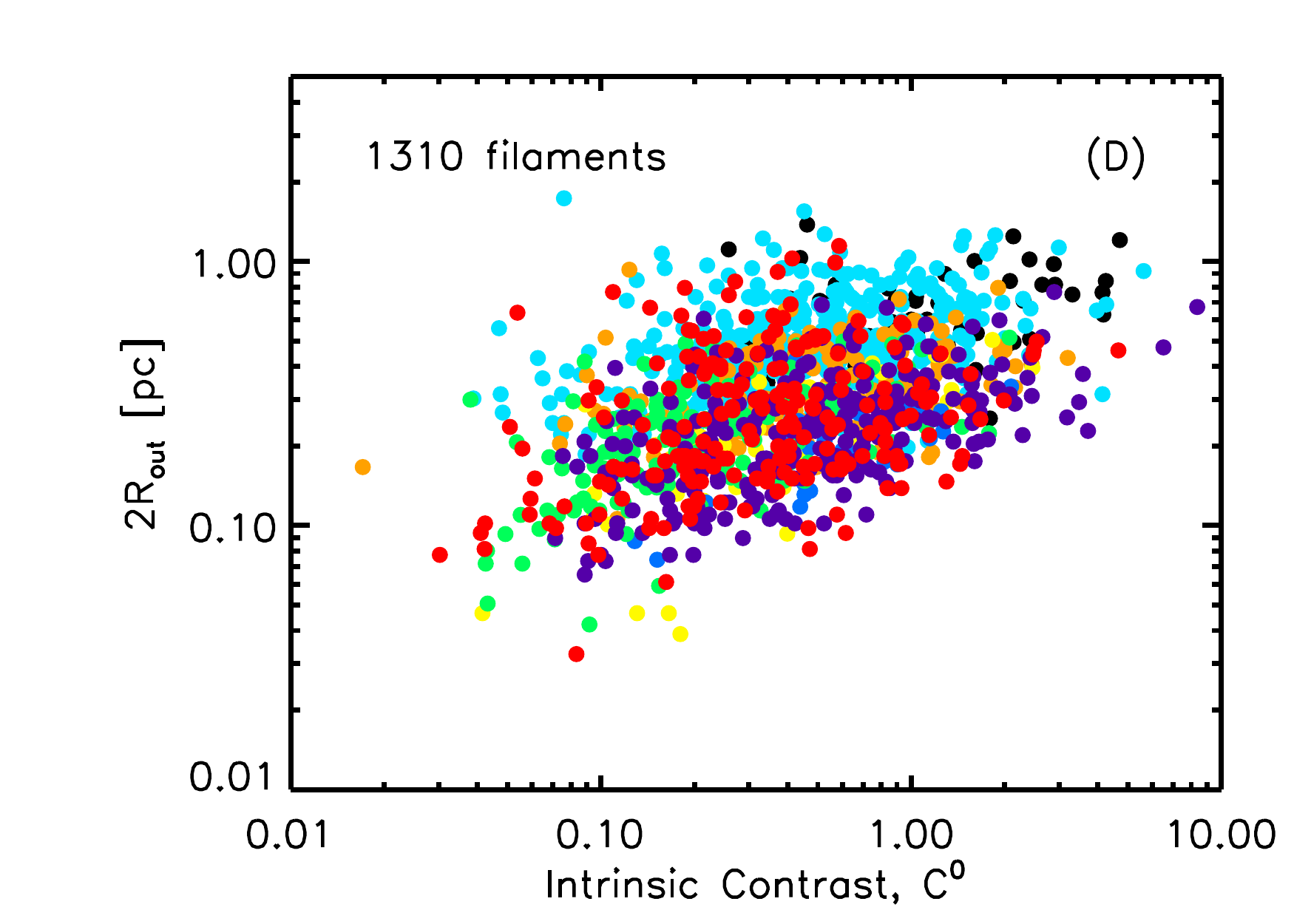}}
  \caption{Plots of estimated {\bf (A)} length $l_{\rm fil}$ against mass per unit length $\ml^{\rm int}$, 
  {\bf (B)} column density contrast $C^0$ against background column density $\nhh^{\rm bg}$, 
   {\bf (C)} deconvolved half-power diameter $hd_{\rm dec}$ against  $\nhh^0$,  {\rev  and {\bf (D)}  outer diameter (2$R_{\rm out}$) against $C^0$},
  for the entire sample of 1310 extracted filaments. 
  In {\bf (B)}, the horizontal line corresponds to $C^0=0.3$, the minimum contrast imposed as a selection criterion in Sect.\,\ref{SelecSample}. 
  In {\bf (C)}, the distribution of  measured $hd_{\rm dec}$ values has a mean of $0.11$\,pc, a standard deviation of $0.06$\,pc, 
  a median of $0.10$\,pc, and an interquartile range of $0.08$\,pc.
        }
  \label{fig:FullSample} 
\end{figure*}

\begin{figure*}[h!]
   \centering
     \resizebox{19.cm}{!}{
       \hspace{-2.cm}
\includegraphics[angle=0]{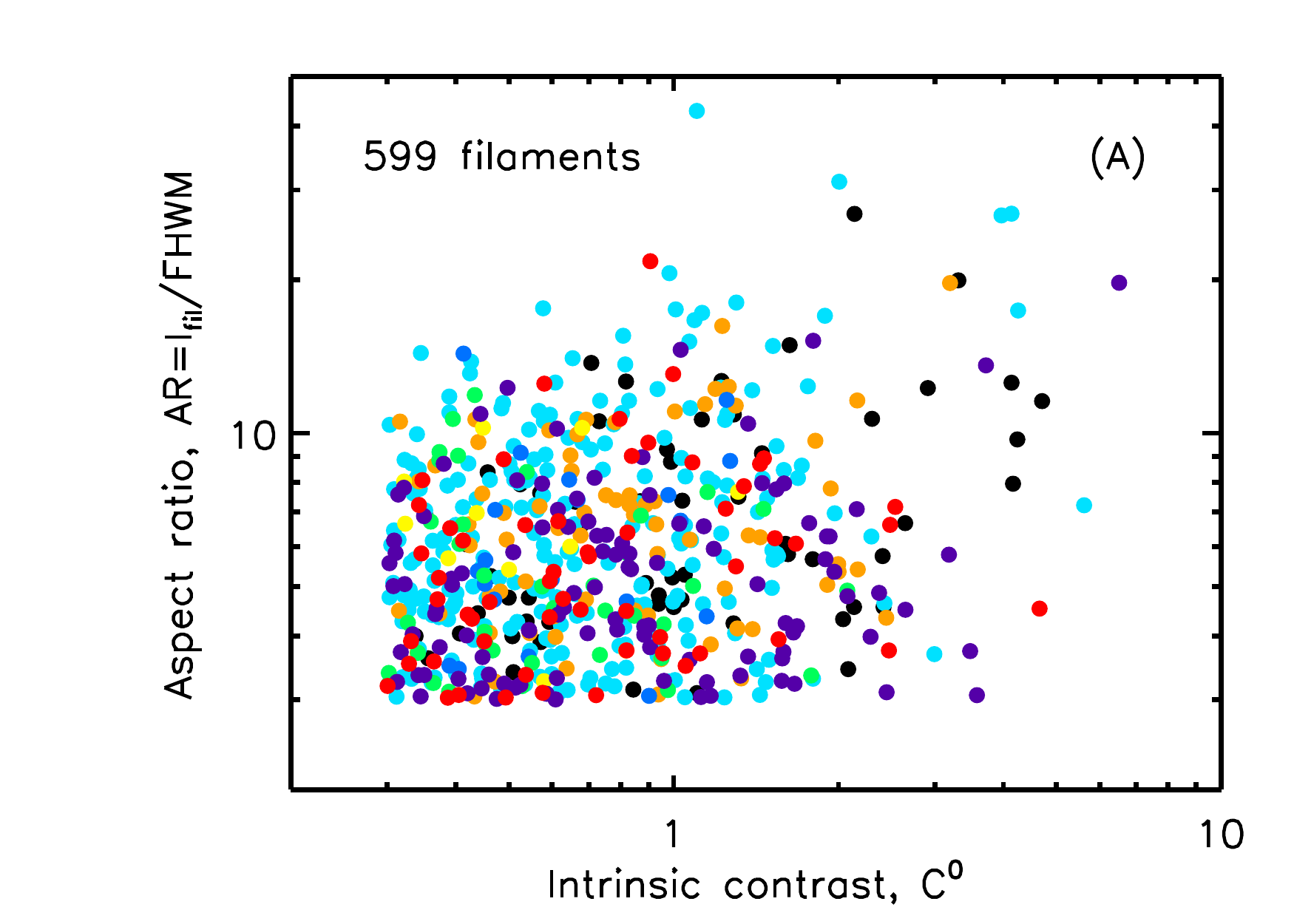}
   \hspace{-1.cm}
 \includegraphics[angle=0]{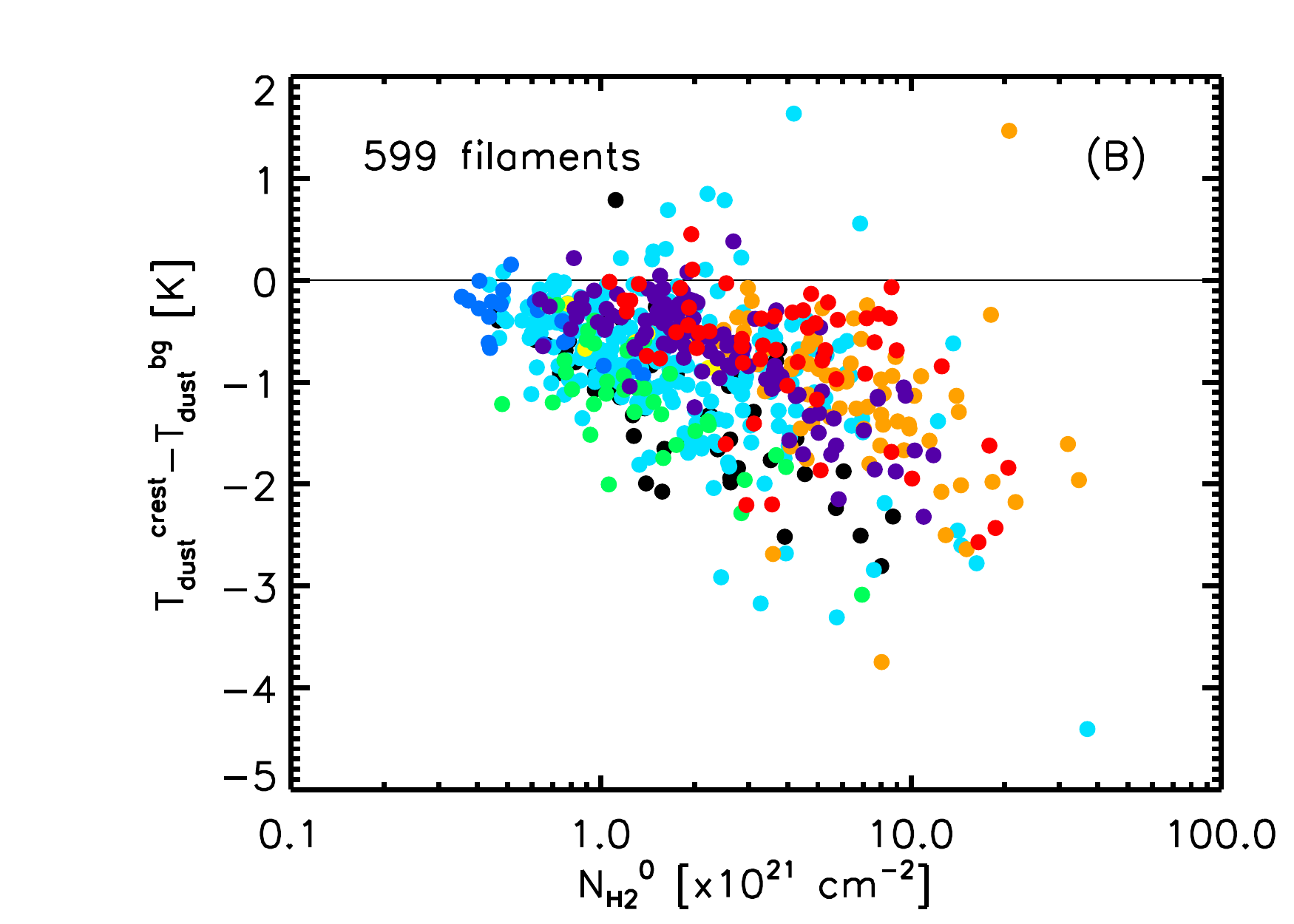}
  \hspace{-1.cm}
\includegraphics[angle=0]{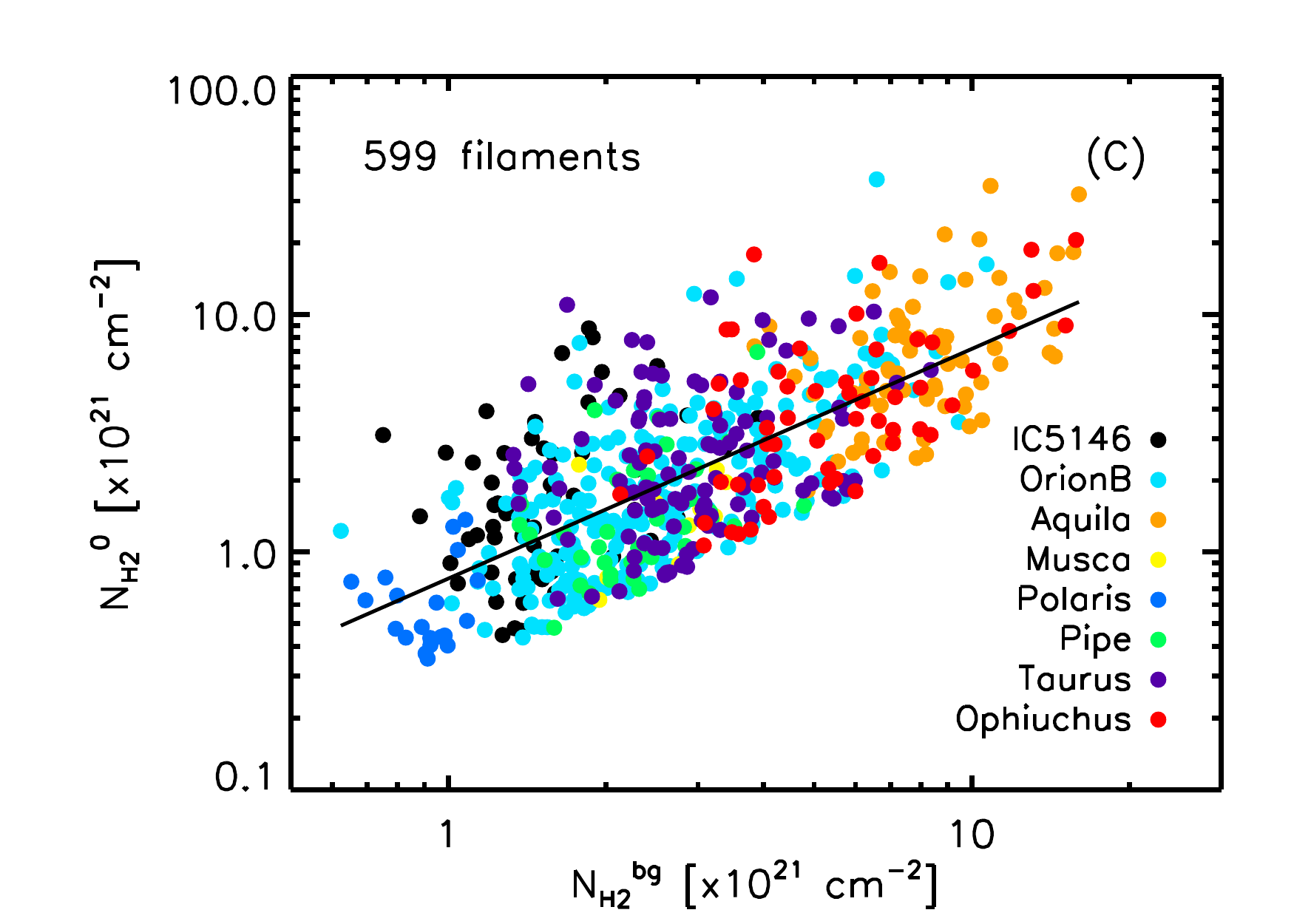}}
  \caption{{\bf (A)} Aspect ratio ($AR$) against intrinsic contrast ($C^0$) for the selected sample of {\rev 599 filaments} with $AR>3$ and $C^0>0.3$ (see Sect.\,\ref{SelecSample}). 
  {\bf (B)} Median dust temperature difference $T_{\rm dust}^{\rm crest}-T_{\rm dust}^{\rm bg}$ against $\nhh^0$, for the selected sample of {\rev 599 filaments}.
   $T_{\rm dust}^{\rm crest}$ and $T_{\rm dust}^{\rm bg}$ are the line-of-sight dust temperatures averaged along the filament crest and the background, respectively.   The horizontal line corresponds to $T_{\rm dust}^{\rm crest} = T_{\rm dust}^{\rm bg}$.
  {\bf (C)} Filament $\nhh^0$ against background column 
  density $\nhh^{\rm bg}$, for the selected sample of {\rev 599 filaments}. 
   {\rev The solid straight line shows the %
 best-fit linear relation
  $ \nhh^0 = (0.95\pm0.15)\, \nhh^{\rm bg} + (-0.15\pm0.39)\times10^{21}$\,cm$^{-2}$.}
          }
  \label{fig:SelSample} 
\end{figure*}

 \subsection{Statistical results on filament-averaged properties}\label{AveProp}

The extended and selected samples comprise 1310 and {\rev 599 filaments}, respectively {\rev (see Figs.~\ref{fig:FullSample} and \ref{fig:SelSample}).}  
Both samples span a broad  range of central column densities $\nhh^0$,  from a few $10^{20}$~cm$^{-2}$ for the faintest  filaments
up to a few $10^{22}$~cm$^{-2}$ for the densest filaments (Fig.\,\ref{fig:FullSample}).
Likewise, the extracted filaments span a wide range in mass per unit length from $\ml<1\,\sunpc$ for the most thermally 
subcritical up to $\ml\gtrsim100\,\sunpc$ for the most thermally supercritical filaments.

The line-of-sight integrated dust temperatures measured toward the filament crests  ($T_{\rm dust}^{\rm crest}$) are found to be typically $\sim$15\,K 
with a dispersion of about 3\,K. 
The dust temperature along the filament crests, $T_{\rm dust}^{\rm crest}$, is generally found 
to be colder than the temperature of the surrounding ambient cloud, $T_{\rm dust}^{\rm bg}$ (see Fig.~\ref{fig:SelSample}B). 
This behavior is consistent with the description of a molecular filament as a 3D  cylindrical structure, where the increase in column density 
corresponds to an increase in density \citep[cf.][]{Li2012,Palmeirim2013}. 
In the absence of luminous embedded protostars, 
the dust in the filament interior is more shielded from the ambient interstellar radiation field  than the dust in  the filament envelope and is therefore colder.

The ranges of background column densities $\nhh^{\rm bg}$ spanned by the filaments in the entire and selected samples can be seen in Fig.\,\ref{fig:FullSample}B and Fig.\,\ref{fig:SelSample}C, respectively.  
Even within the same molecular cloud, the background column density can vary significantly, i.e., by at least an order of magnitude in most regions. 
There is  also a significant correlation between $\nhh^0$ and $\nhh^{\rm bg}$ (Fig.\,\ref{fig:SelSample}C).
These results suggest that environmental properties, such as ambient gas pressure, may change significantly 
from one filament to the other, even within the same cloud, possibly affecting their evolution. 
In each cloud, the observed filaments span a wide range in column density contrast $C^0$ regardless of their  $\nhh^{\rm bg}$ (Fig.\,\ref{fig:FullSample}B). 
 
\begin{figure}
   \centering
 \resizebox{9cm}{!}{
\includegraphics[angle=0]{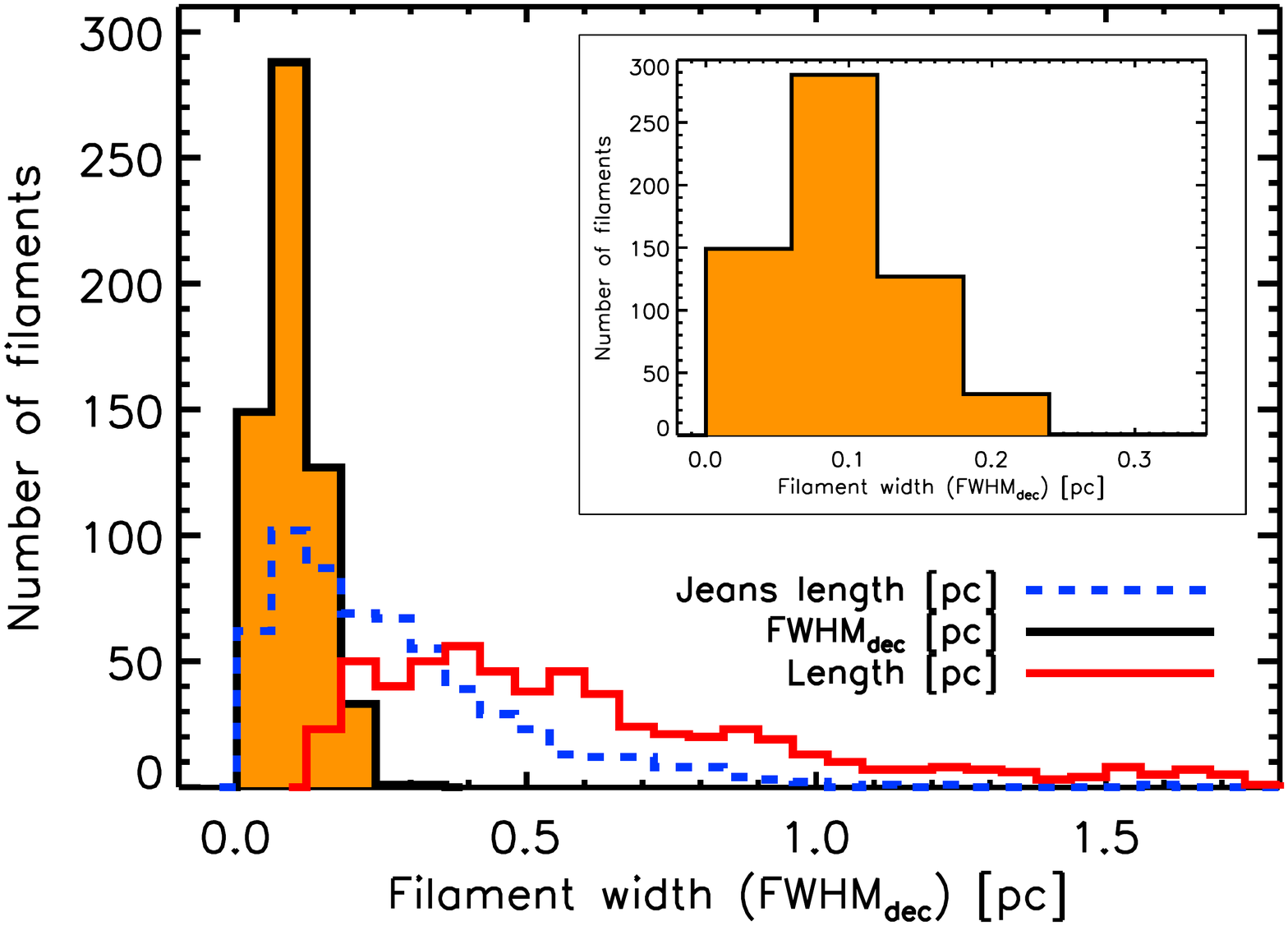}
}
      \vspace{-0.5cm}
  \caption{
Distribution of deconvolved $FWHM_{\rm dec}$ widths
  derived from Gaussian fits for the selected sample of   {\rev 599} filaments (black solid histogram filled in orange). The same histogram is  shown in the top right of the panel with a narrower x-axis range. 
 This distribution has a mean of $0.10$\,pc, a standard deviation of $0.05$\,pc, 
  a median of $0.09$\,pc, and an interquartile range of $0.07$\,pc.
For comparison, the blue dashed histogram shows the distribution of central Jeans lengths corresponding
to the central column densities of the filaments [$\lambda_{J} =  c_{s}^{2}/\left({G \Sigma_{0}}\right) $] for a gas temperature of 10\,K. 
The red histogram shows  the distribution of filament lengths.
The bin size of all three histograms is $0.06$\,pc. 
It can be seen that the distribution of filament widths is much narrower than the other two distributions. 
  }
  \label{fig:histo-width}
\end{figure}

\begin{figure}[!ht]
   \centering
  \resizebox{8.5cm}{!}{
  \includegraphics[angle=0]{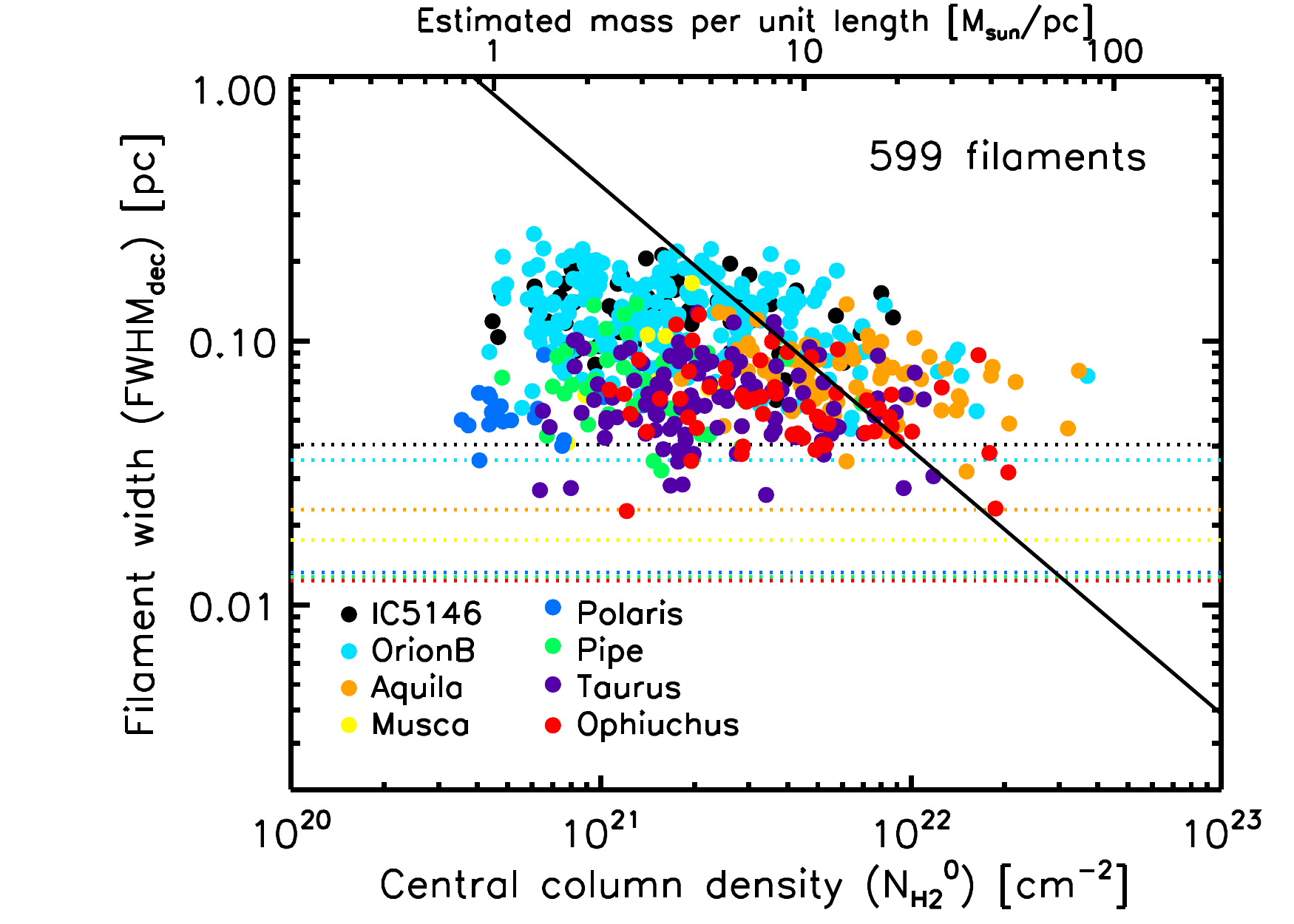}}
  \caption{Plot of deconvolved  $FWHM$ width against central column density $\nhh^0$ for the selected sample of  {\rev 599 filaments.}
The dotted horizontal lines mark the spatial resolutions of the column density maps used in the analysis for each field. 
The solid line running from top left to bottom right shows the central (thermal) Jeans length as a function of central column density 
[$\lambda_{J} =  c_{s}^{2}/\left({G \Sigma_{0}}\right) $] for $T\,=\,10$\,K. 
 {\rev The upper $x$-axis shows an approximate mass per unit length scale derive from the bottom $x$-axis scale as 
$\ml= \mu_{\rm H_2}m_{\rm H}  \nhh^0 \times W_{\rm fil}$ for  $W_{\rm fil}=0.1$\,pc. }
  }
  \label{fig:width-coldens}
\end{figure}

\begin{figure*}
   \centering
     \resizebox{18.5cm}{!}{
     \hspace{-1.cm}
     \includegraphics[angle=0]{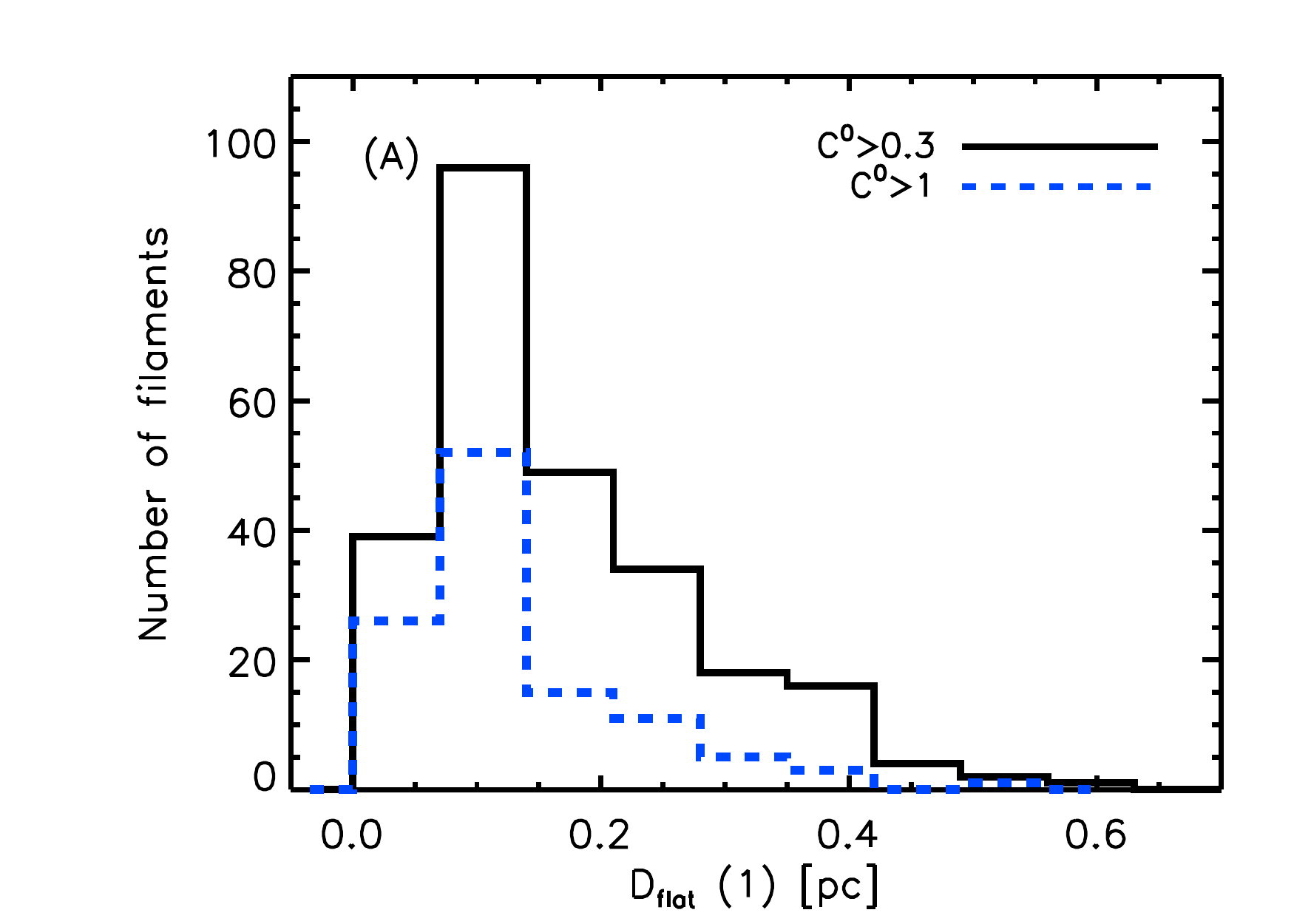}
 \hspace{-1.cm}
\includegraphics[angle=0]{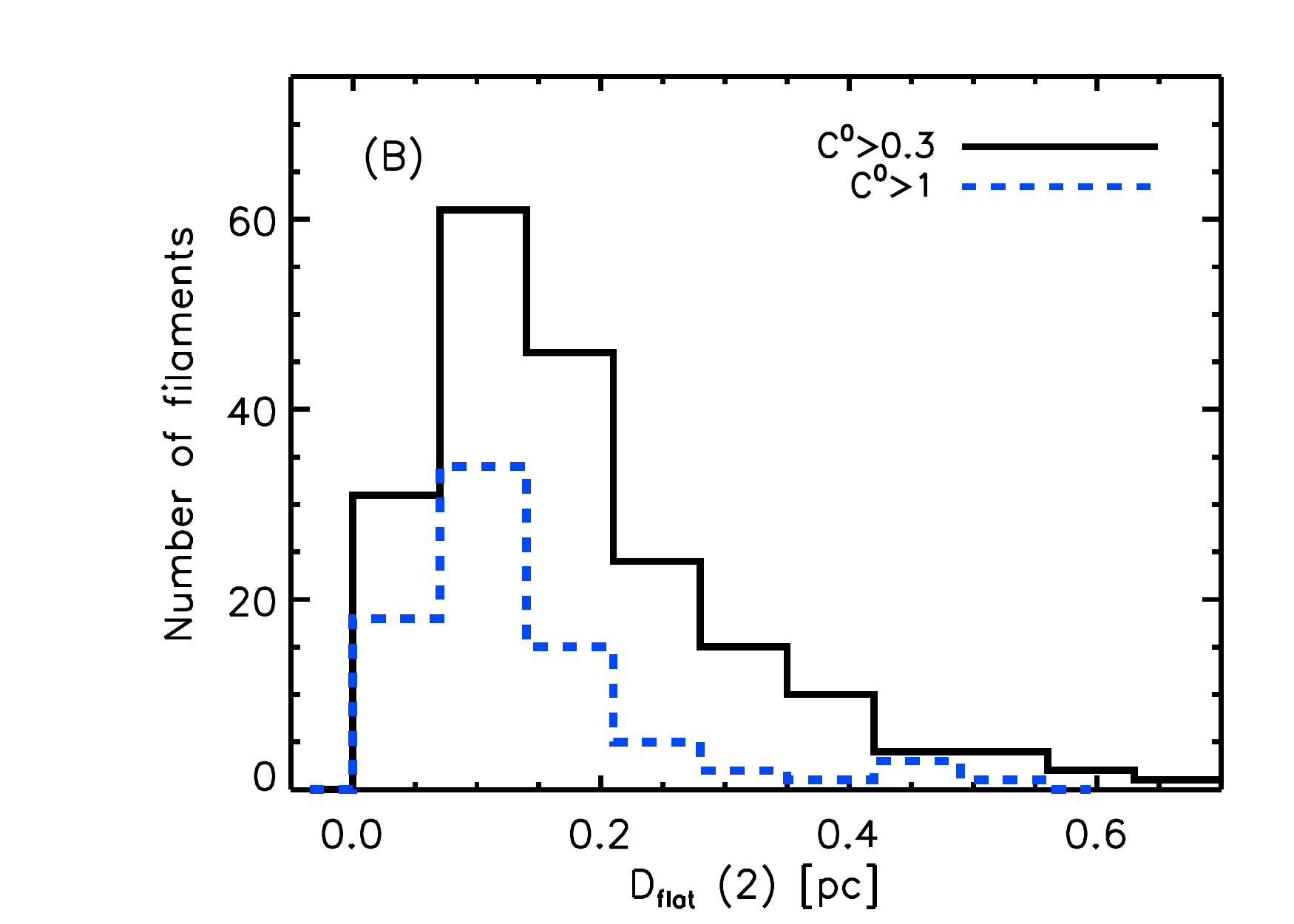}
 \hspace{-0.5cm}
\includegraphics[angle=0]{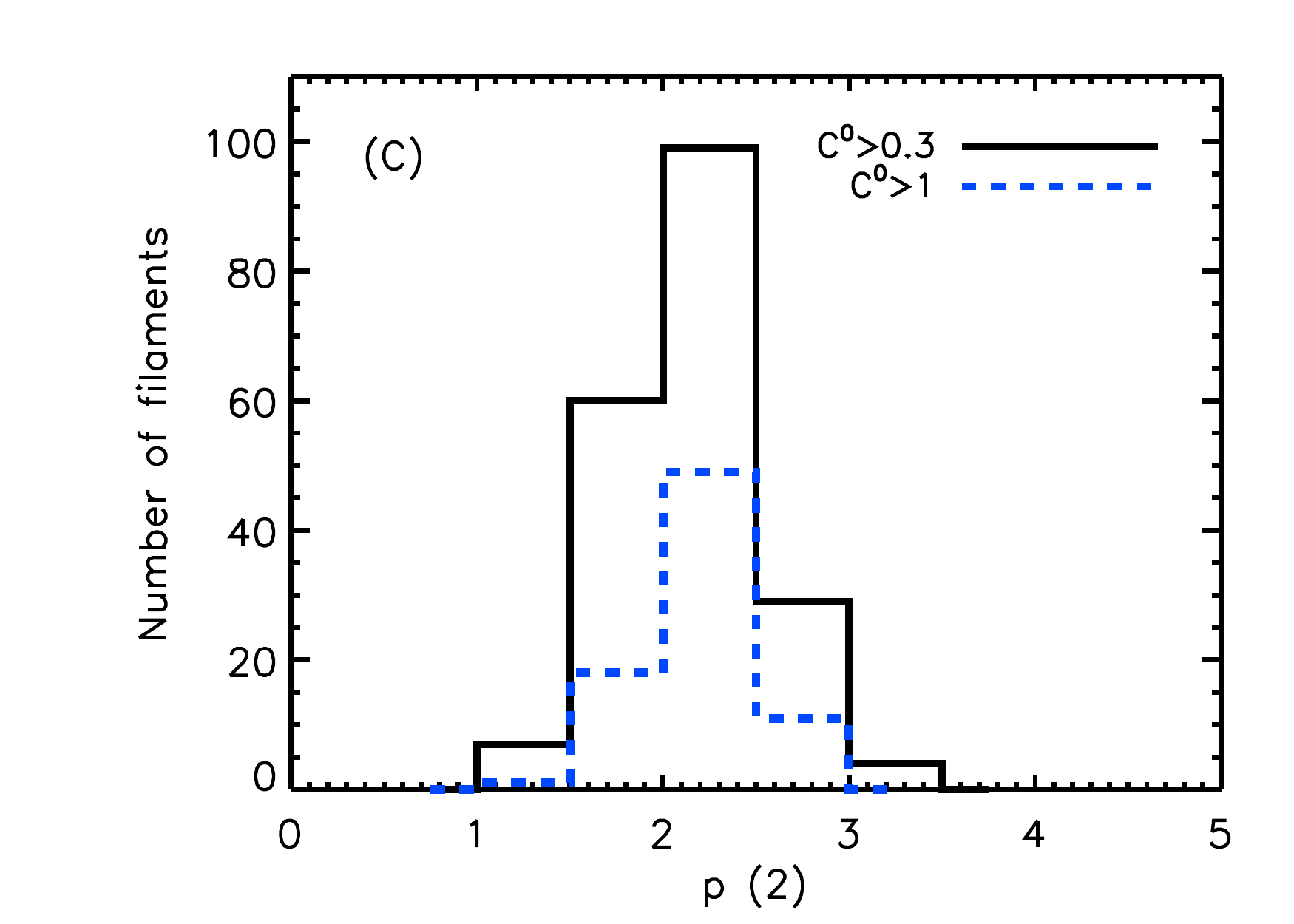}}
  \caption{ Results of Plummer fits to the median radial column density profiles of the filaments in the selected sample with contrasts $C^0>0.3$ and $C^0>1$.
{\bf (A)}  Distributions of flat inner diameters $D_{ \rm flat}$(1) obtained for a fixed value of the Plummer power-law index $p=2$
for  {\rev 260} %
filaments with $C^0>0.3$ (black solid histogram) and {\rev 113} %
filaments with $C^0>1$  (blue dashed histogram). 
The median $D_{ \rm flat}$(1) values are  {\rev  $0.13$\,pc and $0.10$\,pc, and the equivalent standard deviations (estimated from scaling the measured IQRs)  are $0.10$\,pc  and $0.08$\,pc, 
for the subsets of filaments with $C^0>0.3$ and $C^0>1$,}   respectively.
{\bf (B)} Distributions of $D_{ \rm flat}$(2) values obtained when $p$ is left as a free parameter (i.e., fitting both $R_{ \rm flat}$ and $p$)  
for  {\rev 199 %
filaments with $C^0>0.3$ and  79 %
filaments with $C^0>1$, respectively.}
The median $D_{ \rm flat}$(2) values are $0.15$\,pc and $0.12$\,pc, and the equivalent standard deviations are 
{\rev  $0.10$\,pc and $0.07$\,pc, for the subsets of filaments with $C^0>0.3$ and $C^0>1$,   respectively.
(NB: The total number of filaments for which  reliable Plummer fits can be derived is lower when $p$ and $R_{ \rm flat}$ are fitted simultaneously
than  when the fit is performed by fixing $p=2$.) 
}
{\bf (C)} Distribution of power-law index values $p$(2) corresponding to the $D_{ \rm flat}$(2) values in {\bf (B)}.
The median $p$(2) value is 2.2  for both  subsets of filaments (with $C^0>0.3$ and $C^0>1$),   
and the equivalent standard deviations of the distributions are {\rev 0.4 and 0.3}, respectively.
 }
  \label{fig:PlumResults} 
\end{figure*}

{\rev Figure\,\ref{fig:FullSample}D shows a plot of  filament outer diameter  $2\rout$ as a function 
of filament intrinsic column density contrast $C^0$. 
High-contrast filaments tend to have larger outer diameters owing to more developed power-law wings. 
We see similar correlations between $2\rout$ and  $\ml$ or $\nhh^0$ (not shown here).
However,} there is no correlation between the length $l_{\rm fil}$ or the inner width $W_{\rm fil}$
and either the central column density $\nhh^0$ or the mass per unit length $\ml$. 
While the filament lengths span a wide range from  $\gtrsim0.1\,$pc up to a few parsecs, the deconvolved half-power widths  $W_{\rm fil}=hd_{\rm dec}$ 
have a low dispersion around a median value of $0.1$\,pc (see Fig.~\ref{fig:FullSample}C). 
Accordingly, 
the aspect ratios of the filaments in the selected sample %
span a wide range from a minimum of 3 (defined by one of our selection criteria) up to $\sim 30$ for the longest filaments (see Fig.~\ref{fig:SelSample}A). 
{\rev About  59$\%$ and 14$\%$  of the filaments have $AR>5$ and   $AR>10$, respectively.}  
While the filaments detected in the HGBS  images are typically parsec-scale 
structures, {\it Herschel} observations of more massive star-forming regions at  $\sim$kpc distances 
have revealed 
longer filaments in the Galactic Plane  \citep[e.g.,][]{Molinari2010,Schisano2014}.

The deconvolved $FWHM$ widths of the {\rev 599} filaments in our selected sample, as measured from Gaussian fits to their radial column density profiles, 
have a narrow distribution centered around a median value of 0.09\,pc, with an equivalent standard deviation of 0.05\,pc 
(scaled from a measured IQR of 0.07\,pc --
see Fig.\,\ref{fig:histo-width} and Table\,\ref{tab:table_stat}). 
{\rev The  deconvolved $hd$ values also have a narrow distribution about a median value of 0.11\,pc and an equivalent standard deviation of of 0.05\,pc (Table\,\ref{tab:table_stat}).}
The same filaments span more than two orders of magnitude in central column density (cf. Fig.\,\ref{fig:width-coldens}),  
implying a spread of two orders of magnitude in the distribution of central Jeans lengths [$ \lambda_{\rm J}\left({r=0}\right) =  c_{\rm s}^{2}/\left({G \Sigma_{r=0}}\right) $], 
which is much broader than the observed spread in the distribution of filament widths.
 The horizontal dotted lines in  Fig.\,\ref{fig:width-coldens} show the spatial resolution limits of the column density maps used to construct the filament profiles. 
The measured filament widths are always significantly above the corresponding resolution limit. 
Moreover, we stress that only the {\it physical} widths of the filaments  (in pc) show a narrow distribution  (cf. Fig.\,\ref{fig:histo-width}), 
while the {\it angular} widths (in arcsec) vary as a function of the parent cloud distances \citep[see Table\,\ref{tab:table_stat} and also Table\,2 in][]{Arzoumanian2011}.
This strongly suggests that the $\sim0.1\,$pc inner width is an intrinsic physical property of the observed filaments 
and is not  affected by the finite resolution of the $Herschel$ data. 
Our earlier result on the existence of a characteristic filament width (for 90 filaments) in three molecular clouds, IC5146, Aquila and Polaris  \citep{Arzoumanian2011}, 
has therefore been generalized to a much larger filament sample and five additional  clouds. 

\begin{figure*}
   \centering
     \resizebox{18.5cm}{!}{
     \hspace{-3.cm}
     \includegraphics[angle=0]{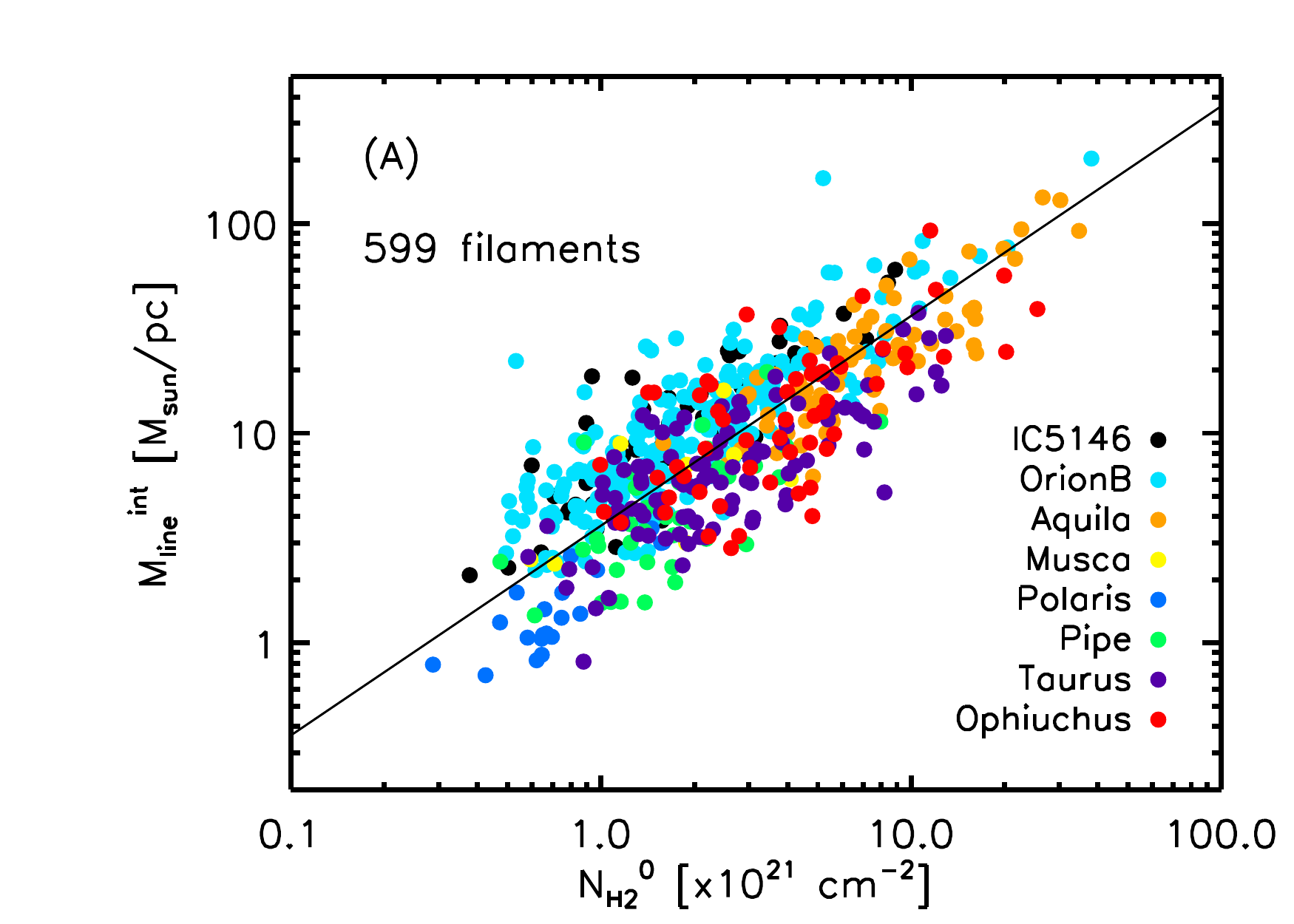}
 \hspace{-1.cm}
\includegraphics[angle=0]{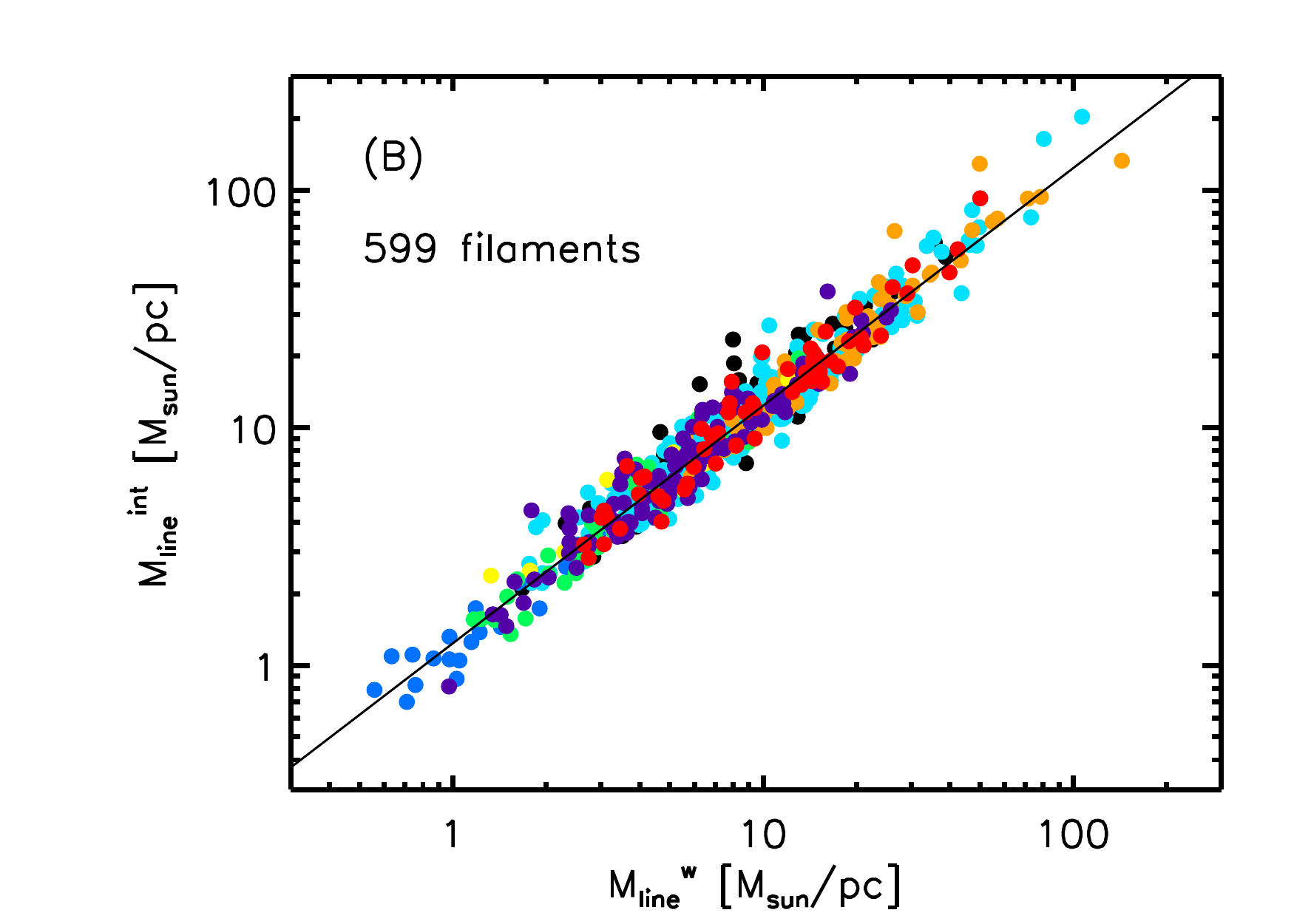}
 \hspace{-0.5cm}
\includegraphics[angle=0]{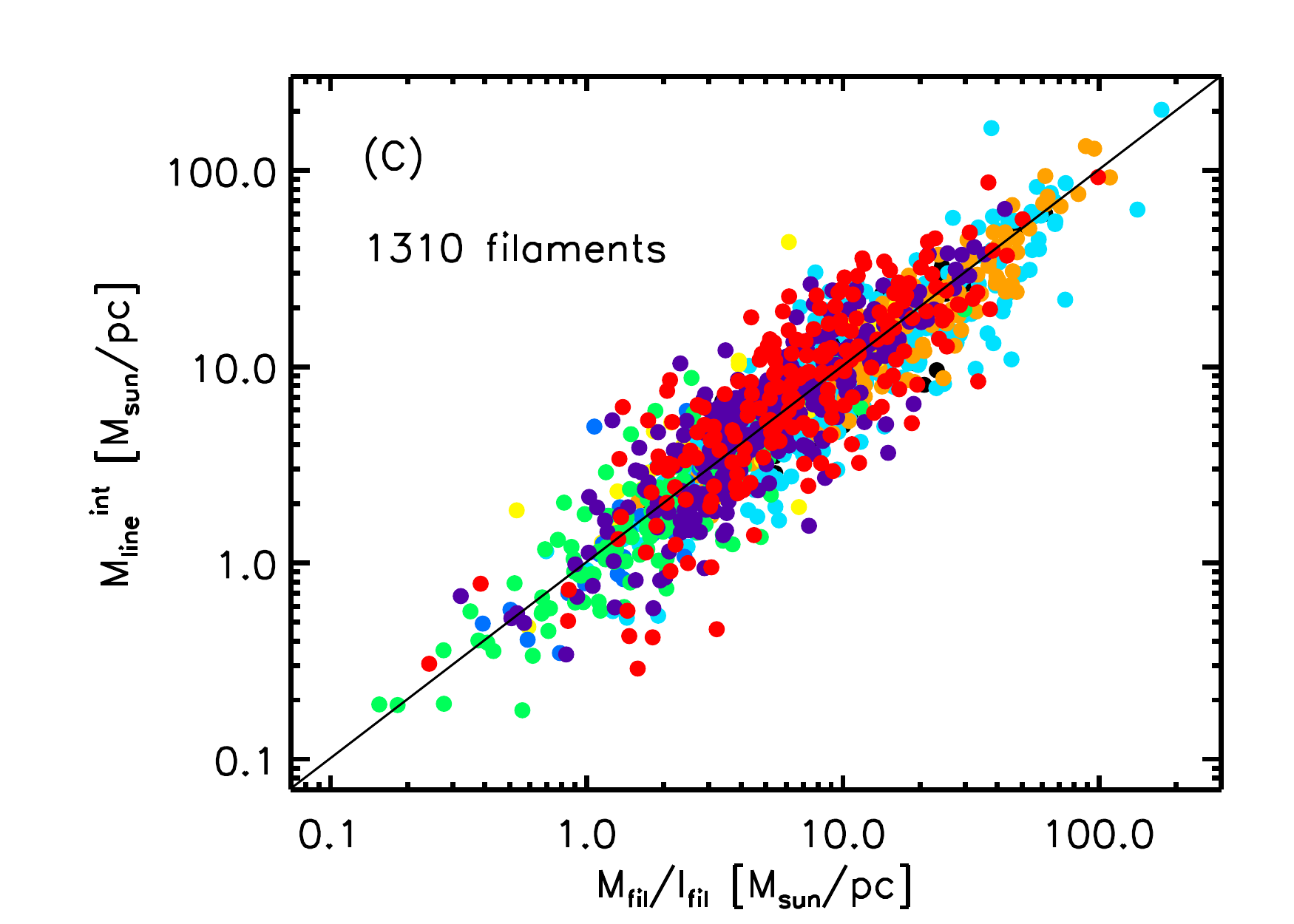}}
  \caption{ {\bf (A)} Mass per unit length integrated over the  radial column density profiles, $\ml^{\rm int}$, against median central column density, $\nhh^0$, for the {\rev 599} filaments  of the selected sample. 
  The solid straight line shows the one-parameter best-fit linear relation
  $\ml^{\rm int}\, (M_\odot /{\rm pc}) = (0.16\pm0.05\, {\rm pc}) \times \Sigma_{\rm fil}^0 \, (M_\odot /{\rm pc}^2) $, where 
  $\Sigma_{\rm fil}^0 = \mu_{\rm H_2}m_{\rm H}  \nhh^0 $ is the central gas surface density. 
  {\bf (B)}  Plot of $\ml^{\rm int}$ against 
  $\ml^{\rm w}\sim 2hr\nhh^0$ for the same sample of filaments. The one-parameter  best-fit linear relation 
  is  $\ml^{\rm int}=(1.34\pm0.03)\,\ml^{\rm w}$.     
    {\bf (C)}  Mass per unit length estimated as $M_{\rm fil }/l_{\rm fil }$ (where $M_{\rm fil }$ is calculated summing the column density over the pixels within the area between the filament crest and  \rout, after subtraction of  $\nhh^{\rm bg}$, see text  of Sect.\,\ref{MlineFilAnalysis}) 
    against $\ml^{\rm int}$ for the 1310 filaments of the ``total'' sample.
    The solid straight line shows the one-parameter  best-fit linear relation $M_{\rm fil }/l_{\rm fil }\sim(0.96\pm0.07)\,\ml^{\rm int}$.
   }
  \label{fig:Mline} 
\end{figure*}

{\rev Our estimates of the physical filament width are somewhat sensitive to uncertainties in cloud distances \citep[see, e.g., discussion in ][]{Arzoumanian2011}. %
The median and equivalent standard deviation values quoted above for the distribution of $FWHM_{\rm dec}$ widths  ($0.09\pm0.05$\,pc)
were obtained  for  the default distances adopted for each region as listed in Table\,\ref{Table1}. 
Assuming alternate distances for the studied clouds, namely 950\,pc for IC5146  \citep{Harvey2008}, 500\,pc for Orion B \citep{Schlafly2015}, 400\,pc 
for Aquila \citep{Bontemps2010,Ortiz-Leon2017}, %
400\,pc for Polaris \citep{Schlafly2014}, and the same distances as the  default values  for Musca (200\,pc), 
Taurus, and Ophiuchus (140\,pc) would lead to a median $FWHM_{\rm dec}$ value and equivalent standard deviation of $0.15\pm0.07$\,pc. 
While the distribution of filament inner widths is somewhat broader with the alternate cloud distances,
the median filament width is only 50\% larger than, and remains consistent with, that derived 
assuming the default distances of  Table\,\ref{Table1}.  
}

Figure\,\ref{fig:PlumResults} presents the distributions of flat inner diameters  $D_{\rm flat} = 2\,R_{\rm flat}$  
and power-law indices $p$ derived from Plummer fits to the median column density profiles of the filaments with contrasts {\rev $C^0>0.3$} (solid line histogram)
and $C^0>1$ (dashed line histogram). 
These plots show that the distribution of $D_{\rm flat}$ diameters
peaks at a median value of about $0.10$\,pc, with an equivalent standard deviation of $0.08$\,pc
(for $C^0>1$, see Table\,\ref{tab:PlummerFit}), 
which is fully consistent with  the $FWHM$ widths derived from Gaussian fits to the inner part of the observed radial profiles 
(see, e.g., Fig.\,\ref{fig:histo-width}). 
{\rev The distributions of median $D_{\rm flat}$ widths exhibit, however, larger dispersions than
the distribution of median  $FWHM_{\rm dec}$ widths,
 especially for filaments with $C^0<1$. 
This larger dispersion may result from 1) larger measurement uncertainties using the Plummer-like function fitting method 
and 2) less well defined power-law profiles and outer radii for low-contrast filaments
(see tests presented in Appendix\,\ref{App2b}).    
}
The radial column density profiles of the selected filaments, especially those with column density contrasts $C^0>1$, 
tend to show a power-law behavior at $r>>hr$ with an exponent $p \approx 2$   \citep[see Figs.\,\ref{fig:Multiprof} and \ref{fig:PlumResults},  Table\,\ref{tab:PlummerFit}, and also][and others]{Arzoumanian2011,Hill2012,Palmeirim2013,Andre2016}.

The presence of  power-law wings with $p\sim2$ in the radial column density profiles of many molecular filaments implies 
that  mass per unit length estimates 
may depend on the outer boundary of the filaments. 
{\rev Interestingly, the radius of this outer boundary, \rout, is observed to increase with the column density contrast (Fig.\,\ref{fig:FullSample}D), 
central column density, or \ml\ of the filaments. 
Such a behavior is indeed expected for a Plummer-like column density profile with fixed $p$ and  \rflat, and increasing central column density.} 
For a Gaussian-like radial \nhh\ profile, 
the mass per unit length is $\ml^{\rm w}\sim  \mu_{\rm H_2}m_{\rm H}  \nhh^0 \times W_{\rm fil}$. 
For filaments with significant power-law wings at $r>>W_{\rm fil}/2$, 
however, 
a non-negligible fraction of the total mass may lie  within the non-Gaussian wings \citep[see also][]{Rivera-Ingraham2016,Rivera-Ingraham2017}.   
This additional mass 
is taken into account when deriving  $\ml^{\rm int}$ by integrating the \nhh\ profile over radii up to  \rout\ (see Sect.\,\ref{MlineFilAnalysis}). 

{\rev Figure\,\ref{fig:Mline} shows correlations between  the mass per unit length (derived in three different ways) and the central column density 
of the observed filaments.}
The correlation between the two estimates of the mass per unit length  shows that   $\ml^{\rm int}\sim1.3\ml^{\rm w}$ on average (cf Fig.\,\ref{fig:Mline}B), 
suggesting that most filaments have non-Gaussian \nhh\ profiles, which typically contribute $30\%$ of their total mass on average.
Figure\,\ref{fig:Mline}A shows a very good correlation between $\ml^{\rm int}$ and $\nhh^0$. 
{\rev The correlation between these two (partly independent) quantities is consistent with  
the presence 
of a common width shared by all  filaments of the selected sample. 
}
The  $\ml^{\rm int}$ estimate is also well correlated with another estimate of the mass per unit length derived 
as  $M_{\rm fil}/l_{\rm fil}$ (see Sect.\,\ref{MlineFilAnalysis}). 
In addition, a good correlation is observed 
between $\ml^{\rm int}$ and background column density $\nhh^{\rm bg}$ (cf., Sect.\,\ref{Discussion}).

\begin{figure}
   \centering
 \resizebox{8.5cm}{!}{
\includegraphics[angle=0]{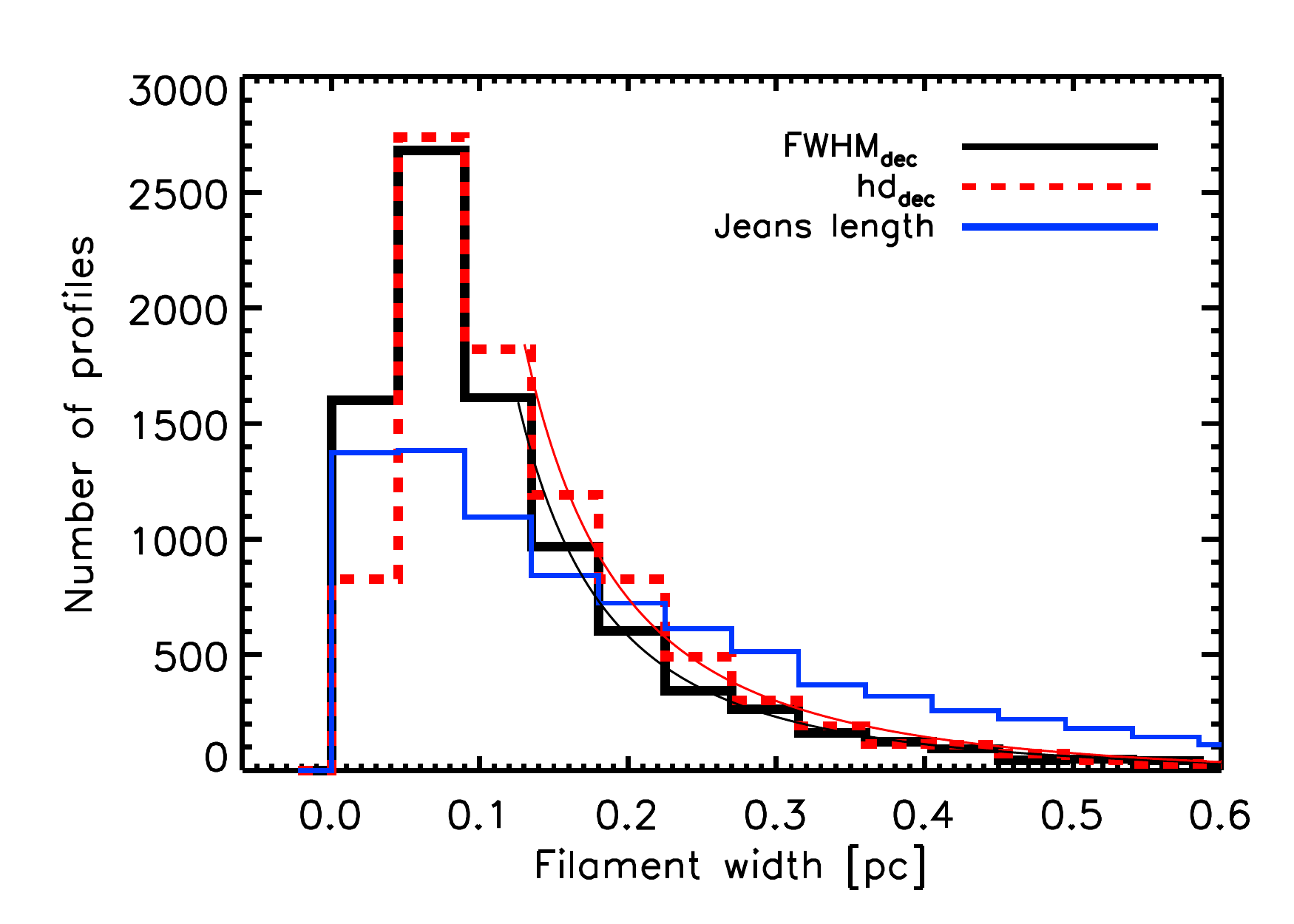}
}
  \caption{
     {\rev  Distributions of $FWHM_{\rm dec}^{pix,\pm}$ (black) and $hd_{\rm dec}^{pix,\pm}$ (red) widths derived from  individual profiles along  and on either side of the  crests for  the selected sample of 599 filaments.    
 The median $FWHM_{\rm dec}^{pix,\pm}$  and $hd_{\rm dec}^{pix,\pm}$  
  values are $0.09$\,pc and $0.11$\,pc, respectively,  and  the  interquartile range is $0.11$\,pc for both distributions. 
  The  black and red curves 
  show power-law fits to the tails of the $FWHM_{\rm dec}^{pix,\pm}$ and $hd_{\rm dec}^{pix,\pm}$ distributions, respectively ($\Delta N/\Delta W\propto W^{-2.6\pm0.5}$, for both distributions).
     {\Newrev For comparison, the distribution of Jeans lengths,  $\lambda_{J}^{pix,\pm}$, is also shown (blue histogram, cf. Fig.\,\ref{fig:histo-width})}.  
  }}
  \label{fig:histo-width-alongfil}
\end{figure}

\subsection{Properties of filaments along and on either side of their crests}\label{AlongFilProp}

{\rev In this section, we discuss the statistical distributions of filament inner widths and background column densities derived from individual column density profiles, 
taken along and on either side of the filament crests. 
In practice, the filament properties discussed here were measured on spatially-independent radial profiles averaged 
over $2\times HPBW$-long segments  along the filament crests (see Sect.\,\ref{RadProf}). 
Figure\,\ref{fig:histo-width-alongfil} shows 
the distributions of individual $FWHM_{\rm dec}^{pix,\pm}$  and $hd_{\rm dec}^{pix,\pm}$ widths measured along and on either side of the crests  for the selected sample of 599 filaments
(see Sect.\,\ref{RadProf}). 
{\Newrev For comparison, the distribution of individual  Jeans lengths $\lambda_{J}^{pix,\pm}$  corresponding
to the  background subtracted central column densities along and on either side of the crests is also shown.
The  $FWHM_{\rm dec}^{pix,\pm}$  and $hd_{\rm dec}^{pix,\pm}$ distributions} are based on 8682 and 8827 independent measurements, 
respectively (for $\sim2\%$ of the individual profiles, reliable Gaussian fits could not be derived). 
They have median values of $0.09$\,pc and $0.11$\,pc, respectively, an interquartile range of $0.11$\,pc in both cases, 
mean values of  $0.13$\,pc and $0.14$\,pc,  and standard deviations of $0.12$\,pc and $0.11$\,pc, respectively. 
These %
distributions, which measure the statistical importance of possible variations of the inner width along and on either side of the filament crests, 
peak essentially at the same 
values as the distributions of median $FWHM_{\rm dec}$ and $hd_{\rm dec}$ widths (e.g. Fig.~\ref{fig:histo-width}).
The distributions of individual widths (Fig.\,\ref{fig:histo-width-alongfil}) exhibit power-law-like tails of 
values significantly larger than $\sim0.1$\,pc \citep[see also][]{Panopoulou2017}. %
Since the distribution of median widths along the filaments  (Fig.\,\ref{fig:histo-width}) has a much narrower dispersion around its peak value, 
it appears that the local inner width of a filament significantly exceeds $\sim0.1$\,pc at most at a few positions along the filament crest.
Such large local excursions of the measured inner width significantly beyond $\sim0.1$\,pc are averaged out when computing the median filament width.
They may be due to several reasons. For example, 
individual profiles along the crest of a filament may be contaminated by 1) the  presence of prestellar cores on or slightly off the crest \citep[e.g.,][]{Malinen2012}, 
2) the presence of fiber-like substructures \citep[e.g,.][]{Hacar2013,Hacar2018}, 
3) the intersection points between distinct but overlapping filaments, 
4) bad measurements due to disturbed individual profiles, 
5) excursions of the \disperse\,-traced crest connecting two neighboring filaments. 
Such excursions about the median width, 
  are also seen, {\Newrev albeit less prominently than in Fig.\,\ref{fig:histo-width-alongfil},} in the distributions of measured individual widths for the tests described in Appendix\,\ref{App2}, even when the  input width is constant along the filament crest (see, e.g., Fig.\,\ref{histo_GaussW_mock}).  
Further analyses and tests would be required to estimate the relative contribution of the various possible factors 
to the shape of the observed distributions of individual $FWHM_{\rm dec}^{pix,\pm}$  and $hd_{\rm dec}^{pix,\pm}$ widths (Fig.\,\ref{fig:histo-width-alongfil}). 
For a given filament in the selected sample, the median absolute deviation of the individual inner widths measured along and on either side of the filament crest  
ranges between  $\sim0.02$\,pc and $\sim0.06$\,pc, which is similar to the dispersions of the distributions of median filament widths 
measured in each region (cf.Table\,\ref{tab:table_stat}). 
} 

{\rev
To investigate potential asymmetries in the column density profiles of the filaments, 
we plot in Fig.\,\ref{fig:WidthbgDiff}  the distributions of  differences in $FWHM$ width ($FWHM_{\rm dec}^--FWHM_{\rm dec}^+$) and  background column density ($\nhh^{ \rm bg-} -\nhh^{ \rm bg+}$) 
between the two sides of each filament crest. 
Independent estimates of $FWHM$ and $\nhh^{\rm bg}$ on either side of the crest were not possible for 71 filaments. 
The strongly peaked shape of the distributions in Fig.\,\ref{fig:WidthbgDiff} indicates that a large majority of the filaments in our selected sample exhibit fairly symmetric profiles. 
The presence of clear wings in the distributions of Fig.\,\ref{fig:WidthbgDiff} nevertheless implies that 
significant asymmetries in width and $\nhh^{ \rm bg}$ exist for some filaments between the two sides of their crests. 
Interestingly, there is no obvious correlation between an asymmetry detected  
in filament width and an asymmetry in background column density $\nhh^{ \rm bg}$. 
In other words, a difference in $\nhh^{ \rm bg}$ between the $+$ and $-$ sides of a filament crest 
does not necessarily imply a difference in  $FWHM$ width, and vice versa.
A dedicated analysis of filaments with asymmetric profiles would be useful  
as it may provide valuable constraints on the formation and evolution of filamentary structures in general, 
as well as their interaction with the parent molecular clouds 
\citep[see, e.g.,][]{Peretto2012}.
}

\begin{figure}[]
   \centering
     \resizebox{8cm}{!}{
     \hspace{-0.5cm}
\includegraphics[angle=0]{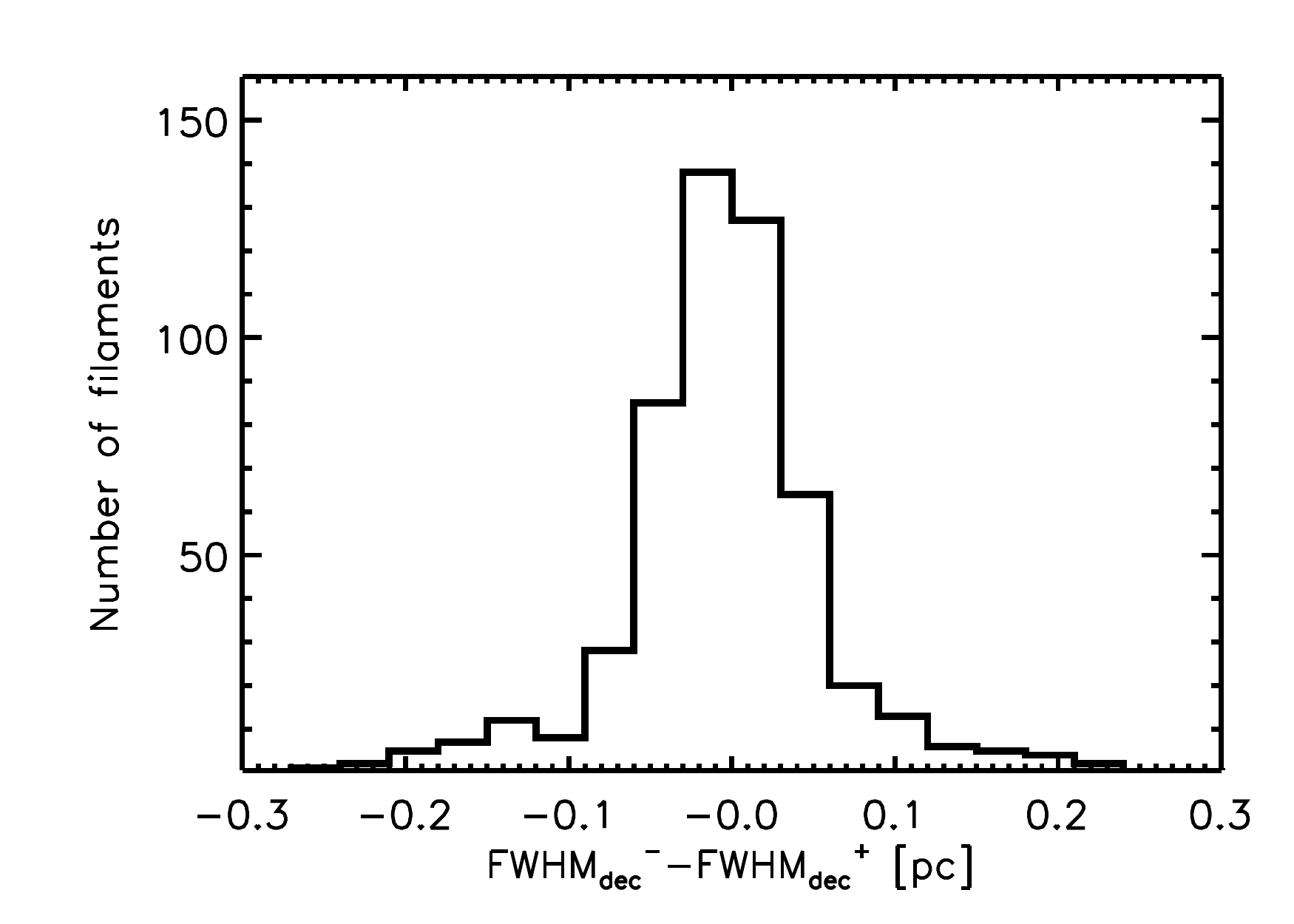}}
         \resizebox{8cm}{!}{
\includegraphics[angle=0]{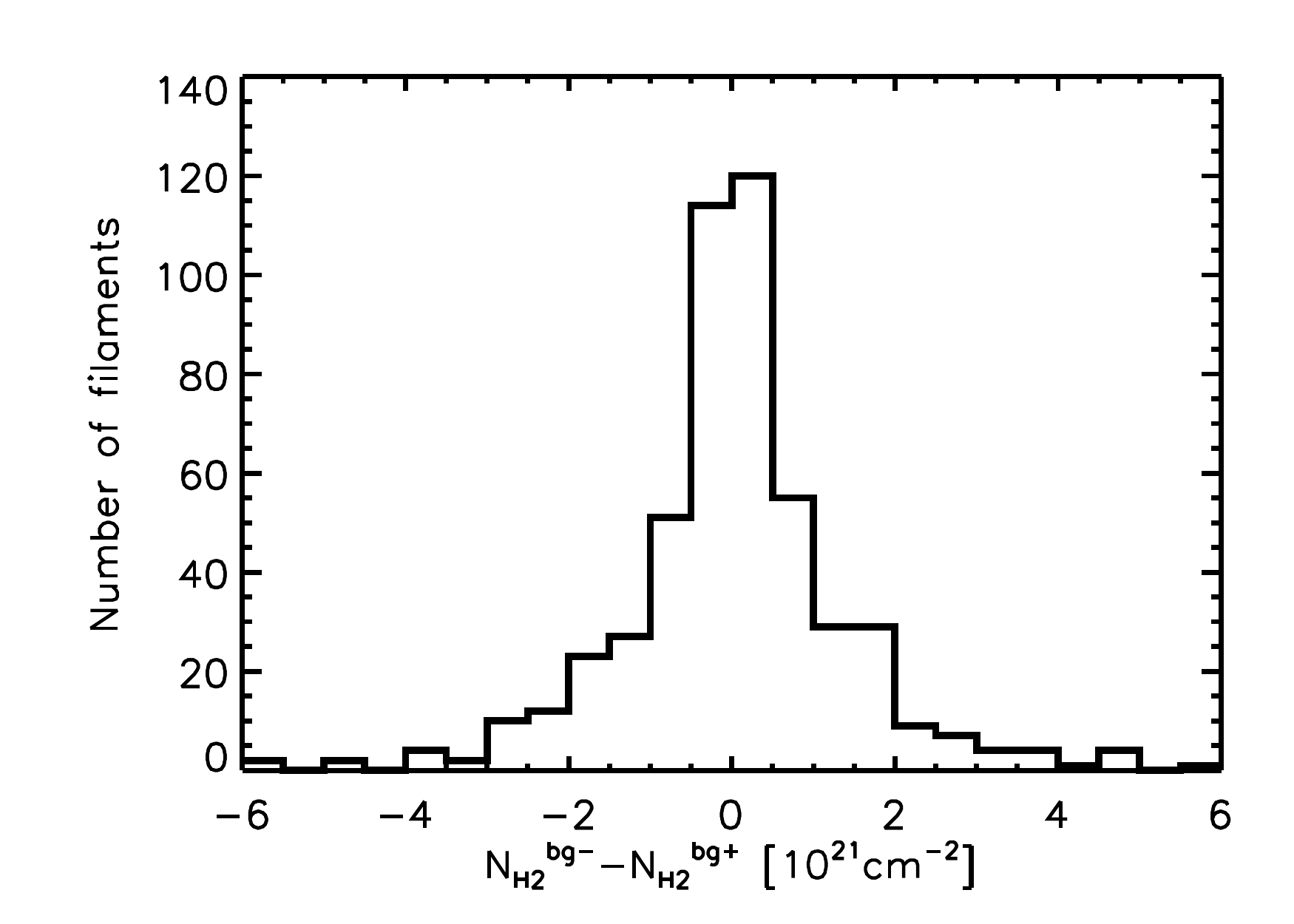}
}
  \caption{{\rev Distributions of  the differences in median width ($FWHM_{\rm dec}^--FWHM_{\rm dec}^+$, top)  and median background column density  ($\nhh^{ \rm bg-} -\nhh^{ \rm bg+}$,  bottom), derived 
      independently on either side of the 599 filament crests in the selected sample. 
      These two distributions have median $\pm$ equivalent standard deviation values of ($0.003\pm0.043$)\,pc and $(0.04\pm0.93)\times10^{21}$\,cm$^{-2}$, respectively. 
  }
          }
  \label{fig:WidthbgDiff} 
\end{figure}

\begin{table*}[!ht]  
\centering
 \caption{Fractions of subcritical, transcritical, and supercritical filaments, along with filament mass and area filling factors, in each cloud}
\begin{tabular}{|c|c|ccc|cccc|}   
\hline\hline   
Field & $N^{\rm fil}_{\rm tot}$&$N^{\rm fil}_{\rm sub}/N^{\rm fil}_{\rm tot}$&$N^{\rm fil}_{\rm crit}/N^{\rm fil}_{\rm tot}$&$N^{\rm fil}_{\rm sup}/N^{\rm fil}_{\rm tot}$& $M^{\rm fil}_{\rm tot}$& $M^{\rm fil}_{\rm tot}/M^{\rm cloud}_{\rm tot}$ & $M^{\rm fil}_{\rm dense}/M^{\rm cloud}_{\rm dense}$ &$A^{\rm fil}_{\rm tot}/A^{\rm cloud}_{\rm tot}$    \\
&  $\#$&$\%$&$\%$&$\%$&M$_\sun$&$\%$&$\%$  &$\%$   \\
  (1)&(2)& (3) & (4) &(5)&(6)&(7)&(8) &(9) \\  
\hline
IC5146&67&40&54&6&966&26&69&7  \\
OrionB&410&52&41&6&4280&16&60&7  \\
Aquila&137&21&61&18&1509&6&33&3  \\
Musca&47&79&19&2&109&11&89&4  \\
Polaris&32&100&0&0&19&4&$<0.1$&2  \\
Pipe&148&95&5&1&135&7&96&7  \\
Taurus&266&67&31&2&581&25&91&11  \\
Ophiuchus&204&50&44&6&577&17&87&7  \\
 \hline
All &1311&58&37&6&8175&13&49&6  \\
Median &148&67&41&6&581&16&89&7  \\
                           \hline  \hline
                  \end{tabular}\label{tab:Tab_MassFrac} 
                  
\begin{list}{}{}
 \item[]{{\bf Notes:} {} 
  {\bf Col. 2:}  Total number of filaments in each field (same as Col. 5 of Table\,\ref{tab:SumParamDisp}).  \\
  \hspace{1.cm}  {\bf Col. 3:} Fraction of subcritical filaments with $\ml^{\rm int} < M_{\rm line,crit}/2$, i.e., $\ml^{\rm int} < 8\sunpc$ in each field.   \\
\hspace{1.cm}  {\bf Col. 4:} Fraction of transcritical filaments with $M_{\rm line,crit}/2\le\ml^{\rm int} \le2M_{\rm line,crit}$, i.e., $ 8\sunpc$$ \le \ml^{\rm int}\le32\sunpc$.  \\
\hspace{1.cm}  {\bf Col. 5:} Fraction of supercritical filaments with $\ml^{\rm int} > 2M_{\rm line,crit}$, i.e., $\ml^{\rm int} > 32\sunpc$.  \\
\hspace{1.cm} {\bf Col. 6:}  Total gas mass of extracted filaments in each cloud, where the mass of a filament of length $l_{\rm fil}$ was estimated as $M^{\rm fil}=\ml^{\rm int}\,l_{\rm fil}$.\\
\hspace{1.cm}  {\bf Col. 7:} Fraction of cloud mass in the form of filaments,  $M^{\rm fil}_{\rm tot}/M^{\rm cloud}_{\rm tot}$, where $M^{\rm cloud}_{\rm tot}$ is given in Col.~5 of Table\,\ref{Table1}.\\
 \hspace{1.cm} {\bf Col. 8:}  Fraction of dense gas mass ($\nhh>7\times10^{21}$\,cm$^{-2}$) in the form of filaments, $M^{\rm fil}_{\rm dense}/M^{\rm cloud}_{\rm dense}$. 
 To estimate $M^{\rm fil}_{\rm dense}$, a mask corresponding to 0.1\,pc-wide filaments was first constructed from the \disperse\ skeleton derived  in each region. 
 $M^{\rm fil}_{\rm dense}$ was then computed within this filament mask from the corresponding column density map.  
 $M^{\rm cloud}_{\rm dense}$ corresponds to the product of Col.\,6 and Col.\,7 in Table\,\ref{Table1}.  \\ 
  \hspace{1.cm} {\bf Col. 9:} Area filling factor of the entire sample of extracted filaments in each field, $A^{\rm fil}_{\rm tot}/A^{\rm cloud}_{\rm tot}$, 
  where the area of a filament was estimated as $A^{\rm fil}=2hr \times l_{\rm fil}$ and $A^{\rm cloud}_{\rm tot}$ is given in Col.\,5 of Table\,\ref{Table1}.
}
 \end{list}      
  \end{table*}

\section{Discussion}\label{Discussion} %

Our detailed analysis of the radial column density profiles of nearly $600$ filamentary structures imaged in eight molecular clouds as part of the HGBS survey
confirms and strengthens the result of \citet{Arzoumanian2011}
that nearby molecular filaments share a common {\rev mean (crest-averaged)} inner width of about 0.1\,pc, while they span a wide range in central column density $\nhh^0$ 
and mass per unit length $\ml$\footnote{The $\nhh^0$ and $\ml$ values derived here overestimate the intrinsic central column densities and masses per unit length 
of the observed filaments by $\sim60\%$ on average, assuming random inclination angles of the filaments with respect to the plane of the sky \citep[cf.][]{Arzoumanian2011,Arzoumanian2013}. 
The plots and values given in the tables of this paper are {\it not} corrected for this inclination effect.} 
(cf. Fig.~\ref{fig:width-coldens} and Fig.\,\ref{fig:Mline}). 
{ \rev Independent measurements of filament widths in nearby molecular clouds have generally been consistent with our result 
when obtained through submillimeter dust continuum observations \citep[e.g.][]{Malinen2012,Ysard2013,Koch2015,Salji2015,Rivera-Ingraham2016,Rivera-Ingraham2017}. 
Measurements obtained using molecular line tracers have been less consistent with our finding, 
with observations in dense gas tracers such as N$_2$H$^+$ or NH$_3$ typically leading to filament widths significantly smaller than $0.1$\,pc 
\citep[e.g.,][]{Pineda2011,Fernandez-Lopez2014, Hacar2018}, 
and studies in low-density tracers such as $^{13}$CO finding widths significantly larger than $0.1$\,pc \citep[e.g.,][]{Panopoulou2014}.
We stress, however, that the dynamic range achieved in (column) density by observations in any given molecular line tracer 
is lower than that achievable by submillimeter continuum observations, especially from space with, e.g., {\it Herschel}.
This has direct implications for the reliable characterization of the intrinsic column density profiles of filaments. 
High-density line tracers are typically only sampling gas above a certain critical density and 
may not probe well the low-density outer parts of filament profiles. 
Conversely, due to depletion and optical depth effects, low-density gas tracers such as $^{13}$CO 
do not probe well the dense inner parts of many molecular filaments.
}

The filaments analyzed in the present paper span a wide range of  
column density contrasts $C^0$ from $\sim0.1$ for the most (thermally) subcritical filaments in our sample 
to $C^0\sim10$ for the most thermally supercritical filaments.
The filaments with contrasts $C^0>1$ span a range of central volume densities from a few $10^2$\,cm$^{-3}$ to $\sim10^5$\,cm$^{-3}$  
as inferred from the Plummer fits to their radial column density profiles. 

The strong linear correlation between $\ml$ and $\nhh^0$ (Fig.\,\ref{fig:Mline}) suggests that  
the filaments in our sample are characterized by an effective 
width given by $W_{\rm fil}^{\rm eff}\sim(\ml^{\rm int}/\ml^{\rm w})W_{\rm fil}\sim1.3W_{\rm fil}\sim0.13$\,pc (see Fig.\,\ref{fig:Mline}).
This result implies that the effective mass (and surface area coverage) of a molecular filament is typically enclosed within a strip of width $1.3\,W_{\rm fil}$ around the central crest. 
This value is also consistent with the observed power-law wings for $r>W_{\rm fil}/2$ (see Fig.\,\ref{fig:Multiprof}) and the results derived from Plummer fits of the column density profiles 
(see Fig.\,\ref{fig:PlumResults} and Table\,\ref{tab:PlummerFit}), suggesting that 
many molecular filaments  have non-Gaussian power-law wings with $\rho \sim r^{-2}$
beyond a flat inner plateau \citep[cf. also, e.g.,][]{Palmeirim2013,Cox2016}.  
On average, however, the non-Gaussian wings contribute little ($\sim 30\%$) additional mass. 
When they are well-developed, 
the observed power-law wings are shallower than the \citet{Ostriker1964} model (with $p=4$) of isothermal filaments in hydrostatic equilibrium. 

As already discussed in  \citet{Arzoumanian2013},  the observational distinction between thermally subcritical and thermally supercritical filaments 
does not correspond to a sharp boundary but to a range of masses per unit length 
around the theoretical value of $M_{\rm line, crit}$. 
For the purposes of this paper, we divide our sample of extracted filaments 
into three families: thermally subcritical with  $M_{\rm line}^{\rm obs} \lesssim M_{\rm line,crit}/2$, 
``transcritical'' filaments with $M_{\rm line,crit}/2 \lesssim M_{\rm line}^{\rm obs} \lesssim 2\,M_{\rm line,crit}$, 
and thermally supercritical filaments with $M_{\rm line}\gtrsim2\,M_{\rm line,crit}$ (see Table\,\ref{tab:Tab_MassFrac}). 
The three families  (subcritical, transcritical, and supercritical) contribute  
{\rev 20\,\%, 51\,\%, and 29\,\%}
of the total gas mass in the whole sample of (1310) filaments, respectively, 
{\rev 
with significant variations 
from cloud to cloud (cf. Table\,\ref{tab:Tab_MassFrac}), as a function of, e.g.,  cloud total dense gas and star-formation activity.}

{\rev  In the selected  sample of (599) filaments, $57\,\%$ of the filaments with  column density contrast  $C^{0}<1$ 
are  thermally subcritical, with a small fraction  ($4\,\%$) being thermally  supercritical, while $100\,\%$ of the filaments with  column density contrast  $C^{0}>2$ are thermally transcritical or supercritical, 
 (Fig.\,\ref{fig:Mline_bg} for the selected sample).} 
 
Given that our census of filamentary structures is essentially complete to transcritical and supercritical filaments 
(more than $95\,\%$ complete for $C^{0}\gtrsim1$ -- see Sect.~\ref{comp} and Appendix~\ref{App1}),
we can estimate the fraction of cloud mass in the form of such filaments. 
Table\,\ref{tab:Tab_MassFrac} gives the total mass in form of filaments in each of the 8 clouds. 
On average, $16\,\%$ of the total gas mass 
and about $7\,\%$ of the total surface area of each cloud 
is in the form of filaments. 
A companion paper \citep{Roy2018} discusses the implication of the relatively low area filling factor of filamentary structures 
for the power spectrum of cloud images.
The above estimate of the fraction of cloud mass in the form of filaments should be considered a lower limit due 
to the incompleteness of  our sample to low-contrast 
filaments, {\rev which may also be contaminated by spurious structures} (cf. Appendix\,\ref{App1}).  
Incompleteness to subcritical filamentary structures 
may not have a strong impact on the mass budget in the clouds, however, 
since the total gas mass in the form of filaments appears to be dominated 
by transcritical filaments with 
 $M_{\rm line}\sim M_{\rm line,crit}$,
which belong to the regime of central column density contrasts $C^{0}\gtrsim1$  where the extracted filament sample is mostly complete
{\rev ($56\,\%$ and $67\,\%$  of the filaments with $M_{\rm line}\gtrsim M_{\rm line,crit}$ and $M_{\rm line}\gtrsim 2\,M_{\rm line,crit}$, respectively, have $C^{0}>1$, cf. Fig.\,\ref{fig:Mline_bg}).}
We refer to Andr\'e, Arzoumanian et al., in prep., for a detailed discussion 
of the observed filament mass and $\ml$ distributions. 

A  dominant fraction $\gtrsim 80\%$ of the dense molecular gas in the clouds is in the form of (mostly supercritical) filaments, 
where we define dense gas based on column density, i.e., 
$N_{\rm H_{2}}^{\rm dense}\ge7\times10^{21}$~cm$^{-2}$.

 \begin{figure}[!ht]
   \centering
  \resizebox{9.cm}{!}{
      \hspace{-1.5cm}
   \includegraphics[angle=0]{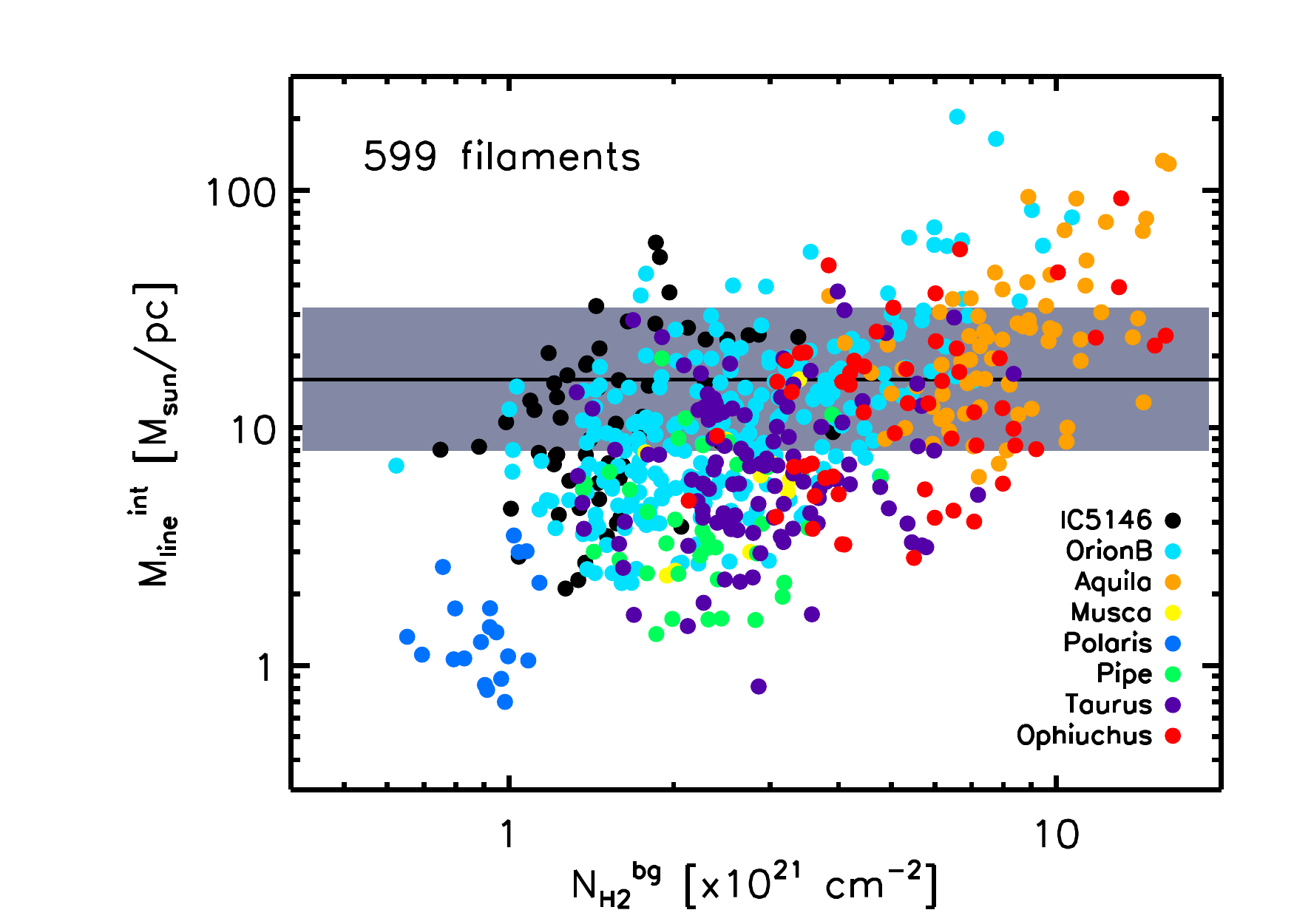}}
   \resizebox{9.cm}{!}{
      \hspace{-1.5cm}
   \includegraphics[angle=0]{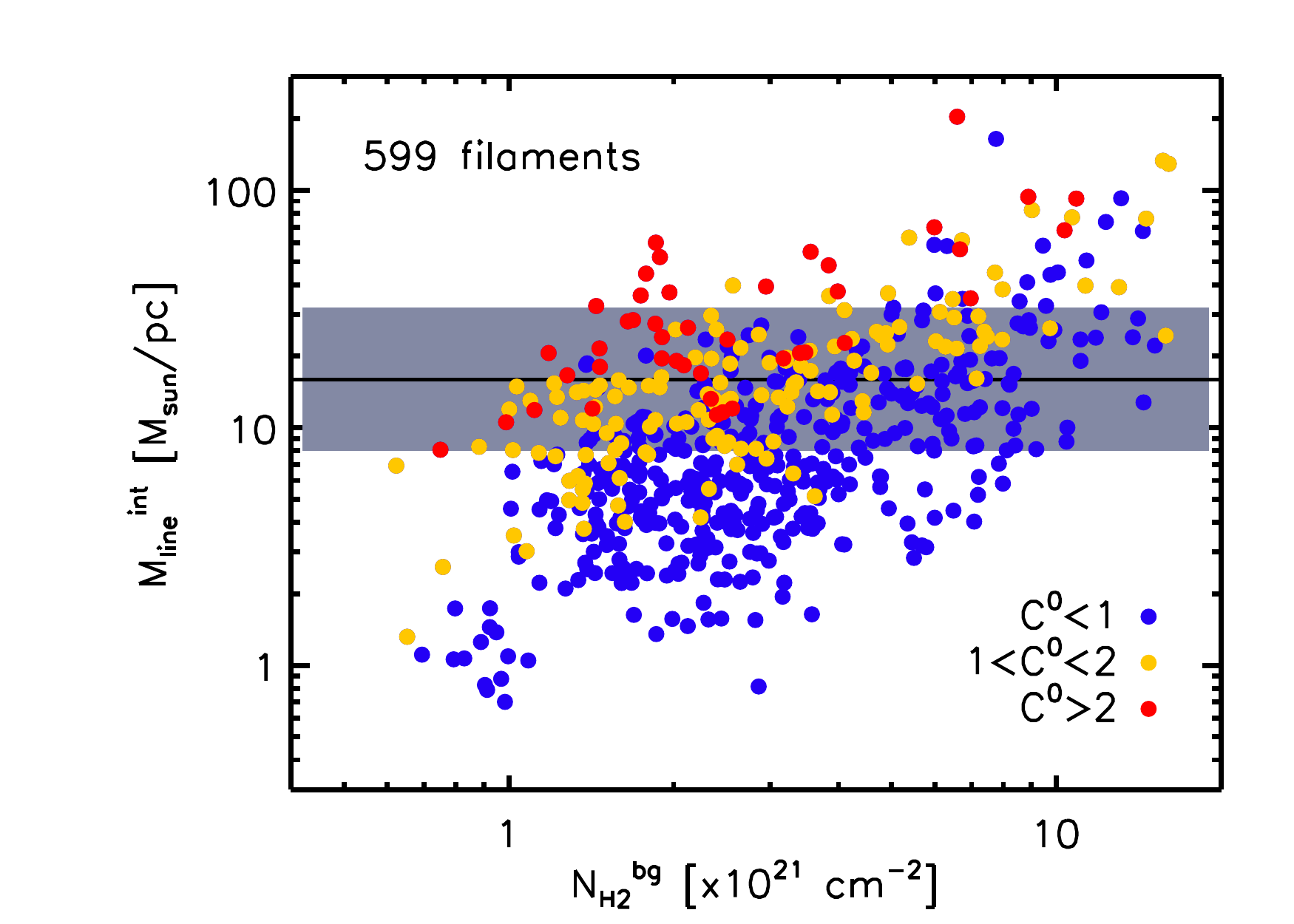}}
    \caption{{\bf Top:} Mass per unit length $M_{\rm line}^{\rm int}$ against 
    background column density $\nhh^{\rm bg}$ for the selected sample of 599 filaments. The horizontal black line  corresponds to $M_{\rm line}=M_{\rm line,crit}=16\,\sunpc$, and the grey area shows the transcritical filament regime with $M_{\rm line,crit}/2\le\ml^{\rm int} \le2M_{\rm line,crit}$ (cf. Table\,\ref{tab:Tab_MassFrac} and Sect.\,\ref{Discussion}).
    {\bf Bottom:} Same as the top panel, where each dot representing a filament is color coded here as a function of column density contrast $C^0$ (the color code is shown on the bottom right). 
    The percentages of  filaments with $C^0<1$, $1<C^0<2$, and $C^0>2$ are about $70\%$, $23\%$, and $7\%$, respectively, in the selected sample.
  }
  \label{fig:Mline_bg}
\end{figure}

The three families of filaments, defined above on the basis of their \ml\ as derived from \herschel\ dust continuum observations, 
exhibit marked differences in terms of their velocity dispersion 
and dust polarization properties.
Thermally subcritical filaments have sonic or transonic non-thermal velocity dispersions, and 
are gravitationally unbound with virial parameters ($\alpha_{\rm vir} \equiv M_{\rm line, vir}/M_{\rm line}$) larger than 2  \citep[e.g.,][]{Arzoumanian2013,Hacar2016}.
Their orientations on the plane of the sky are well aligned with the local magnetic field lines \citep[e.g.,][]{planck2016-XXXII}, and without any star forming activity. 
In contrast, thermally supercritical filaments tend to have supersonic velocity dispersions and are self-gravitating with $\alpha_{\rm vir} \sim 1$. 
They tend to be mostly perpendicular to the magnetic field lines observed on large scales in the surrounding ambient cloud \citep[][]{Sugitani2011,Palmeirim2013},
and are the main sites of prestellar core formation \citep[][]{Andre2010,Konyves2015,Marsh2016}.
Transcritical filaments have properties in common with the other  two groups. Namely,  they tend to have 
transonic velocity dispersions \citep[][]{Arzoumanian2013}, but are mostly observed perpendicular to the local magnetic field 
orientation \citep[][]{Cox2016}, and show indications of some star formation activity \citep[][]{Kainulainen2016}.  
Indeed, transcritical filaments with  \ml\  close to the thermal value of the critical mass per unit length 
may be the most appropriate cloud structures for the investigation of  the initial conditions of  core and star formation \citep[][]{Roy2015}.

Despite  marked differences,  the three families of filaments all appear to share a common {\rev mean} inner width.  
The lack of anti-correlation between filament width and central column density  is
surprising, since one  would naively expect  filament widths to scale with  local Jeans lengths, 
represented by the solid line running from top left to bottom right in Fig.\,\ref{fig:width-coldens}. 
Such an anti-correlation would indeed be expected for isothermal filaments in hydrostatic equilibrium \citep{Ostriker1964}. 
Figure\,\ref{fig:width-coldens} implies that the molecular filaments in our sample are not well described by models 
of isothermal cylinders in hydrostatic equilibrium\footnote{{\rev Significantly steeper power-law density profiles 
than the $p \sim 2$ profiles found here  for $r>>\rflat$ have been observed in N$_2$H$^+$ or NH$_3$ by, e.g.,  \citet{Pineda2011} and \citet{Hacar2018} 
who reported $p\sim4$, in agreement with the isothermal equilibrium 
model of \citet{Ostriker1964}. As pointed out earlier, however, high-density gas tracers 
such as N$_2$H$^+$ or NH$_3$ lines may not be sensitive to the low-density outer parts of filament profiles, 
biasing the derived $p$ index toward larger values.}}.

The existence of a common {\rev crest-averaged} width $\sim 0.1\,$pc 
shared by all filaments (at least by the filaments analyzed here) may provide hints on the formation process of filaments. 
\citet[][]{Arzoumanian2011} suggested that the characteristic filament width may be linked to the sonic scale of turbulence 
in the cold ($\sim10$\,K) ISM, observed to be around 0.1\,pc \citep[][]{Larson1981,Goodman1998}. 
The latter appears to be roughly the scale at which supersonic MHD turbulence dissipates and interstellar turbulence turns  
from supersonic and magnetohydrodynamic on larger scales to subsonic and 
hydrodynamic on smaller scales  \citep[][]{Federrath2010,Vazquez-Semadeni2003}.
Together with the observed subsonic to transonic velocity dispersions of subcritical/critical filaments \citep{Arzoumanian2013,Hacar2013,Hacar2016}, 
the common width of filaments suggests that the dissipation of large-scale shock waves (turbulent or not)  in the ISM 
may be important for the formation of the observed filamentary web 
\citep[see also][]{Inutsuka2015,Inoue2017}.

The observed correlation between filament mass per unit length and background column density (Fig.\,\ref{fig:Mline_bg}) 
indicates some link between 
filament properties and local conditions in the parent cloud \citep[see also][]{Rivera-Ingraham2016,Rivera-Ingraham2017}. 
Transcritical and subcritical filaments are observed toward a wide range of  background column densities ($\sim10^{21}$\,cm$^{-2}\lesssim \nhh^{\rm bg} \lesssim10^{22}$\,cm$^{-2}$ for transcritical filaments).
Only subcritical filaments are found at $\nhh^{\rm bg}\lesssim10^{21}$\,cm$^{-2}$.
At the other extreme, thermally supercritical filaments are mostly observed in high column density portions of the clouds ($\nhh^{\rm bg}\gtrsim5\times10^{21}$\,cm$^{-2}$). 
This high-density background may represent the reservoir of gas mass from which 
supercritical filaments appear to accrete while evolving \citep[e.g.,][]{Schneider2012,Hennemann2012,Arzoumanian2013,Palmeirim2013}.
The physical properties of the ambient cloud 
may possibly also change during the typical lifetime of filaments 
due to  (single or multiple) interactions with interstellar shock waves   \citep[see, e.g.,][]{Inutsuka2015,Arzoumanian2018}.

Thermally supercritical filaments with  $\ml>2M_{\rm line,crit}$ are expected to be gravitationally unstable to radial contraction
and  thus should nominally  collapse to spindles \citep{Inutsuka1997}, which is apparently inconsistent with their characteristic inner width $\sim 0.1\,$pc.
A possible explanation is that the non-thermal motions observed toward and along supercritical filaments \citep{Schneider2010,Arzoumanian2013,HKirk2013,Peretto2014,Hacar2017} 
may provide additional support against gravity and effectively prevent radial contraction. 
The presence of non-thermal motions is also consistent with models of accretion-driven turbulence 
within evolving supercritical filaments \citep{HennebelleAndre2013,Clarke2016,Inoue2017}. 
Another key ingredient may be the ambient magnetic field which 
may act against complete collapse of supercritical filaments,  
creating ribbon-like, tri-axial structures \citep[][Iwasaki et al. in prep.]{Tomisaka2014,Auddy2016}.
One limitation of the magnetized ribbon model comes from observational constraints on the maximum aspect ratio between the two 
short axes\footnote{A ribbon-like filamentary structure  \citep[e.g.,][]{Auddy2016} has one long axis, the main axis of the filament, and two short axes: 
one is set by hydrostatic equilibrium and corresponds to the Jeans length, and the other two by magnetohydrostatic equilibrium, resulting in an hourglass shape.} 
of observed filamentary structures, set to first order by the dispersion of measured filament widths. 

The very organized magnetic field structure revealed by \planck\ polarization observations \citep{planck2016-XXXIII} 
around supercritical filaments 
suggests that magnetic fields play a dynamically important role in shaping the cold ISM \citep[see also][]{planck2016-XXXV}. 
They may be channeling flows of low-density ambient gas 
onto star-forming supercritical filaments, which may be growing in mass per unit length and central column density 
while fragmenting into prestellar cores \citep[][Shimajiri et al.,  in prep.]{Palmeirim2013,Arzoumanian2013,Cox2016}.  
Subcritical filaments, on the other hand, are not self-gravitating and may be only transient density enhancements, 
dispersing in a turbulence crossing time 
$\sim 0.3\,$Myr (for 0.1\,pc wide filaments with internal total velocity dispersion of $\sim0.3\,$km\,s$^{-1}$) unless confined 
by external pressure \citep[][Inutsuka et al., in prep.]{Fischera2012}.

While a complete theoretical scenario for the formation and evolution of molecular filaments 
and their role in the star formation process is still lacking, 
the observational results presented in this paper
set strong constraints on possible models 
and represent a challenge for numerical simulations aiming to reproduce realistic filament properties.
The existence of a common mean inner 
width $\sim 0.1\,$pc for filamentary structures 
also provides 
hints on the physics at play in the cold ISM 
and may have strong implications for our understanding of the star formation process \citep[see][]{Andre2014,Roy2015,Shimajiri2017,Lee2017}.

\section{Summary and conclusions}\label{Summary}

In this paper,  we presented a census of filamentary structures observed in {\it Herschel} Gould Belt survey images 
of eight  nearby 
molecular clouds: IC5146, Orion B, Aquila, Musca, Polaris, Pipe, Taurus L1495, and  Ophiuchus. 
Our method of analysis and 
main results may be summarized as follows:

 \begin{enumerate}
\item The highly filamentary structure of a nearby molecular cloud 
imaged with \herschel\ can be well traced 
using the \disperse\ algorithm with a persistence threshold $PT={\rm rms}_{\rm min}$ and a robustness threshold $RT=1.5\nhh^{\rm bg,min}$, 
where ${\rm rms}_{\rm min}$ and $\nhh^{\rm bg,min}$ are the minimum rms level of background fluctuations and the minimum background column column density
in the input  \herschel\ column density map, respectively.
These values for the two main parameters $PT$ and $RT$ of \disperse\ were selected from multiple tests performed on synthetic maps,  
including realistic populations of mock filaments (see Appendix\,\ref{App1} and Appendix\,\ref{App2}).\\

\item Using \disperse\ with the above persistence and robustness thresholds, as well as a well-defined procedure (see Sect.~\ref{FilIdentification}), we identified
a total of 1310 filamentary structures in our 8 target fields and a selected sample of {\rev 599 filaments} with aspect ratio $AR \ge 3$ 
and central column density contrast  $C^0>0.3$. 
Based on extraction tests performed on synthetic maps (see Appendix\,\ref{App1}), this selected sample is estimated to be more than $95\,\%$ complete
to filaments with column density contrast $C^0\ge1$ {\rev (contaminated by 
$\sim 5\,\%$ of spurious detections)}, corresponding mostly to thermally transcritical and supercritical filaments with $\ml \gtrsim M_{\rm line,crit}$.\\
  
 \item 
Filament properties were derived by constructing radial column density radial profiles perpendicular to the filament  crests.
Without any fitting of the radial profiles, and for each extracted filament, we obtained estimates of the median column density along the filament crest, 
the outer radius on either side of the crest, the background column density at the outer radius, the filament width from the deconvolved half-power diameter ($hd_{\rm dec}$), and
the filament length.\\

\item Our sample of nearby molecular filaments is characterized by a narrow distribution of  {\rev crest-averaged inner  widths},  
with a median $hd_{\rm dec}$ diameter of 0.10\,pc and an interquartile range of 0.08\,pc, 
but spans wide ranges of lengths, column densities, and column density contrasts. 
The background column densities around the sampled filaments also span a wide range, reflecting  
significant variations in environmental properties (e.g., external gas pressure).
The line-of-sight dust temperatures observed along the filament crests are generally lower than the dust temperature in the background cloud, 
and the difference in dust temperature is larger for higher column density filaments.\\

 \item  In addition, the radial column density profiles observed on either side of the filament crests were fitted with Gaussian functions for radii $r \in [0,1.5hr]$, 
where $hr$ is the half-power radius directly measured on each profile (see point 3 above). 
The resulting distribution of deconvolved, {\rev crest-averaged} $FWHM$ widths 
{\rev for the 599 filaments in the selected sample} 
is sharply peaked around a median value of 0.10\,pc 
with an interquartile range of 0.07\,pc. 
This distribution is in stark contrast to the much broader distributions of filament lengths and local central Jeans lengths 
(inversely proportional to the central column densities of the filaments).\\ 

\item As the observed filaments often feature power-law wings that cannot be well reproduced by Gaussian fits, the radial profiles 
were also fitted with Plummer-like model functions (Sect.~\ref{Plumfit}). 
The resulting distribution of flat inner diameters $D_{\rm flat} = 2\, R_{\rm flat}$ has a median value of 0.10\,pc and an interquartile range of 0.11\,pc, 
in excellent agreement with both the distribution of deconvolved $FWHM$ widths found from Gaussian fitting 
and the distribution of  deconvolved half-power diameters ($hd_{\rm dec}$) directly measured on the filament profiles.
The power-law exponent $p$ of the underlying density profiles at large radii $r >> R_{\rm flat}$
is found to have a median value of  2.2 and an equivalent standard deviation of 0.3. Finally, filaments with central column density contrasts $C^0>1$ are found to span a range of central volume densities from a few $10^2$\,cm$^{-3}$ to $\sim10^5$\,cm$^{-3}$.  \\ 

\item  Our method of estimating filament properties from the radial column density profiles was validated by performing tests on synthetic cloud maps,  
including populations of mock filaments with both Gaussian and Plummer-like input radial profiles (see Appendix\,\ref{App2}).\\

{\rev
 \item The distributions of  individual $hd_{\rm dec}^{pix,\pm}$ and $FWHM_{\rm dec}^{pix,\pm}$ widths, 
derived from independent profiles taken along and on either side of the filament crests,
have the same median and  mean values as the distributions of 
$hd_{\rm dec}$ and $FWHM_{\rm dec}$ widths averaged along and on either side of the filament crests. 
They exhibit larger dispersions about their mean, however, with power-law-like tails of values significantly larger than 0.1\,pc.  
These unusually large values of the $hd_{\rm dec}^{pix,\pm}$ and $FWHM_{\rm dec}^{pix,\pm}$ widths may result from 1) physical variations of 
properties along each filament crest, 2) local variations in the background cloud, and/or 3) bad measurements due to locally disturbed radial profiles. 
Further analysis and tests would be needed to assess the relative contribution of these different factors to the observed distributions.       
} \\

 \item Our filament sample can be divided into three families: 
thermally subcritical filaments with $M_{\rm line} \lesssim 0.5\,M_{\rm line,crit}$, transcritical filaments with $ 0.5\,M_{\rm line,crit} \lesssim M_{\rm line} \lesssim 2\,M_{\rm line,crit}$, 
and thermally supercritical filaments with $M_{\rm line} \gtrsim 2\,M_{\rm line,crit}$. 
Transcritical filaments contribute more than half of the total gas mass of the whole sample of extracted filaments. 
On average, $16\,\%$ of the total mass in the target clouds is in the form of filaments and about $80\,\%$ of the dense gas mass (at $\nhh>7\times10^{21}$\,cm$^{-2}$)  
is in the form of (mostly transcritical and supercritical) filaments.\\
 
 \item  The masses per unit length derived for the sampled filaments correlate very well with their central column densities, 
which {\rev is consistent with} 
the existence of a characteristic filament width.\\

 \item These results suggest that, contrary to expectations, dense supercritical filaments with $\ml>2M_{\rm line,crit}$ do not  collapse radially to spindles 
but  somehow maintain a crest-averaged width of 
 $\sim0.1$\,pc while evolving and fragmenting into prestellar 
cores.\\

 \end{enumerate}

\begin{sidewaystable*}   
\centering
 \caption{Summary of measured properties for the selected sample of {\rev 599 filaments} (see Sect.\,\ref{SelecSample}) identified in eight regions of the HGBS.}
 \hspace{-1.cm}
\begin{tabular}{|c|c|c|c|c|c|c|c|c|c|c|c|c|}   
\hline\hline   
Field &beam& $N^{\rm fil}_{\rm select}$& $N_{\rm H_{2}}^{\rm bg}$& $N_{\rm H_{2}}^{\rm 0}$& $T^{\rm bg}_{\rm dust} $& $T^{0}_{\rm dust} $&$l_{\rm fil}$& $hd_{\rm dec}$&$\theta_{FWHM_{\rm dec}} $ & $FWHM_{\rm dec} $  &2\rout&$\ml^{\rm int}$ \\ 
 &[pc]& \#&[$10^{21}$~cm$^{-2}$]&[$10^{21}$~cm$^{-2}$]&[K]&[K]&[pc]&[pc]  &["]  & [pc]& [pc] &M$_{\odot}$/pc\\
  (1)&(2)& (3) & (4) &(5) &(6) &(7) &  (8) &(9)&(10)&(11)&(12)&(13)\\
    \hline
IC5146& 0.041& 59& 1.5 {\tiny$\pm$}  0.5 & 3.2 {\tiny$\pm$}  1.6 & 15.9 {\tiny$\pm$}  1.0 & 14.9 {\tiny$\pm$}  1.3 & 0.80 {\tiny$\pm$}  0.38 & 0.16 {\tiny$\pm$}  0.06 & 57 {\tiny$\pm$}  20 & 0.13 {\tiny$\pm$}  0.04 & 0.59 {\tiny$\pm$}  0.20 & 10 {\tiny$\pm$} 11 \\
Orion B& 0.035& 234& 2.3 {\tiny$\pm$}  1.3 & 4.1 {\tiny$\pm$}  2.7 & 16.7 {\tiny$\pm$}  1.5 & 15.9 {\tiny$\pm$}  1.6 & 0.82 {\tiny$\pm$}  0.45 & 0.15 {\tiny$\pm$}  0.04 & 64 {\tiny$\pm$}  22 & 0.12 {\tiny$\pm$}  0.04 & 0.48 {\tiny$\pm$}  0.20 & 9 {\tiny$\pm$} 8 \\
Aquila& 0.023& 71& 7.8 {\tiny$\pm$}  2.9 & 14.1 {\tiny$\pm$}  7.3 & 15.0 {\tiny$\pm$}  1.1 & 14.0 {\tiny$\pm$}  1.1 & 0.46 {\tiny$\pm$}  0.21 & 0.09 {\tiny$\pm$}  0.03 & 59 {\tiny$\pm$}  15 & 0.08 {\tiny$\pm$}  0.02 & 0.34 {\tiny$\pm$}  0.12 & 23 {\tiny$\pm$} 16 \\
Musca& 0.018& 10& 2.9 {\tiny$\pm$}  0.9 & 4.2 {\tiny$\pm$}  1.3 & 14.7 {\tiny$\pm$}  0.3 & 14.1 {\tiny$\pm$}  0.3 & 0.54 {\tiny$\pm$}  0.23 & 0.08 {\tiny$\pm$}  0.04 & 73 {\tiny$\pm$}  36 & 0.07 {\tiny$\pm$}  0.04 & 0.26 {\tiny$\pm$}  0.07 & 6 {\tiny$\pm$} 4 \\
Polaris& 0.013& 20& 0.9 {\tiny$\pm$}  0.1 & 1.4 {\tiny$\pm$}  0.2 & 15.1 {\tiny$\pm$}  0.2 & 14.8 {\tiny$\pm$}  0.4 & 0.35 {\tiny$\pm$}  0.19 & 0.07 {\tiny$\pm$}  0.02 & 76 {\tiny$\pm$}  14 & 0.06 {\tiny$\pm$}  0.01 & 0.19 {\tiny$\pm$}  0.05 & 1 {\tiny$\pm$} 1 \\
Pipe& 0.013& 38& 2.3 {\tiny$\pm$}  0.7 & 3.4 {\tiny$\pm$}  1.3 & 16.7 {\tiny$\pm$}  0.6 & 15.5 {\tiny$\pm$}  0.7 & 0.35 {\tiny$\pm$}  0.22 & 0.08 {\tiny$\pm$}  0.03 & 101 {\tiny$\pm$}  36 & 0.07 {\tiny$\pm$}  0.03 & 0.30 {\tiny$\pm$}  0.12 & 4 {\tiny$\pm$} 3 \\
Taurus& 0.012& 110& 2.8 {\tiny$\pm$}  1.0 & 5.2 {\tiny$\pm$}  2.3 & 13.6 {\tiny$\pm$}  0.3 & 12.9 {\tiny$\pm$}  0.5 & 0.30 {\tiny$\pm$}  0.13 & 0.07 {\tiny$\pm$}  0.03 & 86 {\tiny$\pm$}  32 & 0.06 {\tiny$\pm$}  0.02 & 0.24 {\tiny$\pm$}  0.10 & 6 {\tiny$\pm$} 6 \\
Ophiuchus& 0.012& 57& 5.3 {\tiny$\pm$}  2.4 & 9.8 {\tiny$\pm$}  4.3 & 16.6 {\tiny$\pm$}  0.9 & 16.1 {\tiny$\pm$}  1.4 & 0.32 {\tiny$\pm$}  0.15 & 0.07 {\tiny$\pm$}  0.02 & 83 {\tiny$\pm$}  24 & 0.06 {\tiny$\pm$}  0.02 & 0.26 {\tiny$\pm$}  0.15 & 13 {\tiny$\pm$} 11 \\
   \hline    \hline
Range & 0.012 $-$ 0.041 & &0.6 $-$ 16.0 & 1.3 $-$ 48.1 & 12.6 $-$ 30.2 & 11.5 $-$ 30.3 & 0.13 $-$ 3.52 & 0.03 $-$  0.29 & 33 $-$ 157 & 0.02  $-$ 0.30 & 0.09  $-$  1.24 & $<1$  $-$ 204 \\
Median & 0.018& & 2.6 & 4.8 & 16.0 & 15.0& 0.53 & 0.11 & 128 & 0.09 & 0.38 & 9 \\
IQR & $-$ &&1.8 $-$ 4.5 & 3.1 $-$ 8.5 & 14.8 $-$ 17.0 & 13.7 $-$ 16.3 & 0.34 $-$  0.84 & 0.08 $-$  0.15 & $-$ & 0.06  $-$ 0.13 & 0.26  $-$  0.52 & 5  $-$ 17 \\
Mean & 0.021 & &3.6 & 7.0 & 16.2 & 15.3& 0.66 & 0.12 & 141 & 0.10 & 0.41 & 14 \\
Stdev & 0.011 & &2.8 & 6.2 & 2.2 & 2.5& 0.46 & 0.05 & 67 & 0.05 & 0.21 & 18 \\
    \hline
                           \hline  
                  \end{tabular}
\vspace*{-0.45ex}
 \begin{list}{}{}
 \item[]{{\bf Notes:} {} 
The two values given in each column from {\bf Col.\,4} to {\bf Col.\,13} are the median and equivalent standard deviation (scaled from the interquartile range) 
 of the corresponding filament property in each field.\\   
  {\bf Col.\,1:} Field name.  
 {\bf Col.\,2:} Effective spatial resolution (in pc) of the column density map used in each field. 
 {\bf Col.\,3:} Number of filaments in each field from the selected sample. 
{\bf Col.\,4,\,5,\,6,\,7:}  Median  background column density, central column density, background dust temperature, and central dust temperature
of the selected filaments in each field, respectively. 
 {\bf Col.\,8:} Median filament length in each field, derived from integrating the filament crests traced with \disperse.
 {\bf Col.\,9:} Median deconvolved diameter ($hd_{\rm dec} $) of the filament column density profiles in  each field.
  {\bf Col.\,10:} Median deconvolved angular width ($\theta_{FWHM_{\rm dec}}$) of the selected  filaments in  each field.
  {\bf Col.\,11:} Median deconvolved width ($FWHM_{\rm dec}$ in pc)  of the selected  filaments in  each field.
{\bf Col.\,12:} Median outer diameter ($2\rout$) of the selected filaments in each field. 
{\bf Col.\,13:} Median mass per unit length ($\ml^{\rm int}$)  of the selected filaments in each field.\\   
The bottom five raws of the table give the observed range (i.e., minimum and maximum values), the median value, the interquartile range (IQR), the mean value, and the standard deviation (Stdev) 
of each property in the whole sample of 599 selected filaments.
  }
 \end{list}      
   \label{tab:table_stat} 
         \end{sidewaystable*}

   \begin{table*}[!ht]  
\centering
 \caption{Filament properties derived from Plummer fits.}  
 \label{tab:PlummerFit}  
\begin{tabular}{|c|cc|ccc|}   
\hline\hline   
Field & $N^{\rm fil}(1)$ & $D_{\rm flat}(1)$ & $N^{\rm fil}$(2) & $D_{\rm flat}(2)$ & $p(2)$ \\
  &  \#&[pc] & \# & [pc] & \\
  (1)&(2)& (3) & (4) &(5)&(6) \\
\hline
IC5146&17&0.14   $\pm$ 0.10&14&0.17   $\pm$ 0.21&2.2   $\pm$ 0.4\\
OrionB&31&0.21   $\pm$ 0.14&18&0.17   $\pm$ 0.15&2.2   $\pm$ 0.6\\
Aquila&10&0.12   $\pm$ 0.07&6&0.13   $\pm$ 0.08&2.4   $\pm$ 0.6\\
Musca&1&0.08    &$-$ &   $-$ &   $-$ \\
Polaris&2&0.10   $\pm$ 0.01&3&0.13   $\pm$ 0.08&2.1   $\pm$ 0.7\\
Pipe&5&0.08   $\pm$ 0.03&4&0.10   $\pm$ 0.04&2.2   $\pm$ 0.4\\
Taurus&33&0.07   $\pm$ 0.04&25&0.10   $\pm$ 0.06&2.3   $\pm$ 0.1\\
Ophiuchus&14&0.10   $\pm$ 0.04&9&0.12   $\pm$ 0.04&2.3   $\pm$ 0.4\\
\hline
All &113&0.10  $\pm$ 0.08&79&0.12   $\pm$ 0.07&2.2   $\pm$ 0.3\\
                           \hline  \hline
                  \end{tabular}
\vspace*{-0.45ex}
 \begin{list}{}{}
 \item[]{{\bf Notes:} {}
 {\bf Col.\,1:} Field name.  
 {\bf Col. 2:} 
 Number of filaments with contrast $C^0>1$ for which a Plummer fit with fixed power-law index $p=2$ was possible in each field.  We derived reliable Plummer fits for  {\rev 108} filaments with $C^0>1$ corresponding to $73\%$ of the selected filament sample with $C^0>1$. 
 It was possible to  derive reliable Plummer fits for  $61\%$ of the filaments of the selected sample with $C^0>1$.  
 {\bf Col. 3:} Median flat inner diameter $D_{ \rm flat}$(1)$=2R_{ \rm flat}$(1), and equivalent standard deviation (scaled from the measured interquartile range), derived from Plummer fits with fixed $p=2$ in each field. 
 {\bf Col. 4:}  Number of filaments with $C^0>1$ for which a Plummer fit with free power-law index $p$ was possible in each field. Reliable fits were possible for a  total number of {\rev 79}  filaments with  $C^0>1$. 
 {\bf Col. 5:} Median $D_{ \rm flat}$(2)$=2R_{ \rm flat}$(2) value and equivalent standard deviation derived from Plummer fits with free power-law index $p$ in each field. 
{\bf Col. 5:} Median $p$ value and equivalent standard deviation derived from Plummer fits with free power-law index $p$ in each field. 
}
 \end{list}      
  \end{table*}

\begin{acknowledgements}
DA acknowledges an International Research Fellowship from the Japan Society for the Promotion of Science (JSPS).
This work has received support from the European Research Council under the European Union's Seventh Framework Programme 
(ERC Advanced Grant Agreement no. 291294 - ORISTARS).
A.R. and N.S. acknowledge support by the french ANR and the german DFG through 
the project "GENESIS" (ANR-16-CE92-0035-01/DFG1591/2-1).
The present study has made use of data from the Herschel Gould Belt survey (HGBS) project (http://gouldbelt-herschel.cea.fr). 
The HGBS is a Herschel Key Programme jointly carried out by SPIRE Specialist Astronomy Group 3 (SAG 3), scientists 
of several institutes in the PACS Consortium (CEA Saclay, INAF-IFSI Rome and INAF-Arcetri, KU Leuven, MPIA Heidelberg), 
and scientists of the Herschel Science Center (HSC). 
\end{acknowledgements}

\bibliographystyle{aa}
\bibliography{aa} 

\newpage
\begin{appendix}

\section{Completeness of the extracted filament sample}\label{App1}

In this Appendix, we discuss several tests performed  on 
synthetic data to estimate the completeness of the extracted filament sample following the method described in Sect.\,\ref{FilIdentification}.

 To test our method of tracing filaments, synthetic column density maps of filamentary molecular clouds were generated by adding realistic populations
of mock filaments 
to a \herschel\ column density map where most, if not all, compact sources and filamentary structures had been removed using
the \textsl{getsources} and \textsl{getfilaments} algorithms \citep[][]{Men'shchikov2012,Men'shchikov2013}.

{\rev We built the background column density map (see Fig.\,\ref{SyntMap}a), by periodically duplicating  
the lower half of the  \herschel\ column density image of the Aquila cloud \citep[cf. Fig.~1 of][]{Konyves2015} resulting} in a $4^\circ\times2^\circ$ map corresponding 
to $\sim \,$18.5\,pc$\,\times\,9.2$\,pc at a distance $d = 260\,$pc.
Synthetic filaments were generated using 
{\rev a real filament crest traced with \disperse\ in the Aquila cloud.} 
{\rev Starting from this observed crest used as a template}, the synthetic filaments were given a Gaussian column density profile with a fixed width $FHWM=0.1$\,pc 
at $d = 260\,$pc  {\rev and random orientations with respect to the Cartesian $x$-- and $y$--axes of the background map. }  
For each realization, 
a group of {\rev 100} such synthetic filaments were distributed randomly over the background column density map,
controlling the  
contrast $C^0$ between the median column density along the crest of each synthetic filament and the local background column density (see Fig.\,\ref{SyntMap}c).  
A large number of such realizations were generated and the same method as described in Sect.~\ref{FilIdentification} for identifying filamentary structures 
in the real data was applied in each case.

As mentioned in Sect.\,\ref{SkelDisperse}, the completeness of our filament extractions was estimated 
by investigating the variation of the fraction of extracted synthetic filaments as a function of input filament contrast, $C^0$, and aspect ratio, $AR$, 
for several values of the persistence threshold, $PT,$ and robustness threshold, $RT$, of the \disperse\ run (see Sect.\,\ref{SkelDisperse}).\\ 

{\rev Ten}  synthetic maps, each including {\rev100} Gaussian-shaped filaments, i.e., a total of 1000 synthetic filaments  
were generated for each combination of the above mentioned four parameters ($C^0,\,AR,\,PT,\,RT$). 
Figure\,\ref{SyntMap}c shows an example of one of these realizations.

We then applied the filament extraction method of Sect.~\ref{FilIdentification} to each realization.
Since, as mentioned in Sect.\,\ref{SkelDisperseHerschel}, a good choice for $PT$ is on the order of the background column density fluctuations, 
$PT$ was varied as follows: 
${\rm rms}_{\rm min}/2$,  ${\rm rms}_{\rm min}$, $2\,{\rm rms}_{\rm min}$, $4\,{\rm rms}_{\rm min}$, where ${\rm rms}_{\rm min}$ 
is the minimum ${\rm rms}$ in the background column density map (see Fig.\,\ref{SyntMap}b).
As to the robustness threshold, it was varied between $RT = 0.75 \nhh^{\rm bg,min}$ and $RT = 2.25\nhh^{\rm bg,min}$, where  $\nhh^{\rm bg,min}$ is the minimum $\nhh^{\rm bg}$ 
in the  background column density map, since
it is linked to the minimum column density contrast of the filaments to be extracted, $RT\,\sim\,C^0 \nhh^{\rm bg}$ (cf., Sect.\,\ref{SkelDisperseHerschel}). 
Finally, 
for each realization, we estimated the number of ``true'' mock filaments that were recovered using our method of filament tracing,  
as well as the number of ``spurious''\footnote{A ``true'' detection corresponds to a synthetic filament added to the background column density map, 
while a ``spurious'' detection is a structure that is traced by \disperse\ but does not correspond to any synthetic filament. } extracted structures.
{\rev To do so, we compared the crests traced by \disperse\ with the input skeleton maps. When a  \disperse\ crest matched 
a crest from the input skeleton (numbered from 1 to 100 in each of the realizations) for at least 30 pixels (corresponding to $5\times HPBW$), the \disperse\ crest was considered a 
 ``true'' detection. When the \disperse\ crest did not have a counterpart in the  input skeleton for at least 30 pixels, 
 the \disperse\ crest  was flagged as a  ``spurious'' detection (see for an example Fig.\,\ref{SyntMap}e).}
 
{\rev We define the fraction of extracted  filaments as the ratio of  the number of ``true'' detections to the number of input filaments, i.e., 1000, for each combination of the  four studied parameters ($C^0,\,AR,\,PT,\,RT$). 
The fraction of ``spurious'' detections is defined as the ratio of the number of ``spurious'' detections to the total number of  (``true''$+$``spurious'') detections. 
}

    \begin{figure*}
   \centering
     \resizebox{15cm}{!}{
\includegraphics[angle=0]{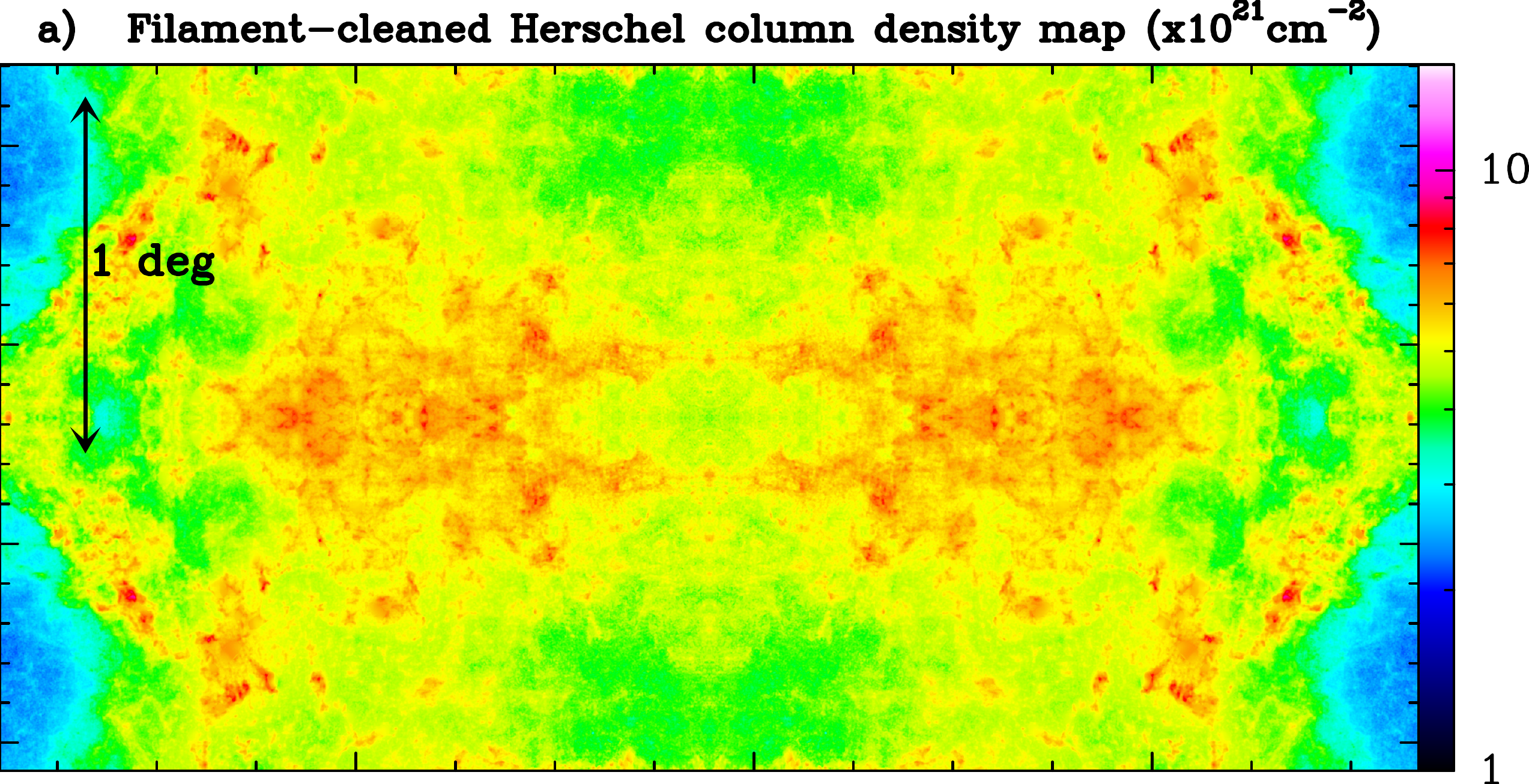}
\hspace{1.cm}   \vspace{2.cm}
\includegraphics[angle=0]{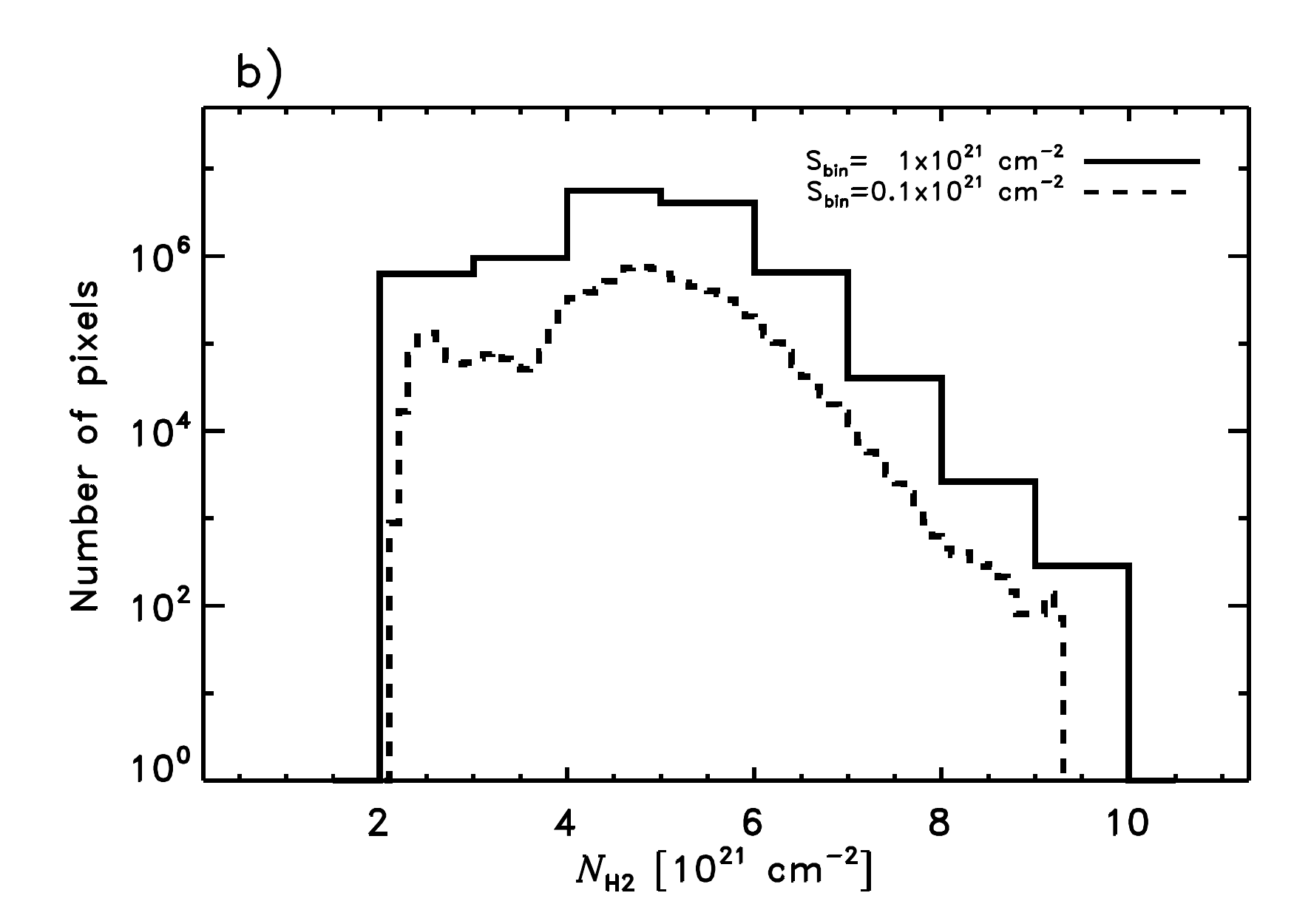}}
  \resizebox{10cm}{!}{  \vspace{2.cm}
 \includegraphics[angle=0]{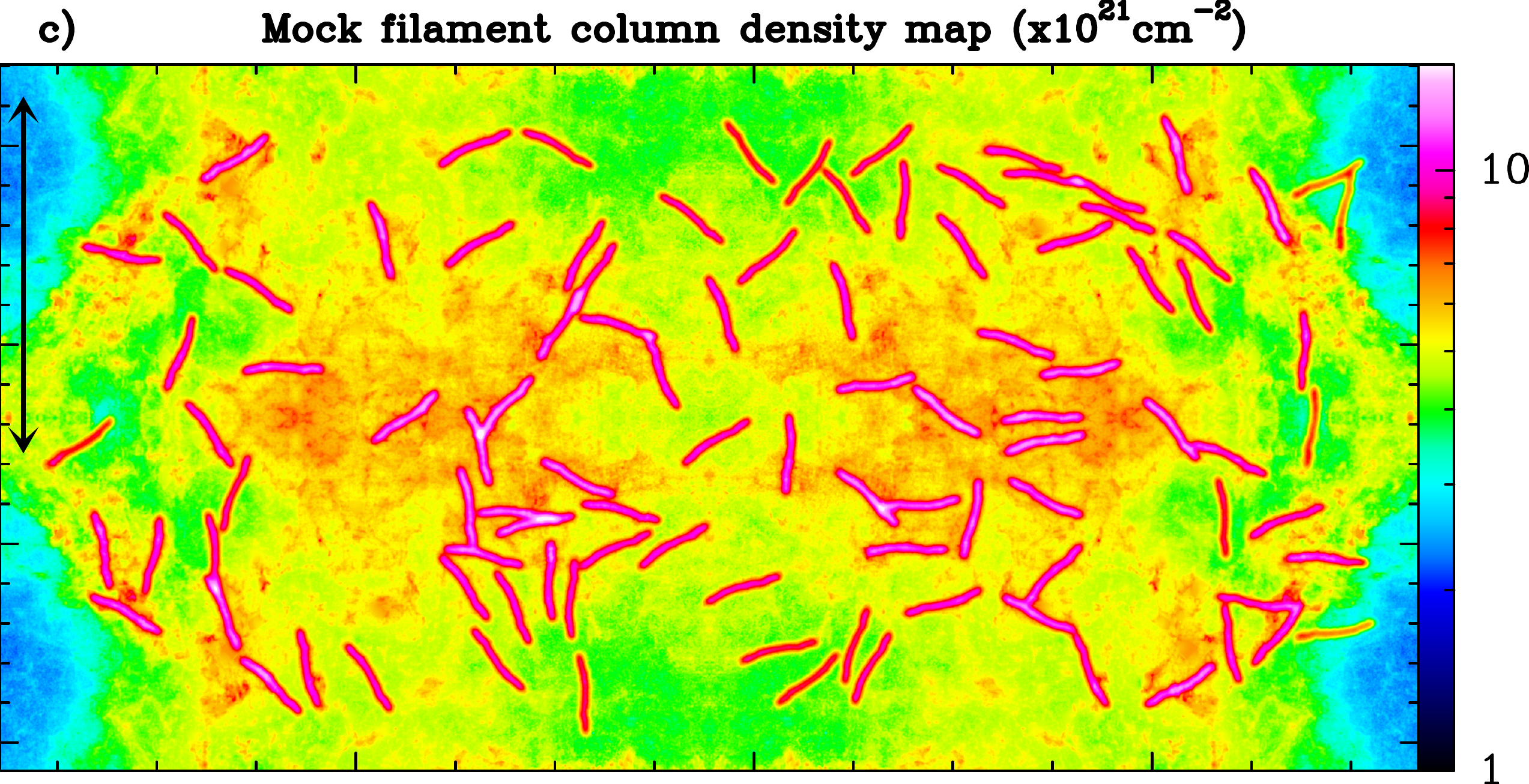}}
  \resizebox{19cm}{!}{
     \vspace{2.cm}
 \includegraphics[angle=0]{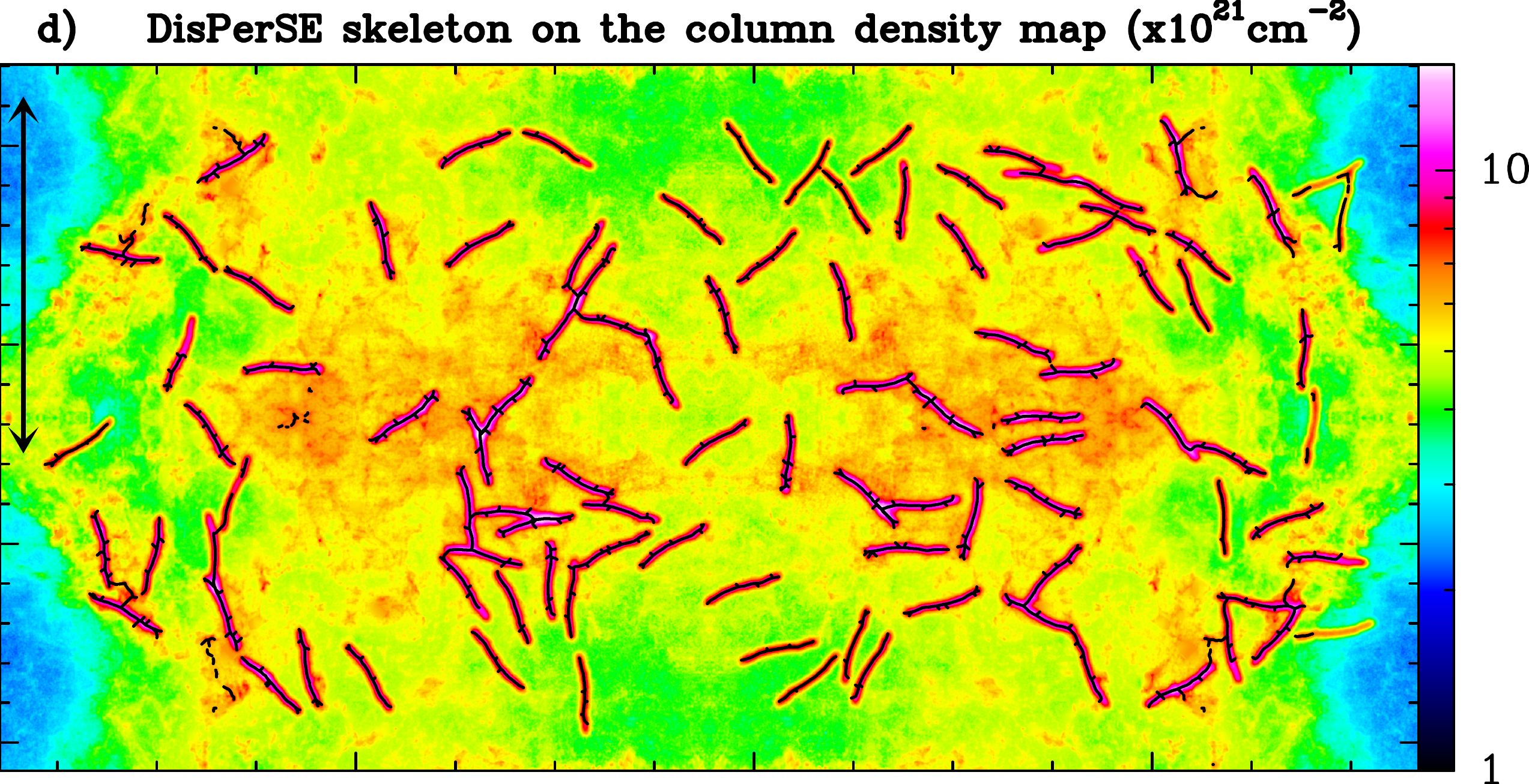}
  \hspace{1.cm}
\includegraphics[angle=0]{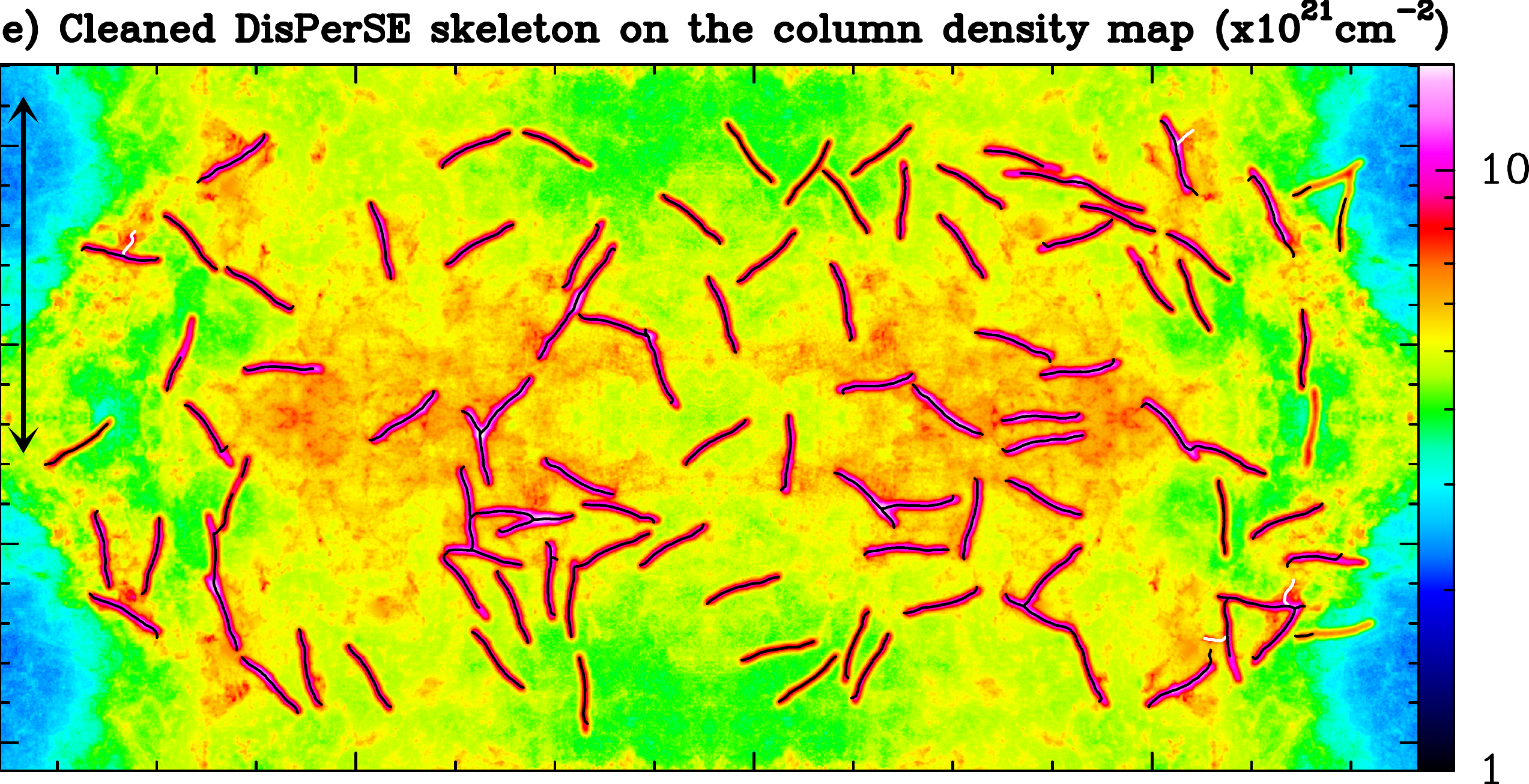}}
  \caption{{\bf a)}  ``Background'' map used in our tests of filament extraction, corresponding to a $4^\circ\times2^\circ$ portion of  
a   \herschel\  column density  map 
  after subtraction of compact sources and filamentary structures with the 
  \textsl{getsources} and \textsl{getfilaments} algorithms \citep[][]{Men'shchikov2012,Men'shchikov2013}. 
  {\rev The vertical double arrow indicates the $1^\circ$ scale of the map and is the same on all the maps of this figure. The spatial resolution of the maps is 0.023\,pc, corresponding to $18\arcsec $ 
 at a distance of 260\,pc. }
   {\bf b)} Column density histograms of the ``background map'' in {\bf a)} using two different bin sizes, S$_{\rm bin} = 10^{21}\, {\rm cm}^{-2}$ (solid histogram) and 
S$_{\rm bin} = 10^{20}\, {\rm cm}^{-2}$ (dashed histogram). 
The minimum background column density, $\nhh^{\rm bg,min}$, and the minimum level of background fluctuations, ${\rm rms}_{\rm min}$, 
used to adjust the \disperse\ parameters $RT$ and $PT$ (see text)
were estimated from the median and standard deviation of column density values in the first bin of the solid histogram, respectively: 
$\nhh^{\rm bg, min}=2.5\times10^{21}$cm$^{-2}$ and ${\rm rms}_{\rm min}=0.19\times10^{21}$cm$^{-2}$.
 {\bf c)} Example of a synthetic column density map obtained by distributing {\rev100} synthetic filaments with contrast $C^0=1$ and aspect ratio $AR=10$ over the  ``background map'' of panel {\bf a)}.
Here, all synthetic filaments were given a Gaussian profile with $FWHM=0.1$\,pc and a length of $\sim\,1$\,pc at the distance of 260\,pc. 
{\bf d)}  Same as panel {\bf c)} with the filament crests traced with \disperse\  overlaid in black.  \disperse\ was run with a persistence threshold $PT={\rm rms}_{\rm min}$ 
and a robustness threshold $RT=1.5\,\nhh^{\rm bg,min}$.
{\bf e)}  Same as panel {\bf d)} after removing filament segments shorter than 10 times the $HPBW$ beam, i.e., after ``cleaning'' the  \disperse\ skeleton (cf. Sect.\,\ref{SkelDisperseHerschel}). {\rev The white crests   indicate  the structures identified as "spurious" (four in this case), while the black crests correspond to "true" filaments}. 
}          
  \label{SyntMap}
    \end{figure*}
    
    \begin{figure*}
   \centering
     \resizebox{8cm}{!}{
\includegraphics[angle=0]{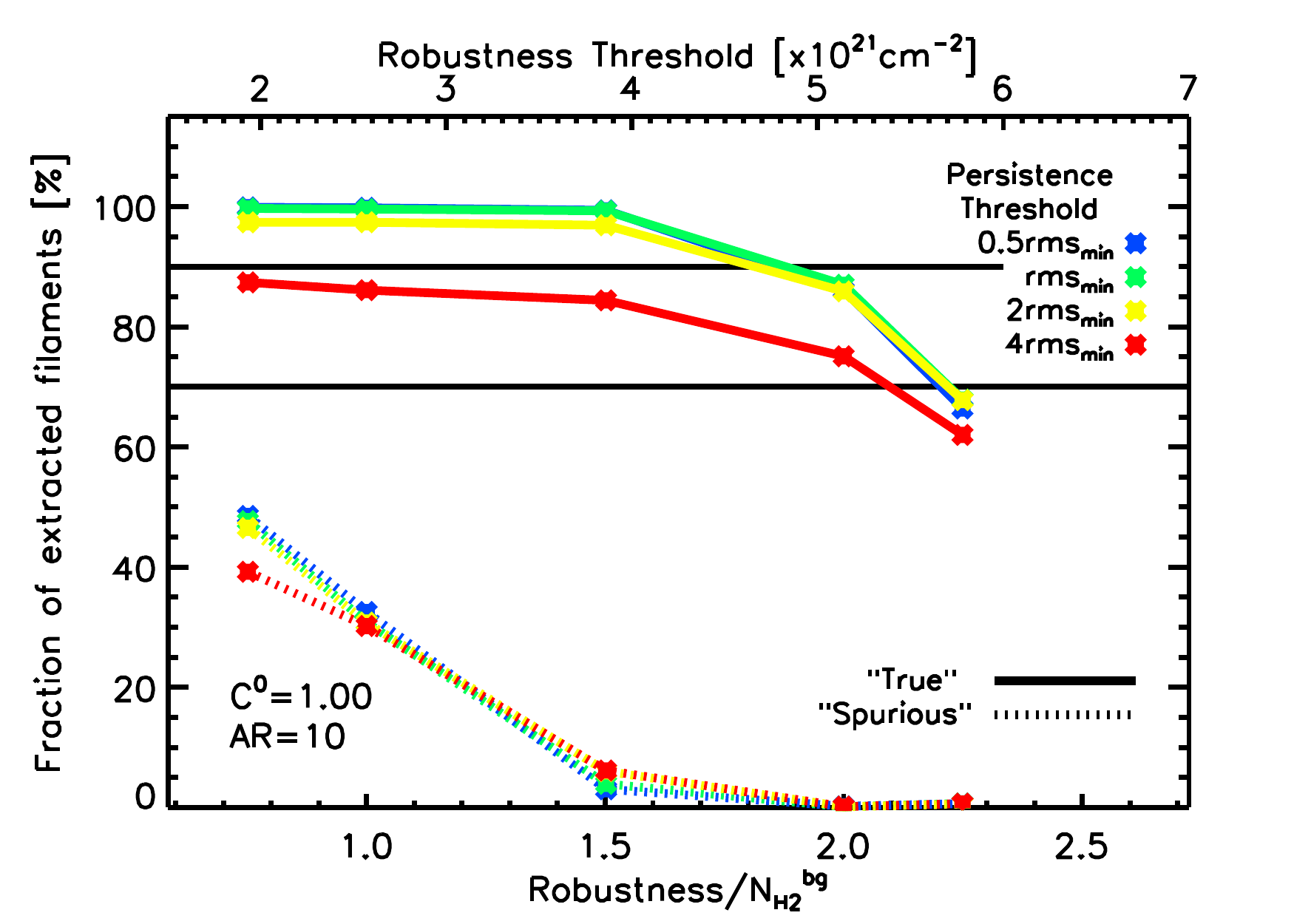}}
     \resizebox{8cm}{!}{
\includegraphics[angle=0]{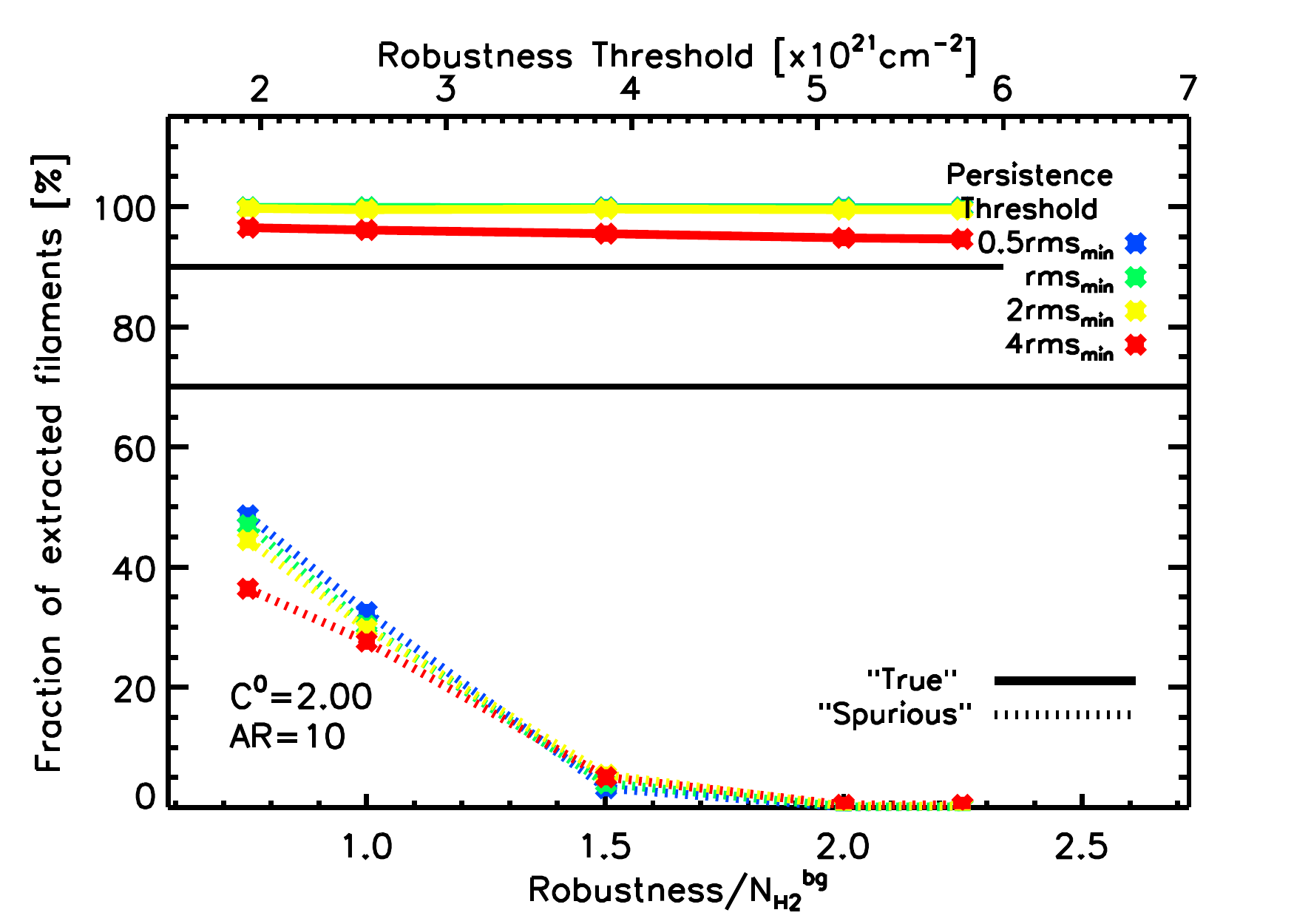}}
  \caption{{\bf Left:} Fractions of ``true'' detections (solid color lines) and ``spurious'' detections (dotted color lines) in tests of filament extractions as a function of the robustness threshold $RT$ 
used with \disperse\ (in units of  $ \nhh^{\rm bg}$ on the bottom x-axis and $10^{21}$cm$^{-2}$ on the top x-axis), for four values of the persistence threshold $PT$ ($0.5\, {\rm rms}_{\rm min}$, 
${\rm rms}_{\rm min}$, $2\, {\rm rms}_{\rm min}$, $4\, {\rm rms}_{\rm min}$, color coded as shown at the top right of the plot). 
The synthetic filaments used in these tests all had Gaussian profiles with $FWHM=0.1$\,pc, an aspect ratio $AR=10$, a column density contrast $C^0=1$ over the background,  
and were distributed in the map shown in Fig.~\ref{SyntMap}a. 
The statistics shown in the plot are based on a total of 1000 such synthetic filaments for each set of the two  \disperse\ parameters $RT$ and $PT$. 
The two horizontal black  lines indicate completeness levels of 70$\%$ and 90$\%$, respectively. 
{\bf Right:} Same as left panel but for synthetic filaments with intrinsic column density contrast $C^0=2$ over the background.
 }          
  \label{Comp_randW}
    \end{figure*}

    \begin{figure*}
   \centering
     \resizebox{8cm}{!}{
\includegraphics[angle=0]{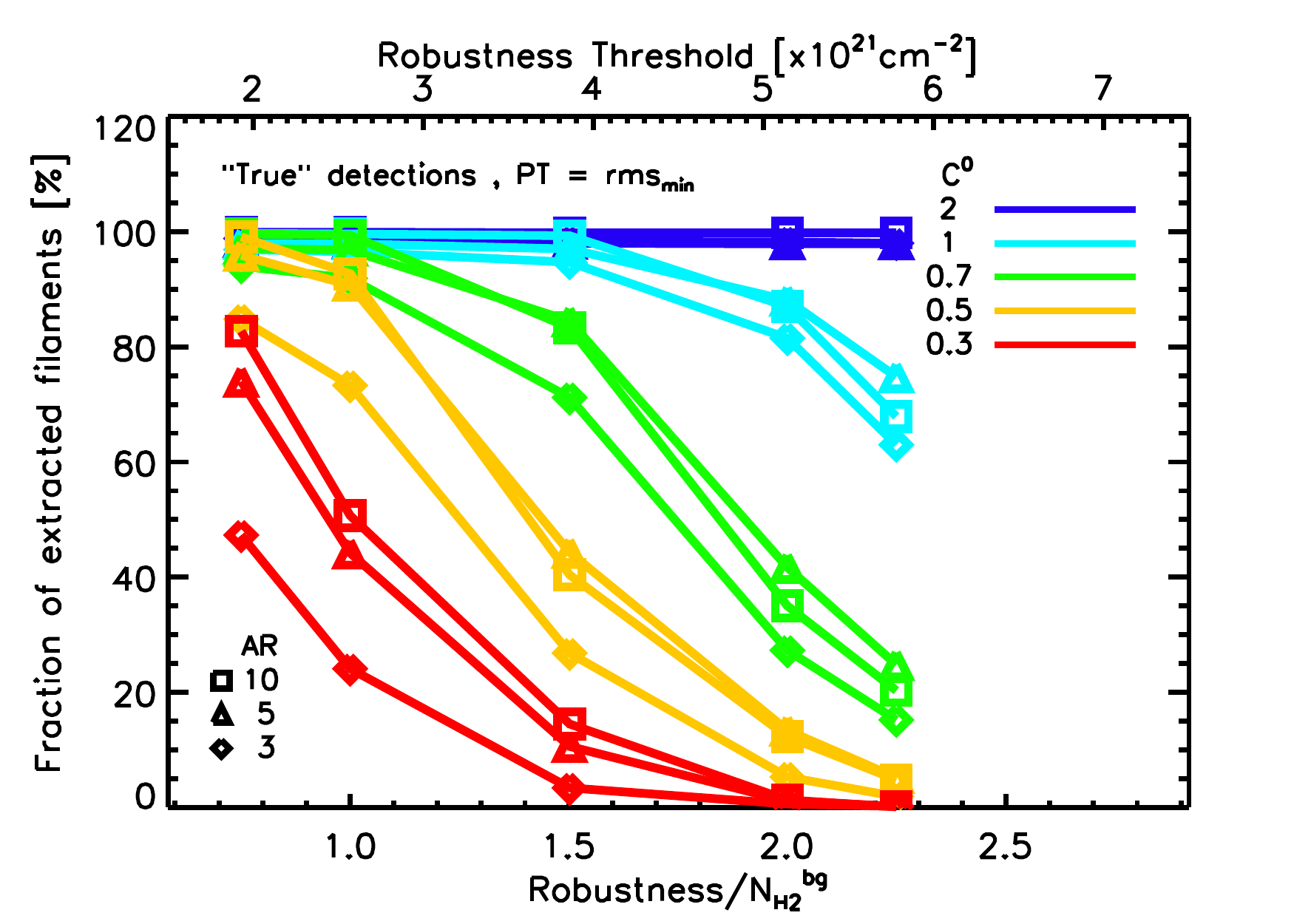}}
\hspace{1.cm}
\resizebox{8cm}{!}{
\includegraphics[angle=0]{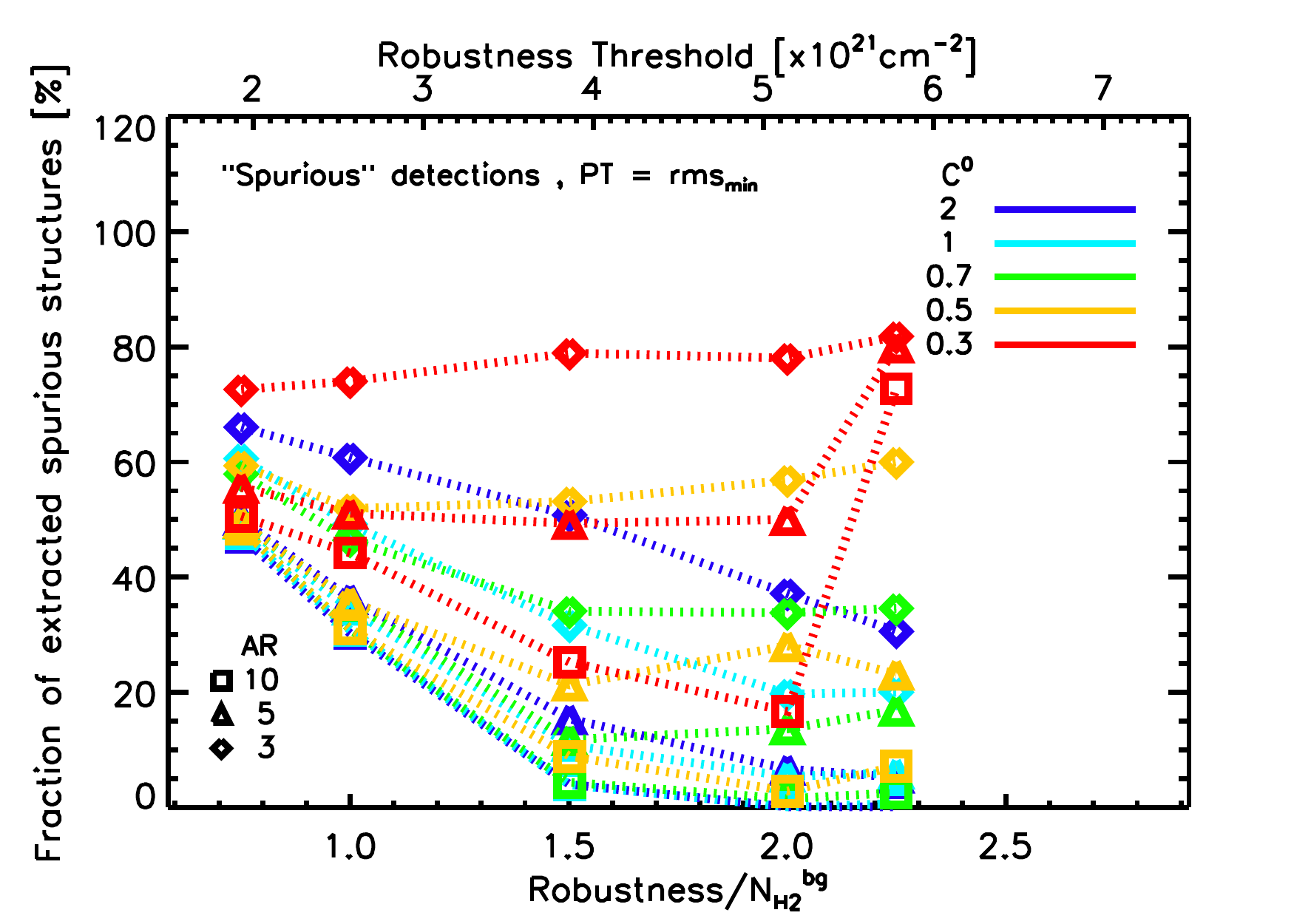}}
  \caption{ {\bf Left:} Fractions of ``true'' detections (solid color lines) in tests of filament extraction as a function of the robustness threshold $RT$ 
used with \disperse\ (in units of  $ \nhh^{\rm bg}$ on the bottom x-axis and $10^{21}$cm$^{-2}$ on the top x-axis), 
for three values of the input aspect ratio ($AR = 3$, 5, 10 -- see symbols at the botttom left of the plot) and four values 
of the input column density contrast ($C^0 = 0.3$, 0.5, 0.7, 1, 2 -- color coded as shown at the top right of the plot).
The statistics shown in this plot are based on a total of 1000 synthetic filaments with Gaussian profiles of $FWHM=0.1$\,pc 
for each set of ($AR$, $C^0$) input values,  added to the background map shown in Fig.~\ref{SyntMap}a. 
The persistence threshold used in \disperse\ was fixed to $PT={\rm rms}_{\rm min}$. 
 {\bf Right:} Same as left panel but for the fractions of ``spurious'' detections (dotted color lines). 
 }          
  \label{Comp_randW_AR_C}
    \end{figure*}

Figure\,\ref{Comp_randW} shows the statistics 
of ``true'' and ``spurious'' 
detections, based on a total of 1000 synthetic filaments for each set of  $PT$ and $RT$ values. 
Here, the synthetic filaments 
all have an aspect ratio $AR=10$ and column density contrasts $C^0=1$ (left panel) or $C^0=2$ (right panel). 

Thermally subcritical filaments 
are expected to have column density contrasts $C^0 \lesssim 1$, while thermally transcritical/supercritical filaments should have $C^0 > 1$ (see  Sect.~\ref{comp}).
As an illustration, filaments with intrinsic column density contrast $C^0=1$ and Gaussian width  
$FWHM=0.1$\,pc, embedded in a background cloud with  $\nhh^{\rm bg}=2.5\times10^{21}$cm$^{-2}$ (for $\mu_{\rm H_{2}}=2.8$), 
have a mass per unit length $\ml=11\,\sunpc$. %

For all input filament contrasts $C^0$, 
the fractions of extracted synthetic filaments were found to be almost independent of the persistence threshold 
for $PT$ values within a factor of two of the minimum ${\rm rms}$ in the background column density map, ${\rm rms}_{\rm min}$. 
The fraction of extracted synthetic filaments with $C^0=2$ 
is larger than $95\%$ for robustness threshold values $RT\le2.25\,\nhh^{\rm bg,min}$. 
The same level of completeness is achieved for synthetic filaments with $C^0=1$ when $RT\le1.5\,\nhh^{\rm bg,min}$.
The fraction of extracted synthetic filaments with $C^0=1$ decreases below $95\%$ when $RT>1.5\,\nhh^{\rm bg,min}$ 
but nevertheless remains  $\geq 80\%$ for $RT\le2\,\nhh^{\rm bg,min}$.

The fraction of ``spurious'' detections behaves in the opposite way, i.e., it decreases when $RT$ increases. 
``Spurious'' detections correspond to 
cirrus-like  column density structures in the background column density map. 
These structures are traced by \disperse\ but are not ``true'' filamentary structures. 
For $RT\ge1.5\,\nhh^{\rm bg,min}$, the fraction of spurious extracted structures is less than $10\%$, 
but it increases rapidly for $RT<1.5\,\nhh^{\rm bg,min}$.

We also investigated the effect of the robustness parameter $RT$
on the fractions of extracted synthetic filaments and ``spurious'' detections for aspect ratios $AR<10$ and column density contrasts $C^0<1$. 
The plots of Fig.\,\ref{Comp_randW_AR_C} show the fractions of extracted synthetic filaments and ``spurious'' structures for 
mock filaments with $0.3\le\,C^0\le2$ and $3\le\,AR\le10$, as a function of $RT$ for a fixed persistence threshold $PT={\rm rms}_{\rm min}$. 
It can be seen that the fraction of extracted synthetic filaments is almost independent (within $10\%$) of filament aspect ratio $3\le\,AR<10$, 
but decreases significantly when $C^0$ decreases for $RT<\nhh^{\rm bg,min}$. 
The fraction of ``spurious'' detections is almost constant (within $15\%$) for all $C^0>0.5$ and $AR>3$ values, 
and decreases when $RT$ increases. 
For $C^0\le\,0.5$ and $AR\le\,3$, the fraction of ``spurious'' extracted structures exceeds 
$\sim50\%$ of the total number of detections.

\section{Reliability of derived filament properties}\label{App2}

 \begin{figure*}
   \centering
      \resizebox{19cm}{!}{
         \hspace{-1cm}
\includegraphics[angle=0]{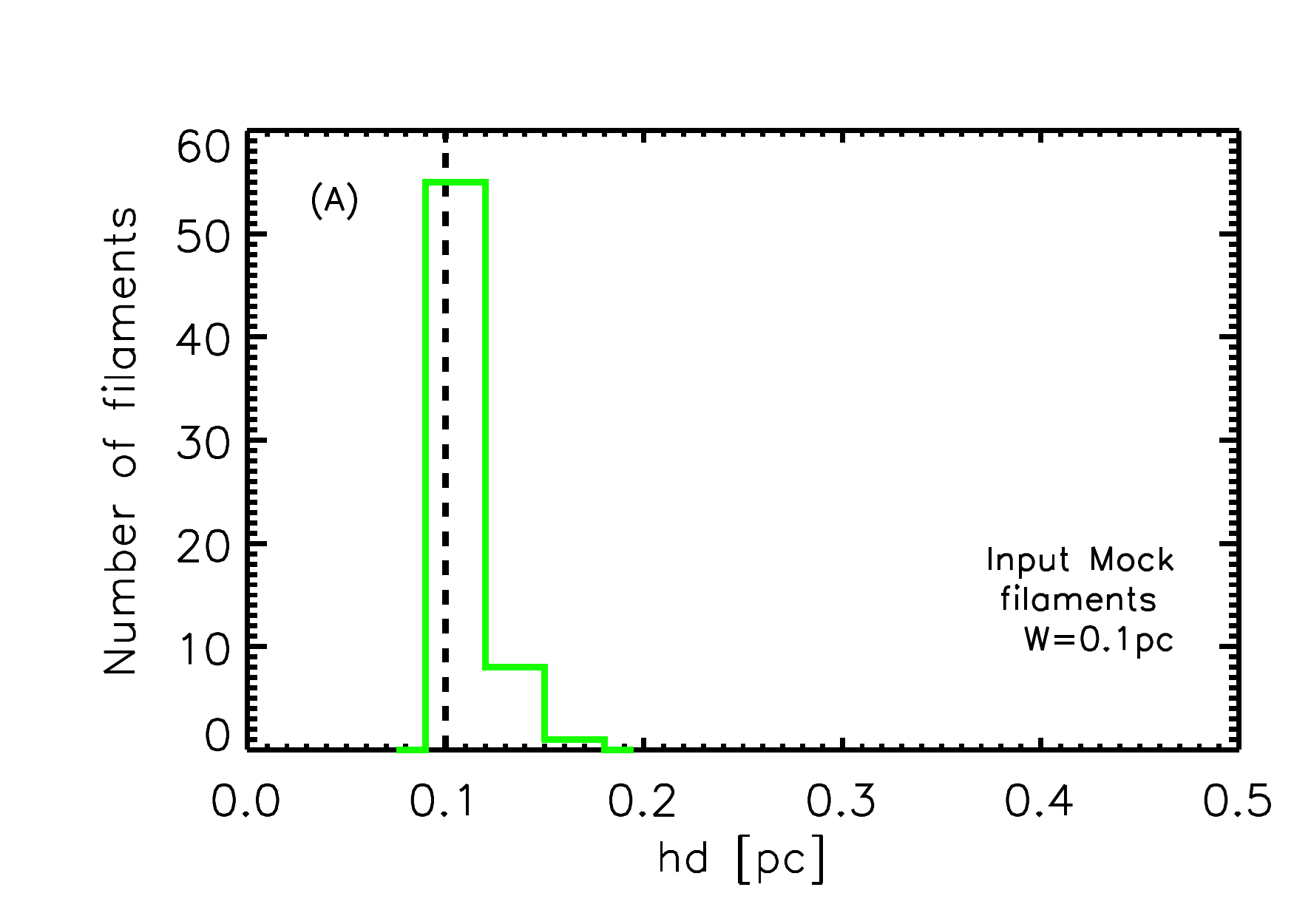}
\includegraphics[angle=0]{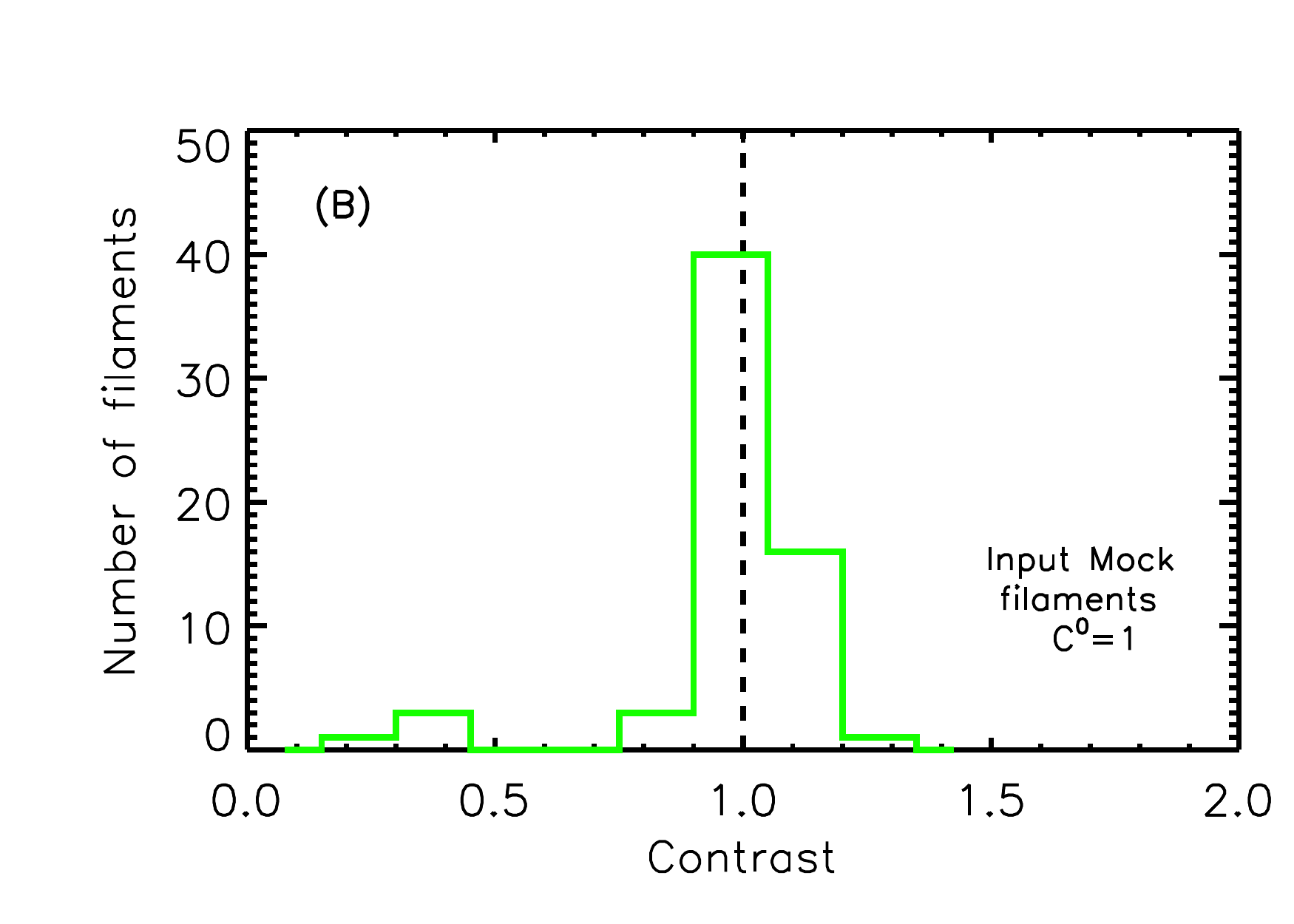}
\includegraphics[angle=0]{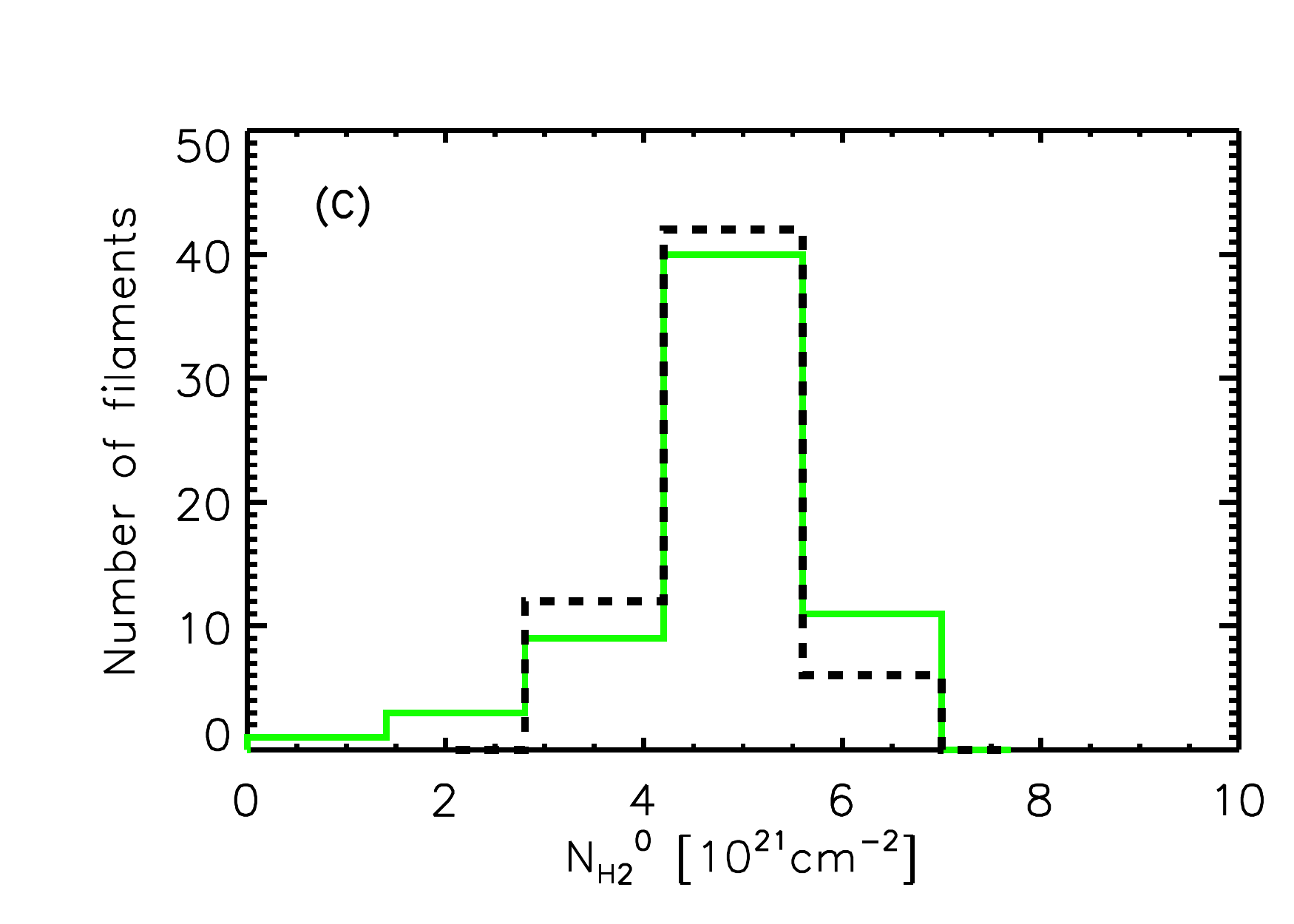}
}
   \vspace{-0.5cm}
  \caption{Histograms of derived crest-averaged properties  [half-power diameter, hd, in panel (A); filament contrast, $C^0$, in panel (B); background-subtracted
central column density, $\nhh^0$, in panel (C)] in measurement tests using a population of  Gaussian mock filaments.
{\rev The vertical dashed lines in panels (A) and (B) indicate the fixed input properties of the mock filaments: $FWHM$ width ($W$=0.1\,pc) and column density contrast ($C^0$=1).
In panel  (C), the  black and green histograms show the distribution of input and derived background-subtracted column densities 
averaged along the filament crests, respectively. 
}
 }          
  \label{histo_mock_nofit}
    \end{figure*}

In this Appendix, we assess the reliability of our method of estimating filament properties using 
the measurement steps described in  Sect.\,\ref{FilMeasure}.
{\rev To this end, synthetic maps including mock filaments with Gaussian and Plummer-like column density radial profiles, 
and various distributions of input properties, such as  inner width, power-law slope at large radii, and filament contrast, were constructed.  
Several sets of synthetic maps were generated by distributing between {\rev 60} and {\rev 180} mock filaments 
with given input properties ($AR$, $C^0$, $p$)
within the background column density map of Fig.\,\ref{SyntMap}a, which has a spatial resolution of 0.023\,pc corresponding to $\sim 18\arcsec $ at a distance of 260~pc.
The mock filaments were distributed in an ordered way and at well separated locations  to reduce measurement uncertainties due to possible intersections or blending between filaments.   
 The filament crests of the synthetic maps were traced using \disperse\ with $PT={\rm rms}_{\rm min}$ and  $RT=\,1.5\,\nhh^{\rm bg,min}$ following the steps described in Sect.\,\ref{FilIdentification}, and   analyzed in the same way as the real data as explained in Sect.\,\ref{FilMeasure}. 

Figure\,\ref{histo_mock_nofit} compares the  distributions of input and measured filament properties 
derived without any fitting of the column density radial profiles and after determination of the filament crest, outer radius, and background column density values, as explained in 
Sect.\,\ref{FilIdentification} and Sect.\,\ref{FilMeasure}. 
The distribution of measured half-power diameters, $hd$, for filaments with Gaussian profiles has a median value consistent with 
the input $FWHM$ width of the mock 
filaments (Fig.\,\ref{histo_mock_nofit}A). 
The input filament contrasts and central/background column densities are also well recovered. 

In the following two subsections (Sect.\,\ref{App2a} and Sect.\,\ref{App2b}), we discuss the reliability of inner width measurements 
derived from Gaussian  
and Plummer 
fits to the filament radial profiles, respectively. We summarize the results of these tests in Sect.\,\ref{App2c}. 
}

 \begin{figure*}
   \centering
      \resizebox{19.5cm}{!}{
      \hspace{-2.5cm}
\includegraphics[angle=0]{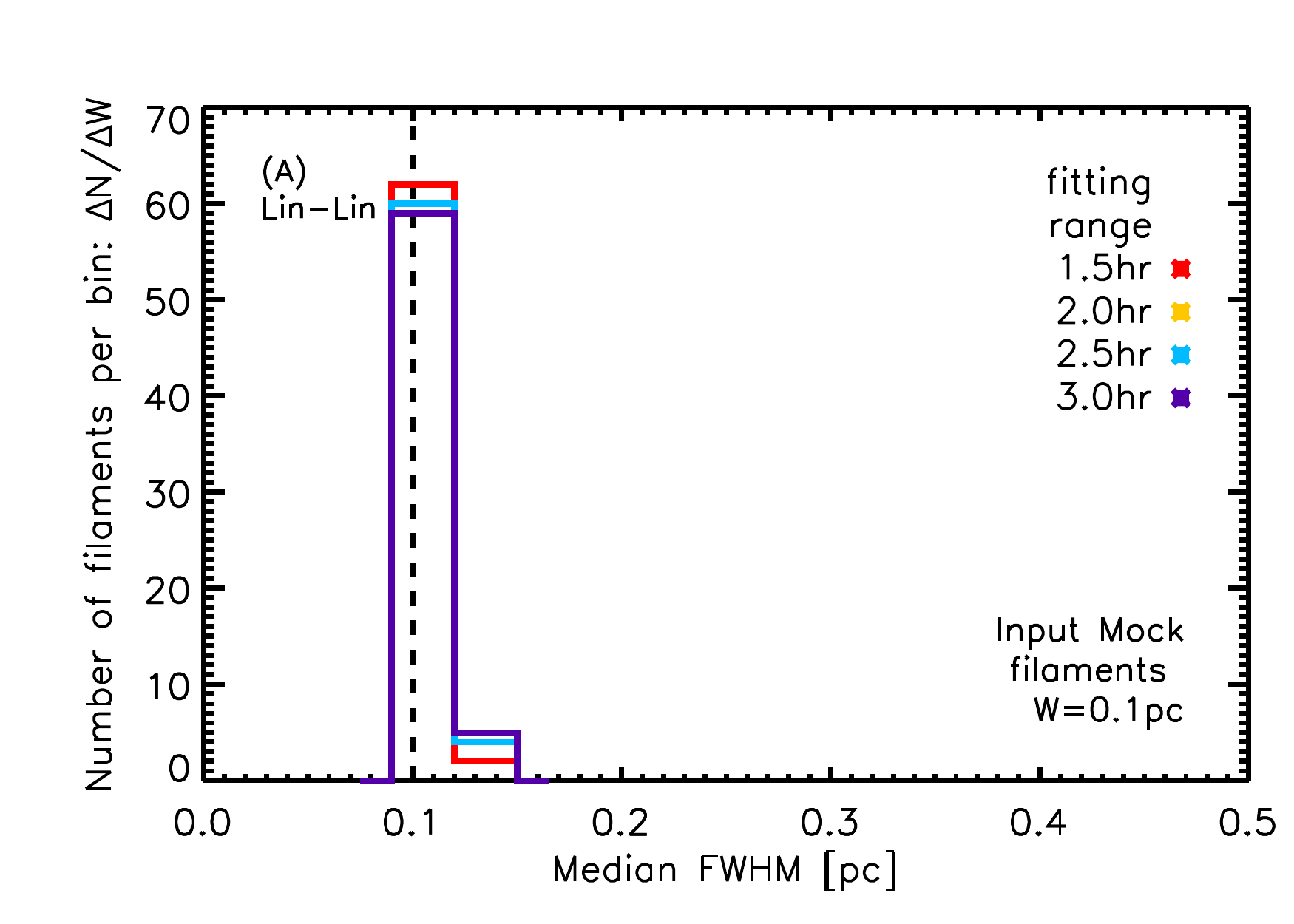}
\hspace{-.5cm}
\includegraphics[angle=0]{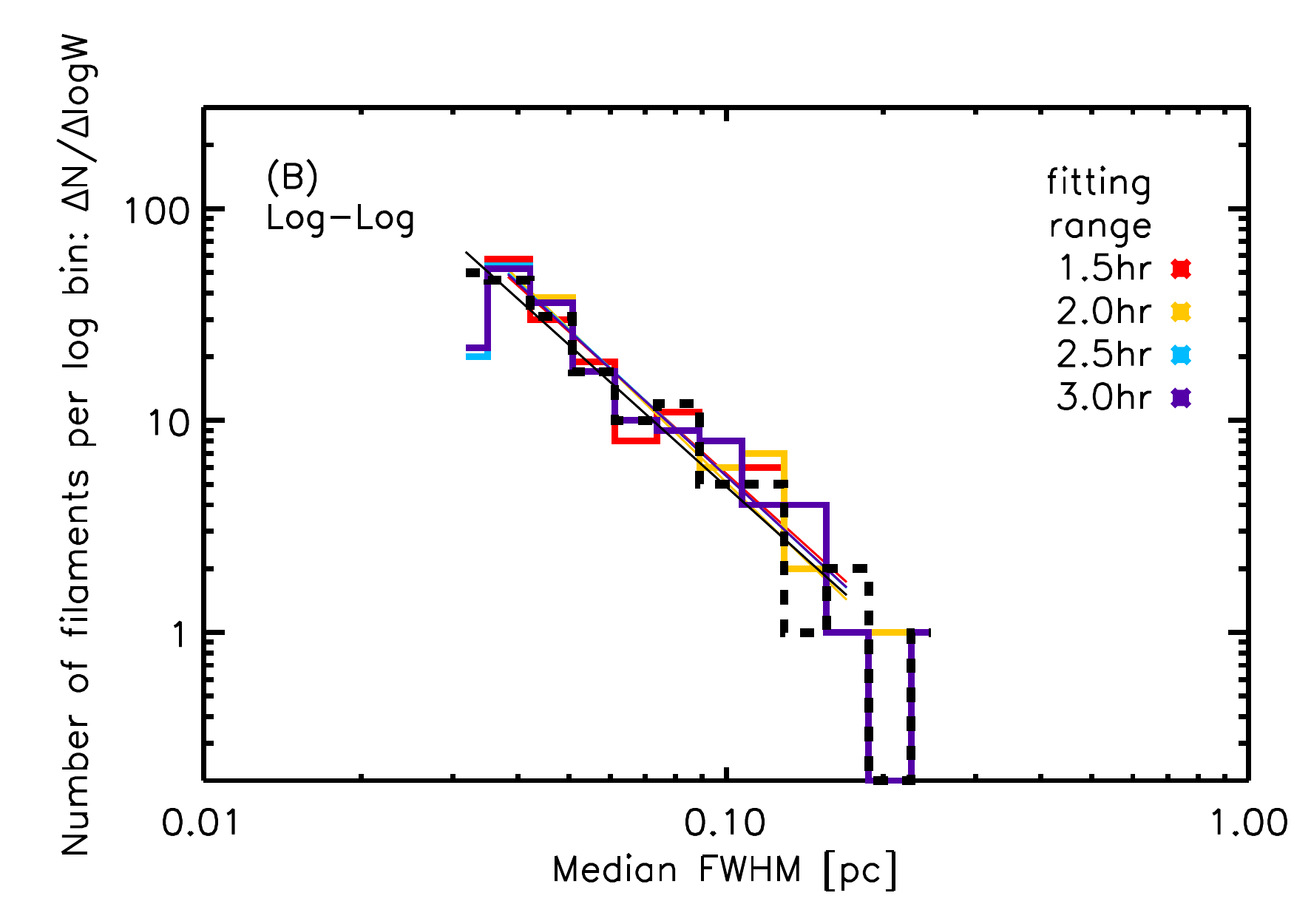}
 \hspace{-1.5cm}
   \includegraphics[angle=0]{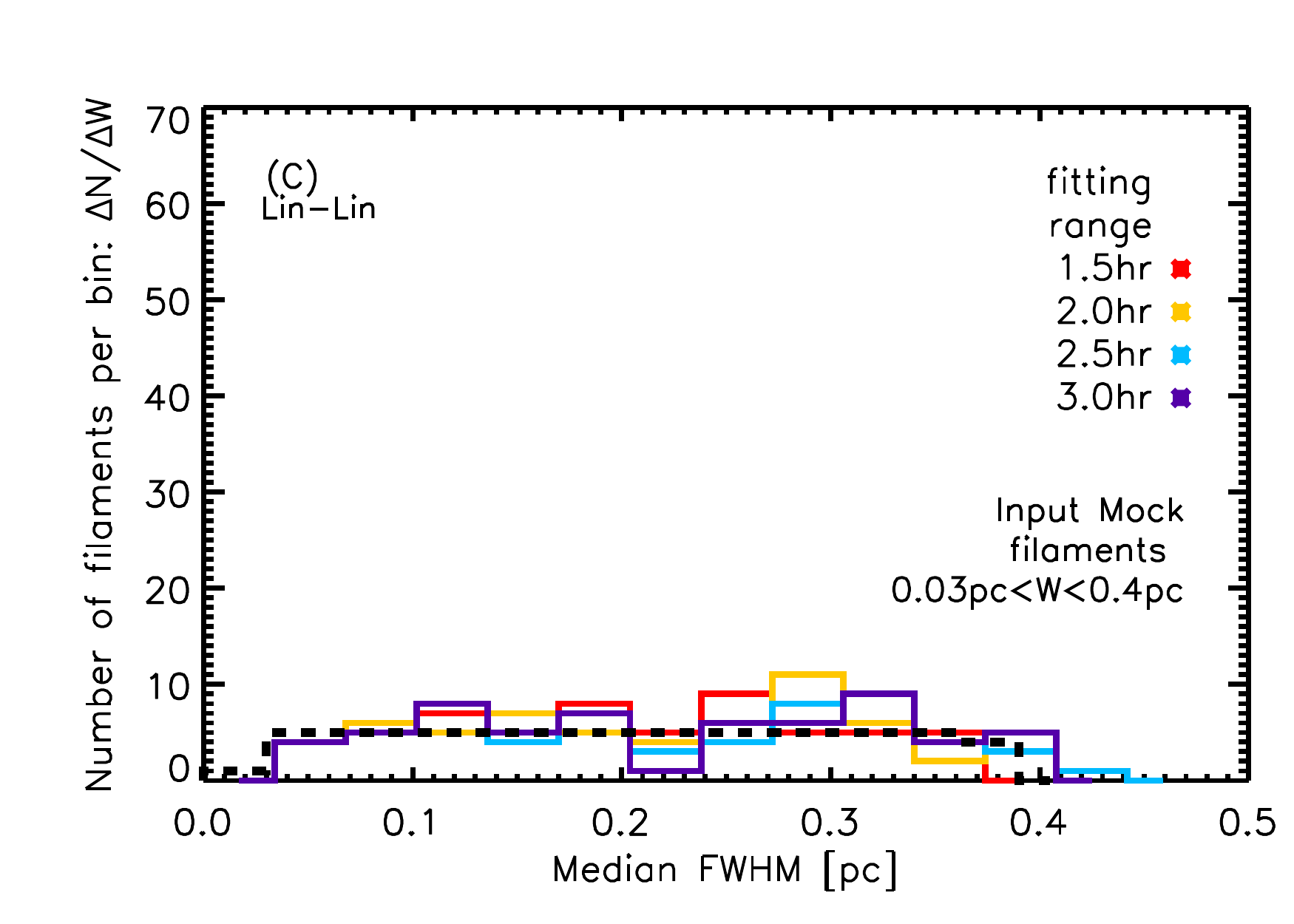}}
\vspace{-0.5cm}
  \caption{Histograms of FWHM widths derived from Gaussian fitting to the radial column density profiles of synthetic filaments 
  with Gaussian input profiles and   fixed input $FWHM$ width ($W=0.1$\,pc) (A), 
  {\rev  a  power law distribution of $FWHM$ widths between 0.03\,pc and 0.25\,pc (B), } 
 and a flat input distribution of $FWHM$ widths between 0.03\,pc and 0.4\,pc (C). 
    {\rev The input $FWHM$ width distributions are shown with the black dashed lines on all panels.}
  Results are presented for four choices of the fitting range ($1.5hr$, $2.0hr$, $2.5hr$, $3.0hr$), color coded as shown on the right hand side of each plot (cf. Sect.\,\ref{Linkfit}).
 Note how the measured distributions of filament widths are almost independent of the fitting range. 
      }          
  \label{histo_width_mock}
    \end{figure*}

   \begin{figure*}[]
   \centering
      \resizebox{19.5cm}{!}{
    \hspace{-2.5cm}
     \includegraphics[angle=0]{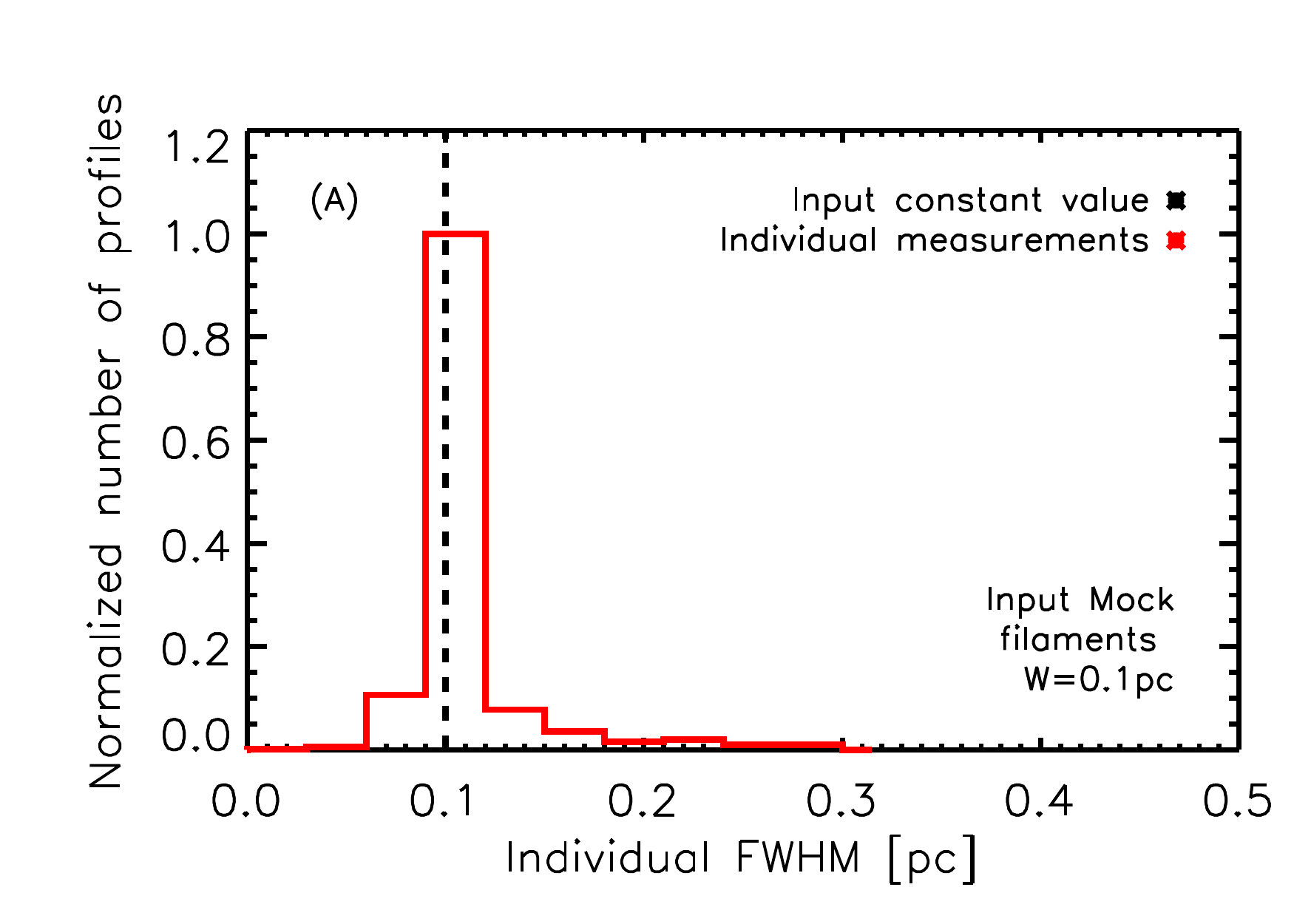}
      \hspace{-1.5cm}
   \includegraphics[angle=0]{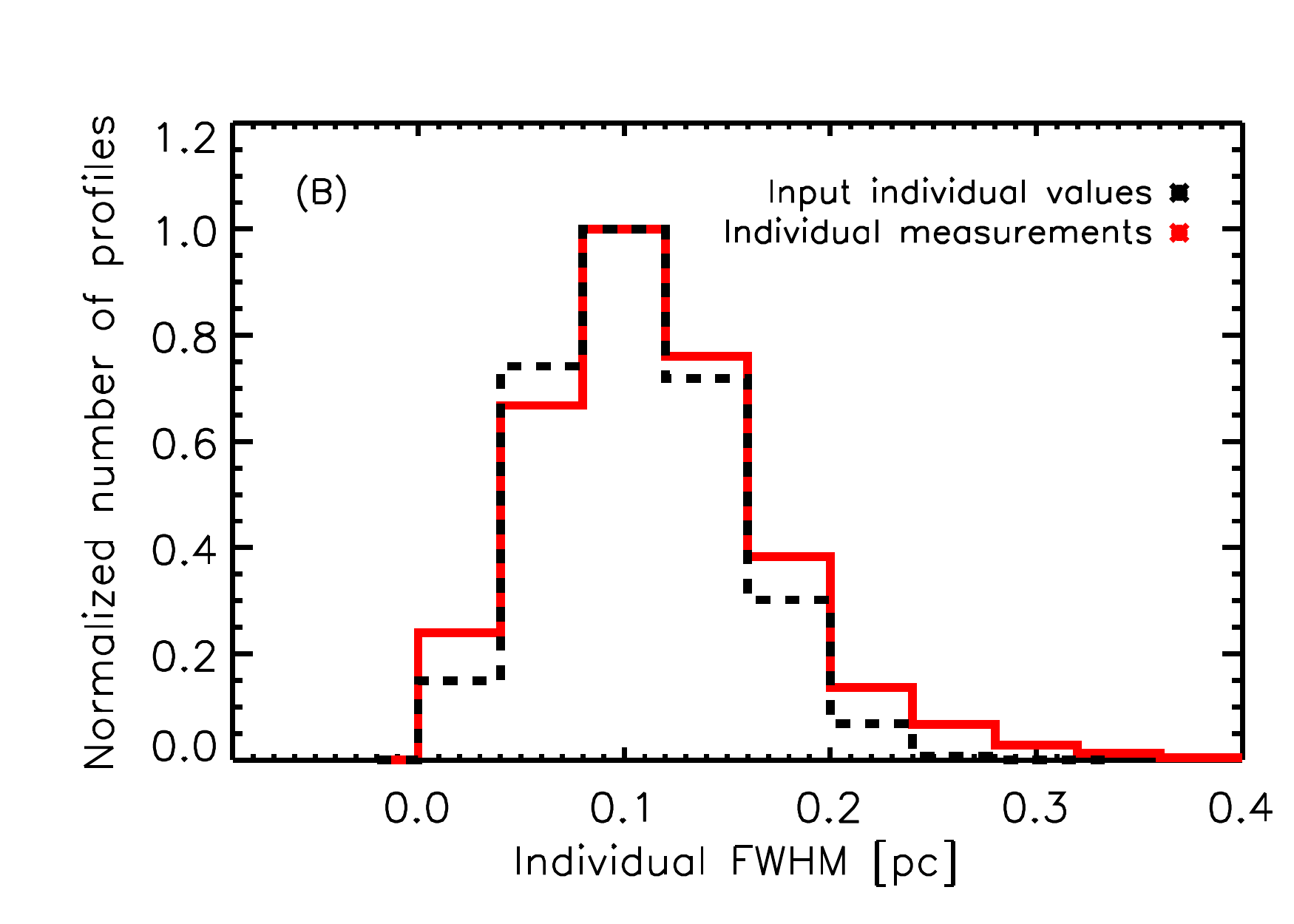}
    \hspace{-1.5cm}
   \includegraphics[angle=0]{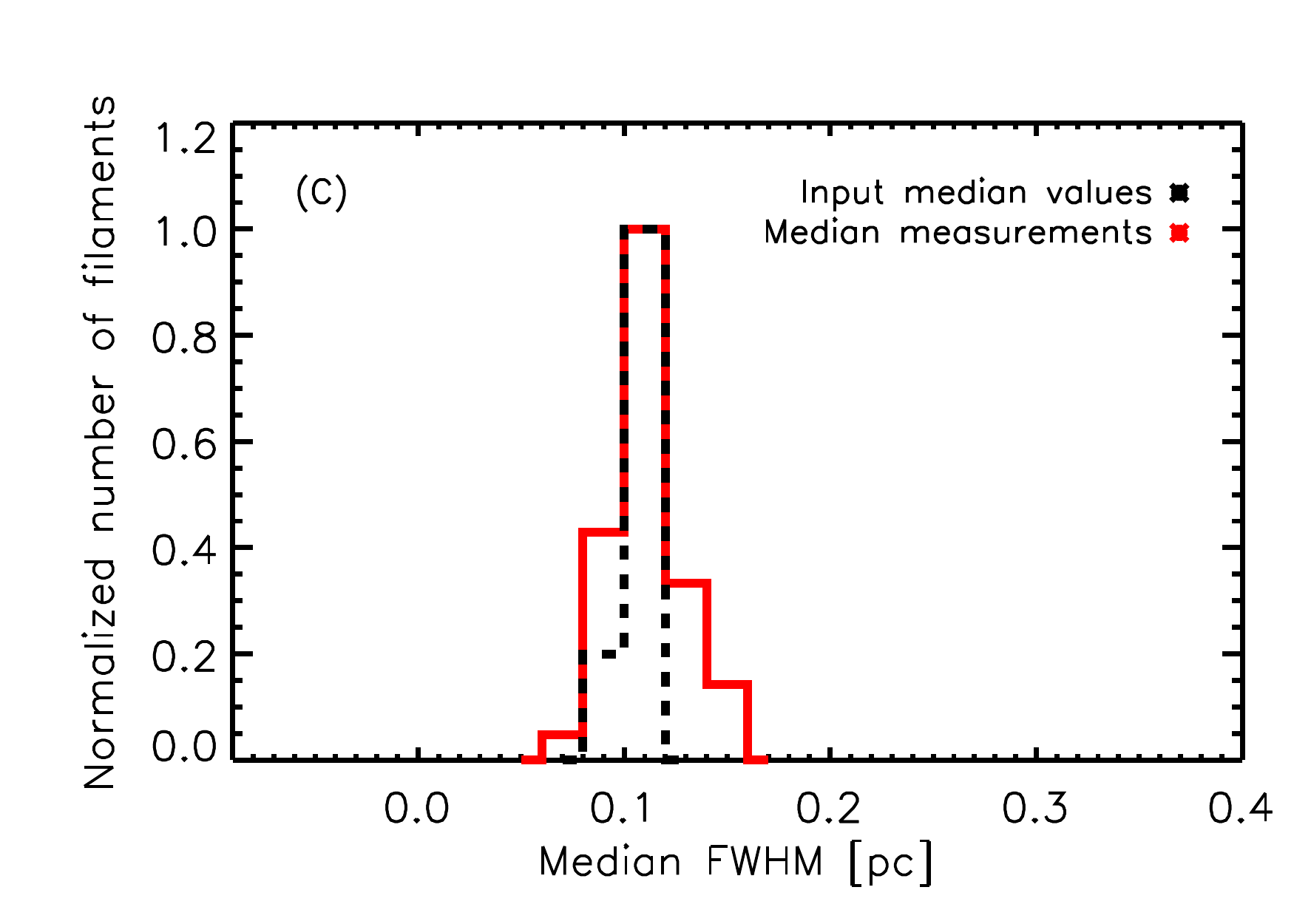}
}
\vspace{-0.5cm}
  \caption{{\rev Histograms of FWHM widths derived from Gaussian fits to the column density profiles of Gaussian-shaped 
synthetic filaments. The distributions of input and measured $FWHM$ widths are displayed in black and red, respectively.
The range radii used for the Gaussian fits was [0, $1.5hr$].   
Panel (A) shows the distribution of individual $FWHM$ widths measured along and on either side of the filament crests for input $FWHM$ width ($W=0.1$\,pc) constant along the filament length (same as Fig.\,\ref{histo_width_mock}A).
Panel (B) shows the distribution of individual $FWHM$ widths measured along and on either side of the filament crests, for  input individual $FWHM$ widths with a Gaussian  distribution  centered at  0.1\,pc with a standard deviation of $0.05$\,pc. 
Panel (C) shows the distribution of median $FWHM$ widths ``averaged'' over  each filament crest for the same input $FWHM$ distribution as in   panel (B).
        }         } 
  \label{histo_GaussW_mock}
    \end{figure*}

\subsection{Synthetic filaments with Gaussian input radial column density profiles}\label{App2a}

{\rev As illustrated in Fig.\,\ref{fig:Multiprof} (panels A and B), some filaments are observed to have 
radial column density profiles that are reasonably well described by a Gaussian profile. 
To discuss the accuracy of deriving the inner widths of such filaments,  
} 
we generated a number of synthetic maps by distributing 
identical Gaussian-shaped mock filaments 
with aspect ratio $AR=10$ and column density contrast $C^0=1$,
within the background column density map of Fig.\,\ref{SyntMap}a. {\rev The effective spatial resolution of all synthetic maps was 0.023\,pc (i.e., $\sim 18\arcsec $ at 260\,pc).}

{\rev We used four  distributions of Gaussian-shaped synthetic filaments:}

\noindent
a) All mock filaments had the same input $FWHM$ width, {\rev  uniform along each filament crest (Fig.\,\ref{histo_width_mock}A and Fig.\,\ref{histo_GaussW_mock}A).}

{\rev
\noindent 
b) The sample of mock filaments had a power-law distribution of $FWHM$ widths between 0.03\,pc and 0.25\,pc, {\Newrev  and a slope of $-2$ in 
$\Delta N/\Delta {\rm log}(FWHM)$}
(180 input mock filaments were used in this case -- Fig.\,\ref{histo_width_mock}B).  }

\noindent
c) The number of mock filaments per linear bin of width was constant in the range $0.03\, {\rm pc} \le FWHM \le 0.4\, {\rm pc}$. i.e., flat distribution of input widths,  {\rev (Fig.\,\ref{histo_width_mock}C)}.

{\rev
\noindent
d) The mock filaments had a Gaussian distribution of $FWHM$ widths along their crests, with a mean value of 0.1\,pc and a standard deviation of $0.05$\,pc (Fig.\,\ref{histo_GaussW_mock}B and Fig.\,\ref{histo_GaussW_mock}C).}

{\rev Figure\,\ref{histo_width_mock} shows the distribution of measured $FWHM$ widths derived from Gaussian fitting to the radial column density profiles 
of Gaussian-shaped mock filaments 
for three input distributions of filament $FWHM$ widths (constant, power-law, and flat, respectively).}
A peaked distribution of measured $FWHM$ widths  was obtained solely when the input mock filaments had {\rev the same  $FWHM$ value constant along their crest (Fig.\,\ref{histo_width_mock}A)}. 
Likewise, 
{\rev a  power-law} distribution of measured $FWHM$  widths was obtained only when the input mock filaments had  {\rev a power-law}  distribution of $FWHM$ widths (Fig.\,\ref{histo_width_mock}B), 
and a flat  distribution of measured $FWHM$ widths  was obtained only when the input distribution was flat (Fig.\,\ref{histo_width_mock}C). 

{\rev 
We can also see that reliable  filament widths can be measured down to $\sim0.03$\,pc, corresponding to the smallest input width considered,  
close to the 0.023\,pc resolution of the synthetic column density map. 
The slope of the input power-law distribution of mock filament widths was recovered with an accuracy better than $\sim10\%$  (for all fitting ranges). 
}

  {\rev Figures\,\ref{histo_GaussW_mock}A shows the distribution of individual $FWHM$ widths measured  along and on either  sides of the filament crests for an input  width  of 0.1\,pc constant along the crests (same as Fig.\,\ref{histo_width_mock}A). The median value of the measured widths is consistent with the input constant width, but a tail of larger values can also be seen on the distribution. These excursions from the constant input width  may be attributed to bad measurements and/or spurious structures. }

{\rev Figures\,\ref{histo_GaussW_mock}B and \ref{histo_GaussW_mock}C show the results of Gaussian width measurements  
when the input $FWHM$ widths are not strictly  constant along each filament crest, 
but have a Gaussian distribution with a mean value of 0.1\,pc and a standard deviation of 0.05\,pc. 
Figure\,\ref{histo_GaussW_mock}B  shows the distribution of individual  $FWHM$ widths derived from Gaussian fitting to the individual profiles  along and on either side of the filament crests,  
while Fig.\,\ref{histo_GaussW_mock}C 
shows the distribution of median  $FWHM$ values derived after ``averaging'' the independent measurements obtained along and on either side of each filament crest 
(see Sect.\,\ref{RadProf} and Sect.\,\ref{FilBack}). 
The derived distributions of individual $FWHM$  widths and of median (crest-averaged) $FWHM$  widths 
are both consistent with the corresponding input  distributions. 
We also note that the standard deviation of the derived distribution of median $FWHM$ widths  is larger 
when the input individual $FWHM$ widths have a Gaussian distribution about a mean value (Fig.\,\ref{histo_GaussW_mock}C) 
than when the input $FWHM$ width is constant along each filament crest (Fig.\,\ref{histo_width_mock}A), i.e., $0.02$\,pc and  $<0.01$\,pc, respectively.
}

The $FWHM$ widths derived from Gaussian fitting to the profiles of Gaussian-shaped synthetic filaments 
are {\it not} affected by the range of radii used for fitting the radial column density profiles. 
As expected and as already mentioned in Sect.\,\ref{Linkfit}, however, the fitting range does matter when Gaussian fits are used to estimate the $FWHM$ widths 
of filaments with Plummer-like input column density profiles. {\rev This is discussed in the next section}.

\subsection{Synthetic filaments with Plummer input radial density profiles}\label{App2b}

The 
{\rev logarithmic} 
radial column density profiles of many observed filaments are characterized by a flat inner plateau up to a radius $R_{\rm flat}$ 
and non-Gaussian wings for $r>>R_{\rm flat}$  {\rev (see, e.g., Fig.\,\ref{fig:Multiprof}).} 
Using a Plummer-like model profile (see Eq.\,\ref{Eq:EqPlum}), 
one can in principle reproduce the behavior of the radial  column density distribution for both $r\le R_{\rm flat}$ and $ r>>R_{\rm flat}$, 
which is not possible with a Gaussian fit.

\begin{figure*}
   \centering
      \resizebox{16cm}{!}{
\includegraphics[angle=0]{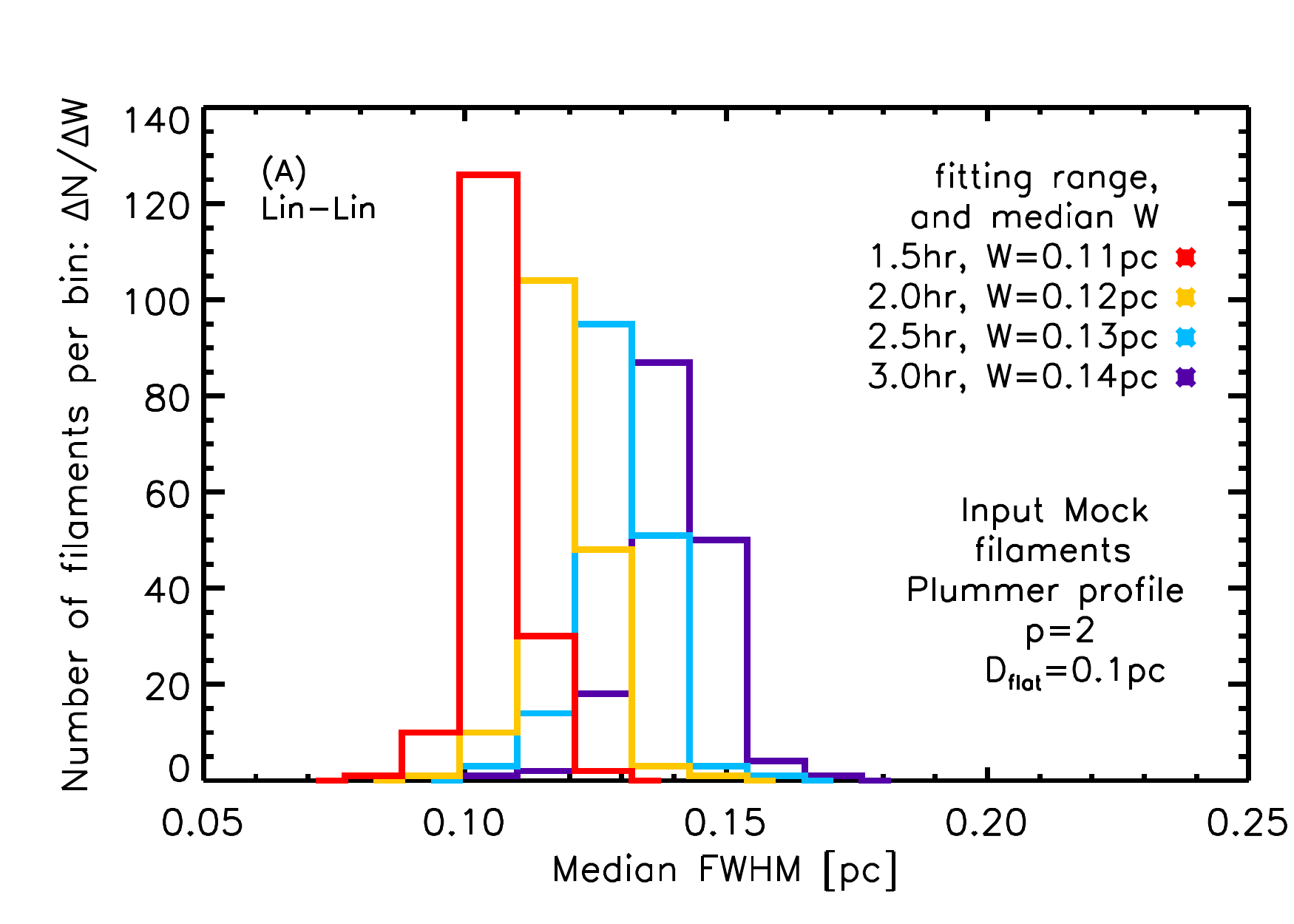}
\includegraphics[angle=0]{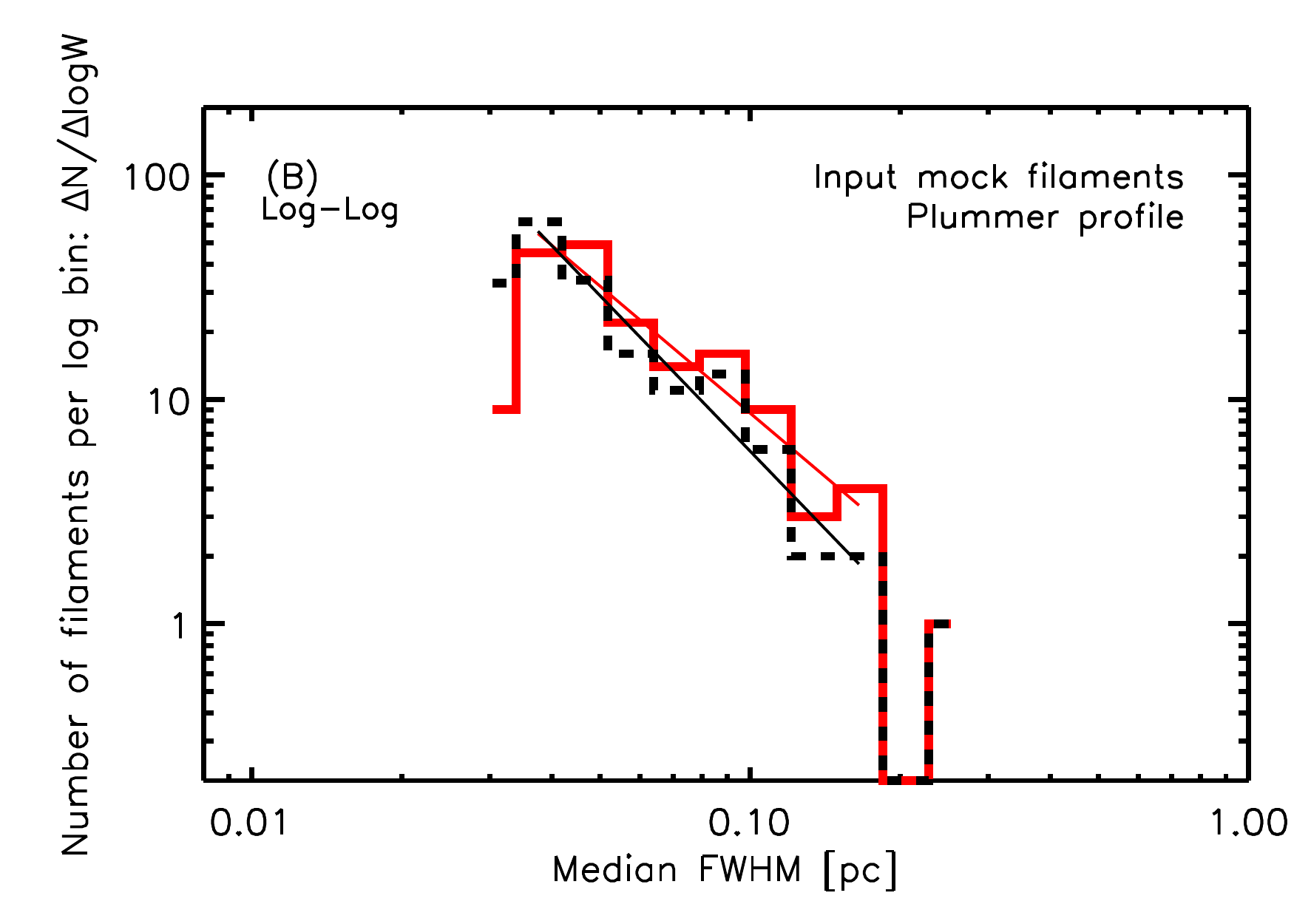}}
  \caption{ {\rev Histograms of $FWHM$ widths derived from Gaussian fitting to the radial column density profiles of synthetic filaments  with Plummer radial profiles and input contrast $C^0$=1. 
 In panel {\bf (A)}, all input 180 mock filaments had Plummer radial profiles with $2\rflat=D_{\rm flat}=0.1$\,pc and $p=2$.
  Results  
 are presented for four choices of the fitting range, [0,$1.5hr$],  [0,$2.0hr$], [0,$2.5hr$], and [0,$3.0hr$], color coded as shown at the top right of the plot (cf. Sect.\,\ref{Linkfit}).
Note how the measured distribution of filament widths depends on the fitting range (the median filament width derived for each fitting range 
is given at the top right of the plot). 
In panel {\bf (B)},  the input 180 mock filaments had Plummer radial profiles with a power-law distribution of $D_{\rm flat}$ diameters between 0.03 and 0.25\,pc 
 (same distribution as the distribution of $FWHM$ widths in Fig.\,\ref{histo_width_mock}B)  
 and a Gaussian distribution of $p$ values with a mean of 2 and a standard deviation of 0.3. 
 The $FWHM$ widths were derived from Gaussian fits to the radial column density profiles with a fitting range of [0,1.5$hr$]. 
 The solid black and red straight lines show  power-law fits to the input and measured distributions of widths, respectively.
     } }          
  \label{histo_widthPlum_mock}
    \end{figure*}

{\rev  In this section, we present  measurement tests  to assess the reliability 
of deriving the inner widths of filaments with Plummer-like radial column density profiles, using both Gaussian and Plummer-like function fits (as described in Sect.\,\ref{Gaussfit} and Sect.\,\ref{Plumfit}, respectively). 
For this purpose, several 
synthetic maps 
were constructed by distributing, within the background column density map of Fig.\,\ref{SyntMap}a  at a spatial resolution of 0.023\,pc,
a population of
Plummer-shaped mock filaments with 
input column density contrasts of  $C^0=1$ and $C^0=0.5$, and 
various input  distributions of  $D_{\rm flat}=2R_{\rm flat}$ and $p$ values.}

   \begin{figure*}
   \centering
      \resizebox{8cm}{!}{
\includegraphics[angle=0]{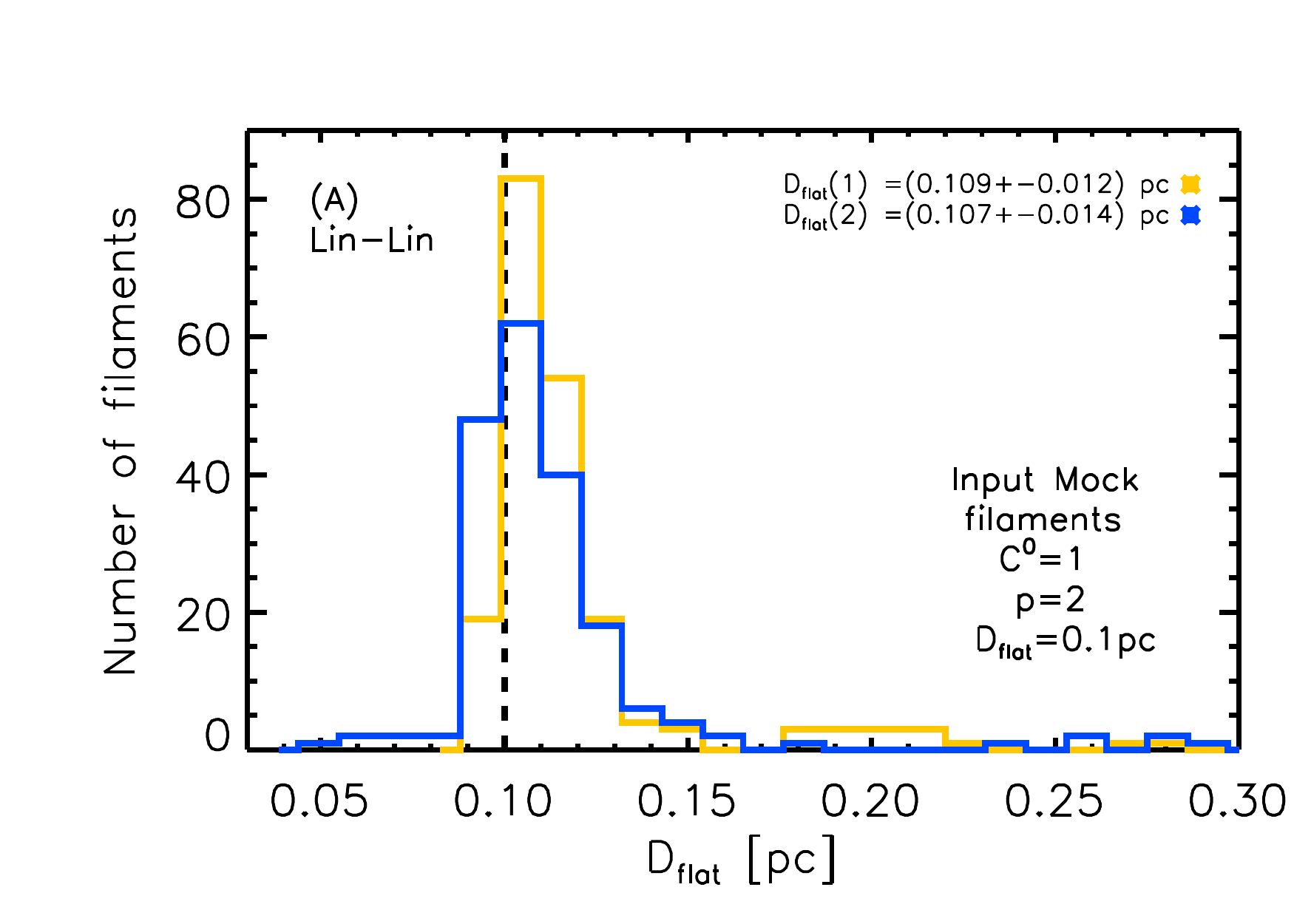}}
     \resizebox{8cm}{!}{
\includegraphics[angle=0]{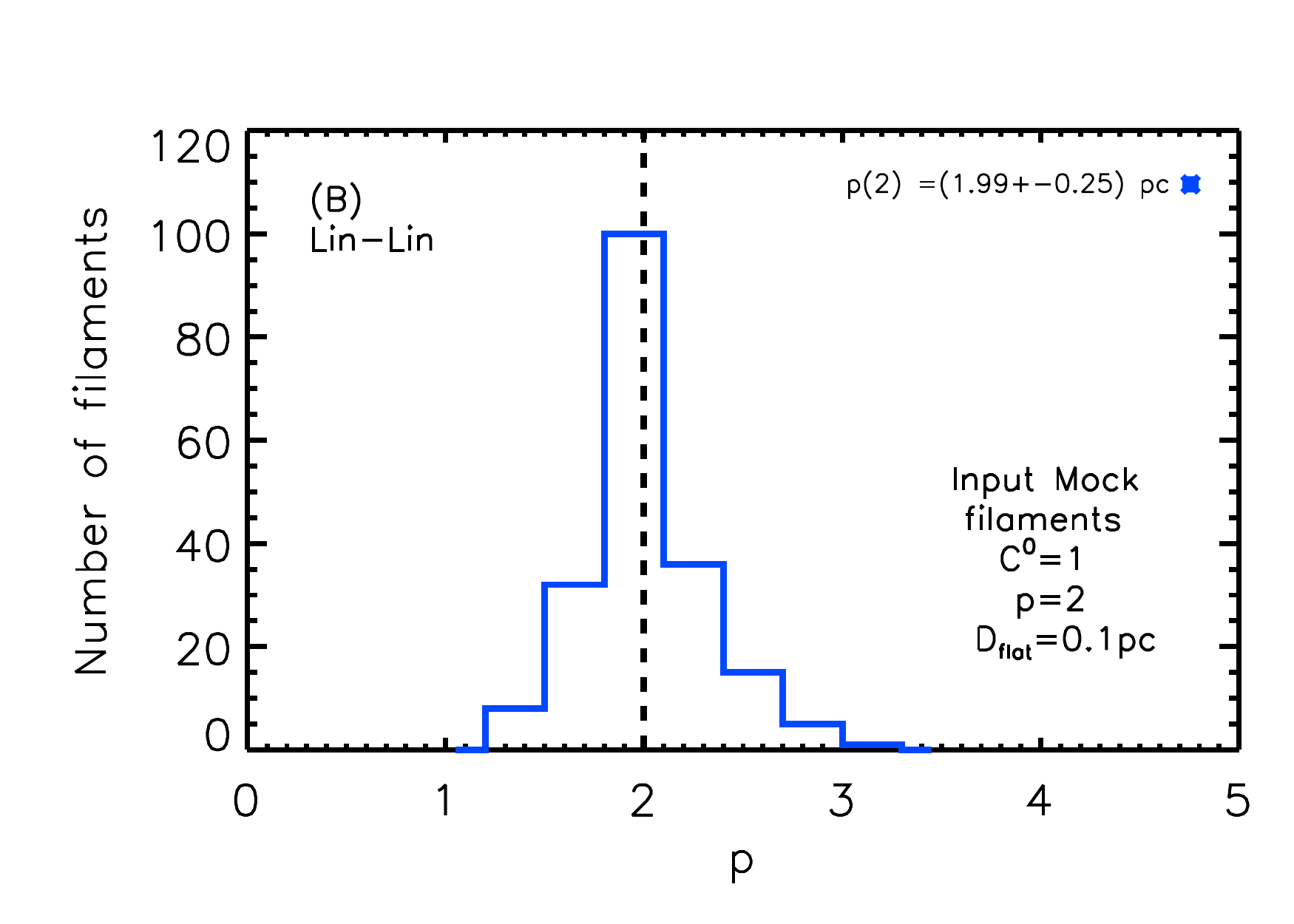}}
  \caption{Results of measurements performed to test the reliability of  Plummer  fits 
  to the radial column density profiles of Plummer-shaped synthetic filaments with fixed input central column density contrast 
$C^0=1$, inner diameter $D_{\rm flat}=0.1$\,pc, and power-law index $p=2$. The input $D_{\rm flat}$ and $p$ values are marked by vertical dashed lines on the plots. 
The synthetic filaments are the same as those used in Fig.\,\ref{histo_widthPlum_mock}.   
Two sets of results are provided for $D_{\rm flat}$ in panel {\bf (A)}, depending on whether the power-law index $p$ 
of the Plummer model profile was fixed to $p=2$ [(1), yellow histogram] or left as a free parameter [(2), blue histogram]. 
The blue histogram of fitted $p$ values in  panel {\bf (B)} corresponds to the blue distribution of $D_{\rm flat}$(2) values in panel {\bf (A)}.
The median values of the derived parameters as well as their equivalent standard deviations (scaled from the measured IQRs)
are provided at the top right of the panels.}          
  \label{fig:histo_mockParamPlum_Cont1}
    \end{figure*}

   \begin{figure*}
   \centering
      \resizebox{15cm}{!}{
\includegraphics[angle=0]{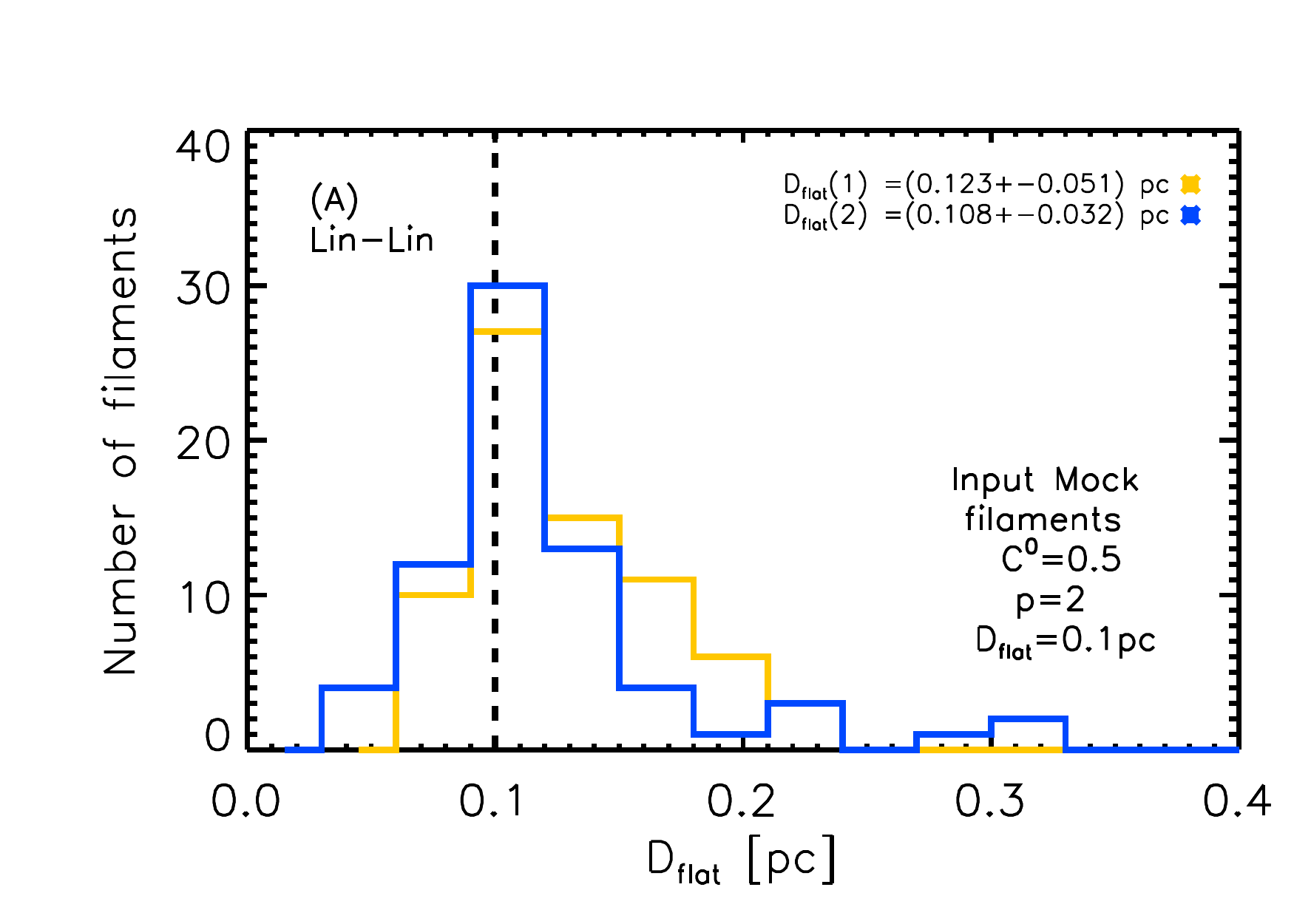}
\includegraphics[angle=0]{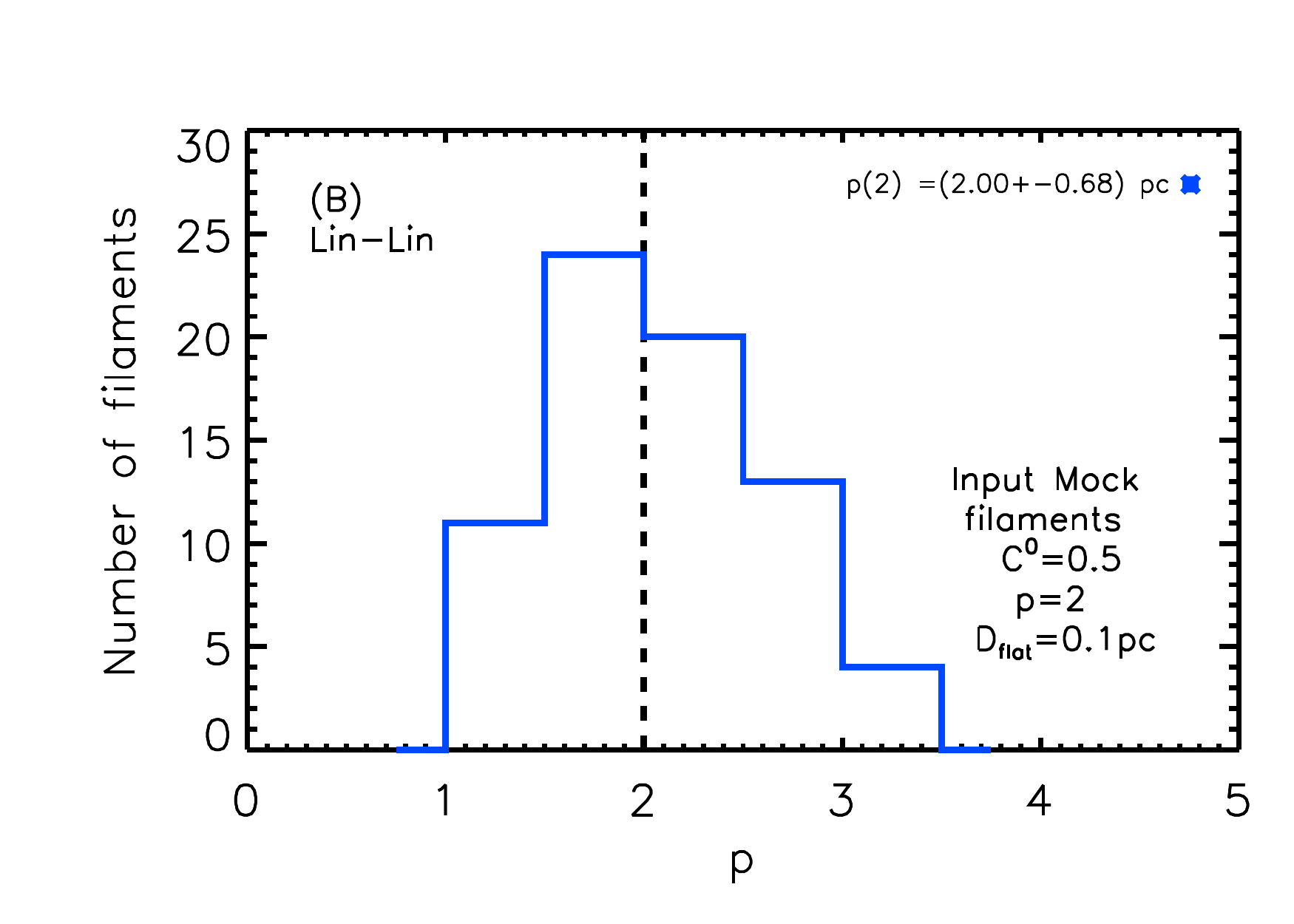}}
  \caption{{\rev Same as Fig.\,\ref{fig:histo_mockParamPlum_Cont1} for Plummer-shaped synthetic filaments with fixed input central column density contrast $C^0=0.5$. 
The input Plummer parameters were fixed to $D_{\rm flat}=0.1$\,pc and $p=2$, as marked by vertical dashed lines on the plots.
} 
}         
  \label{fig:histo_mockParamPlum_Cont05}
    \end{figure*}
    
{\rev
We used the following distributions of Plummer-like synthetic filaments:
\noindent
a) All mock filaments had the same  $\rflat$ and $p$ values, constant along each filament crest. Three sets of synthetic maps were considered, 
with mock filaments having $D_{\rm flat}=0.1$\,pc  and $p=1.5$, 2, and 3, respectively (Fig.\,\ref{histo_widthPlum_mock}A, Fig.\,\ref{fig:histo_mockParamPlum_Cont1}, Fig.\,\ref{fig:histo_mockParamPlum_Cont05}, Fig.\,\ref{fig:histo_mockParamPlum23}).

\noindent
b) All mock filaments had the same power-law index $p=2$, but a flat distribution of $D_{\rm flat}$  (i.e., a constant number of mock filaments per linear bin of $D_{\rm flat}$) in the range $0.03\, {\rm pc} \le \rflat \le 0.34\, {\rm pc}$ (Fig.\,\ref{fig:histo_mockParamPlumFlatR}A).

\noindent
c) All mock filaments had the same  $D_{\rm flat}=0.1$\,pc diameter, but a flat distribution of $p$ values (i.e., a constant number of mock filaments per linear bin of $p$) in the range $1.5\le p \le3.2$ (Fig.\,\ref{fig:histo_mockParamPlumFlatp}). 

\noindent
d) The mock filaments had a power-law distribution of $D_{\rm flat}$ diameters between 0.03\,pc and 0.25\,pc, with a 
 {\Newrev power-law index of  $-2$ (in $\Delta N/\Delta {\rm log}(D_{\rm flat})$})  and a Gaussian distribution of $p$ values 
with a mean of 2 and a standard deviation of 0.3 (Fig.\,\ref{histo_widthPlum_mock}B, Fig.\,\ref{fig:histo_mockParamPlumFlatR}B). 

\noindent
e) The mock filaments had the same $p=2$ index and a power-law distribution of  individual $D_{\rm flat}$ diameters along their crests with a %
 {\Newrev power-law index of  $-2$ (in $\Delta N/\Delta {\rm log}(D_{\rm flat})$})  between 0.022\,pc and 0.19\,pc (Fig.\,\ref{fig:histo_mockParamPlum_plw}).

}

 {\rev 
Figure\,\ref{histo_widthPlum_mock}A  shows the distributions of $FWHM$ widths derived from Gaussian fits to the radial column density profiles of 
Plummer-shaped mock filaments with uniform input $D_{\rm flat}=0.1$\,pc and  $p=2$ parameters.   
For the same input $D_{\rm flat}$ and $p$ values, the derived $FWHM$ widths increase for increasing fitting range, e.g., from [0,1.5$hr$] to [0,3$hr$] 
(cf. also, Sect.\,\ref{Linkfit} and  Fig.\,\ref{ProfFit}). 
This is also the case when the input mock filaments have $p=1.5$ and $p=3$ (not illustrated here with a figure). 
For given fitting range and  input $D_{\rm flat}$ diameter, 
the measured $FWHM$ widths from Gaussian fits to Plummer-like input column density profiles 
decrease for increasing  $p$: e.g., for a fitting range of $0\le r\le1.5hr$, $FWHM=1.4\,(D_{\rm flat}^{p=1.5})$,  $FWHM=1.1\,(D_{\rm flat}^{p=2})$, and $FWHM=0.8\,(D_{\rm flat}^{p=3})$, 
when $p=1.5$, $p=2$, and $p=3$, respectively.   
The logarithmic slope of the power-law profile, $p$,  also influences the derived values of the  half-power diameter, $hd$. 
Our tests show that,  for the same input $D_{\rm flat}$ radius, the  median value of the $hd$  distribution  decreases when the input $p$ index increases: 
$hd=1.8\,(D_{\rm flat}^{p=1.5})$,  $hd=1.4\,(D_{\rm flat}^{p=2})$, and $hd=0.98\,(D_{\rm flat}^{p=3})$, for $p=1.5$, $p=2$, and $p=3$, respectively. 
}

 {\rev Figure\,\ref{histo_widthPlum_mock}B shows the  distribution of $FWHM$ widths derived from Gaussian fitting to the radial profiles 
 of 180 Plummer-shaped mock filaments with a power-law distribution of input $D_{\rm flat}$ diameters between 0.03\,pc and 0.25\,pc 
 and a Gaussian distribution of $p$ values with a mean of 2 and a standard deviation of 0.3. 
For a fitting range of $0\le r\le1.5hr$, the derived distribution of  $FWHM$ widths is in  good agreement 
with the input distribution of filament inner widths down to 0.03\,pc, close to the spatial resolution of the synthetic background map. 
The logarithmic slope of the fitted power-law distribution of $FWHM$ widths is only $\sim15\%$ shallower than that of the input distribution of $D_{\rm flat}$ diameters.
  }

      \begin{figure*}
   \centering
      \resizebox{14.5cm}{!}{
\includegraphics[angle=0]{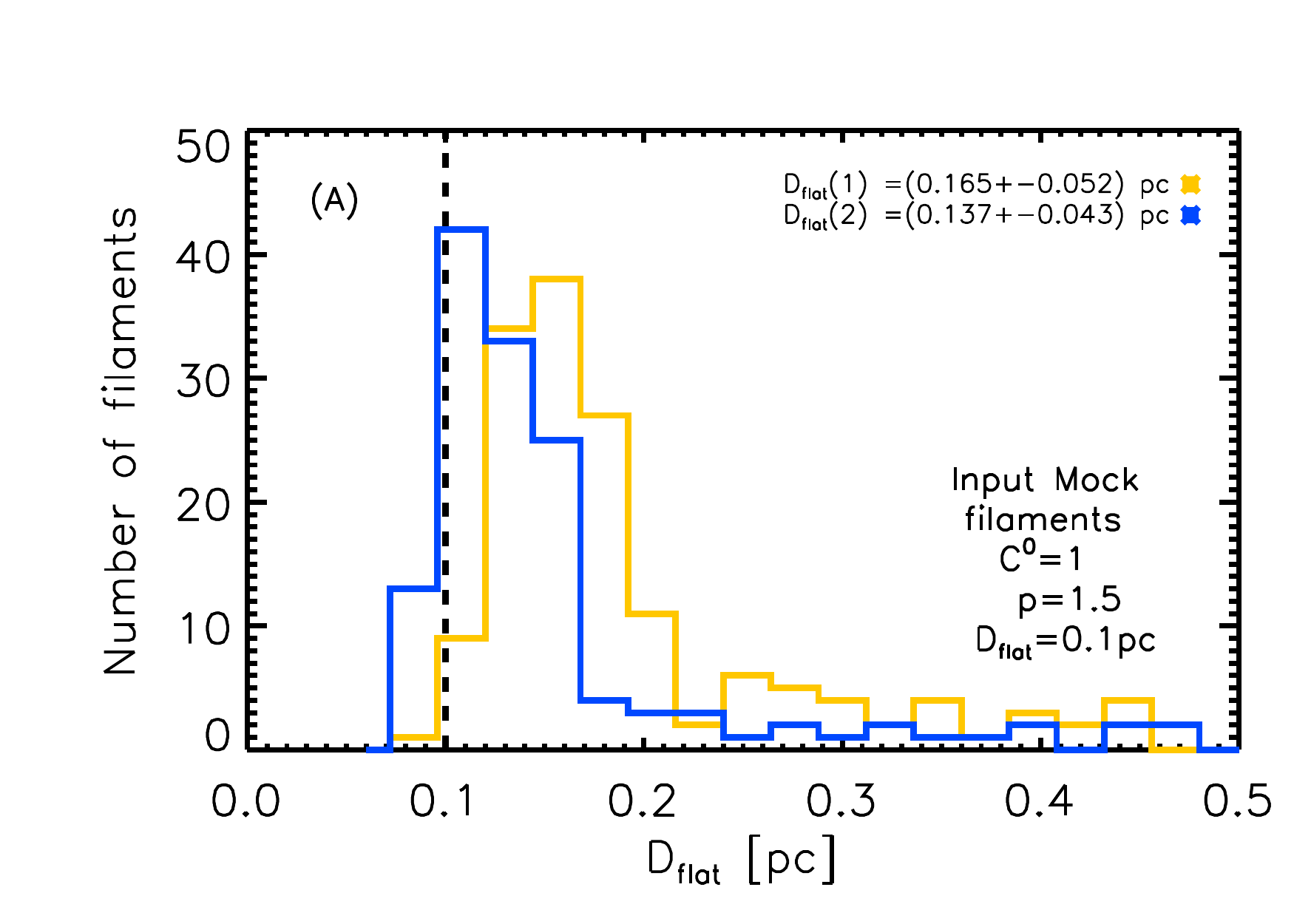}
\hspace{1.cm}
\includegraphics[angle=0]{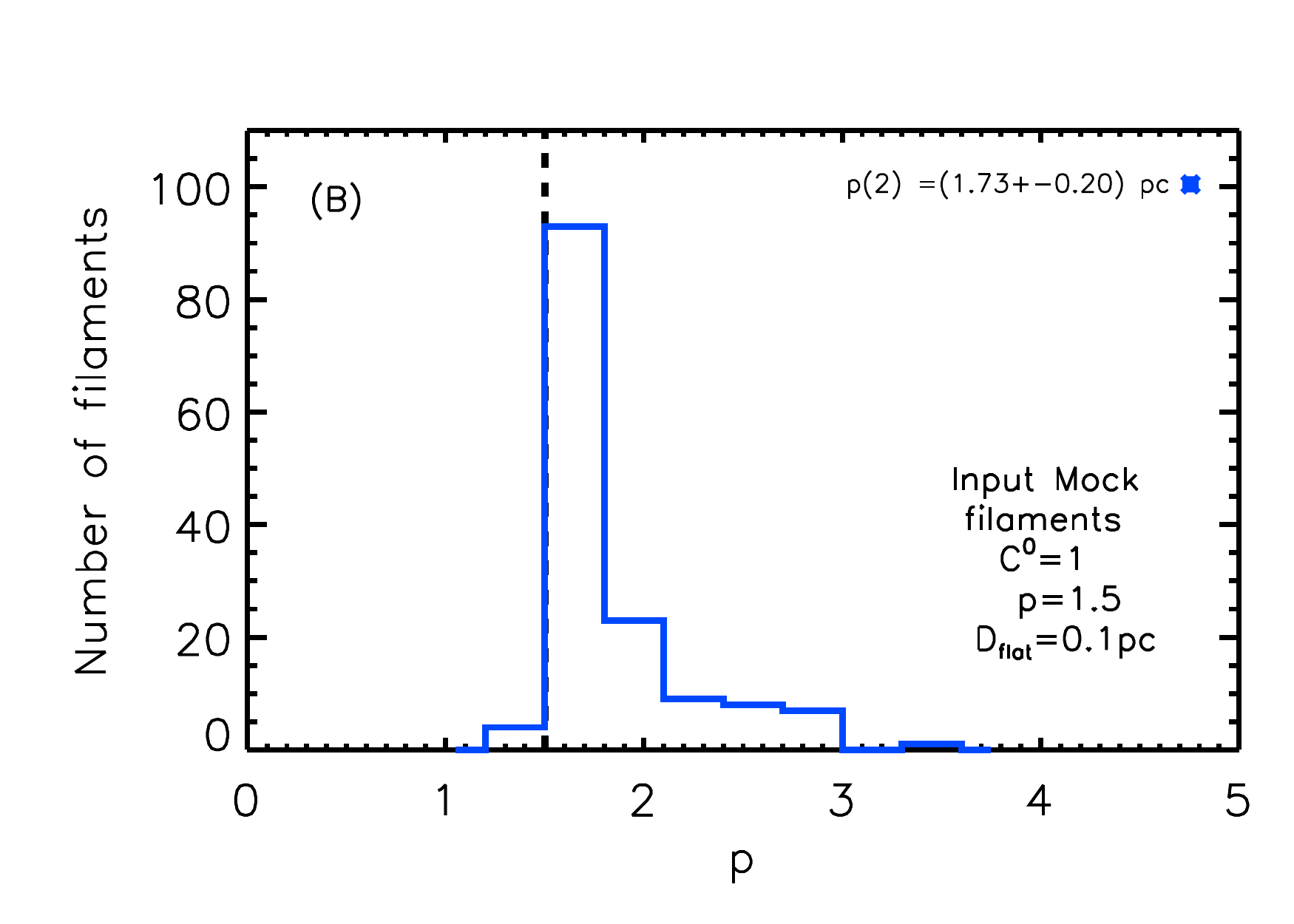}}
    \resizebox{14.5cm}{!}{\vspace{-1.cm}
\includegraphics[angle=0]{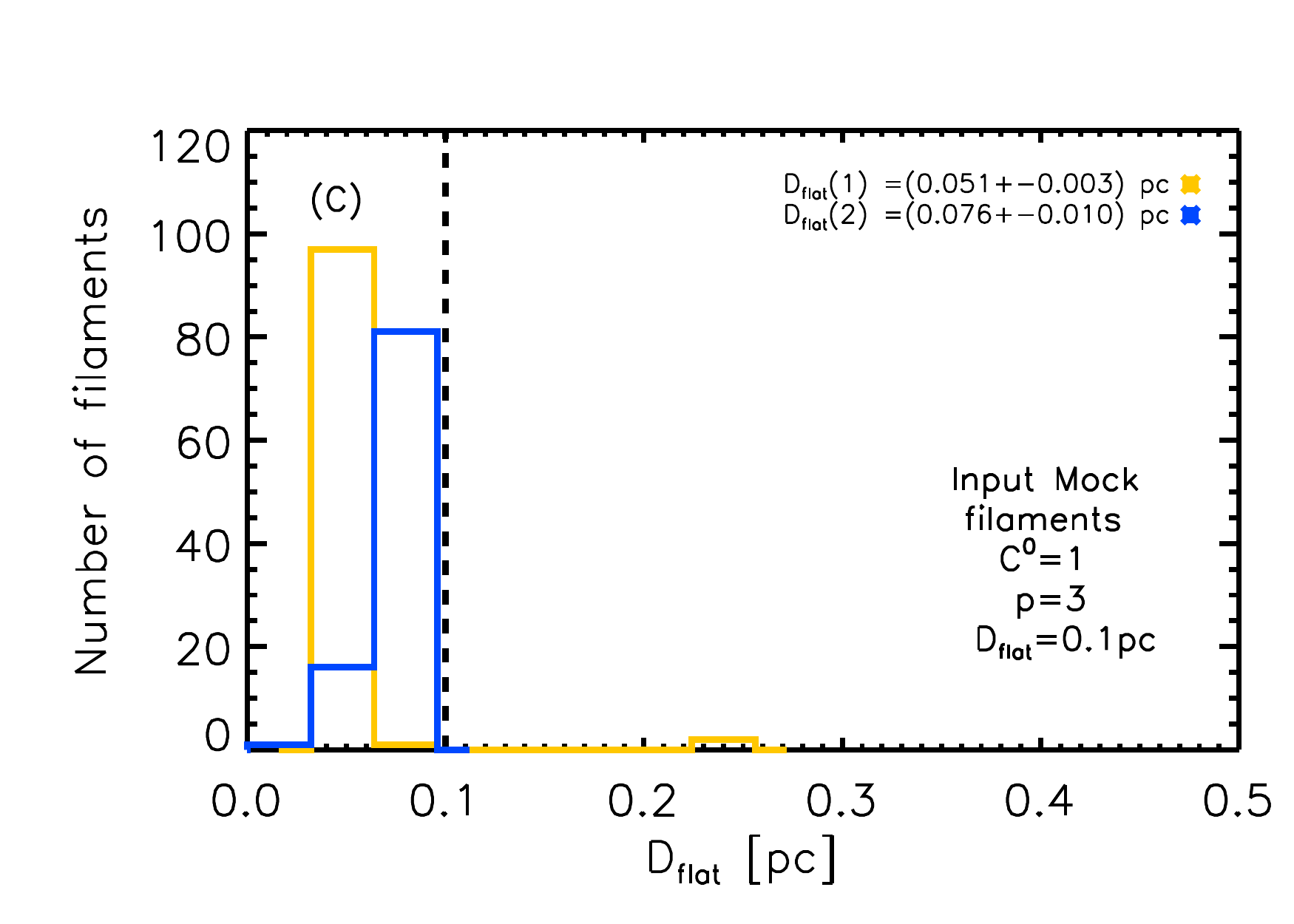}
\hspace{1.cm}
\includegraphics[angle=0]{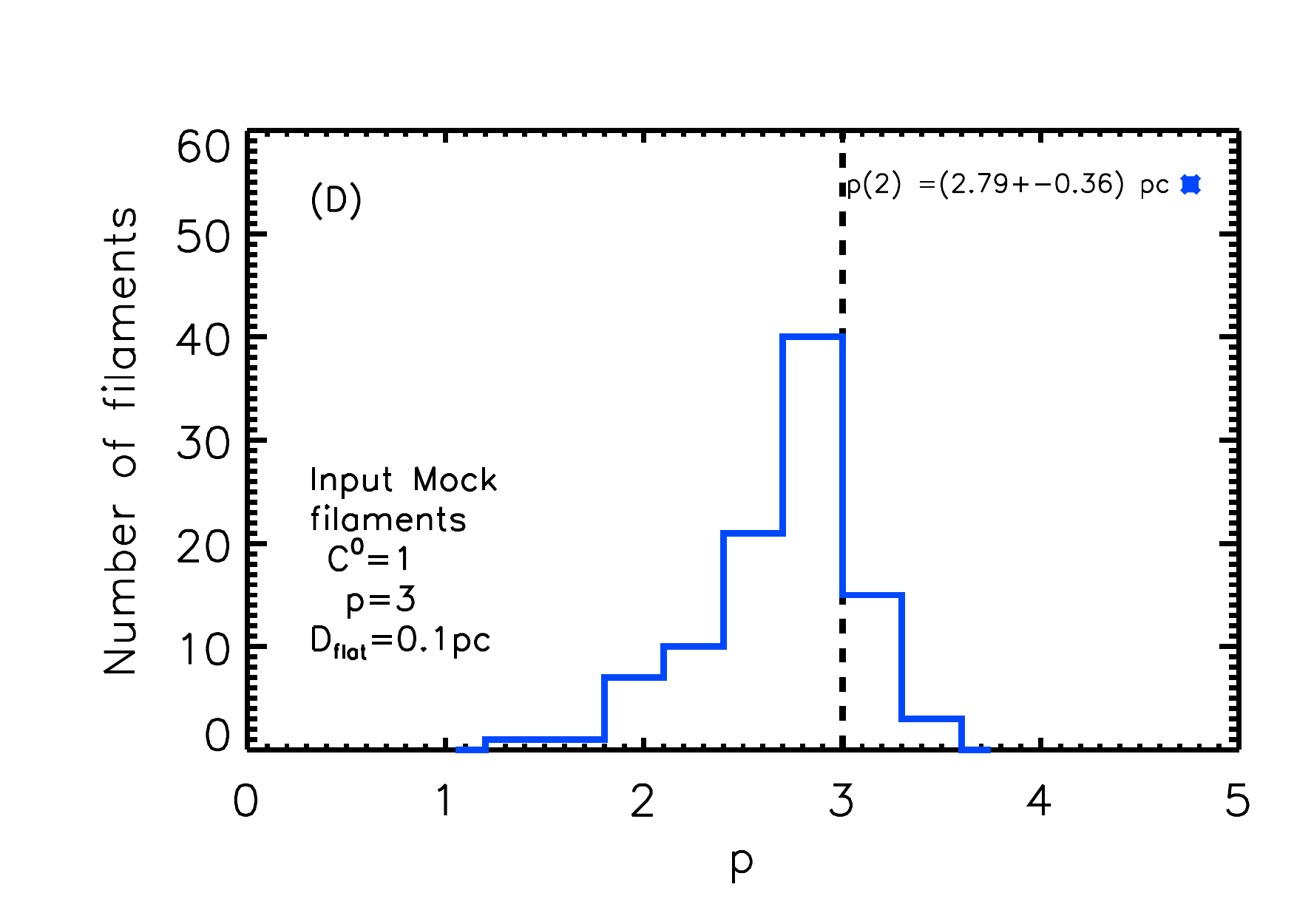}}
  \caption{{\rev Same as Fig.\,\ref{fig:histo_mockParamPlum_Cont1} for Plummer-shaped synthetic filaments with fixed input central column density contrast  $C^0=1$, inner diameter $D_{\rm flat}=0.1$\,pc, 
and power-law index $p=1.5$ [top panels, {\bf (A)} and {\bf (B)}] or $p=3$ [bottom panels, {\bf (C)} and {\bf (D)}].  The input $D_{\rm flat}$ and $p$ values are marked by vertical dashed lines in the plots.
The left panels, {\bf (A)} and {\bf (C)}, show histograms of measured $D_{\rm flat}$ values, 
while the right panels, {\bf (B)} and {\bf (D)}, show histograms of measured $p $ values. 
}   }       
  \label{fig:histo_mockParamPlum23}
    \end{figure*}

      \begin{figure*}
   \centering
      \resizebox{15cm}{!}{
      \hspace{-2.cm}
\includegraphics[angle=0]{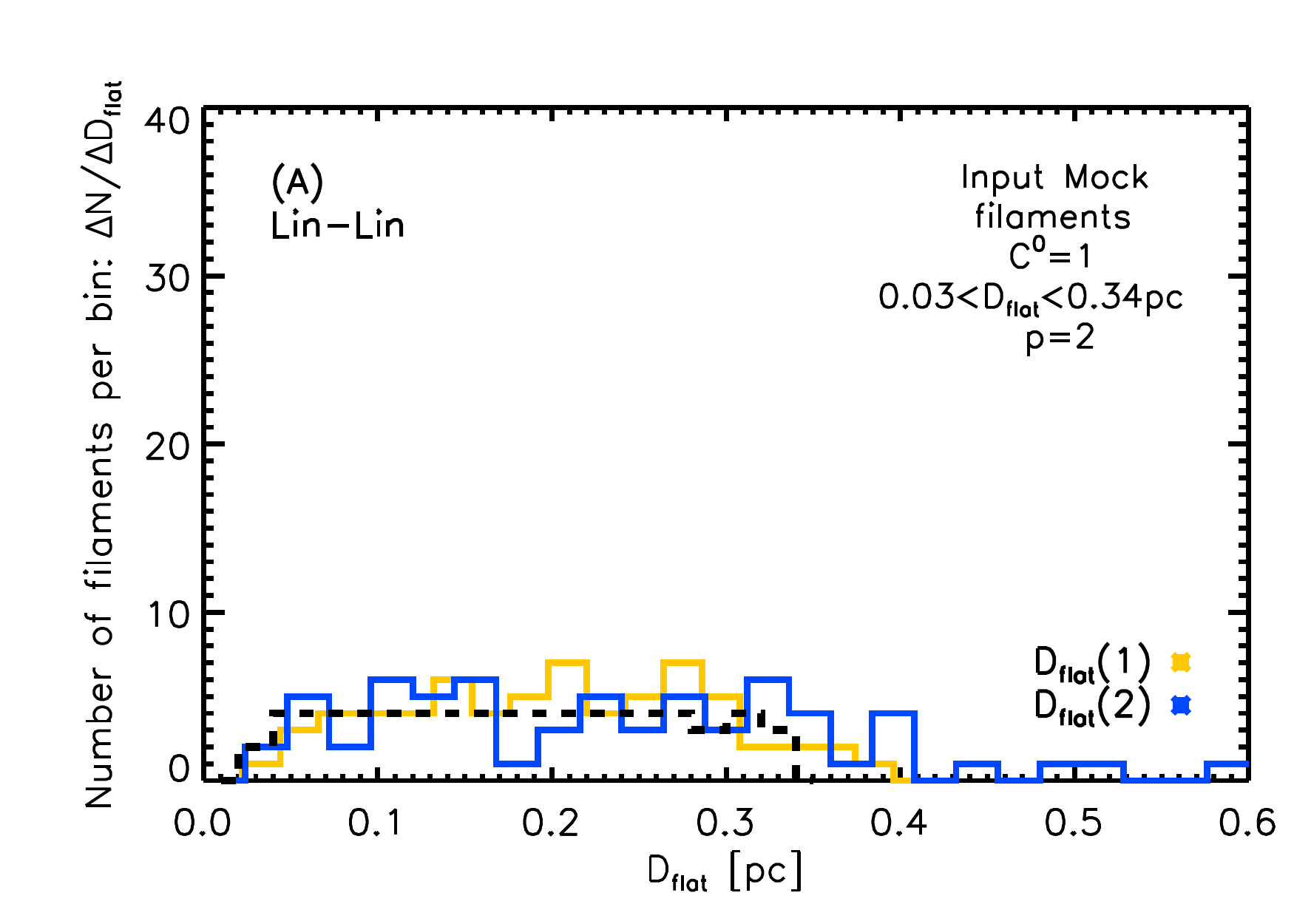}
\hspace{1.cm}
\includegraphics[angle=0]{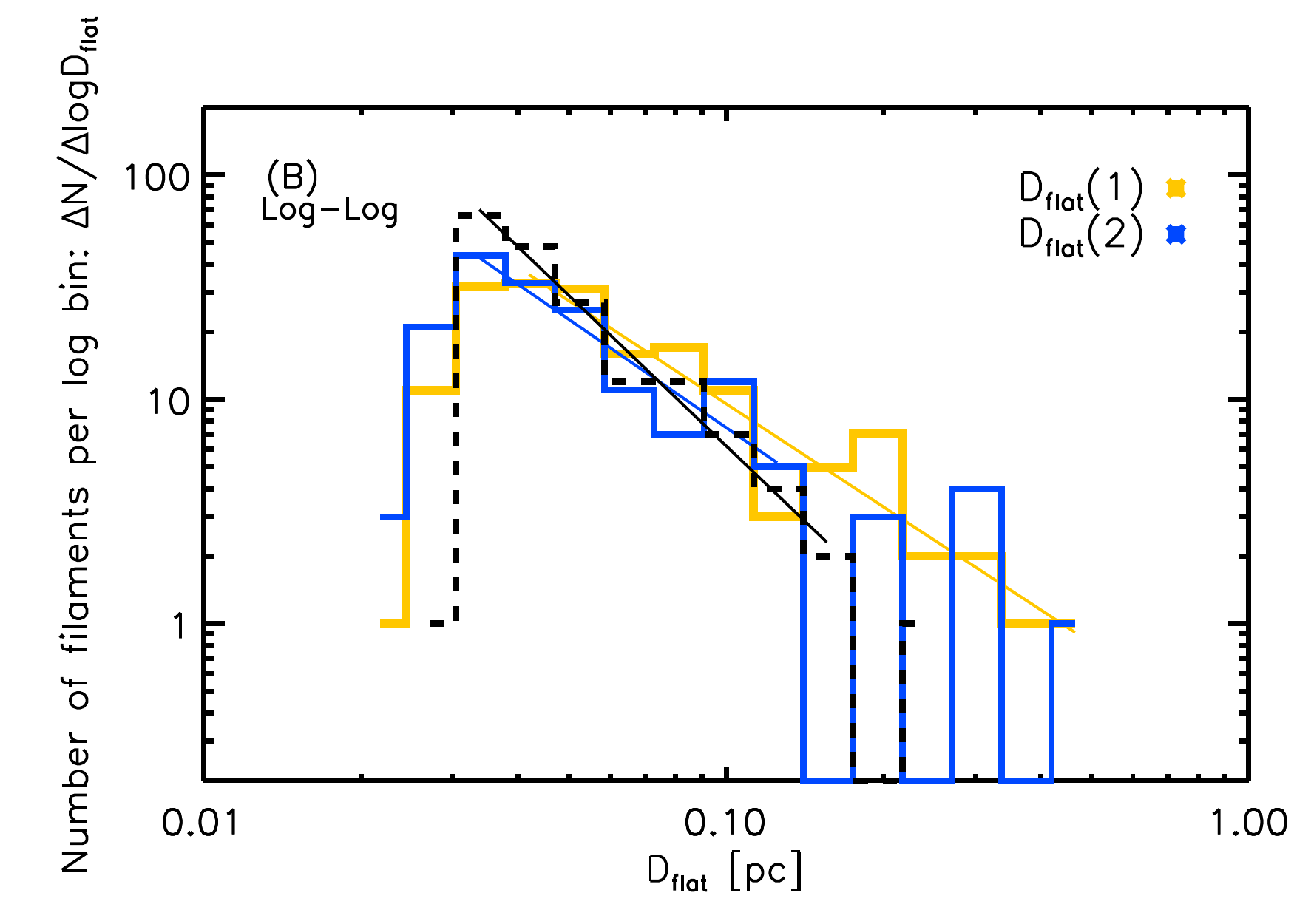}}
  \caption{{\rev Results of Plummer fits to Plummer-shaped  synthetic filaments 
 with a central column density contrast  $C^0=1$ and two distributions of $D_{\rm flat}$ inner diameters [flat in panel (A);  power law in (B)]. 
Two sets of $D_{\rm flat}$ measurements are provided  
depending on whether the power-law index $p$ 
of the Plummer model profile was fixed to $p=2$ [(1), yellow histogram] or left as a free parameter [(2), blue histogram]. 
 The input distribution of $D_{\rm flat}$ values are shown by the black dashed histogram in  both panels. 
 In the left panel {\bf (A)}, the input 
  mock filaments had  a fixed input  $p=2$ and a flat input distribution of $D_{\rm flat}$ between 0.03\,pc and 0.34\,pc. 
In the right panel {\bf (B)},  the input 
mock filaments had a power-law distribution of $D_{\rm flat}$ diameters between 0.03 and 0.25\,pc, similar to the distribution of $FWHM$ widths  in Fig.\,\ref{histo_width_mock}B\,and\,\ref{histo_widthPlum_mock}B, 
and a Gaussian distribution of $p$ values with a mean of 2 and a standard deviation of 0.3. 
The power-law slopes of the measured distributions of $D_{\rm flat}$ diameters (shown as yellow and blue straight  lines for the $D_{\rm flat}$(1) and $D_{\rm flat}$(2) estimates, respectively) 
are only $\sim30\%$ shallower than the slope of the input distribution (black straight line). 
  }     }  
  \label{fig:histo_mockParamPlumFlatR}
    \end{figure*}

    \begin{figure*}
   \centering
      \resizebox{14cm}{!}{
      \hspace{-2.cm}
\includegraphics[angle=0]{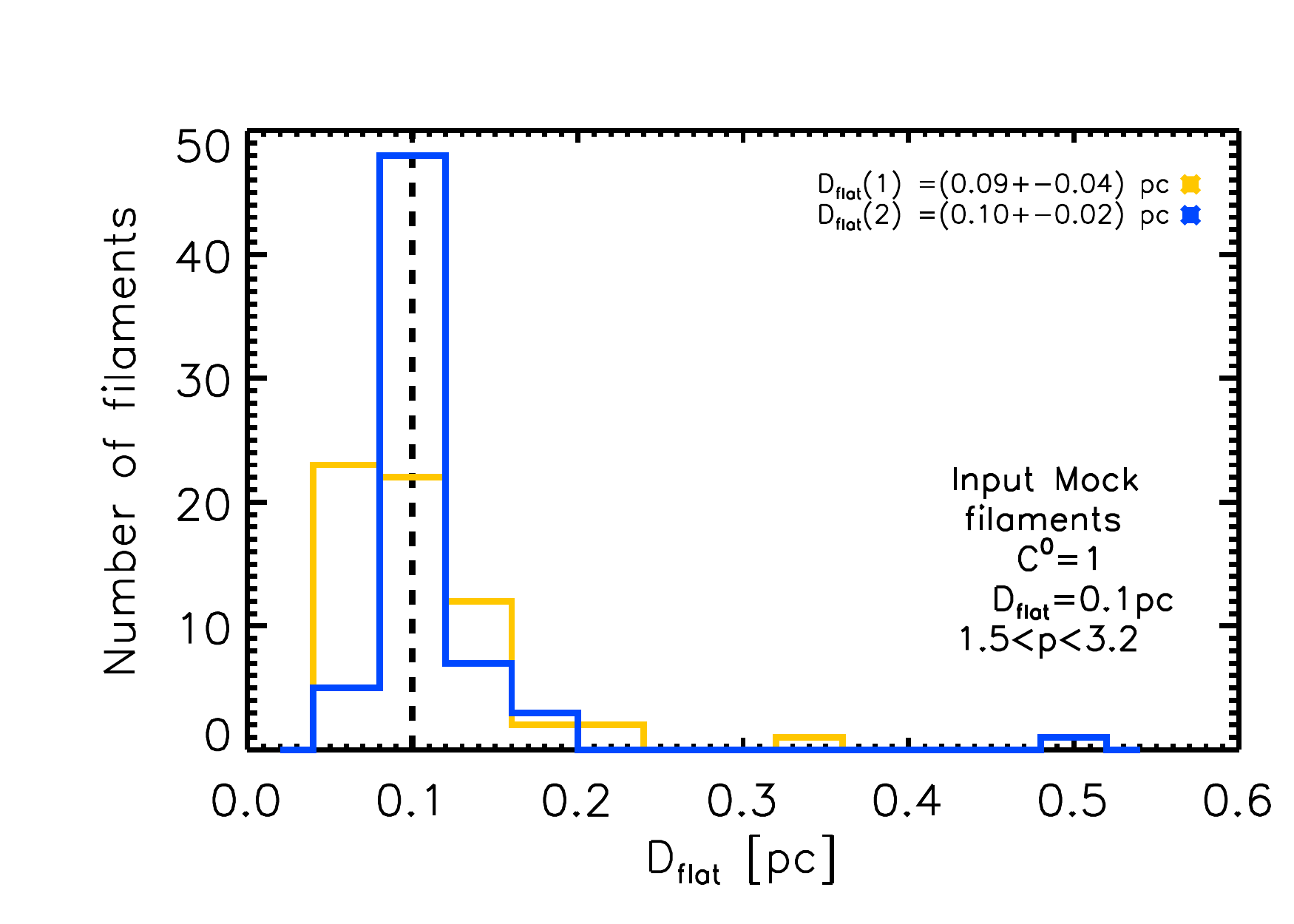}
\hspace{1.cm}
\includegraphics[angle=0]{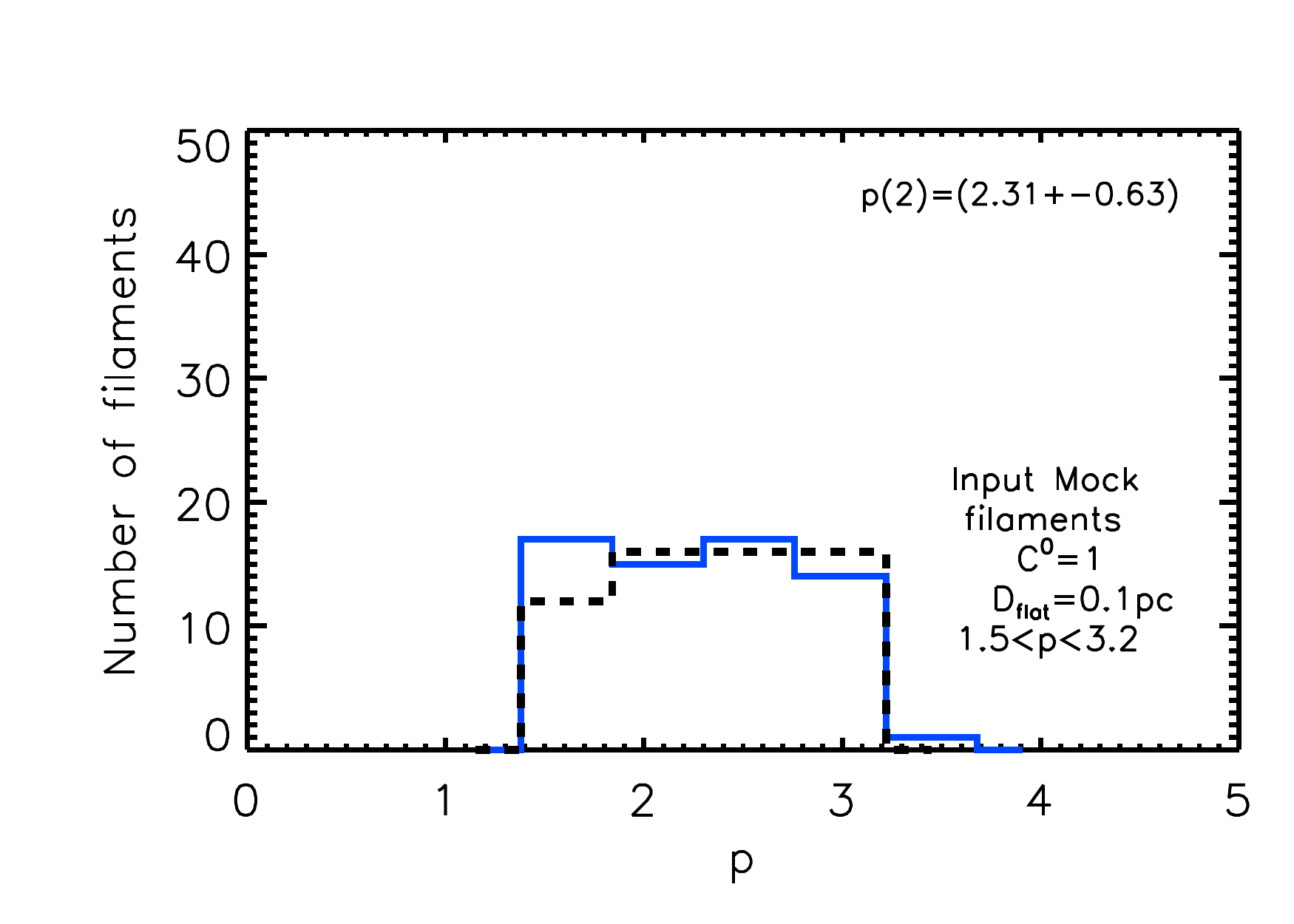}}
  \caption{{\rev Same as Fig.\,\ref{fig:histo_mockParamPlum_Cont1} for  Plummer-shaped synthetic filaments with a fixed central column density contrast  $C^0=1$, a fixed input $D_{\rm flat}=0.1$\,pc diameter 
 (dashed black straight line in the left panel),  and a flat distribution of input $p$ values  between 1.5 and 3.2 (dashed black histogram in the right panel).
  }     }  
  \label{fig:histo_mockParamPlumFlatp}
    \end{figure*}

  Figures\,\ref{fig:histo_mockParamPlum_Cont1} and \ref{fig:histo_mockParamPlum_Cont05} show the results of Plummer fitting 
to the profiles of Plummer-shaped 
filaments with input contrasts $C^0=1$ and $C^0=0.5$, respectively. 
In the  $C^0=1$ case, 
the derived $D_{\rm flat}$ and $p$ parameters 
have peaked distributions 
with median values consistent with the input parameters. 
In the lower contrast case ($C^0=0.5$),  the median values  of the derived $D_{\rm flat}$ and $p$ distributions tend to be slightly larger than the corresponding input values, and 
the dispersions of $D_{\rm flat}$ and $p$ values are about $30\%$ and $40\%$ larger, respectively, than the dispersions measured for $C^0=1$. 
These larger uncertainties in the results of Plummer fitting for input filaments with $C^0=0.5$ may be attributed 
to less well-defined power-law profiles at $r>>R_{\rm flat}$ for low-contrast filaments leading to larger errors in the fitted parameters.

 Figure\,\ref{fig:histo_mockParamPlum23} shows the results of Plummer-fitting measurements
for an input population of Plummer-shaped mock filaments with fixed input parameters:  $D_{\rm flat}=0.1$\,pc and $p=1.5$ [panels (A) and (B)]
or $p=3$  [panels (C) and (D)].
It can be seen that when  $p$ and $D_{\rm flat}$  are fitted simultaneously, the input $p$ and $D_{\rm flat}$ values are reasonably well recovered. 
When the fitting is performed with fixing the power-law exponent to $p=2$, different from the input value $p=1.5$ or $p=3$,  the derived $R_{\rm flat}$ values are larger 
or smaller (by up to $\sim 50\% $) than the input $D_{\rm flat}=0.1$\,pc radius, 
respectively.

{\rev Figure\,\ref{fig:histo_mockParamPlumFlatR}A shows that a flat distribution of $D_{\rm flat}$ values is derived 
when the input mock filaments have a flat distribution of $D_{\rm flat}$ diameters.
Likewise, Fig.\,\ref{fig:histo_mockParamPlumFlatR}B shows that a power law distribution of $D_{\rm flat}$ values 
is derived when the input mock filaments have a  power-law distribution of $D_{\rm flat}$ diameters. 
Correct estimates of the $D_{\rm flat}$ diameter are derived down to 0.03\,pc, close to the spatial resolution of the synthetic map. 
Furthermore,  when both  $D_{\rm flat}$ and $p$ are fitted simultaneously, the median value of the derived $p$ indices is compatible with the input $p$ value in both sets of experiments. }

{\rev Figure\,\ref{fig:histo_mockParamPlumFlatp} shows that a flat distribution of $p$ values is obtained when the input mock filaments have 
a flat distribution of $p$ indices from 1.5 to 3.2 and a fixed  $D_{\rm flat}$ diameter. 
When $p$ and $D_{\rm flat}$ are fitted simultaneously in this case, the distribution of derived $D_{\rm flat}$ values peaks at the constant input $D_{\rm flat}$ value. 
The derived distribution of $D_{\rm flat}$ radii  has a larger dispersion when the fitting is performed while fixing $p$ to 2, different from the input $p$ values. 
 }
 
{\rev Figure\,\ref{fig:histo_mockParamPlum_plw} shows the results of measurement tests 
obtained from both Plummer [panel (A)] and Gaussian [panels (B) and (C)] fitting 
for a sample of Plummer-shaped mock filaments  
with variable input $D_{\rm flat}$ values along their crests. In these tests, 
each synthetic filament was given a power-law distribution of $D_{\rm flat}$ values along its crest. 
(The distribution of median, crest-averaged $D_{\rm flat}$ diameters for the sample of 180 input filaments also followed a power law). 
It can be seen that the distribution of derived median $D_{\rm flat}$ values (Fig.\,\ref{fig:histo_mockParamPlum_plw}A) and the distribution of derived median $FWHM$ widths (Fig.\,\ref{fig:histo_mockParamPlum_plw}B) 
are both consistent with the input distributions down to $\sim0.025$\,pc, 
close to the resolution of the background column density map. 
Figure\,\ref{fig:histo_mockParamPlum_plw}C shows that the distribution of  individual $FWHM$ widths measured along the filament crests (before averaging) 
reproduces well the input distribution of individual $D_{\rm flat}$ diameters.
}

 \begin{figure*}
   \centering
      \resizebox{19cm}{!}{
      \hspace{-2.cm}
\includegraphics[angle=0]{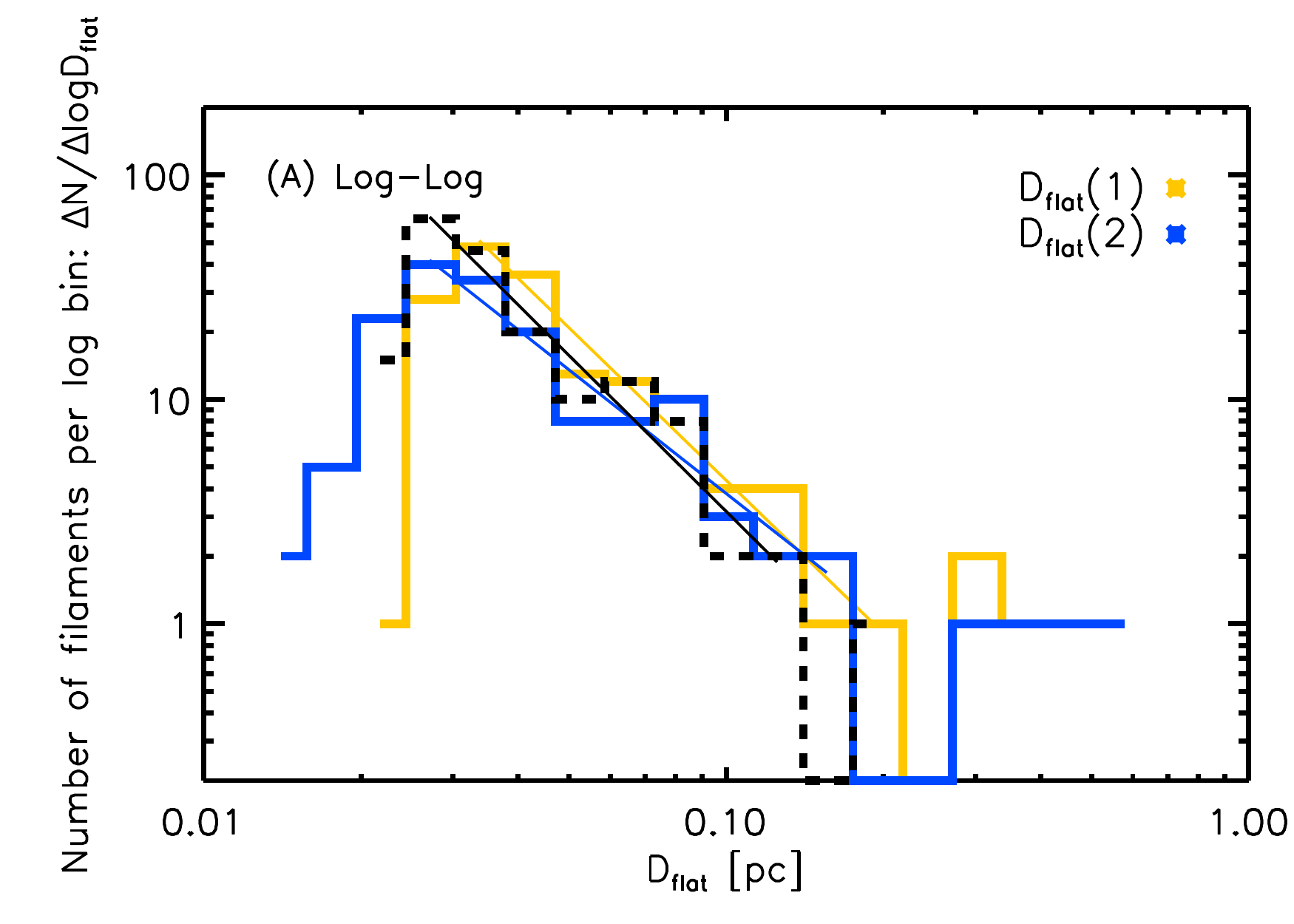}
\includegraphics[angle=0]{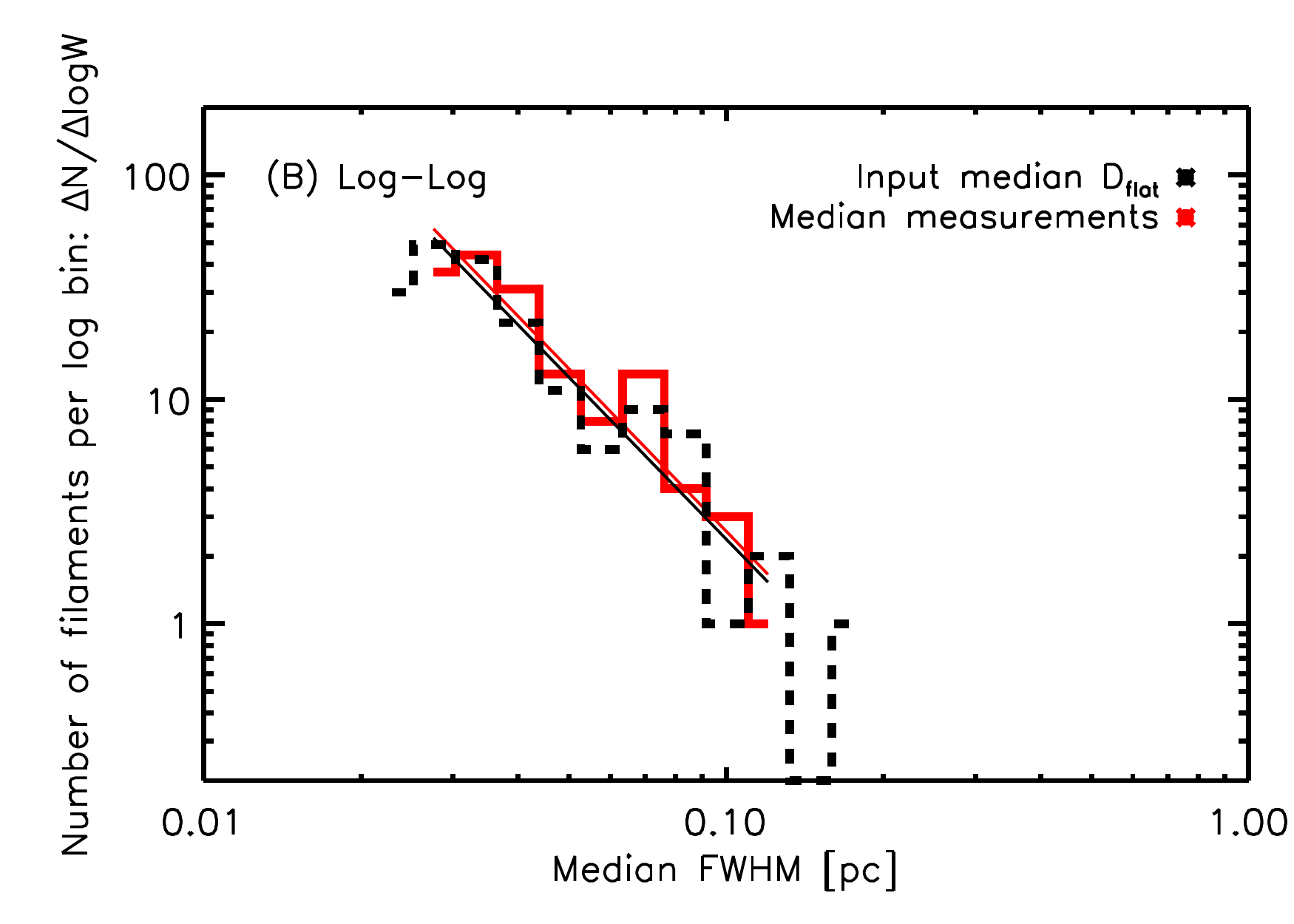}
\includegraphics[angle=0]{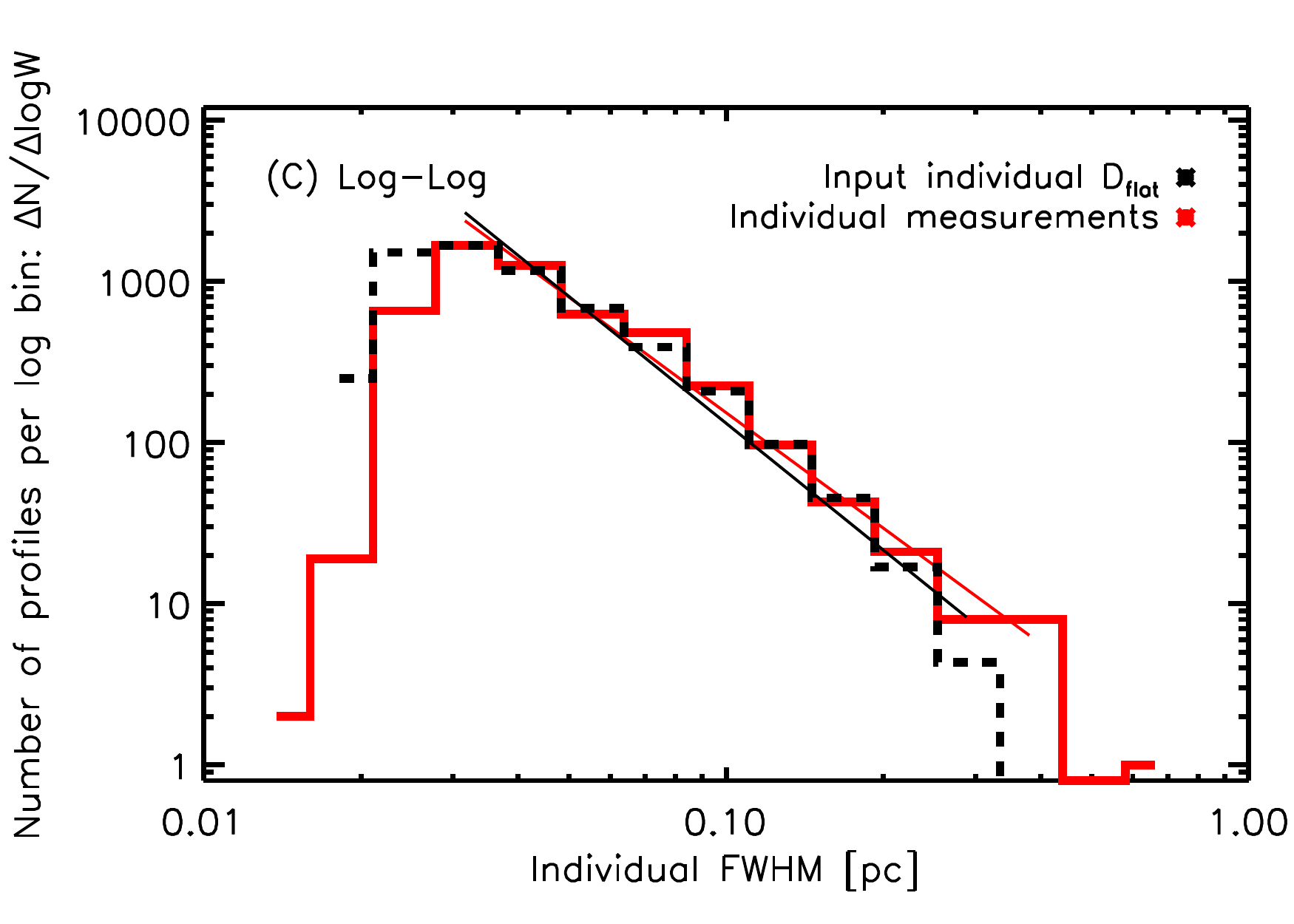}}
  \caption{{\rev
Results of measurements performed to test the reliability of inner-width estimates from both Plummer and Gaussian fits to the profiles of 180 
Plummer-shaped synthetic filaments with variable $D_{\rm flat} $ diameters along the crests.
The synthetic filaments had 
a fixed input central column density contrast $C^0=1$, a fixed index $p=2$, and a power-law distribution of individual input $D_{\rm flat}$ diameters between 0.022\,pc and 0.19\,pc with an 
{\Newrev exponent of   $-2$ (in $\Delta N/\Delta {\rm log}W$}, similar to that in Fig.\,\ref{histo_widthPlum_mock}B) along each crest.
 The distribution of  input inner diameters 
 is shown by a dashed black histogram in all  three plots. 
  {\bf Left:} Histograms of measured $D_{\rm flat}$ diameters from Plummer fitting. Two sets of results are provided for $D_{\rm flat}$ as explained in the caption of Fig.\,\ref{fig:histo_mockParamPlum_Cont1}.  
 {\bf Middle:} Histogram of median $FWHM$ values derived from Gaussian fits after averaging along each filament crest. 
  {\bf Right:} Histograms of individual $FWHM$ values measured along the filament crests. In this plot, 
the dashed black histogram of input  $D_{\rm flat}$ values has been rescaled so as to have the same peak as  
the red histogram of measured $D_{\rm flat}$ values.} 
}          
  \label{fig:histo_mockParamPlum_plw}
    \end{figure*}

\subsection{Summary and conclusions regarding 
the reliability of filament width estimates}\label{App2c}
{\rev
The 
inner widths of filaments with Gaussian-like intrinsic profiles 
are estimated reliably by both 1) our method of deriving the half-power diameter without any fitting (Sect.~\ref{FilBack})
and 2) the $FWHM$ widths derived from  Gaussian fits to the observed profiles (Sect.~\ref{Gaussfit}). 
The latter $FWHM$ estimates are not affected by the fitting range for Gaussian-shaped filaments.  
For a fixed  input  width  constant along the filament crest, the distribution of individual $FWHM$ widths derived along and on either side of the filament crests show some excursion from the median value of the distribution (consistent with the input fixed value), that may be attributed to bad measurements and/or spurious structures. Considering  crest-averaged values result in median filament widths that are robust  in the presence of measurement errors. 

The inner widths of filaments with Plummer-like intrinsic profiles can be estimated 
1) without any fitting ($hd$ estimates), 2) with Gaussian function fitting ($FWHM$ estimates), 
and 3) Plummer function fitting ($D_{\rm flat}=2R_{\rm flat}$ estimates). 
The first two estimates, $hd$ and $FWHM$, are affected by the power-law index $p$ of the Plummer profile: 
for the same input $D_{\rm flat}$ diameter, $hd$ and $FWHM$ both increase for decreasing intrinsic $p$ values. 
For the same $p$ index, the dispersion of the derived $FWHM$ widths is smaller than that of the derived $hd$ values. 
While the derived $FWHM$ widths depend on the fitting range,  they are closest to the input $D_{\rm flat}$ diameters,  
and thus provide satisfactory estimates of the filament inner widths of filaments, when the  fitting range is chosen to be $0\le r \le 1.5hr$. For this fitting range, 
the derived $FWHM$ widths provide accurate estimates of the $D_{\rm flat} $ diameters 
to better than $\sim 50\% $ (for filaments with $C^0 > 1$) when $1.5 < p < 3$. 
The $D_{\rm flat}$ diameters derived from Plummer fitting 
provide satisfactory estimates of the intrinsic inner diameters of Plummer-shaped filaments 
when the filament contrasts are large enough (e.g. $C^0 \ga 0.5$). 
Furthermore, such $D_{\rm flat}$ estimates do not have significant biases in the sense 
that the input distributions of $D_{\rm flat}$ diameters are correctly reproduced 
in tests performed on populations of Plummer-shaped synthetic filaments, independently 
of the shape of the input distributions (e.g. constant, flat, or power law). 
For constant input  $D_{\rm flat}$ diameters, the dispersion of measured values 
increases 1) when the column density contrast of the input filaments decreases 
and 2) when  $D_{\rm flat}$ and $p$   are fitted simultaneously. 
}

\section{Column density maps and filament skeletons}\label{App4}

This appendix presents  the column density maps of the entire fields analyzed in this paper for seven of the eight target clouds (see Table\,\ref{Table1} for an overview of the target clouds). 
The column density map of the entire field analyzed in the eighth target cloud (IC5146) is shown in the right panel of Fig.\,\ref{ColdensMapsSkel}.  
These column density maps were derived from HGBS data (see http://gouldbelt-herschel.cea.fr/archives)
as explained in Sect.\,\ref{SkelDisperseHerschel}.
The  skeletons of the entire sample of  filaments traced in each cloud as explained in Sect.\,\ref{SkelDisperse} are shown as  {\rev cyan or dark blue} solid curves in the column density maps. 
{\rev The cyan curves trace the crests of the selected sample of filaments derived as explained in  Sect.\,\ref{SelecSample}  (cf., Table\,\ref{tab:SumParamDisp}).}

    \begin{figure*}
   \centering
  \hspace{0.1cm}
         \resizebox{8.cm}{!}{
     \hspace{0.1cm}
        \vspace{-0.9cm}
\includegraphics[angle=0]{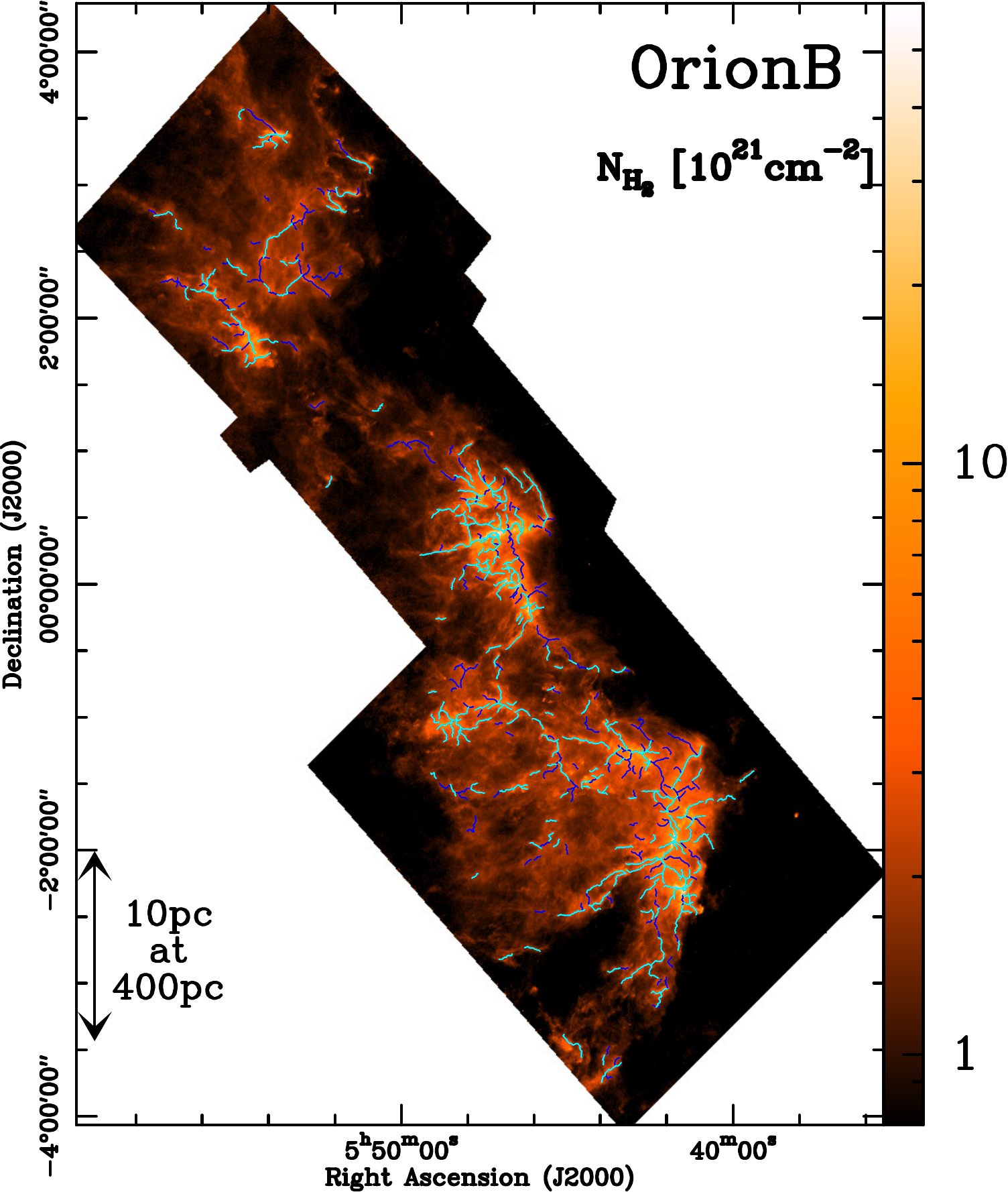}}
    \resizebox{9.9cm}{!}{
\includegraphics[angle=0]{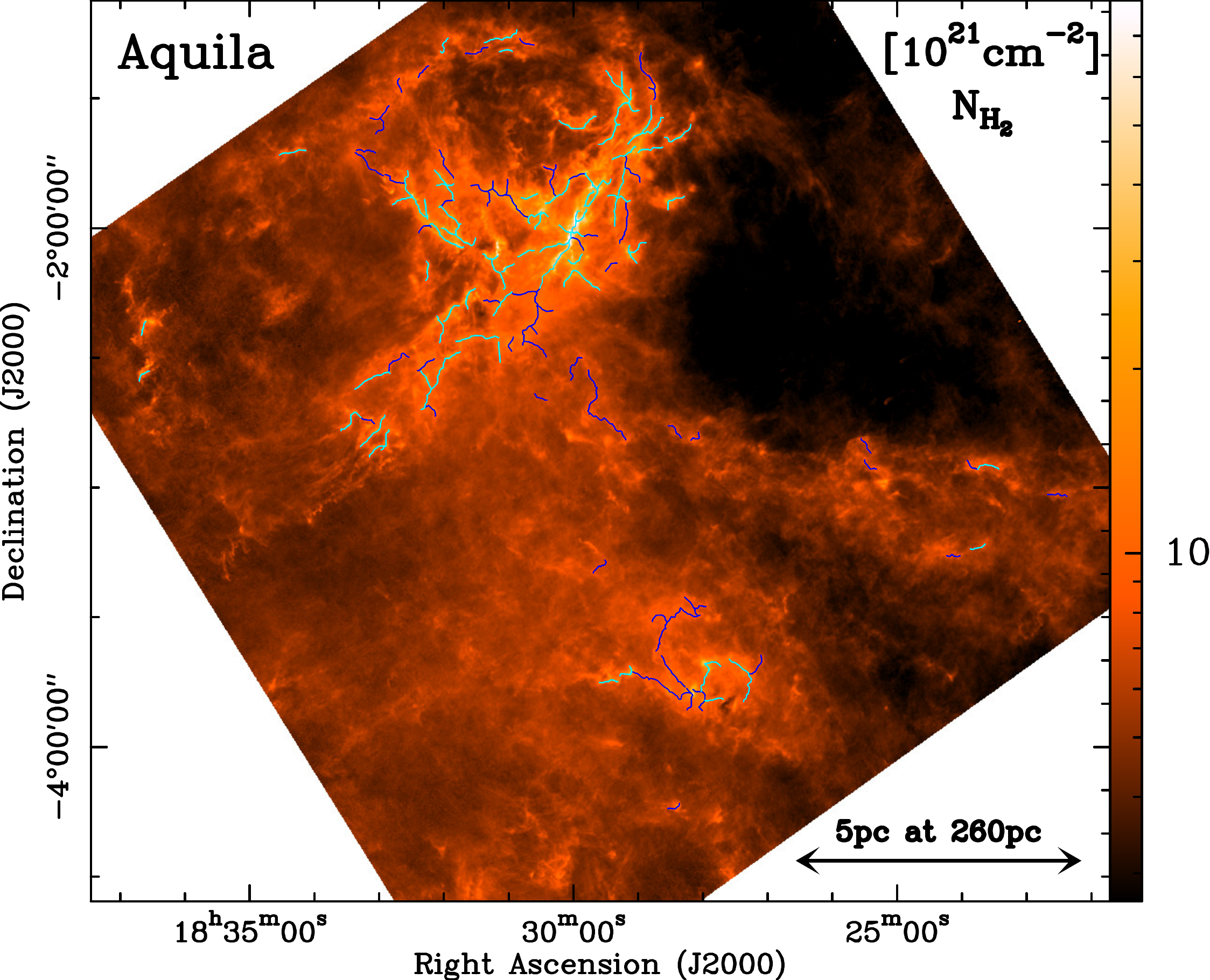}}
 \caption{{\it Herschel} column density maps of the Orion B field (left panel -- K\"onyves et al. 2018) 
and Aquila field (right panel -- \citealp{Konyves2015}) analyzed in this paper, as derived from HGBS data 
\citep[http://gouldbelt-herschel.cea.fr/archives,][]{Andre2010}. 
The effective HPBW resolution 
is $18\parcs2$.    
The crests of the filamentary structures traced in the two clouds using \disperse\ (see Sect.\,\ref{SkelDisperse}) are overlaid as  solid curves.
{\rev The cyan curves trace the filament crests of the selected sample, and the dark blue curves trace the additional filament crests in the extended sample  (cf. Sect.\,\ref{SelecSample}).}
See Table\,\ref{tab:SumParamDisp} for the absolute values of the persistence and robustness thresholds of the \disperse\ runs, 
as well as the number of extracted filaments in each field.  
}          
  \label{Maps1}
    \end{figure*}
    
\begin{figure*}[!ht]
   \centering
  \hspace{0.1cm}
         \resizebox{7.cm}{!}{
     \hspace{0.1cm}
        \vspace{-0.9cm}
\includegraphics[angle=0]{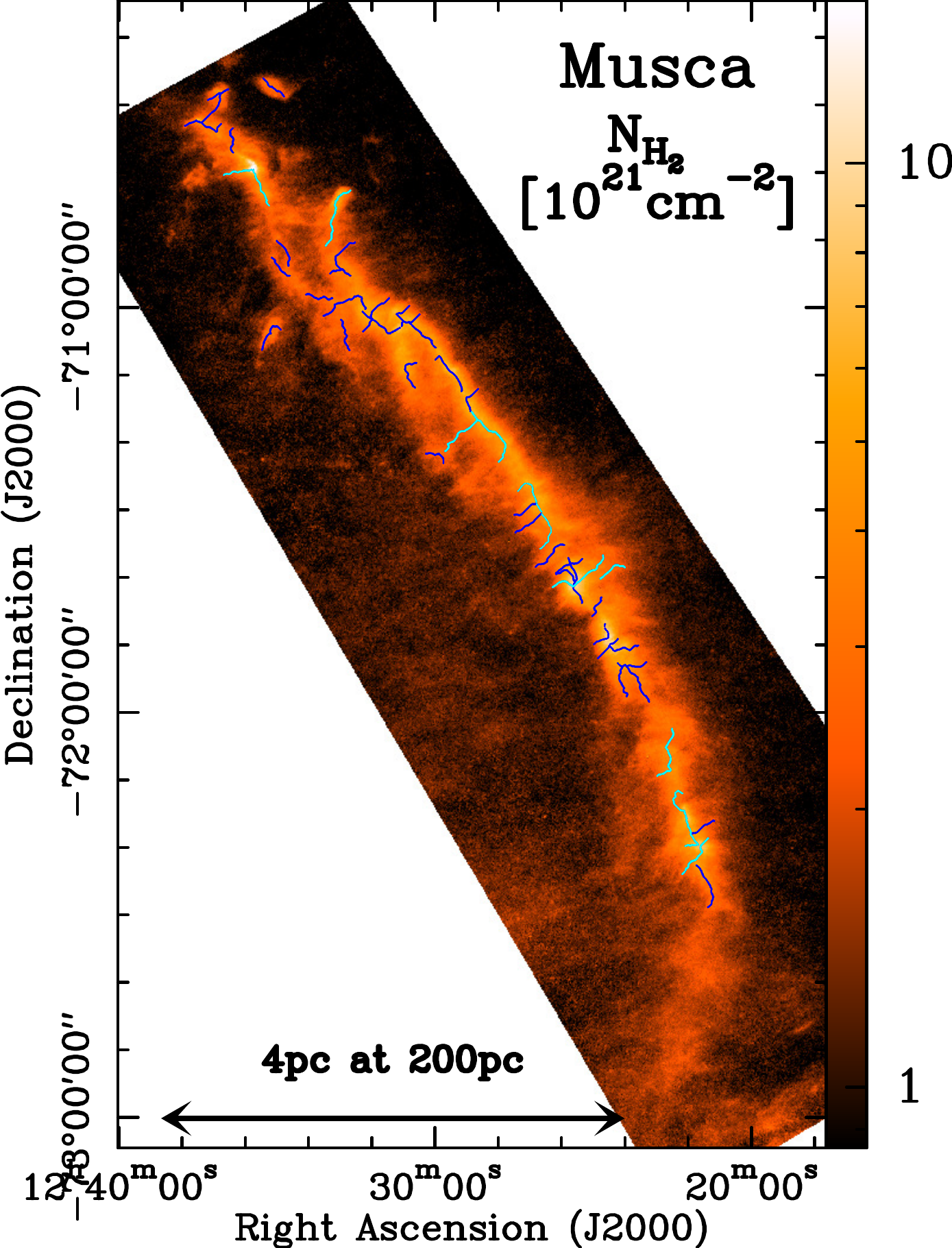}}
    \resizebox{9.9cm}{!}{
\includegraphics[angle=0]{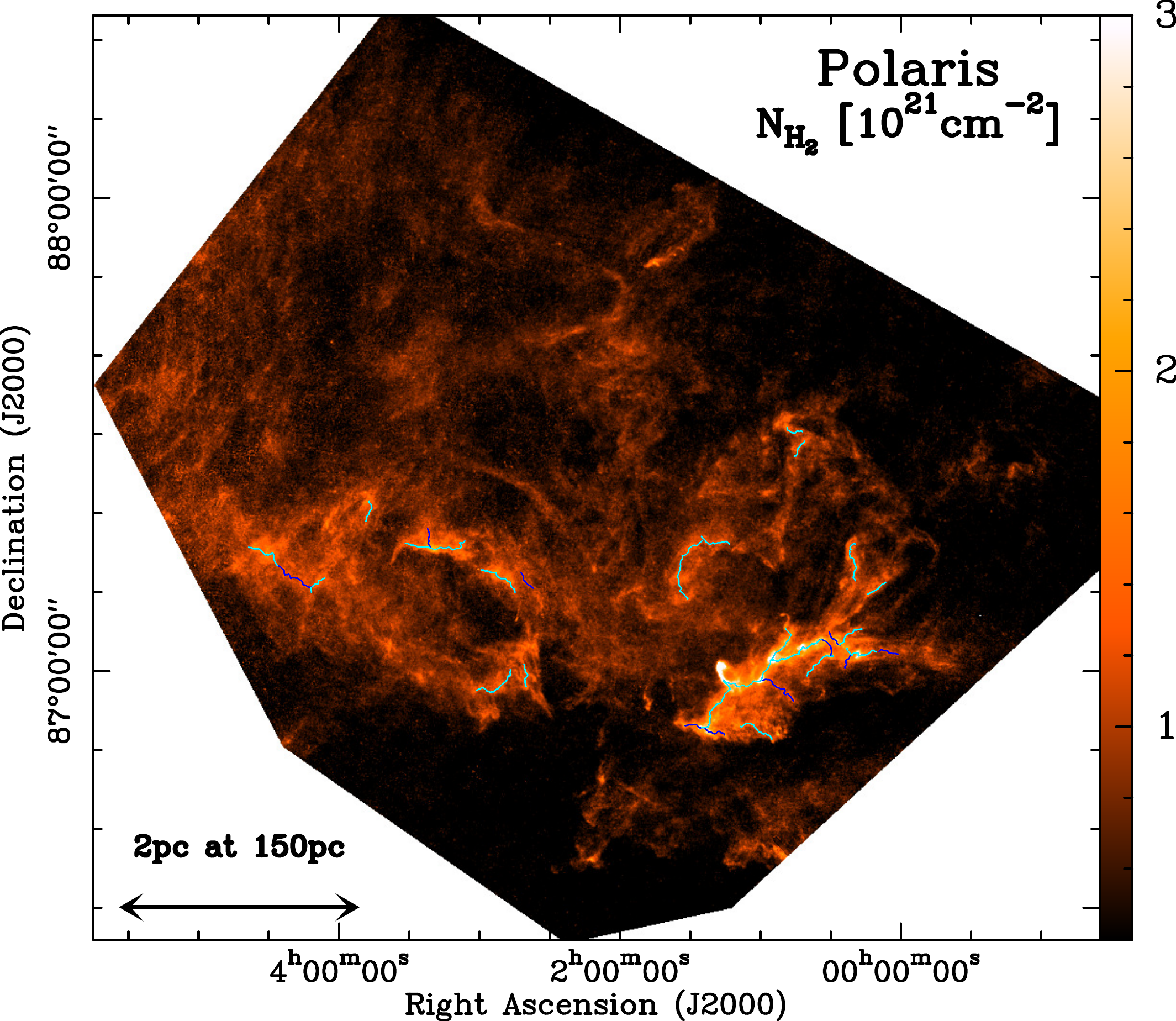}}
 \caption{Same as Fig.\,\ref{Maps1} for the Musca cloud (left panel -- \citealp[see][]{Cox2016}) and Polaris cloud (right panel -- \citealp[see][]{Ward-Thompson2010, Miville2010}). 
}          
  \label{Maps2}
    \end{figure*}
    
\begin{figure*}[!ht]
   \centering
  \hspace{0.1cm}
         \resizebox{9.cm}{!}{
     \hspace{0.1cm}
        \vspace{-0.9cm}
\includegraphics[angle=0]{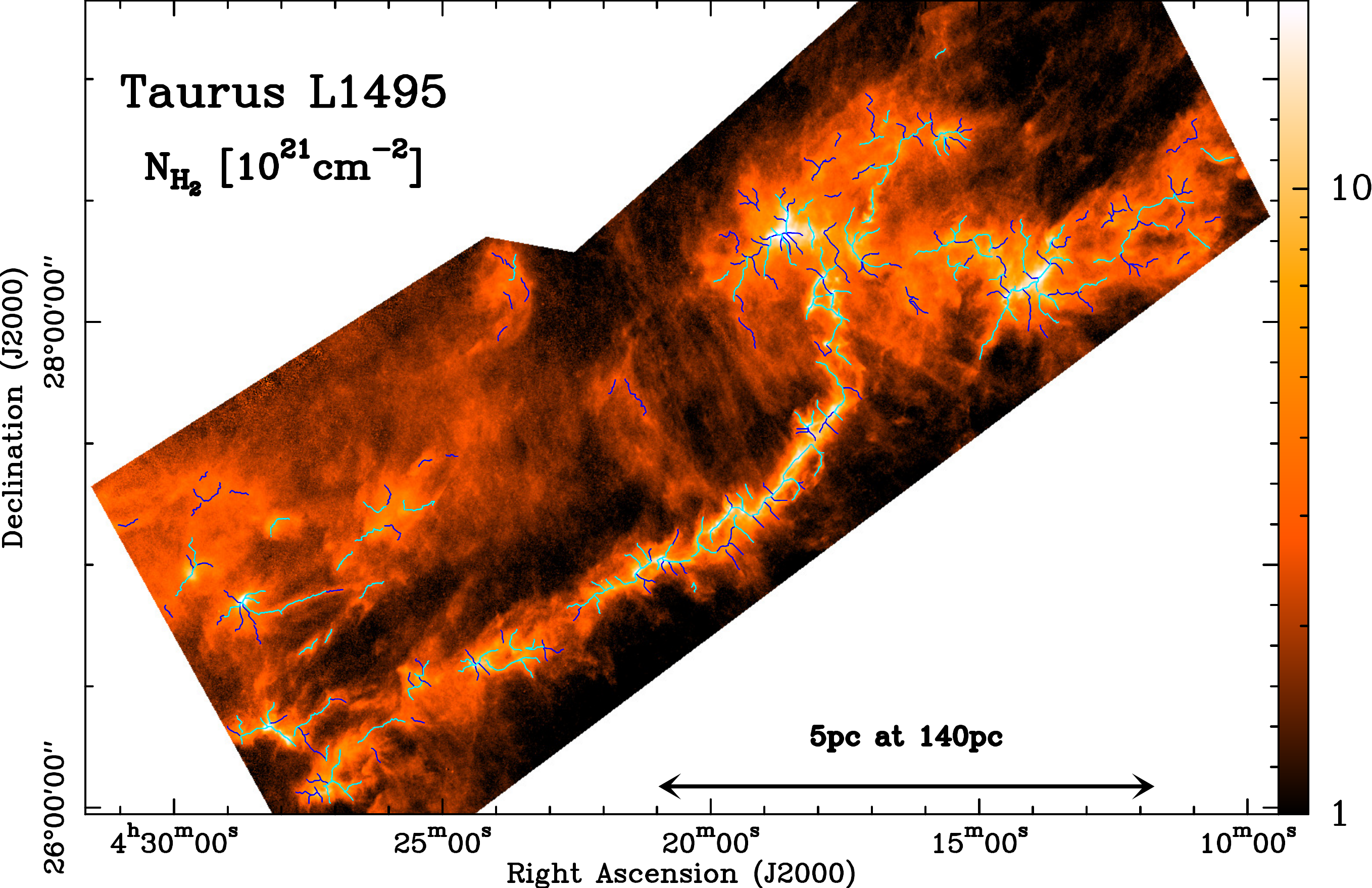}}
    \resizebox{7.5cm}{!}{
\includegraphics[angle=0]{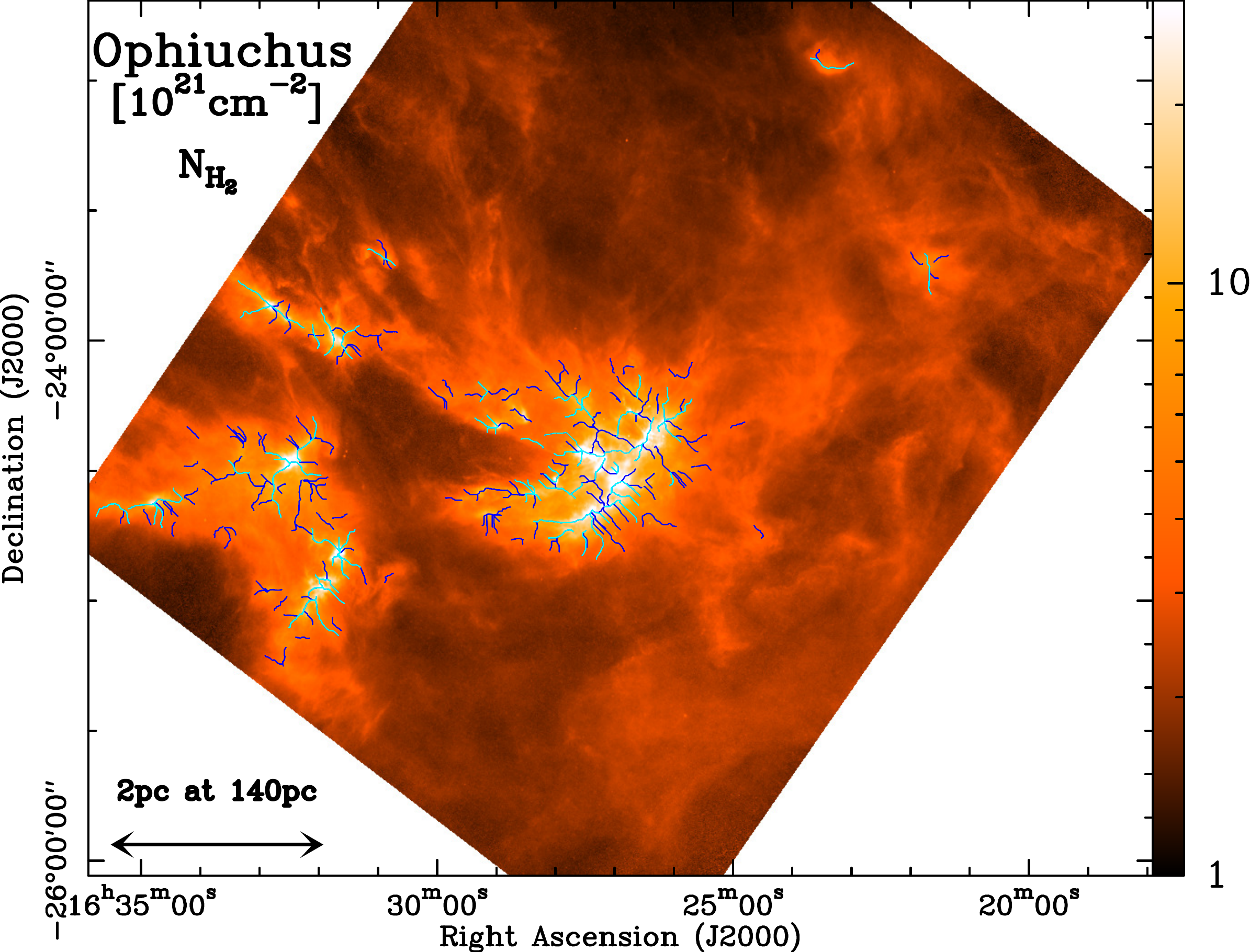}}
 \caption{
Same as Fig.\,\ref{Maps1} for the Taurus/L1495 cloud (left panel -- \citealp[][]{Palmeirim2013,Marsh2016}) and the Ophiuchus L1688/L1689 field (right panel -- Ladjelate et al., in prep.). 
}          
  \label{Maps3}
    \end{figure*}
    
    \begin{figure*}[!ht]
   \centering
  \hspace{0.1cm}
         \resizebox{15.cm}{!}{
     \hspace{0.1cm}
        \vspace{-0.9cm}
\includegraphics[angle=0]{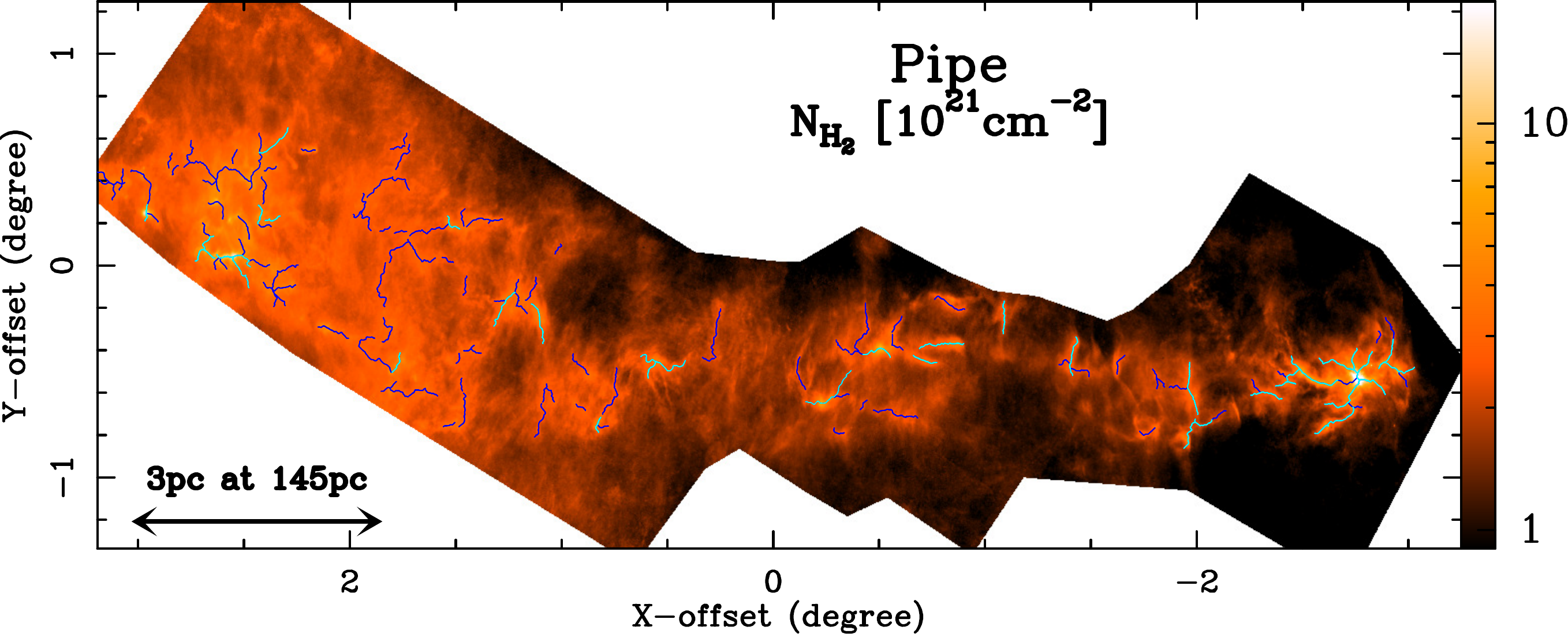}}
 \caption{Same as Fig.\,\ref{Maps1} for the Pipe molecular cloud \citep[][Roy et al., in prep.]{Peretto2012}. 
The labels indicate the offsets in degrees relative to the center of the map at RA(J2000) = 17:38:20 and Dec(J2000) = 27:02:25. 
The X--Y reference frame is rotated by $-37^\circ$ with respect to the RA--Dec reference frame (counting positive angles east of north).
}          
  \label{Maps4}
    \end{figure*}

\end{appendix}

\end{document}